\documentclass[twoside]{sta-thesis}
\usepackage[dvips]{graphics}
\usepackage[applemac]{inputenc}
\usepackage{graphicx,amsmath,amsfonts,amssymb}
\usepackage{ifthen}
\usepackage{rotating}
\usepackage{setspace}
\newcommand\beq{\begin{equation}}
\newcommand\eeq{\end{equation}}
\newcommand\bea{\begin{eqnarray}}
\newcommand\eea{\end{eqnarray}}
\newcommand\nn{\nonumber}

\newcommand\TTS{Sr$_3$Ru$_2$O$_7$}
\newcommand\TOF{Sr$_2$RuO$_4$}
\newcommand\FTT{Sr$_4$Ru$_3$O$_{10}$}
\newcommand\OOT{SrRuO$_3$}

\newcommand\bs{\mathbf}
\newcommand\PchapterI{./ChapterI/}
\newcommand\PchapterII{./ChapterII/}
\newcommand\PchapterIII{./ChapterIII/}
\newcommand\PchapterVI{./ChapterVI/}
\newcommand\PchapterVII{./ChapterVII/}
\newcommand\PAppendixB{./AppendixB/}
\newcommand\PAppendixE{./AppendixE/}
\newcommand\PAppendixF{./AppendixF/}
\newcommand\sinc{\,\textrm{sinc}}
\newcommand\erf{\,\textrm{erf}}

\author{Jean-Fran\c{c}ois Mercure}
\title{The de Haas van Alphen effect\\ near a quantum critical end point\\ in Sr$_3$Ru$_2$O$_7$}
\date{September 2008}

\wordlength{42000}
\rschstartdate{October 2004}
\phdstartdate{October 2004}
\workstart{2004}
\workfinish{2008}

\acknowledgementfile{Acknowledgements}
\abstractfile{Abstract}

\begin{document}

\maketitle

\setcounter{page}{1}
\pagenumbering{roman}

\singlespacing
\declaration
\thesisabstract
\acknowledgement

\markright{Contents}
\tableofcontents
\markright{List of figures}
\listoffigures
\listoftables
\newpage
\setcounter{page}{1}
\pagenumbering{arabic}

\onehalfspacing

\addcontentsline{toc}{chapter}{\sffamily Introduction}
 
 \chapter*{Introduction}
 \markright{Introduction}
 
 Correlated electron materials are systems in which many exotic quantum states can emerge. These are phases of the electronic complex among many others that are predicted by the fundamental expression of quantum mechanics, the Schrödinger equation which, when one considers a specific arrangement of atoms, is difficult to solve. This equation gives rise to complexity from which these phases emerge. It is imperative for the condensed matter physicist to find other ways to understand this phenomenon than starting from the Schrödinger equation, and one can, for instance, identify conditions that are favorable to their formation. Such phases of the electron liquid have important technological potential, for instance in the case of high temperature superconductivity (HTS), where room temperature superconductivity can potentially be achieved by understanding better how it is formed. But they are also of considerable intrinsic fundamental interest, and better knowledge regarding their nature could lead to the discovery of new types of emergent phenomena.
 
 One particular situation brings together the right conditions for the formation of many exotic phenomena,  and is called quantum criticality. It corresponds to a region in parameter space, surrounding a so-called quantum critical point (QCP), which is dominated by quantum fluctuations and where in many cases superconductivity was found. Such a point may arise when the critical temperature of a second order phase transition is varied using a specific parameter, for example pressure or doping, and is made to reach zero temperature. Neighbouring regions in parameter space reveal unusual physical properties, described as non-Fermi liquid behaviour. Theory predicts several of these properties, notably an enhanced specific heat, which may also be expressed as an increase of the quasiparticle masses.
 
\begin{figure}[t]
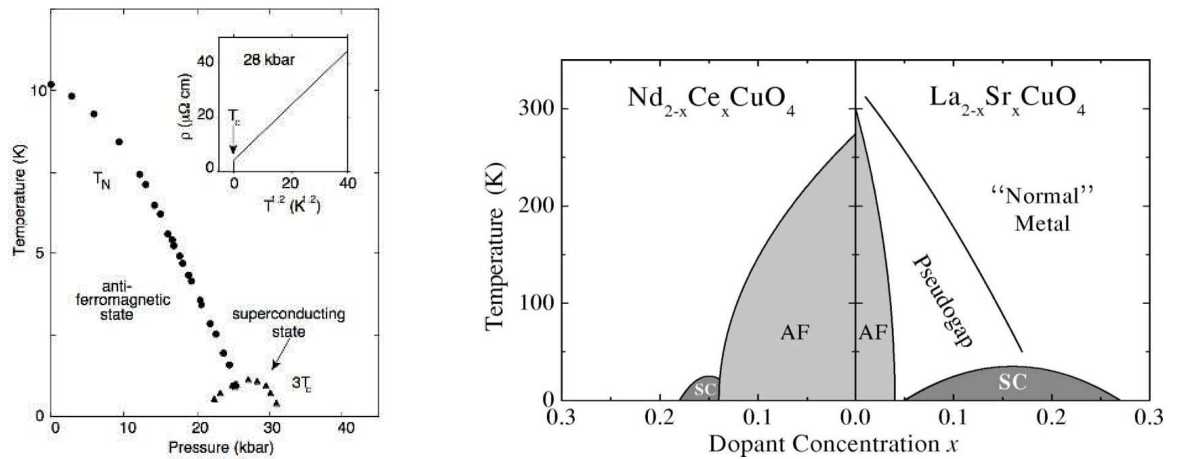

	\begin{minipage}[t]{5cm}
		\begin{center}
		\includegraphics[width=5cm]{\PchapterI Mathur1.epsf}
		\end{center}
	\end{minipage}
	\hfill
	\begin{minipage}[t]{9cm}
		\begin{center}
		\includegraphics[width=9cm]{\PchapterI HTCDiagram.epsf}
		\end{center}
	\end{minipage}
	\caption[Superconductivity near a QCP]{$Left$ Work of Mathur $et$ $al.$ reporting superconductivity near a QCP in CePd$_2$Si$_2$\cite{mathur}. $Right$ Typical phase diagram of high temperature superconductors (taken from Damascelli $et$ $al.$\cite{damascelli}). }
	\label{fig: mathur1}
\end{figure}

 The work of Jaccard $et$ $al.$ in 1992 and that of Mathur $et$ $al.$ in 1998 demonstrated the existence of superconductivity surrounding a QCP in heavy fermion materials, CeCu$_2$Ge$_2$ \cite{jaccard}, CePd$_2$Si$_2$ and CeIn$_3$ \cite{mathur}, where a Néel transition was suppressed to zero temperature using pressure. Figure \ref{fig: mathur1}, left, presents the phase diagram of the first of these systems, the second being very similar. This research has become very influential in the field of condensed matter, and many more systems have exhibited similar behaviour \cite{stewart1,stewart2}. It has moreover been thought that quantum criticality could potentially be part of an explanation of the phenomenon of HTS, where a QCP would lie at the centre of the superconducting dome \cite{vanderMarel}. Figure \ref{fig: mathur1}, right, shows a typical phase diagram for high temperature superconductors, where superconductivity is found either by doping electrons or holes. 
 
An important problem in quantum criticality in metals is to find out how the Fermi surface (FS) evolves near a QCP. There are two main methods to study the FS of a material, which are by using angle resolved photoemission spectroscopy (ARPES) or the de Haas van Alphen effect (dHvA). It is however difficult to perform any of these near a QCP for the following reasons. ARPES is an experiment that is performed at ambient pressure and zero magnetic field, excluding these two parameters for tuning a QCP, and one is required to find a QCP that exists at zero pressure and field, which is not common\footnote{Chemical doping can still be used with ARPES, and high temperature superconductors have been extensively studied in this regard. It is, however, not yet demonstrated whether a QCP exists in those materials.}. dHvA is a probe of the FS that can be used more easily with tuning parameters, but requires that a magnetic field be applied, which may drive the system away from criticality, and very few QCPs tuned with magnetic field have been found at values high enough to observe the dHvA effect.\footnote{Most known QCPs in clean, undoped materials arise at low magnetic field values, for instance in YbRh$_2$Si$_2$, CeCoIn$_5$ and CeAuSb$_2$ \cite{gegenwart,paglione,balicas}.} However, using such a material is an attractive option as it allows the analysis to be performed continuously near the QCP, an aspect not possible when using pressure or chemical doping. 

A quantum critical end point (QCEP) was reported in \TTS\ \cite{science1}, located at a field high enough to perform comfortably the dHvA experiment on both low and high field sides of the non-Fermi liquid region. Such a situation provides a unique opportunity for the observation of the changes occuring in a FS that evolves from a normal Fermi liquid towards a non-Fermi liquid, and is the subject of this thesis. Moreover, this QCEP features, similarly to the cases reported by Mathur and co-workers, an unusual emergent phase of the electron liquid which is not superconductivity \cite{science2}. It was reported more recently to feature a nematic type of electron ordering, which produced an additional wave of interest in the community, motivating many theoretical studies. Understanding the nature of this new phase also required information about the FS of \TTS, and part of our attention in this project was given to this issue. Finally, the dHvA effect also measures the quasiparticle effective mass, which in turn can be expressed as the electronic specific heat, and is expected to be highly enhanced near the QCEP.
 
 The goals of this project were very clear. We were determined to study and resolve three issues related to the FS of \TTS. The first was to determine the three dimensional topology of the FS, which has been achieved in collaboration with ARPES experts. The second was to study the FS of \TTS inside the new nematic phase, and was successfully performed. The last was to revisit and analyse in detail the evolution of the quasiparticle properties near the QCEP, a study that revealed an unexpected result.
 
 This thesis is divided in four chapters. We present in the first chapter a review of the current knowledge surrounding \TTS, its QCEP and nematic phase, but also the standard theory of dHvA. Emphasis is given to all theoretical aspects of relevance for the correct analysis and interpretation of the measured dHvA data, as well as to the previous research that makes up what is known about the electronic structure of \TTS. The second chapter introduces all the experimental methods that were used in this project along with the numerical methods for the analysis of dHvA data that were employed. The dHvA experiment has required the isolation of ultra-pure samples of \TTS, and this procedure and its results are described in detail in this chapter as well. Chapter three presents all the dHvA data measured by the author, raw and processed, with a basic analysis and interpretation. A detailed analysis follows in chapter four, where additional information is incorporated from other experiments, notably ARPES. A complete model for the FS is introduced, and the consequence of the nematic phase and quasiparticle mass measurements are discussed. The properties of the zero, low and high field FS are described, as well as those of the nematic phase region. Finally, we conclude with a brief review of the information gained in this project.

\chapter{Scientific Background\label{chapter:background}}
\markright{Chapter~\ref{chapter:background}: Scientific Background}

\section{Strongly correlated metals}

We introduce in this section the physical properties of strongly correlated electron systems, relevant to this work. Electrons in a correlated system are usually well described by Landau's Fermi liquid theory, a framework which enables condensed matter physicists to understand and predict the properties of most metals. However, we are interested in this project by a case which features characteristics that diverge from the normal properties predicted by the Fermi liquid theory, and we are therefore required to review the theory of non-Fermi liquids. This introduction to the subject is very brief, as this work is mainly of experimental nature, but we provide references for further reading.

\subsection{The Fermi Liquid \label{sect:FermiLiquid}}
Electrons in simple metals are usually well described by the Sommerfeld theory of the electron gas. Electron states are represented by an orthogonal set of plane waves. Consideration of the effects of an underlying periodic potential on the electron gas leads to dispersion relations, featuring band gaps, that can be more or less complex, and to a wide variety of different Fermi surface shapes. However, the basic physics of the Sommerfeld model remain relatively simple: the Pauli exclusion principle is respected and at zero temperature, electrons fill the lowest states in $k$ space up to a surface of constant energy, the Fermi energy. For instance, in simple metals like the alkali elements, the Fermi surface is close to a perfect sphere. In this case, the dispersion relation is close to quadratic, like that of a gas of free electrons. In other cases, the Fermi surface can be very complex.

In this framework a basic physical quantity is usually derived from the dispersion relation $\epsilon(k)$ of an electron gas, the velocity \cite{ashcroft},
\beq
\bs{v}(\bs{k}) = {1\over \hbar}\bs{ \nabla}_\bs{k} \epsilon(\bs{k}).\nn
\eeq
This expression leads to the concept of the effective mass $m^*$ of the electron,
\beq
\label{eq:effmass}
\bs{v} = {\hbar \bs{k} \over m^*}.
\eeq

The Sommerfeld model has been extremely successful in the study of condensed matter, but it neglects a fundamental property of the electron gas, the Coulomb interaction and hence electronic correlations. Consequently, the question arises as to why the model works so well. In the 1950s, Landau constructed a theory that became the standard model for the physics of metals, called Landau's Fermi liquid theory. It is based on a critical assumption, now called adiabatic continuity, that when turning on the interaction between electrons in a non-interacting gas, the low energy excitations evolve continuously from the Fermi gas to the Fermi liquid, preserving their energy ordering and a one to one correspondence with those of the non-interacting gas. Consequently, in this model the Fermi surface picture of the Sommerfeld model is still correct. The wave functions of the new states are different, but they can be approximated by orthogonal linear combinations of plane waves, and the fundamental unit acting as an electron is called a quasiparticle.


It is surprising that even with strong interactions, the excitations remain relatively long lived. One of the reasons why such a scheme works is because of the screening of the electrons. The coulomb potential has a very long range, but in an electron gas, any fluctuation of charge in space should in equilibrium be compensated by a variation of the density of electrons. The interaction is actually of very short range. An electron moving will affect many others, but the global movement of the charge will be well defined. The wave functions of the quasiparticles are not exact eigenstates of the Hamiltonian featuring the interactions and, even at zero temperature, excitations above the Fermi surface will be scattered. It can be shown \cite{schofield} that the scattering rate of an excitation due to electron-electron interaction depends on its energy as
\beq
{1\over \tau} \sim {\epsilon^2 \over \epsilon_F},\nn
\eeq
which is much smaller than the excitation energy $\epsilon$ itself, when $\epsilon$ is sufficiently small. A quasiparticle is better defined the closer it is to the Fermi surface. 

For isotropic Fermi surfaces, Fermi liquid theory makes the following predictions for the resistivity due to interactions, the specific heat and the magnetic susceptibility of a one band electron liquid at low temperatures:
\bea
\rho_{e-e} &=& A_0{ m^{*2} k_B^2 \over n \hbar^3 e^2 k_F^2}T^2, \label{eq:resistivity}\\
C_v &=& {m^* k_F\over3 \hbar^3}k_B^2 T, \label{eq:specific}\\
\chi &=& {1\over1+F_0^\alpha}{\mu_B^2k_Fm^*\over\pi^2\hbar},
\eea
where $F_0^\alpha$ is one of the so called Landau $F$ parameters, $A_0$ is a dimensionless constant and $n$ is the density of free carriers \cite{ashcroft, schofield, pines}. These quantities all involve the effective mass $m^*$, into which was put part of the effect of interactions. The theory predicts that it should be a constant, determined using specific heat or, as we will see later on, de Haas van Alphen measurements.

\subsection{Non-Fermi liquids and quantum criticality \label{sect:QCtheory}}

In 1986, high temperature superconductivity was discovered by Bednorz and Müller in layered materials involving copper oxide \cite{Bednorz}. It was found to be a type of unconventional superconductivity that could not be understood with previously developed $BCS$ (Bardeen-Cooper-Schrieffer) theory which applies to simple metals. The metallic state in transition metal oxides is very anisotropic and unusual and it was naturally thought that the superconducting state of a material would be understood only with a correct description of its normal metallic state. It was found that these metallic states are not very well described by Fermi liquid theory. For example, power laws for the temperature dependence of the resistivity were observed with exponents lower than 2, contrary to the Fermi liquid theory prediction for electron-electron interaction in highly correlated metals \cite{schofield}. Examples of non-Fermi liquid behaviour include the fractional quantum Hall effect in 2D metallic systems, the Luttinger Liquid in one dimensional metals, the Kondo effect and quantum criticality, which is of interest here. 

\begin{figure}[t]
  \begin{center}
	\includegraphics[width=0.65\columnwidth]{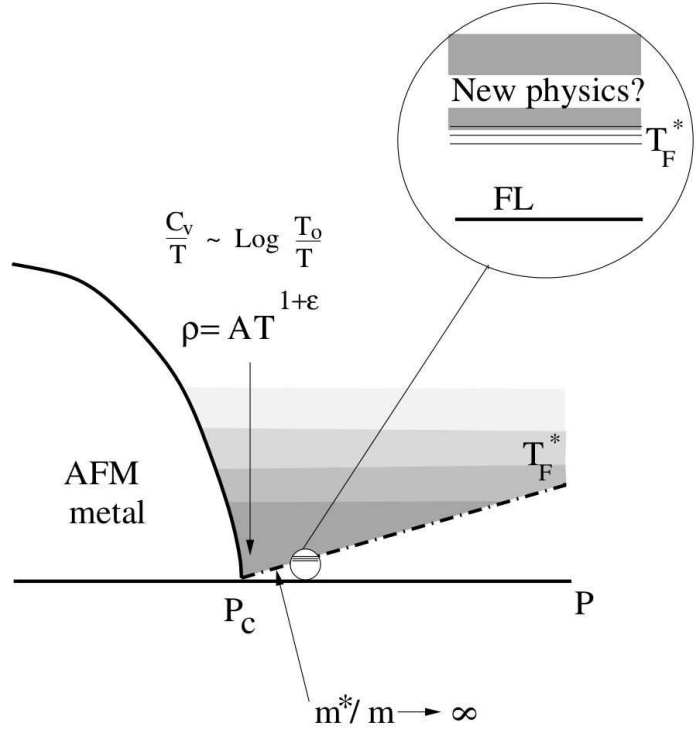}
	\caption[Generic phase diagram near a QCP]{Generic phase diagram for a heavy fermion system exhibiting a quantum critical point (image taken from the work of Coleman $et$ $al.$ \cite{coleman1}). The Néel temperature is tuned to zero Kelvin using the control parameter $p$, to a QCP at $P_c$. The renormalized Fermi temperature $T_F$ is shown under which Fermi liquid behaviour is observed. The region between the line of transition and $T_F$ exhibits non-Fermi liquid behaviour.}
	\label{fig:QCP_coleman}
	\end{center}
\end{figure}

A way towards non-Fermi liquid behaviour is to find a state of the electron liquid where the scattering rate is so great that the quasiparticles cannot exist for long enough to be well defined. One of the models proposed is the proximity to a QCP \cite{Hertz, Millis}, or similarly to a QCEP \cite{science1}\footnote{Described in section \ref{sect:QC}.}.  Proximity to such a point in the phase diagram is dominated by quantum fluctuations. A quantum critical point arises when a second order phase transition approaches absolute zero in any phase diagram. For example, if the critical temperature of a second order transition can be varied using a non thermal control parameter like pressure or chemical doping such that it is depressed to absolute zero, the end point constitutes a QCP. As the entropy of a system approaches zero when lowering the temperature to absolute zero, the transition then occurs between two perfectly ordered states and is dominated by quantum fluctuations. Figure \ref{fig:QCP_coleman} shows the typical phase diagram for heavy fermion quantum critical systems, where the transition in question is the Néel transition, at which the system moves from an antiferromagnetic to a paramagnetic metal, and is tuned with control parameter $p$.

Discovery of a possible close link between QCPs and high $T_c$ superconductivity was made in a variety of systems, for example, by the observation of a linear resistivity at all temperatures outside of the superconducting phase in optimally doped LaSr$_{2-x}$Cu$_x$O$_4$ \cite{gurvitch}, but not at any other doping values \cite{takagi}. Conversely, in a few heavy fermion systems, unconventional superconductivity was found using pressure to tune a transition from antiferromagnetism to paramagnetism towards 0~K, covering the region where a QCP was expected to exist \cite{mathur}. In this view, the proximity to a QCP and the domination of quantum fluctuations are thought to provide a mechanism for types of superconductivity not mediated by phonons.

Regions near a quantum critical point in a phase diagram are characterised by unusual metallic properties. A metallic system containing a QCP to which the itinerant electrons have some coupling usually exhibits the following properties \cite{schofield, coleman1}:

\begin{enumerate}
\item Fermi liquid behaviour away from the QCP at zero temperature,
\item Close to linear temperature dependence of the resistivity, 
\beq
\rho(T)  = \rho_0 + AT^\alpha \label{eq:ExponentRes},
\eeq
with $\alpha$ near 1,
\item Divergent specific heat as a function of temperature at the QCP,
\beq
{C_v(T) \over T} = \gamma_0 \ln({T_0 \over T}), \label{eq:Cvdiv}
\eeq
which suggests, according to equation \ref{eq:specific}, a diverging effective mass $m^*$.
 \end{enumerate}

\section{The strontium ruthenate oxide \TTS}

We present in this section the properties of the material of study, \TTS, that we feel are relevant to this thesis. Numerous papers, both experimental and theoretical, have been written on the subject, and our aim is not to give a complete review on the subject. We rather insist on most aspects related to its crystalline and electronic structure, phase diagram and quantum critical properties. We therefore begin by listing the most important general physical characteristics of \TTS. This is followed by a description of its chemical structure, in order for the reader order become aware of the dimensionality of its electron liquid and the space group symmetry. We then present the electronic structure of \TTS, where using the space group symmetry, we review the shape and size of its Brillouin zone. We moreover construct an approximate Fermi surface starting from arguments related to the various $d$-bands that the quasiparticles populate. Finally, we present a review of the discoveries that were made in the past related to the phase diagram of the electron liquid in \TTS, and the existence of a QCEP.

\subsection{General physical properties}

\TTS\ is a paramagnetic metal, thought to be close to a ferromagnetic instability. It features a peak in the susceptibility, as a function of temperature, near 16~K (see  fig.~1 of the work of Ikeda and co-workers \cite{Ikeda2000}, or equivalently, figure \ref{fig:SquidExample2}, section \ref{sect:SQUID} of this thesis), and Curie-Weiss behaviour above 200~K ($\Theta_W$~=~-39~K). It can be pushed into the ferromagnetic state using pressure, where under 1GPa, magnetic ordering appears around 70~K \cite{Ikeda2000}. It possesses a highly two dimensional character, and the resistivity anisotropy is of a factor 300 at 0.3~K. It is nevertheless metallic, with strong $e-e$ correlations, as the low temperature resistivity follows a $T^2$ behaviour. The specific heat coefficient $\gamma$ was found to be large, 110~mJ/mol~Ru~K$^2$, much higher than in some of the other ruthenates, where in \TOF, $\gamma$ = 38~mJ/mol~Ru~K$^2$ and in \OOT, $\gamma$ = 30~mJ/mol~Ru~K$^2$, but similar to that of \FTT, where $\gamma$~=~109~mJ/mol~Ru~K$^2$. The Wilson ratio of \TTS\ is also high, $R_W$~=~10, due to strong ferromagnetic correlations. Consequently, this material is viewed as a strongly correlated metal on the verge of ferromagnetism.

\subsection{Crystalline structure \label{sect:Cstructure}}

\begin{figure}[t]
  \begin{center}
	\includegraphics[width=0.65\columnwidth]{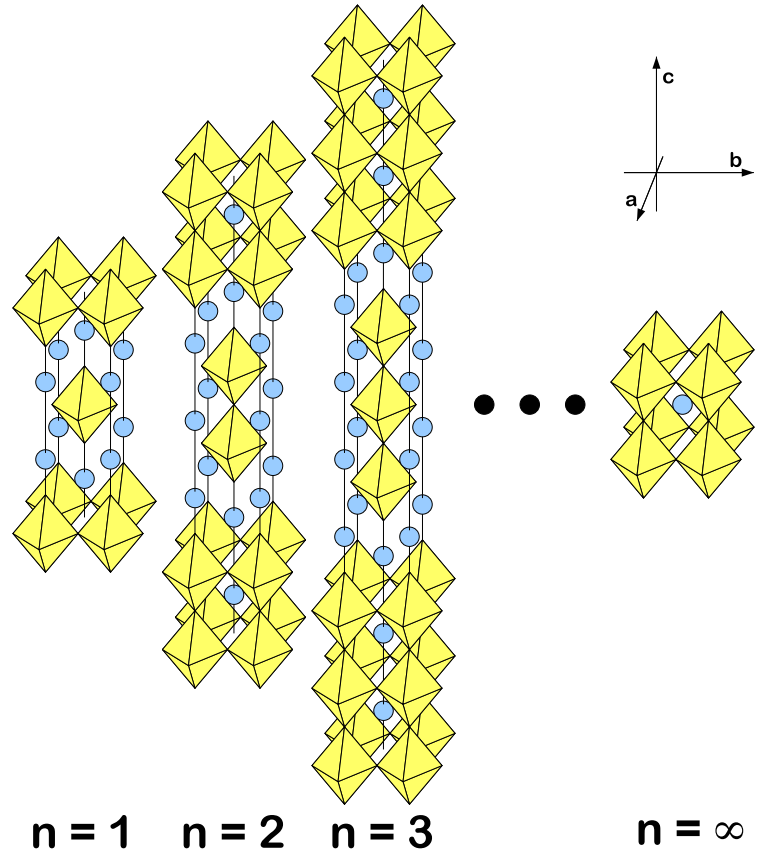}
	\caption[Ruddlesden-Popper crystallographic series]{Schematic representation of the Ruddlesden-Popper series Sr$_{n+1}$Ru$_n$O$_{3n+1}$ with increasing $n$, showing from left to right \TOF, \TTS, \FTT\ and \OOT\ ($n = \infty$). The octahedra represent the Ru ions with six O at their corners, and the spheres correspond to the Sr ions. The crystallographic axes are given in the top right corner.}
	\label{fig: RuthenateCrystal}
	\end{center}
\end{figure}

\TTS\ is member of a family which has been extensively studied, the strontium ruthenate oxide family. Its crystallographic structure has been identified by Ruddlesden and Popper \cite{Popper} as a perovskite  series of the form Sr$_{n+1}$Ru$_n$O$_{3n+1}$. The effective dimensionality of the electron liquid changes as $n$ is varied: it defines the number of consecutive planes of ruthenium oxide, leading to more or less anisotropy of the metallic state. Figure \ref{fig: RuthenateCrystal} presents the crystal structure of a few members of this family schematically. The $n=1$ member \TOF, famous for its unconventional superconductivity, probably of the $p$-wave type, has been extensively described in a review by A. P. Mackenzie and Y. Maeno \cite{mackenzieRMP}, and is isostructural to high T$_c$ superconductor LaSr$_{2-x}$Cu$_x$O$_4$. It is the most strongly two dimensional member of the family. $n=\infty$ \OOT\ is an itinerant three dimensional ferromagnet with a Curie temperature of around 160~K \cite{Kanbayasi, Allen, Cao2}. The $n=2$ compound \TTS\ is an exchange-enhanced paramagnet exhibiting a metamagnetic transition \cite{perry1}. It has one bilayer per formula unit and consequently a dimensionality intermediate between that of \TOF\ and \OOT. Finally, the members with $n>2$ possess physical properties similar to those of \OOT, with a dimensionality that varies with $n$. Of interest here, \FTT, the $n = 3$ member is a ferromagnet with Curie temperature of 105~K \cite{Cao,Crawford}. It is important to note that stacking faults can transform part of a crystal of one member of the ruthenate family into another member. For instance, suppressing RuO planes from \TTS\ leads to inclusions of \TOF. In a similar way, duplicating planes in \TTS\ will produce \FTT. One can also deduce from the similarity between the lattice cell of high $n$ and \OOT\ that such systems should have similar properties.

Strontium ruthenate crystals are usually grown using the floating zone technique. In the case of \TTS\, the crystals used in this project were produced by R. S. Perry using the method designed by Perry and Maeno \cite{perryGrowth}, and has yielded crystals of exceptionally high purity and low disorder. It is well known that each member of the Ruddlesden-Popper family, during growth, requires specific stochiometric ratios of Ru and Sr. A growth performed with intermediate ratios results in a mix of different members (see, for instance, the work of Fittipaldi and co-workers \cite{fittipaldi}). Moreover, if during growth the stochiometry inside the floating zone region drifts slightly, stacking faults appear in order to correct the relative amounts of Ru to Sr, producing inclusions of other compounds. Consequently, the stochiometry needs to be very accurate in order to grow crystals of high phase purity. We will, in this project, present a method for identifying volume fractions of the various phases of the ruthenate family in samples (section \ref{sect:search}).

\begin{figure}[t]
  \begin{center}
	\includegraphics[width=1\columnwidth]{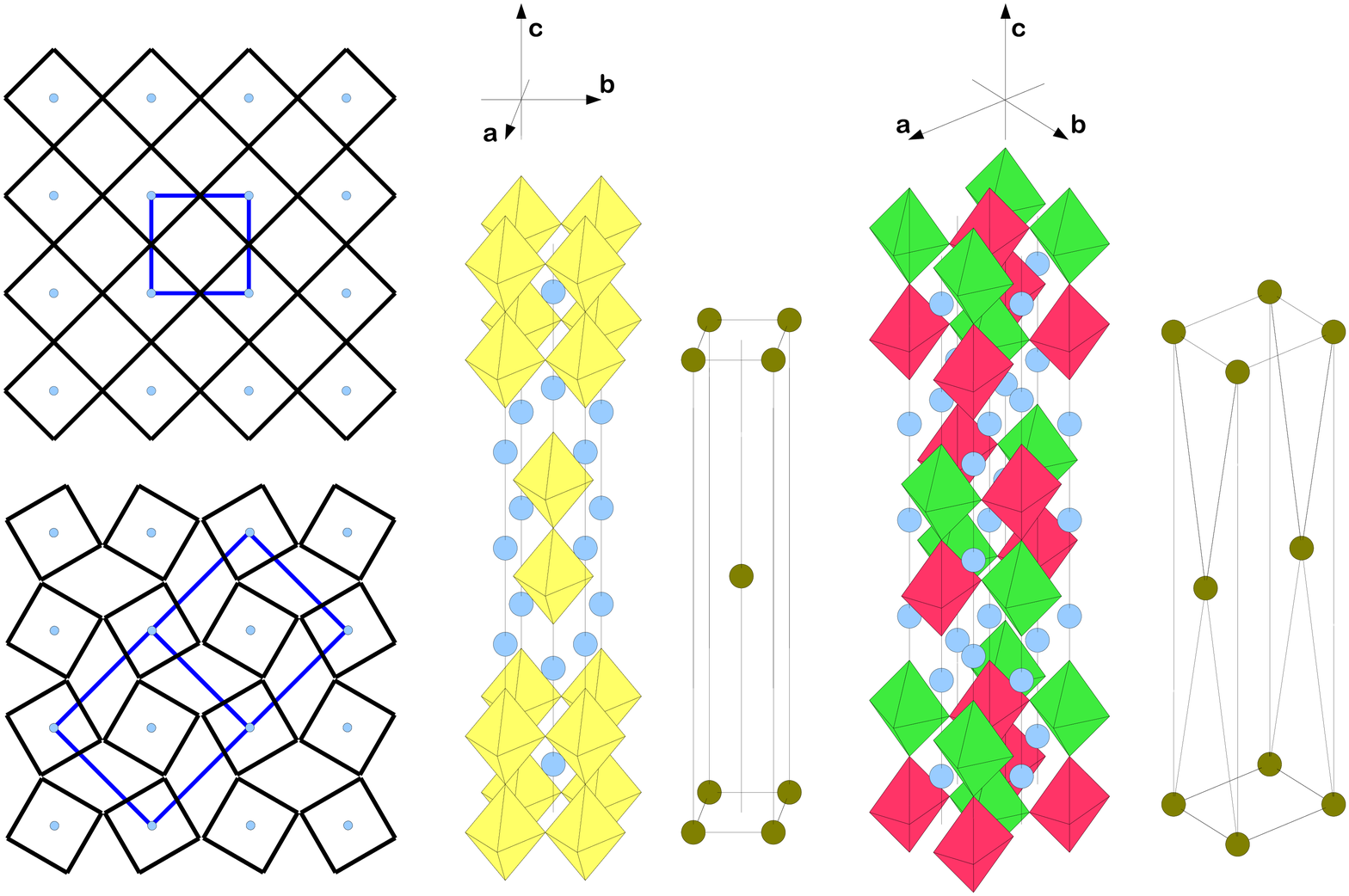}
	\caption[Crystal structure of \TTS]{Schematic representation of the octahedron rotation in \TTS. $Left$ In-plane representation of the octahedra before rotation, top, and after, bottom. The basal unit cell is shown in blue. $Middle$ Undistorted lattice unit cells. $Right$ Distorted unit cell. Yellow octahedra are not rotated, the green are rotated clockwise in the $ab$-plane, and the red counter-clockwise. The respective crystallographic axes are represented, the ones for the distorted lattice rotated by 45$^{\circ}$ around the $c$-axis.}
	\label{fig: RuthenateCrystalRot4}
	\end{center}
\end{figure}

\TOF\ possesses a body-centred tetragonal crystal structure, of space group symmetry $I4mmm$ \cite{mackenzieRMP}. Consequently, it has fourfold rotational and inversion symmetry. Early work on \TTS\ reported it to possess the same space group \cite{ZAAC}, but it was later found using neutron diffraction that this symmetry is broken by a 7$^{\circ}$ rotation of the RuO octahedra \cite{huang,shaked2}. The resulting space group is $Bbcb$, corresponding to single face centred orthorhombic structure with a square base of sides $\sqrt{2}$ larger than those of the undistorted crystal. In this new space group, one no longer has fourfold rotation, even though the lattice parameters $a$ and $b$ are still equal and impossible to distinguish in a Laue experiment. Figure \ref{fig: RuthenateCrystalRot4} presents schematically the rotation of the octahedra. On the left, one can see its effect on the basal plane, where blue lines represent the base of the unit cell. The middle part of figure \ref{fig: RuthenateCrystalRot4} shows the unit cell before rotation, with space group $I4mmm$. The right part shows the effect of the distortion where each neighbouring octahedron is rotated in opposite direction, green shapes indicate those rotated clockwise and red ones counter-clockwise. The primitive cell contains twice the number of atoms compared to the undistorted lattice\footnote{The undistorted primitive cell contains two octahedra, while the distorted one has four.}.

\subsection{Electronic structure \label{sect:estructure}}

\begin{figure}[t]
  \begin{center}
	\includegraphics[width=1\columnwidth]{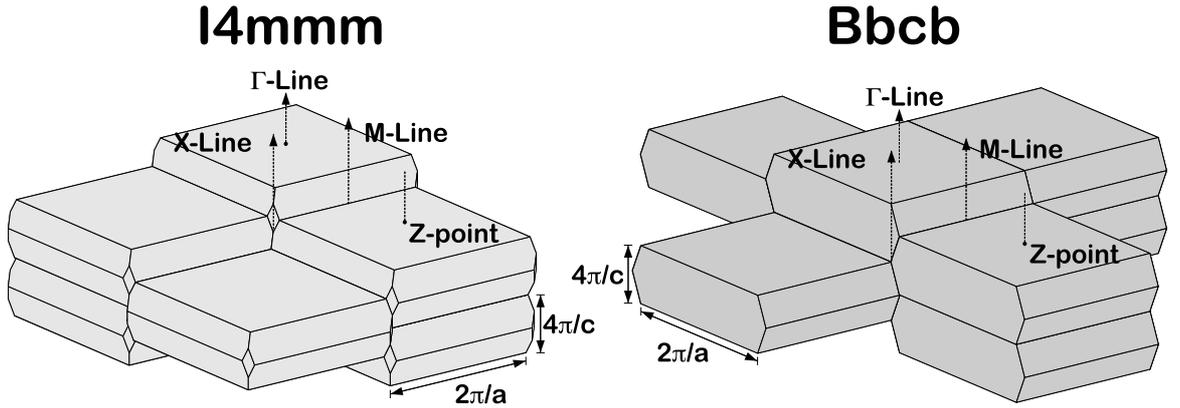}
	\caption[Brillouin zones for space groups $I4mmm$ and $Bbcb$]{Schematic representation of the Brillouin zone for $I4mmm$ , left, and $Bbcb$, right, for the case where the lattice parameters are such that $c > a = b$ (inspired from Bergemann \cite{bergemann}).}
	\label{fig: BZ214and327}
	\end{center}
\end{figure}

The rotation of the octahedra in \TTS\ has a significant effect on the electronic system. The first and obvious modification is the shape of the BZ, due to the change in lattice type. The BZ possesses the same symmetries as the crystalline structure, plus inversion through the centre. Consequently, the new BZ should possess a lower symmetry than the initial one. Figure \ref{fig: BZ214and327} presents BZs for space groups $I4mmm$, left,  and $Bbcb$, right, when the $c$ lattice parameter is longer than both $a$ and $b$, which are equal\footnote{Note that for \TOF\ and \TTS, the shape of the BZ will be flatter and wider than appears in figure \ref{fig: BZ214and327}.}. For $I4mmm$, we have the same BZ as for \TOF, which is fourfold rotational symmetric. In the case of $Bbcb$, the symmetry is not fourfold and the $k_x$ and $k_y$ axes are not equivalent, even though the in-plane lattice parameters are equal. Finally, for \TTS, the base of the BZ has an in-plane area that is twice that of \TOF. Table \ref{tab:BZarea} shows these areas for \TOF\ and \TTS, in units of $k$-space and dHvA frequencies (defined later in section \ref{sect:OscZeroT}), which will be useful in the discussion, chapter~\ref{chap:discussion}.

\begin{figure}[p]
  \begin{center}
	\includegraphics[width=1\columnwidth]{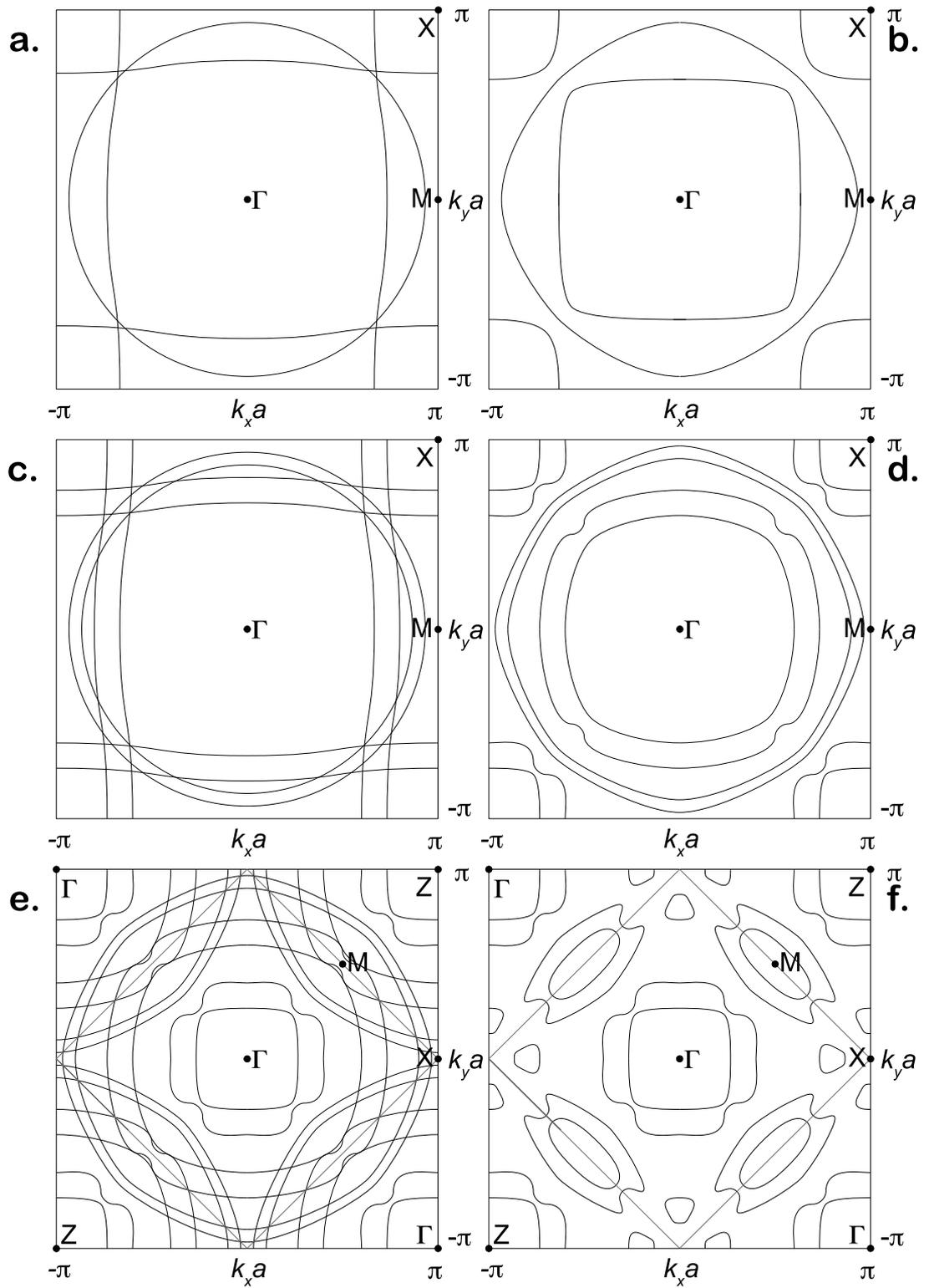}
	\caption[Schematic fermi surface of \TOF\ and of \TTS]{Schematic construction of the Fermi surface of \TOF, $a$ and $b$, and of \TTS, $c$, $d$, $e$ and $f$, where in the last two, the $\sqrt{2} \times \sqrt{2}$ reconstruction of the BZ accompanied by back-folding was included, and the new BZ is represented by the grey diamond. All axes were scaled by the parameters of the undistorted lattice.}
	\label{fig: 327FS2}
	\end{center}
\end{figure}

The FS of \TTS\ can be constructed qualitatively from hand-waving arguments, in the spirit of Bergemann  $et$ $al.$ \cite{bergemann} (see figure \ref{fig: 327FS2}). This does not involve any band structure calculation, but only the simple rule that bands cannot cross one another, and hybridise instead. We start from the situation in \TOF, where three bands are present, $\alpha$, $\beta$ and $\gamma$. These originate from the hybridisation of the various $d$ bands of the Ru atoms, the $d_{xz}$, $d_{yz}$ and $d_{xy}$, which, due to crystal field splitting, are the ones crossing the Fermi level, all the others lying at higher energy values.  The $d_{xz}$ and $d_{yz}$ give rise to almost no dispersion either in the $z$ and, respectively, $y$ and $x$ directions, but rather to quasi one-dimensional hopping. Consequently, the Fermi surface corresponding to these bands should be close to planar in the $k_x k_z$ and $k_y k_z$ direction. These will cross in certain regions of the BZ (see figure \ref{fig: 327FS2}, $a.$), where hybridisation gaps will appear, and the sheets will reconnect into closed surfaces. The $d_{xy}$ orbital, however,  allows hopping in all directions, and the corresponding Fermi surface is close to a perfect circle in the $k_x k_y$ plane. The resulting FS with all bands is shown in figure \ref{fig: 327FS2}, $b$. This construction is consistent the fourfold rotation symmetry of the $I4mmm$ space group.

\begin{table}[t]
	\begin{center}
		\begin{tabular}[h]{|c | c  c |}
			\hline
			 & BZ Area & BZ Area \\ 
			 & \AA$^{-2}$ & kT \\ 
			\hline
			\TOF & 2.61 & 27.74 \\
			\hline
			\TTS & 1.31 & 13.69 \\
			\hline
		\end{tabular}
		\caption[In-plane area of the BZ of \TOF\ and \TTS]{In-plane area of the BZ of \TOF\ and \TTS, in units of $k$-space and dHvA frequencies.}
		\label{tab:BZarea}
	\end{center}
\end{table}

In \TTS, one expects each band to duplicate, due to bilayer splitting \footnote{After distortion, \TTS\ possesses four instead of two non-equivalent Ru atoms, bringing the number of $d$ bands crossing the Fermi level from three to six.}. Consequently, the result is slightly different than for \TOF. One starts from six surfaces (figure \ref{fig: 327FS2}, $c.$), four that originate from the $d_{xz}$, $d_{yz}$ orbitals, which reconnect at the points of crossing, and one should obtain a result similar to that shown in figure \ref{fig: 327FS2}, $d$. This is what the FS could be without the $\sqrt{2} \times \sqrt{2}$ reconstruction due to the octahedral rotation. But since the BZ reconstructs into a square twice smaller, back-folding of the bands occurs, shown in figure \ref{fig: 327FS2}, $e$. In this last plot, many band crossings appear, and the way by which the surfaces reconnect is very complex. However, one can obtain hints from recent ARPES measurements \cite{tamai}. Five orbits are expected, shown in figure \ref{fig: 327FS2}, $f$, which take the form of square and cross shaped hole pockets in the centre, originating from the $d_{xz}$ and $d_{yz}$ orbitals, two lens shaped electron pockets at the $M$ point, and a small pocket near the $X$ point. The complete result from ARPES is more complex and will be discussed in section \ref{sect:ZeroFieldFS}. Note that the result possesses fourfold rotation symmetry although the space group $Bbcb$ does not. This is due to the fact that within a bilayer, the structure is fourfold symmetric, and only the stacking of the layers is not, seen in the right side of figure \ref{fig: RuthenateCrystalRot4}. Since \TTS\ is quasi two-dimensional, we expect the FS to be very close to fourfold symmetric.

\begin{figure}[t]
	\begin{minipage}[t]{7cm}
		\begin{center}
		\includegraphics[width=7cm]{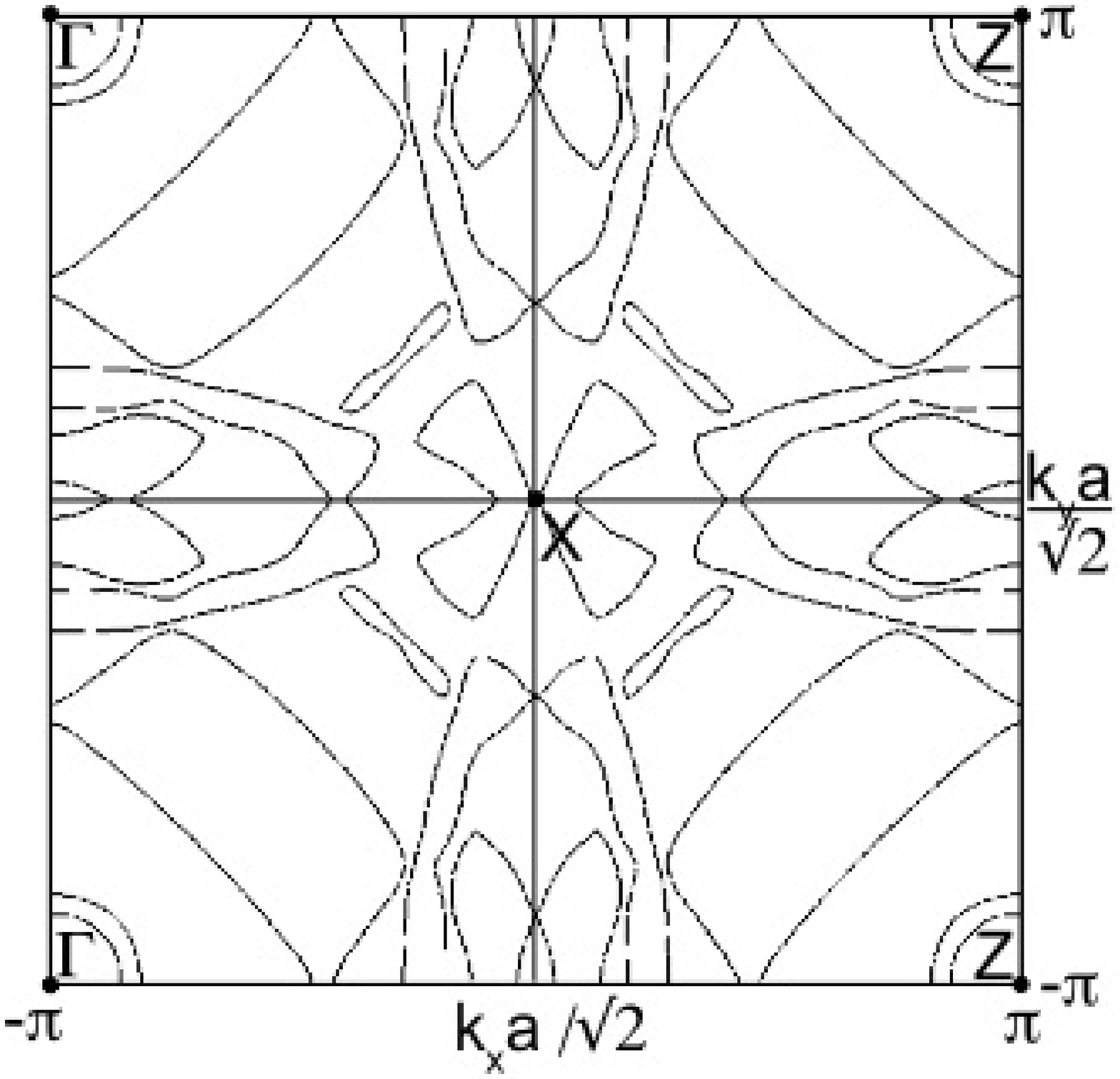}
		\end{center}
	\end{minipage}
	\hfill
	\begin{minipage}[t]{7cm}
		\begin{center}
		\includegraphics[width=7cm]{\PchapterII 3DPhaseDiag1.epsf}
		\end{center}
	\end{minipage}
	\caption[Theoretical FS and phase diagram of \TTS]{$Left$ Fermi surface structure as calculated by Singh and Mazin \cite{singh}, represented in the $\sqrt{2}\times\sqrt{2}$ reconstructed zone with its centre at the corners of the figure. The left part of the figure corresponds to a cut of the FS at $k_z = 0$, while the right part shows the FS at the top (or bottom) of the BZ. $Right$ Three dimensional phase diagram for \TTS, with axes of temperature, magnetic field and magnetic field angle. The green surface corresponds to the metamagnetic transition, and the black line its critical end point.}
	\label{fig: FS327}
\end{figure}

Singh and Mazin calculated the band structure using general potential linearized augmented plane wave method and, taking into account the distortion, suggested the Fermi surface shown in figure \ref{fig: FS327} \cite{singh}, where the FS is plotted with the $X$ point in the centre. In such a representation, the left side is located within a cut at $k_z = 0$, while due to stacking of the BZ (see figure \ref{fig: BZ214and327}), the right side corresponds to a cut at the top of the zone. The calculation features, in contrast to the ideal tetragonal case, lots of small electron and hole pockets, and a FS that breaks slightly the fourfold rotation symmetry. One can see that identifying all the different FS sheets can potentially be a difficult task without ARPES data.

\subsection{Metamagnetism and quantum criticality \label{sect:QC}}

Metamagnetism, for a paramagnetic metal, is defined as a sudden superlinear rise in magnetisation as a function of applied magnetic field. Such phenomena fall into two categories, the first corresponding to an antiferromagnetic transition, and the second to a rise in magnetisation in an itinerant paramagnet, the latter being the one of interest here. In \TTS, it has the characteristics of a first order phase transition, as was inferred from real and imaginary AC magnetic susceptibility, measured by Grigera and co-workers \cite{grigeraPRB}. In these experiments, the real part of the susceptibility ($\chi '$) showed a peak at a critical field, resulting from a jump in magnetisation, while the imaginary part ($\chi ''$) also showed a peak, indicating dissipation and signalling the first order nature of the transition. A field angle study of the complex AC susceptibility revealed a sheet of first order transitions in the 3D phase space defined by magnetic field, angle and temperature ($H$,$\theta$,$T$), terminated by a line of critical points, at which the peak in $\chi ''$ disappeared (shown in figure \ref{fig: FS327}, right). At higher temperatures, the peak in $\chi''$ was not detected. This line was seen to approach absolute zero at a field of 7.85~T and an angle close to 90$^\circ$ with respect to the $ab$ plane.

\begin{figure}[t]
	\begin{minipage}[t]{7cm}
		\begin{center}
		\includegraphics[width=7cm]{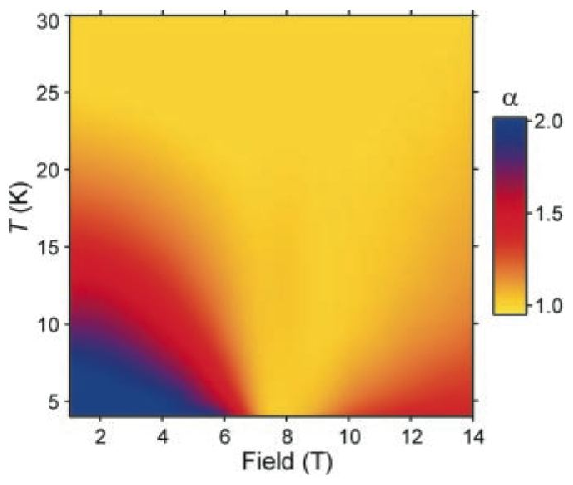}
		\end{center}
	\end{minipage}
	\hfill
	\begin{minipage}[t]{7cm}
		\begin{center}
		\includegraphics[width=7cm]{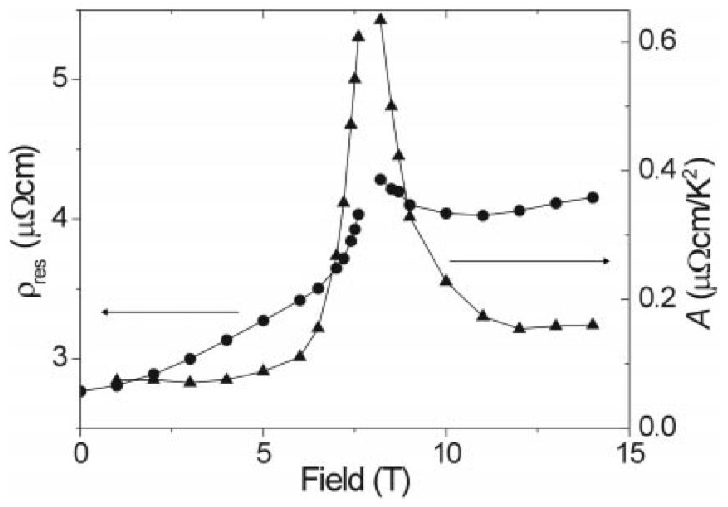}
		\end{center}
	\end{minipage}
	\caption[Power law exponent and $A$ coefficient of the resistivity]{$Left$ Exponent of the temperature dependence of the resistivity (eq. \ref{eq:ExponentRes}) as a function of magnetic field and temperature, obtained using the equation $\alpha = d \ln(\rho-\rho_{res})/ d \ln T$. $Right$ Field dependence of the $A$ coefficient of eq. \ref{eq:ExponentRes} obtained by calculating $(\rho(T) - \rho_{res})/ T^2$, and that of the residual resistivity $\rho_{res}$. Both plots were reproduced from the work of Grigera $et$ $al.$ \cite{science1}. }
	\label{fig:rho_T_c-axis}
\end{figure}

This field angle analysis of the susceptibility in \TTS\ led to the introduction of a new idea, a quantum critical end point (QCEP) arising where the line of critical end points crosses the $T = 0$ plane. As described in section \ref{sect:QCtheory}, a QCP usually arises when a continuous (second order) phase transition reaches 0~K, but a first order phase transition does not lead to a QCP. However, the critical end point terminating a first order transition line in a phase diagram exhibits all the properties of a continuous transition except, in this case, spontaneous symmetry breaking. Grigera $et. al.$ suspected a line of critical end points reaching 0~K to have quantum critical properties, that is, to form a QCEP with field angle as a tuning parameter. 

A careful study of both the specific heat and resistivity in the ($H$, $T$) phase space revealed strong evidence of the existence of such a critical point and of non-Fermi liquid behaviour \cite{science1,perry1}. Figure \ref{fig:rho_T_c-axis} shows the exponent $\alpha$ of the $T$ dependence of the resistivity (eq. \ref{eq:ExponentRes}) for a field aligned in the direction of the $c$-axis, using the equation
\beq
\alpha = {d\ln(\rho-\rho_{res}) \over d\ln T}.\nn
\eeq
Near 8~T, linear $T$ dependence was seen for all temperatures, but away from this field, $T^2$ behaviour was recovered, suggesting that a Fermi liquid existed on both sides of the QCEP. As the resistivity became linear near at 8~T, the parameter $A$ of equation \ref{eq:ExponentRes}, obtained by calculating $(\rho(T) - \rho_{res})/ T^2$ using data measured between 0.2 and 0.9~K, was also seen to increase sharply. This suggested an enhancement of the quasiparticle mass of at least more than one band, as can be seen from equation \ref{eq:resistivity} \footnote{Note that $A$ was obtained from fitting the data $only$ in regions where it follows a $T^2$ dependence, a region that shrinks in length as one comes closer to the metamagnetic transition.}. 

\begin{figure}[t]
	\begin{minipage}[t]{7cm}
		\begin{center}
		\includegraphics[width=7cm]{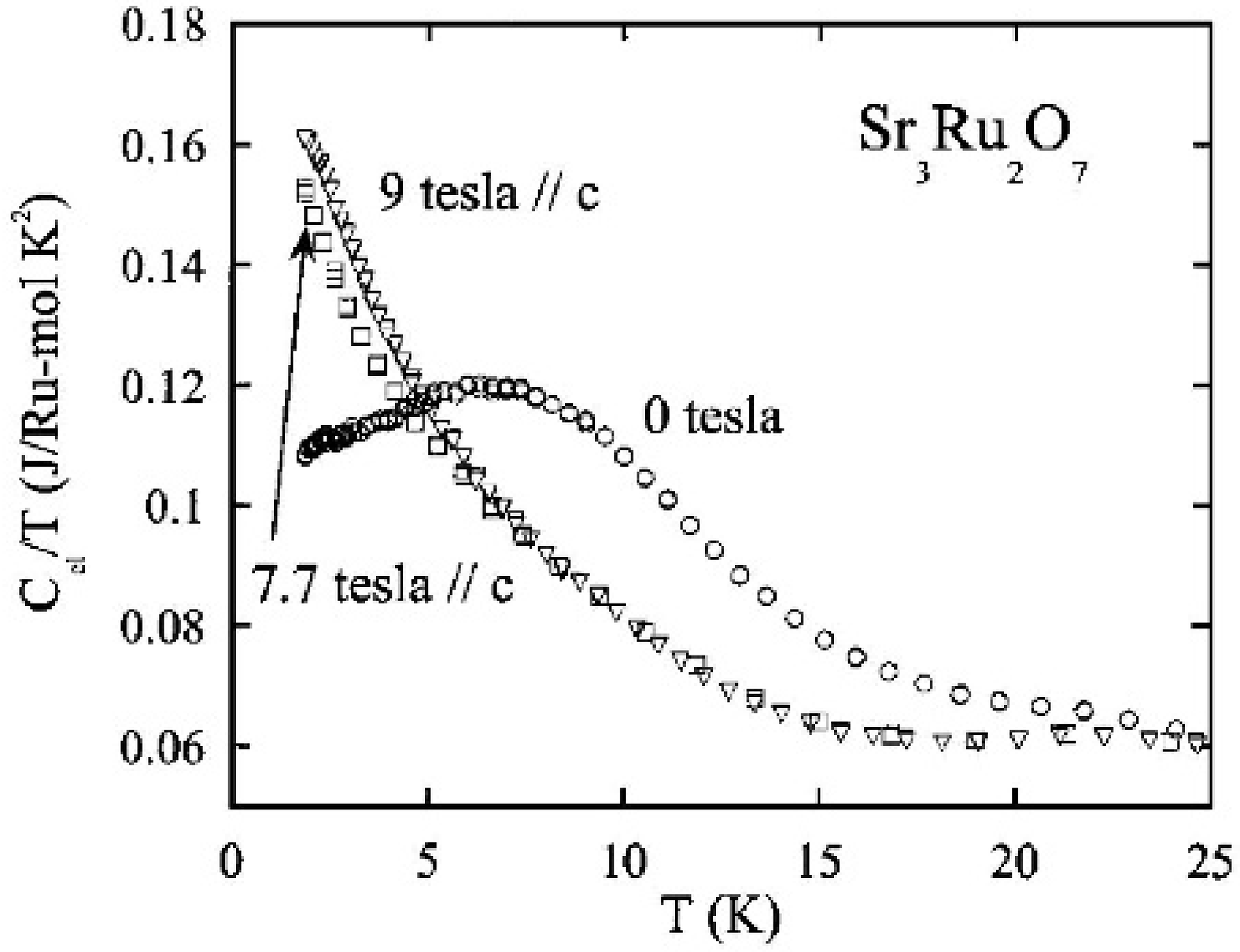}
		\end{center}
	\end{minipage}
	\hfill
	\begin{minipage}[t]{7cm}
		\begin{center}
		\includegraphics[width=7cm]{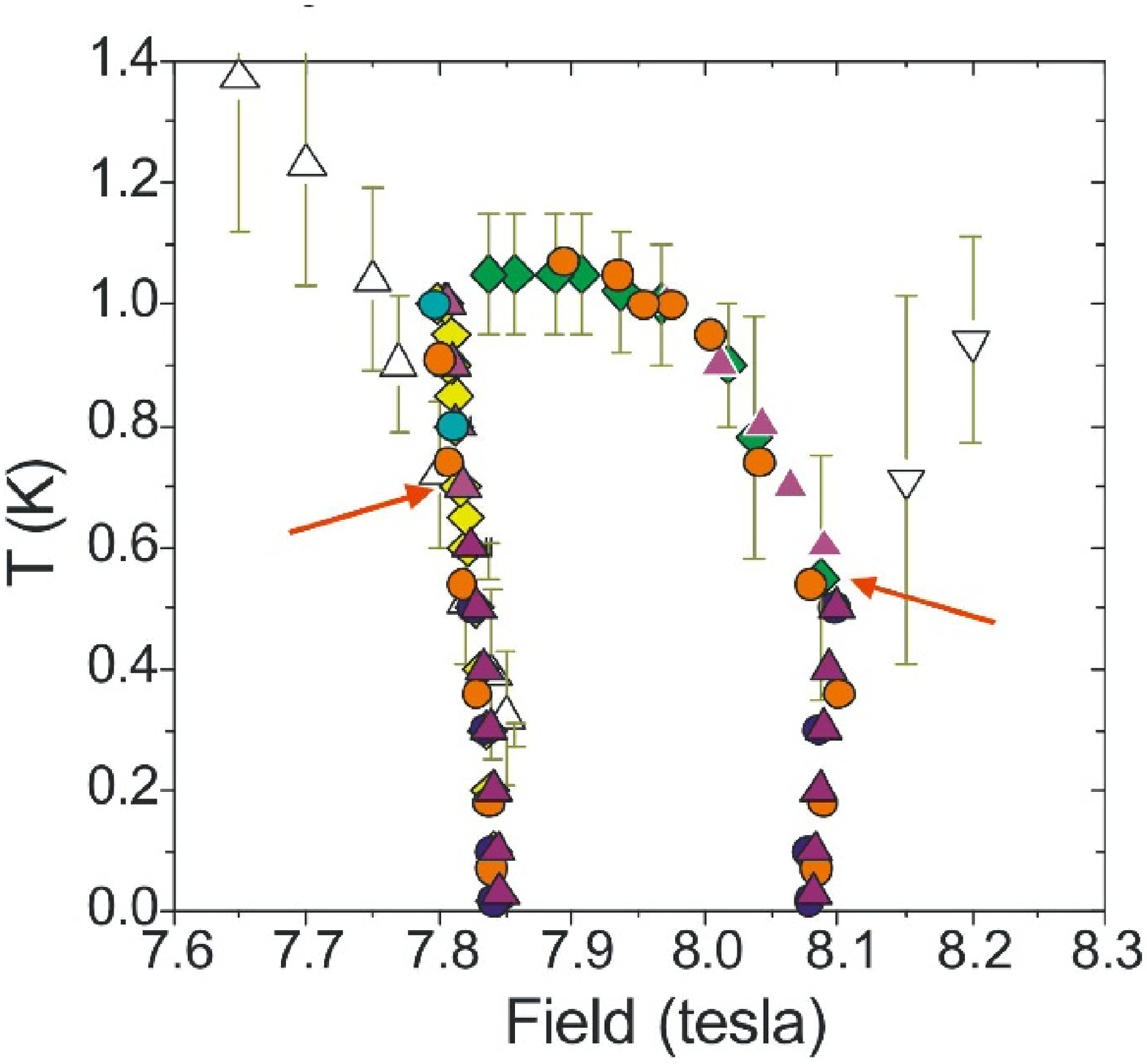}
		\end{center}
	\end{minipage}
	\caption[Electronic specific heat and new phase diagram for \TTS]{$Left$ Electronic specific heat as a function of temperature \cite{perry1}. Logarithmic divergence is observed at 7.7~T. $Right$ New phase reported in \TTS\ surrounding the QCEP, taken from Grigera and co-workers \cite{science2}. Symbols represent features in data taken using various experimental techniques: coloured triangles are the centre of the peaks in the AC susceptibility, white triangles are the loci of peaks in the thermal expansion, orange circles are taken from the DC magnetisation,  yellow diamonds were obtained from $d \rho / d H$, green diamonds are from $d^2 \rho / d T^2$ and the blue circles were extracted from maxima in the magnetostriction. The red arrows indicate the critical points at the end of each of the two lines of first-order transitions.}
	\label{fig:specific_heat}
\end{figure}

The specific heat of \TTS\ was also measured, and reproduced one of the predictions of quantum critical theory. Figure \ref{fig:specific_heat} shows curves of the electronic specific heat $C_{el}/T$ as a function of temperature. Logarithmic divergence as a function of temperature, in accord with eq \ref{eq:Cvdiv}, was seen near the QCEP, which again indicated a divergence of the quasiparticle effective mass $m^*$ (see eq. \ref{eq:specific}). It became desirable to perform an independent measurement of the quasiparticle mass, and quantum oscillations experiments were performed by Perry $et. al.$ \cite{perry2}, and later by Borzi and co-workers near the QCEP \cite{borzi}, where both showed a slight transformation of the FS across the metamagnetic transition and the latter claimed a divergence of the mass of two FS sheets near the QCEP. The work by Borzi $et. al.$ will be described in more details in section \ref{sect:BorzidHvA}.


The system under study met all the requirements (see the list, section \ref{sect:QCtheory}) for quantum critical behaviour, as was demonstrated by Grigera $et$ $al.$ \cite{science1}. However, anomalies were found very near the metamagnetic transition, between fields of 7.8 and 7.9~T, where a decrease in $A$ was observed and a resistivity temperature exponent of 3 was measured between fields of 7.82 and 7.86~T. As new samples were later generated with lower residual resistivities, even more features were discovered near the QCEP, described in the next section.

Finally, we add a brief note about the nature of the magnetic fluctuations present in \TTS, determined with inelastic neutron scattering by Capogna $et$ $al.$ \cite{capogna}. Although at high temperature, the fluctuations are of ferromagnetic nature, as one expects, incommensurate antiferromagnetic fluctuations develop at zero fields as the temperature is lowered. Effectively, a resonance at $\textbf{q} = 0$ was observed at a temperature of 150 K at an energy of 3.1 meV, which vanishes at around 15 K, and corresponds to ferromagnetism. Alongside this excitation were observed resonances at $\textbf{q} = [\pm0.25\; 0\; 0]$ and $[\pm0.09\; 0\; 0]$, which correspond to antiferromagnetism, at temperatures of around 15 K and below. The system is thought to exhibit a competition between the two types of magnetic ground states. 

\subsection{Disorder sensitive phase formation}

As a new generation of samples were grown \cite{perryGrowth}, new experiments were performed, where the metamagnetic transition was revealed to be double\cite{perry2}. From various physical properties\footnote{These included resistivity, magnetic susceptibility, magnetisation, thermal expansion and magnetostriction} of \TTS\ near 8T, strong evidence was found indicating a new ordered phase of the electron liquid surrounding the QCEP \cite{science2}. This phenomenon was reminiscent of famous cases where superconductivity was found surrounding a QCP, for instance in the work on CePd$_2$Si$_2$ and  CeIn$_3$\cite{mathur}. Figure \ref{fig:specific_heat}, right, shows the phase diagram of \TTS\ near 8~T, at low temperatures. In this plot, one can see the bounded phase in the centre, and two first order transition lines extending to higher temperatures away from the phase region. The nature of the new phase was not known but Fermi liquid behaviour was confirmed both to higher and lower magnetic fields. Note that the sides of the phase feature much a stronger thermodynamic signature than the so-called ``roof"\footnote{This subject has been extensively studied by Rost $et$ $al.$, in preparation.}.

\begin{figure}[t]
	\begin{minipage}[t]{6cm}
		\begin{center}
		\includegraphics[width=6cm]{\PchapterII 3DPhaseDiag2.epsf}
		\end{center}
	\end{minipage}
	\hfill
	\begin{minipage}[t]{8cm}
		\begin{center}
		\includegraphics[width=8cm]{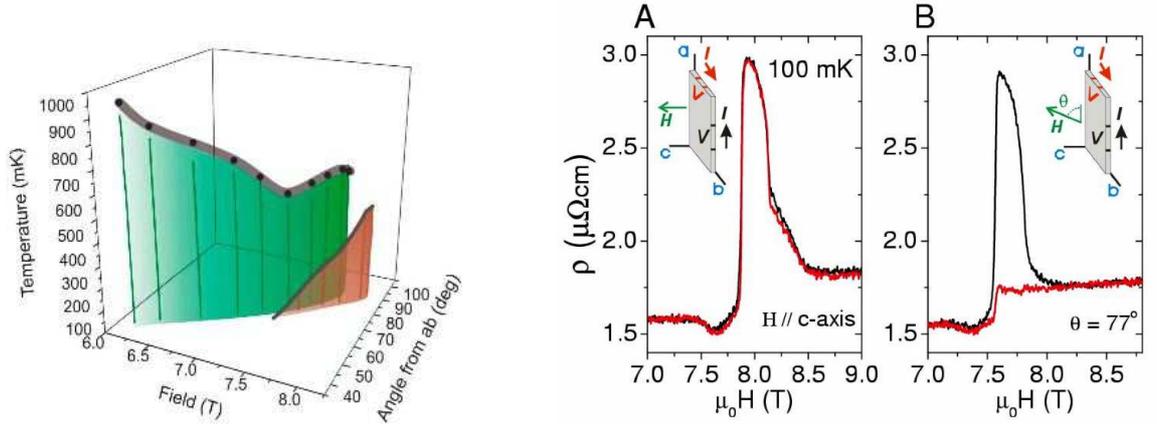}
		\end{center}
	\end{minipage}
	\caption[Three dimensional phase diagram and nematic nature of the new phase]{$Left$ Three dimensional representation of the various phase transitions delimiting the new phase (see text), from Green $et$ $al.$ \cite{PRLgreen}. $Centre$ $and$ $Right$ Nematic nature of the new phase, taken from Borzi $et$ $al.$ \cite{science3}.}
	\label{fig: Nematic}
\end{figure}

The boundaries of the phase were studied in three dimensional space, and it was found to extend in angle to approximately 30$^{\circ}$ from $c$-axis, shown in figure \ref{fig: Nematic}, left. In such a new phase, it was expected that symmetry should be broken, and that domains could be formed. Magnetoresistance experiments at different magnetic field angles were performed, comparing situations where it was aligned or perpendicular to the current, which showed anisotropy, shown in figure \ref{fig: Nematic}, centre and right. It was concluded that the electron liquid possessed nematic properties \cite{science3}. The original article suggested that this could correspond to a deformation of the FS which produces domains that impede the transport of current in a preferential direction that depends on the orientation of the magnetic field. Other studies claim that it may originate from Condon domains \cite{PRLBinz}, or from the appearance of a spontaneous transverse magnetisation\footnote{Work from Green $et$ $al.$, to be published.}. However, the first of these two was ruled out by a careful study of the the system with various crystal geometries, which showed no variation of the critical fields. We found, during this project using dHvA, that the system still possesses metallic properties, by the observation of quantum oscillations inside the phase (see section \ref{sect:nematicCambridge}).

One of the very interesting properties of \TTS\ is the dependence of its properties on disorder. This has motivated us in performing more systematic growth and characterisation of crystals, in the hope of finding samples of outstanding quality. This has been one of the main aspects of this project, and is described in section \ref{sect:search}.

\section{The de Haas van Alphen Effect}

The dHvA experiment is the traditional method for studying the FS of a material. It is normally used for determining the volume and shape of the FS, as well as quasiparticle masses. Accessing a quantum critical point requires a control parameter that is usually not the magnetic field, and when it is, low field values have usually been required (see for example \cite{gegenwart}). Applying a field thus tunes the system away from quantum criticality, so that the dHvA experiment is not possible near a QCP. It is not the case for \TTS, since the QCEP is field tuned between 5 and 8~T with field angle $\theta$. It is the ideal system for exploring a FS near a QCP. Since the dHvA experiment yields the quasiparticle mass for each branch of the FS, it is possible to obtain them as a function of both field and field angle, but moreover, it can show whether there are changes in the FS topology as well. This section thus reviews the basics of the dHvA experiment, along with methods for calculating the quasiparticle masses as a function of magnetic field. For more details, the reader is referred to the main reference on dHvA, the book by Shoenberg \cite{shoenberg}, and a review on \TOF\  by Bergemann $et$ $al.$ \cite{bergemann}.

This section is divided into six parts. We first derive the basic equations and explain the origin of the effect at zero temperature in a perfect lattice. Next, we include the effects of finite temperature and quasiparticle mean free path. Then, a short digression is given about the effect of the spin. We discuss how the two dimensional nature of a material affects dHvA and finally, we review the first measurements that were performed on \TTS\ prior to this project.

\subsection{Oscillations at zero temperature\label{sect:OscZeroT}}
We present here a semi-classical derivation of the quantum oscillations at zero temperature. The source of the quantum oscillations is the Landau quantisation of the orbital movement of electrons in a magnetic field. When a magnetic field is applied to a free electron gas in a box, in the $z$ direction, electrons undergo an orbital or helical movement that corresponds to quantised orbits in $\bs{k}$ space, the $k_z$ quantum number remaining unchanged, with energies
\beq
\epsilon_n = (n + {1\over2})\hbar \omega_c + {\hbar^2 k_z^2 \over 2 m_e},\nn
\eeq 
where $\omega_c = eH/m_e$ is the cyclotron frequency. In such a system, without field, electrons would, at $T = 0$, take the lowest $\bs{k}$ values and form a sphere in $\bs{k}$ space, the highest energy being the Fermi energy $\epsilon_F$, whereas with the field turned on, they will fill so-called Landau $tubes$, concentric hollow cylinders up to the surface of the same sphere. Effectively, for high quantum numbers $n$, the degeneracy of the Landau levels is such that within a $k$-space area delimited by a contour of constant energy, without magnetic field, the number of states equals approximately the same as the sum of the number of states available in all the Landau levels included within that area for a specific field, and that degeneracy is linear in field. Consequently, the Fermi level lies at the same energy value with or without a magnetic field\footnote{High quantum numbers $n$ arise when the Fermi energy is high compared to $n\hbar \omega_c$, which is true in most metals; exceptions to the rule are materials with very small Fermi surfaces, for instance doped semiconductor layers.}.

More often than not, the electronic system forms a Fermi surface that is not isotropic. The orbital movement in $\bs{k}$ is not circular but follows paths along surfaces of constant energy in $\bs{k}$ space. In a magnetic field, the rate of change of momentum of the quasiparticles is
\beq
\dot{\bs{k}} = -{e \over \hbar} \bs{v} \times \bs{H},\nn
\eeq
where
\beq
\bs{v} = \bs{\nabla}_k \epsilon(\bs{k}).\nn
\eeq
$\bs{v}$ is always normal to a surface of constant energy, and the rate of change of momentum is in the direction of constant energy, in a plane perpendicular to $\bs{H}$. It can be integrated in time to give
\beq
\bs{k} - \bs{k}_0 =  -{e \over \hbar}(\bs{r} - \bs{r}_0) \times \bs{H}. \label{eq:kr}
\eeq
The Bohr Sommerfeld quantisation rule for periodic motion states that, for canonically conjugate operators $\bs{p}$ and $\bs{q}$,
\beq
\oint \bs{p}\cdot d\bs{q} = 2 \pi \hbar (n + \gamma),\nn
\eeq
where $n$ is the orbital quantum number and $\gamma$ is a constant. $\bf{p}$ is the canonical momentum $\hbar \bs{k} - e\bs{A}$, $\bs{A}$ being the vector potential and $\bs{q}$ is the component of the position operator $\bs{r}$ in the plane perpendicular to $\bs{H}$, that we will denote $\bs{r'}$. Using (\ref{eq:kr}) for the first term and Stokes' theorem for the second, we obtain:
\bea
\oint \bs{p}\cdot d\bs{q} &=&  e\oint \big[\hbar \bs{k} - e\bs{A} \big]\cdot d\bs{r},\nn\\
&=& e \bigg[ \oint \bs{H} \cdot (\bs{r'} \times d\bs{r'}) - \int_S \bs{H} \cdot d\bs{S} \bigg],\nn
\eea
where $\bs{S}$ is the surface enclosed by the orbit in real space. In this expression, $\bs{r}$ should be decomposed into its components parallel and perpendicular to $\bs{H}$, and only the first will contribute. The second integral is equal to $|\bs{H}|$ times the area $S$, while the first is twice that:
\beq
2 \pi (n + \gamma){ \hbar \over e} = 2HS - HS = HS\nn
 \eeq
 so that
 \beq
 S(n, H) = 2 \pi (n + \gamma){\hbar \over e H}.\nn
 \eeq
This means that it is the $area$ in real space that is quantised, and it is also inversely proportional to the field $H$. We are interested in the area in $\bs{k}$ space, though, and from eq. (\ref{eq:kr}) we note that
\beq
|\bs{k} - \bs{k}_0| = -{e H\over \hbar}|\bs{r'} - \bs{r'}_0|.
\label{eq:scaling}
\eeq
It follows from the scaling factor, ${e H \over \hbar}$, that the area in $\bs{k}$ space is proportional to that in real space through
\beq
S_k(n,H) = \bigg({e H \over \hbar}\bigg)^2S(n,H),\nn
\eeq
so that
\beq
S_k(n,H) = 2 \pi (n + \gamma) {e H \over \hbar},\nn
\eeq
which is proportional to the field $H$. Consequently, for any shape of orbital motion, the Landau tubes have a cross-sectional shape defined by constant energy paths in planes perpendicular to the magnetic field. If the value of $H$ increases or decreases, then the tubes, respectively, $grow$ or $shrink$ in size. When a tube crosses a region of the Fermi surface parallel to $\bf{H}$ (an $extremal$ region of the Fermi surface with respect to the field direction), a sudden change in the DOS at the Fermi surface arises, with a period that is a constant of the inverse field $X = 1/H$. This can be more easily understood by looking at the area of the $n$th tube crossing the Fermi surface at a field $H_1$, and that of the next, $n-1$ at $H_2$, as $H$ increases:
\bea
S_k(n,H_1) &=& (n + \gamma) {2 \pi e \over \hbar} H_1, \nn\\
S_k(n-1,H_2) &=& (n -1 + \gamma) {2 \pi e \over \hbar} H_2.\nn
\eea
These areas are equal, and by isolating $n$, we calculate
\bea
{S_k \hbar \over 2 \pi e H_1} - \gamma &=&  {S_k \hbar \over 2 \pi e H_2} - \gamma + 1,\nn
\eea
The period in inverse field $T_{1/H}$ is a constant of the system, and corresponds to a dHvA frequency $F$:
\bea
T_{1/H} = {1\over H_1} - {1\over H_2} &=& {2 \pi e \over \hbar S_k},\nn\\
\Rightarrow F = \bigg[\Delta \big({1 \over H}\big)\bigg]^{-1} &=& {\hbar S_k \over 2 \pi e}.
\label{eq:period}
\eea
The frequency $F$ is proportional to the cross-sectional area $S_k$ of the Fermi surface. 

The critical consequence of this phenomenon is that all physical quantities that depend on the DOS at the Fermi level exhibit oscillations as the magnetic field is swept \footnote{For instance, oscillations have been measured in the magnetisation, resistivity, susceptibility, magneto-caloric effect and even sample size.}, of the form
\beq
\chi(H) = \sum_p A_p \cos\bigg({2\pi pF \over H} + \phi\bigg),\nn
\eeq
where $p$ stands for harmonics of the fundamental frequency $F$, due to the fact that the oscillations are not sinusoidal. The name de Haas-van Alphen was given to oscillations in the magnetisation, by the name of their discoverers \cite{dHvA}, while Shubnikov-de Haas (SdH) was given to those in the resistivity \cite{SdH}, and these experiments followed a prediction by Landau in his work on diamagnetism \cite{Landau}. Applying a Fourier transform to quantum oscillation data as a function of $X = 1/B$, one extracts the area of all extremal orbits that exist in a field direction, and by rotating the sample, one can in principle construct the full 3D shape of the FS. In 2D systems, frequencies of different extremal orbits may be very close together and give rise to beat patterns, and this is discussed in section \ref{sect:BergemanAnalysis}. 

For later use, we rewrite the energy $\epsilon_n$ of the quasiparticle as a function of the area of the $n$th Landau level and its $k_z$ wave vector:
\beq
\epsilon_n =  {\hbar^2 S_n \over 2 \pi m^*} + {\hbar^2 k_z^2 \over 2 m^*}. 
\label{eq:epsilonS}
\eeq

Finally, we note here that electron densities can be calculated from FS volumes determined with dHvA, and the resulting number should respect the stochiometry of the material under study. This is called Luttinger's theorem \cite{luttinger}. For \TTS, we will use this in two dimensions, where this translates to the sum of in-plane areas, with dHvA frequencies $F_n$ of $n$ different sheets,
\beq
N_e = 2{2\pi e \over \hbar S_k^{BZ}} \sum_n F_n = {2 \over F_{BZ}} \sum_n F_n,
\label{eq:LuttingerSum}
\eeq
with $S_k^{BZ}$ and $F_{BZ}$ the in-plane area of the BZ and corresponding dHvA frequency, and the factor two stands for the two spin species. Note that for hole-like FS pockets, one should use the area of $electrons$, which is the area of the BZ minus that of the cyclotron orbit, or alternatively, $F_n = F_{BZ} - F_{hole}$. The result of this sum should correspond to the number of electrons in the material minus the number of electrons in filled bands.

\subsection{Oscillations at finite temperature \label{sect:LK}}

We discuss in this section the role of temperature. At $T = 0$, the jumps in the DOS are produced by the crossing of the Landau levels through a perfectly sharp Fermi surface, which broadens proportionally to $k_B T$ as temperature increases. This results in smearing of the quantum oscillations, through an average weighted by a probability distribution, $g(\epsilon)$ that corresponds to the negative derivative of the Fermi-Dirac distribution with respect to energy $\epsilon$,
\beq
g(\epsilon) = - {df(\epsilon) \over d\epsilon} = {1 \over 2 k_B T \big[1 + \cosh\big((\epsilon - \mu)/kT\big)\big]}. \label{eq:dfde}
\eeq
The oscillations reduce in amplitude, through a convolution of the signal $\chi(X)$, $X = 1/B$, by $g(\epsilon)$ as a function of dHvA phase $\phi$, which we normalise by an unknown factor $\lambda$:
\beq
\chi(X,T) = \int_{0}^\infty \chi_{T = 0}(2\pi p F X + \phi) g\bigg({\phi \over \lambda}\bigg) d\phi. 
\label{eq:LKconvolution}
\eeq
The relation between $\epsilon$ and $\phi$ comes from slight differences in frequency, or dephasing of
\bea
\phi &=& 2\pi \bigg({p \Delta F \over H}\bigg) = \bigg({2 \pi p \over H}\bigg) {\partial F \over \partial S_k}{\partial S_k \over \partial \epsilon}\Delta \epsilon,\nn\\
&=&   {2\pi p m^* \over e \hbar H} \Delta \epsilon\nn
\eea
with $\Delta \epsilon = \epsilon - \mu$, $\mu$ being the chemical potential, and the effective cyclotron mass $m^*$ is taken from its definition,
\beq
m^* = {\hbar^2 \over 2 \pi} {\partial S_k \over \partial \epsilon}.
\label{eq:dAde}
\eeq
We rewrite $g(\epsilon)$ as a function of $\phi / \lambda$, with $\phi / \lambda = \Delta \epsilon / k_b T$:
\beq
g\bigg({\phi \over \lambda}\bigg) =  {1 \over 2 k_B T \big[1 + \cosh\big({e \hbar H \over 2 \pi m^* k_B T} \phi \big)\big]}\nn
\eeq
and obtain
\beq
\lambda =  {2 \pi p m^* k_B T \over e \hbar H},\nn
\eeq
which corresponds to the half width of the function $g$, and is specific to one Fermi surface sheet with quasiparticle mass $m^*$. Eq. \ref{eq:LKconvolution} is the Fourier transform of $g$:
\beq
Re\bigg[ \int_{0}^\infty A_p \exp\bigg(2\pi i p F_n X\bigg) \exp(i \phi) g\bigg({\phi \over \lambda}\bigg) d\phi\bigg].\nn
\eeq
 It is a calculation done using Cauchy's residue theorem, and is given in appendix~\ref{App:A}. The result is a coefficient to the oscillations of the form
\bea
\tilde{g}(\lambda) = {\pi \lambda \over \sinh(\pi \lambda)}.\nn
\eea
This function is called the Lifshitz-Kosevich relation (LK) \cite{LK1,LK2}, which we can rewrite as a function of the variables of interest, $X$ and $T$:
\beq
LK(X,T) = {Cm^*XT \over \sinh(Cm^*XT)}, \quad C = {2 \pi^2 p k_B \over e \hbar}.
\label{eq:LK}
\eeq
The LK prefactor is different for dHvA frequencies with different masses, but also for all harmonics $p$.  For an orbit of frequency $F$ with mass $m^*$, its harmonics at $pF$ will appear to possess a mass of $pm^*$.

\begin{figure}[t]
  \begin{center}
	\includegraphics[width=.60\columnwidth]{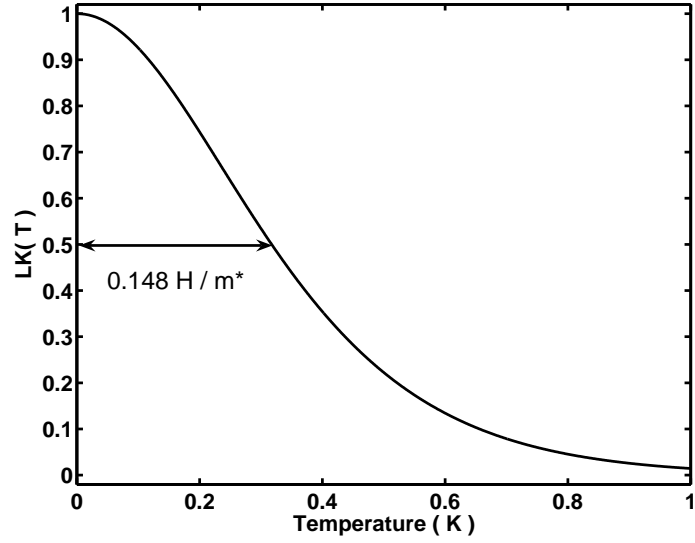}
	\caption[dHvA amplitude reduction factor $LK$]{The factor $LK(X,T)$ as a function of temperature, using typical values for the mass and magnetic field ($m^* = 7 m_e$ and $H$ = 12T respectively).}
	\label{fig: RLK}
	\end{center}
\end{figure}

Figure \ref{fig: RLK} shows the shape of LK(T). One can see that the oscillations die off at a temperature of approximately $0.148 {H\over m^*/m}$, the half-width of the function, which involves the mass $m^*$. Moreover, it is possible to extract $m^*$ from dHvA using a non-linear fit of the amplitude of one frequency as a function of the temperature, which will be shown in section \ref{sect:mass}. The masses are usually different for each FS sheet since these originate from the hybridisation of different atomic orbitals with different dispersions, but can be extracted independently.

One last word should be said about quasiparticle masses. These can be compared with the electronic specific heat of the material under study, in order to help evaluate whether a complete set of Fermi surface sheets has been identified in a dHvA experiment. For two-dimensional materials, one can calculate that the linear specific heat coefficient $\gamma$ relates to the masses through a simple sum,
\beq
\gamma = {\pi k_B^2 N_A a^2 \over 3 \hbar^2} \sum_n m^*_n = 1.48 \, \textrm{mJ/molK}^2 \sum_n m^*_n,
\label{eq:SpecificHeatSum}
\eeq
where $a$ is the in-plane lattice parameter, and the factor 1.48~mJ/molK$^2$ was obtained using the lattice parameter $a$ of \TTS \footnote{For \TTS, the in-plane lattice parameters are equal, $a = b$.}. The values of $m^*_n$ are the masses of each FS sheet. The result of this equation possesses units of mJ/mol~K$^2$ per formula unit. In our case, we will want to quote the result per mole of Ru, expressed as mJ/mol~Ru~K$^2$, and a factor two will be taken out of the coefficient. This relation has been used with success in various systems, notably in Sr$_2$RuO$_4$ \cite{bergemann} and Sr$_2$RhO$_4$ \cite{baumberger}. 

\subsection{The role of disorder and impurities \label{sect:Impurities}}

Crystal disorder and impurities also have an effect on the amplitude of the dHvA oscillations in a very similar way to thermal phase smearing. Disorder is temperature independent, and only affects the field dependence of the oscillations. Scattering of electrons on impurities or defects has the effect of a finite lifetime for Landau levels. If the scattering centres are distributed uniformly and fixed in space, the rate of scattering $\Gamma$ is constant, and the probability of having an electron travelling without scattering during time $t$ is 
\beq
P(t)dt = e^{-\Gamma t}dt, \quad \Gamma = {1\over \tau},\nn
\eeq
In frequency (or energy) space, this corresponds to a lorentzian probability,
\beq
P(\omega)d\omega = {\Gamma \over \Gamma^2 + (\omega-\omega_0)^2}d\omega.\nn
\eeq
The Landau levels are broadened, and the probability of having an electron in level $n$ with energy  between $\epsilon$ and $\epsilon + d\epsilon$ is
\beq
P(\epsilon)d\epsilon = {1 \over 1 + (\epsilon-\epsilon_n)^2/ \hbar^2 \Gamma^2}d\epsilon.\nn
\eeq
The effect is the same as for the Fermi surface thermal broadening, it corresponds to a convolution of the signal with a Lorentzian function, where the phase $\phi$ is defined as 
\beq
{\phi \over \lambda} = {\Delta \epsilon \over \hbar \Gamma}, \nn
\eeq
where this time $\lambda$ is
\beq
\lambda =  {2 \pi p c m^* \Gamma \over e H}.\nn
\eeq
The prefactor to the oscillations is the Fourier transform of the Lorentzian,
\beq
\tilde{g}(\lambda) = \exp(-|\lambda|).\nn
\eeq
This reduction function is called the Dingle factor, after its discoverer \cite{Dingle}. Shoenberg expresses the function as \cite{shoenberg}
\beq
D(X) = \exp(-C m^* T_D |X|),
\label{eq:D}
\eeq
where $C$ is the same constant as in the LK function, and the Dingle temperature $T_D$, assuming all the missing quantities, is expressed as a temperature but relates to disorder and impurities:
\beq
T_D = {\hbar \Gamma \over k_B}\nn
\eeq
A subtlety arises from the fact that the scattering rate does not depend exclusively on the mean free path $\ell$, but also on the size of the Fermi surface and the quasiparticle mass:
\beq
\Gamma = {\bar{v_F} \over \ell} = {\hbar \bar{k_F} \over m^*\ell},\nn
\eeq
where $\bar{v_F}$ and $\bar{k_F}$ correspond to the average Fermi velocity and $k_F$ vector along the cyclotron orbit. The Fermi wave vector in this equation is an average relating to one cyclotron orbit. The Dingle factor can be rewritten in the following ways:
 \beq
 D(X) = \exp\bigg(-{2\pi p\over \ell} \sqrt{{2\hbar F\over e}} X \bigg) = \exp(-{\hbar p\over 2e} {C_F \over \ell} X) = \exp(-{\hbar p\over \omega_c \tau}),
 \label{eq:D2}
 \eeq
 where $F$ is the dHvA frequency for one cyclotron orbit and $C_F$ the circumference of that orbit in $k$ space. The damping is consequently stronger for larger FS cross sections but does not depend on the quasiparticle mass.
 
\subsection{2D systems and the angular dependence of dHvA\label{sect:BergemanAnalysis}}
 
Quasi two dimensional metallic systems have very special characteristics in terms of dHvA. When metals possess a very anisotropic resistivity, the Fermi surface usually connects with itself through the Brillouin zone in the direction defined as the $k_z$ direction, as it is the case in \TOF\ and \TTS. For \TOF, the Fermi surface has been determined with extremely high accuracy and a good review about how it was done was written by C. Bergemann $et$ $al.$\cite{bergemann}. We reproduce here the points that are relevant to this thesis, and the main calculation, which is rather arduous, is presented in detail in appendix~\ref{App:B}. The main characteristic that we observe is that when a material is close to being perfectly two dimensional, the Fermi surface takes the shape of one or many cylinders, oriented in the $k_z$ direction and connecting with the top and bottom of the BZ, with an arbitrary shaped base in the $k_x,k_y$ plane (not necessarily a circle), and only small deviations along $k_z$. This situation is generally accompanied by a real space lattice $c$ parameter that is much larger than the other two, $a$ and $b$, such that the overlap of the orbitals in that direction is small, resulting in a much lower conductivity in the $c$ direction. In such a case, the Brillouin zone is shorter in the $k_z$ direction than in the other two. 

When one performs a dHvA experiment with the magnetic field aligned with the $c$-axis, the frequencies obtained correspond to the area of the base of each cylinder, and they are generally easy to identify to, for instance, $k$ space areas calculated from angle resolved photoemission spectroscopy (ARPES) data. Moreover, if one tries to perform dHvA with the magnetic field aligned in the basal $k_x,k_y$ plane, no oscillations are detected, since no closed Landau orbits exist in that orientation. dHvA frequencies are at a minimum in the $c$-axis direction and, as one rotates the magnetic field from the $c$-axis towards the plane, they gradually increase to infinity, and their amplitude decreases to zero. In quantitative terms, as one rotates the magnetic field the cross-sectional shape becomes stretched in one direction, by a scaling factor which corresponds to the inverse of a cosine:
\beq
F = {F_0 \over \cos(\theta)},
\label{eq:cosinelaw}
\eeq
where $F_0$ is the dHvA frequency in the $c$-axis direction.

Since metallic materials are never truly two-dimensional, and the various Fermi cylinders have to connect to the top of the Brillouin zone at an angle of 90$^{\circ}$ for perdiodicity, it follows that they must possess at least two extremal areas in the $k_z$ direction. If the area difference between those is small compared to the spacing between the Landau levels at the magnetic field that is used, then the extremal area treatment of dHvA is not appropriate, and one has to use a slightly more elaborate approach. Moreover, the oscillations produced possess very special beat patterns, which evolve with field angle. We will now show that these patterns can be identified in order to work back and deduce the deviations of the true FS from simple cylinders. 

When the extremal orbit approximation is not applicable, one must calculate the following relation in order to obtain the fundamental oscillatory part in the magnetisation, for one Fermi surface pocket:
\beq
M \propto \int_{-{\pi\over c}}^{{\pi\over c}} dk_z \sin\bigg({2 \pi F(k_z) \over B}\bigg) \propto  \int_{-\pi}^{\pi} d\kappa_z \sin\bigg({\hbar \over e} A(\kappa_z)X \bigg),
\label{eq:IntOsc}
\eeq
where $X$ is the inverse of the field, $\kappa_z$ corresponds to $k_z \times 2\pi / h_{bz}$, $h_{bz}$ being the height of the Brillouin zone $4\pi/c$, $c$ the lattice parameter, and $A(\kappa_z)$ is the cross-sectional area (not necessarily an extremal area) of the Fermi pocket at $\kappa_z$. For the complete dHvA signal, with the harmonics and the whole Fermi surface, one needs to sum one such term per harmonic, using $p$ times the area $A(\kappa_z)$, and one per Fermi surface piece, using an appropriate area $A_n(\kappa_z)$. This equation produces the beat pattern that is the result of the deviations from a cylinder in the $k_z$ direction, roughly the interference pattern from the difference between the extremal areas. When we rotate the magnetic field away from $c$-axis, these areas change, and so does the beating period. This produces a pattern in the field-angle plane that is unique to the corrugation of the Fermi surface pocket. It can thus be used to identify accurately those deviations. Consequently, what is required is to calculate this area as a function of $\kappa_z$ and field angle $\theta$, and putting it into the integral, one should obtain the full dHvA signal as a function of $X$ and $\theta$.

\begin{figure}[t]
	\begin{minipage}[t]{7cm}
		\begin{center}
		\includegraphics[width=7cm]{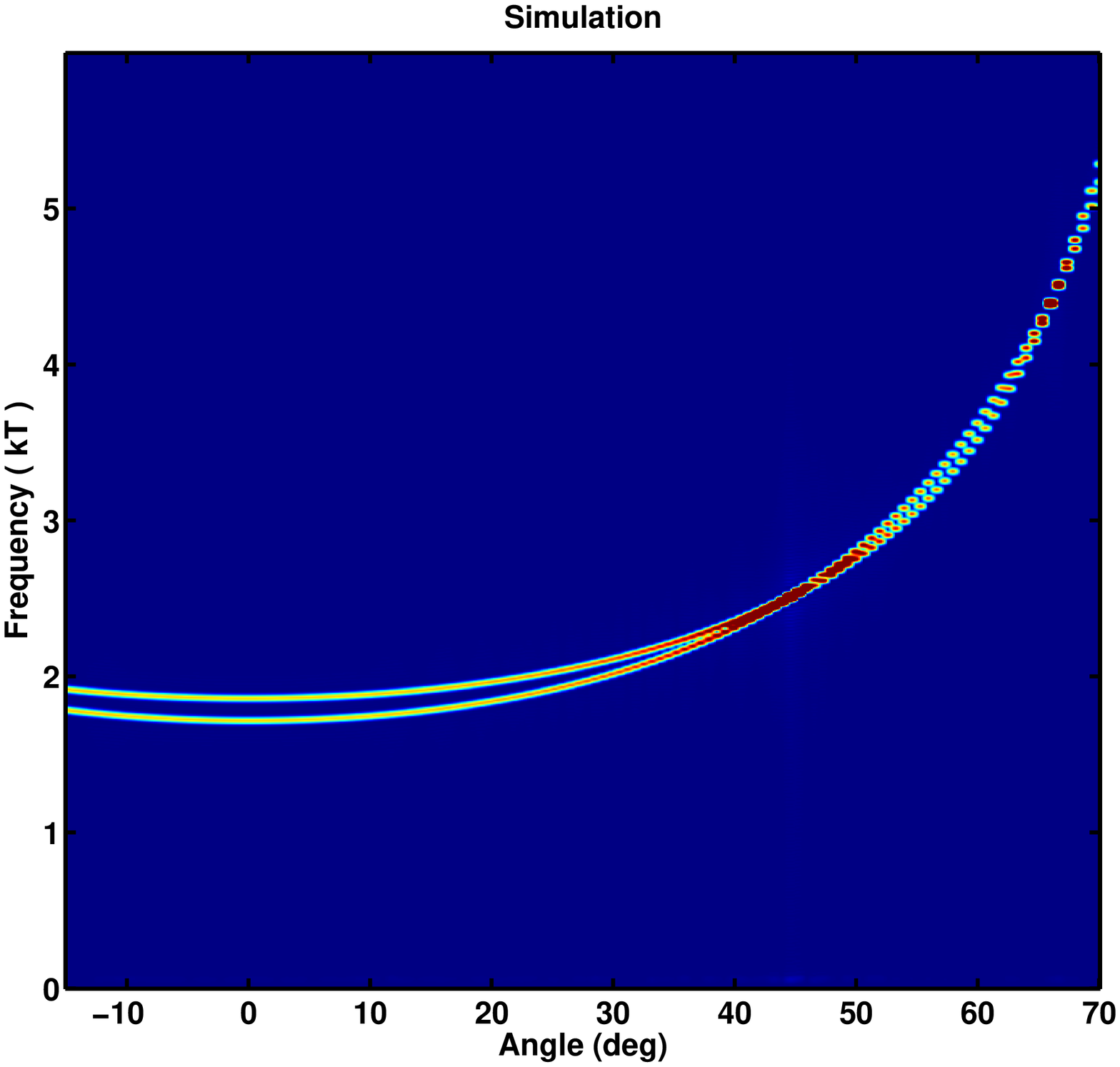}
		\end{center}
	\end{minipage}
	\hfill
	\begin{minipage}[t]{7cm}
		\begin{center}
		\includegraphics[width=7cm]{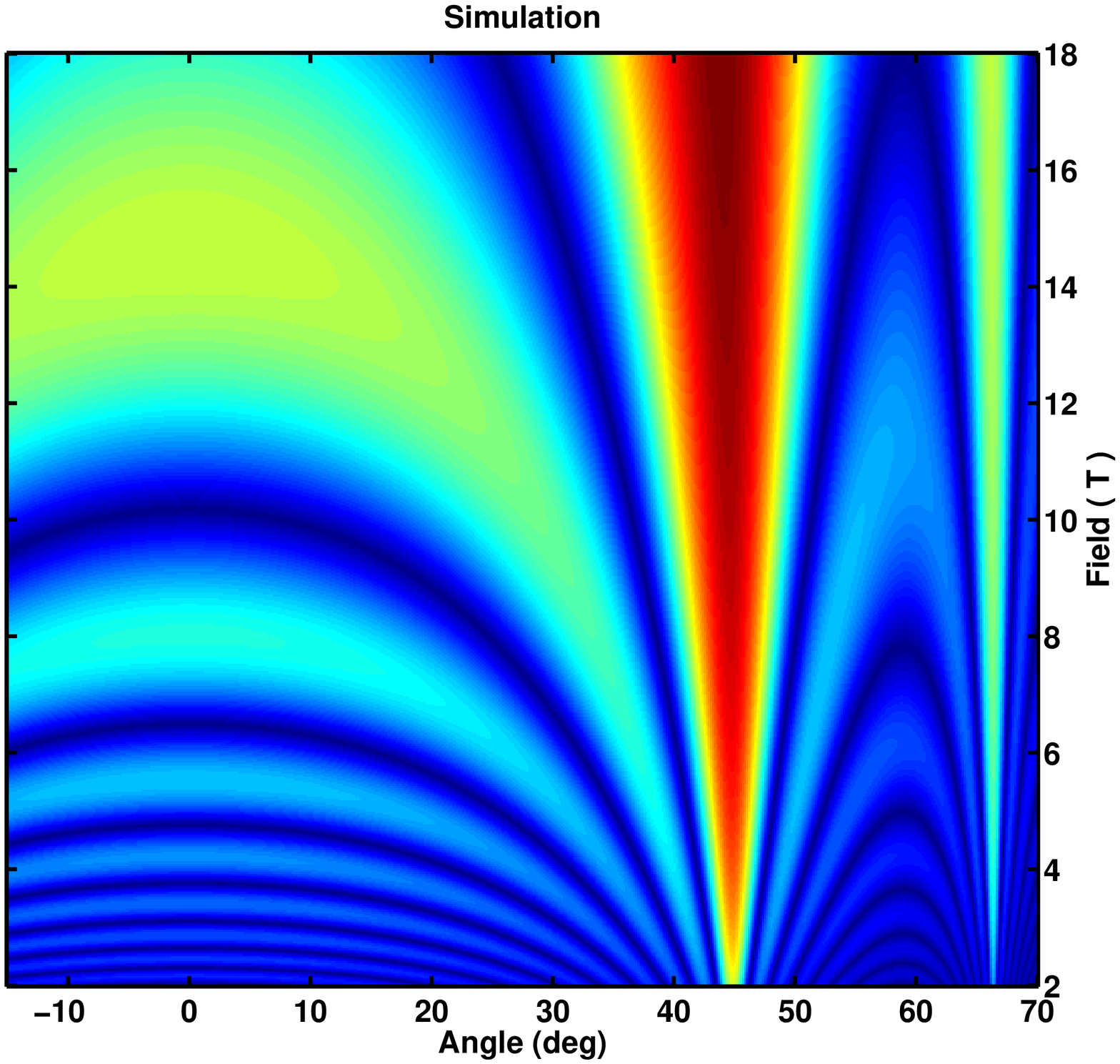}
		\end{center}
	\end{minipage}
	\caption[Angular dependence of dHvA in a 2D material.]{$Left$ Angular dependence of a dHvA frequency produced by a simple corrugation of type $k_{01}$. For both figures, red corresponds to high values, green to intermediate and dark blue to low values. $Right$ Field-angle beat pattern produced by a simple corrugation of type $k_{01}$. }
	\label{fig: QuarterPercentSim}
\end{figure}

The deviation of $A(\kappa_z)$ from a cylinder is arbitrary, but it is a periodic function of $\kappa_z$, and one can expand it in a Fourier series. Moreover, since the in-plane shape of the Fermi surface is not known, it can be expanded in polar harmonics. Since this in-plane shape has a specific orientation in $k$ space, one more parameter is required, which we express as an in-plane angle $\phi_0$. We then write this cylindrical expansion with subscripts, $\mu$ for the polar expansion and $\nu$ for the $\kappa_z$ Fourier series :
\beq
k_F(\kappa_z, \phi, \phi_0) = k_{00} + \sum^{\infty}_{\mu = 1, \nu = 1} k_{\mu \nu} \cos \nu \kappa_z \cos (\mu\phi + \phi_0),
\label{eq:kfexpansion}
\eeq
where the parts sinusoidal in $\nu \kappa_z$ were dropped for reasons of symmetry\footnote{In the Brillouin zone, electrons at $\vec{\bf{k}}$ and $-\vec{\bf{k}}$ possess the same energy.}. One is required to calculate  the area difference  between the real area and that averaged along $k_z$, $\Delta A_{\mu \nu}$, as a function of $\kappa_z$ and $\theta$. This is a difficult calculation that may be of interest to some readers, and it is presented in appendix~\ref{App:B}. The result is
\bea
\Delta A_{\mu \nu} &=& {2 \pi k_{00}k_{\mu \nu} J_\mu(\nu \kappa_F \tan \theta) \over \cos \theta} \cos \mu \phi_0 \times(-1)^{\mu \over 2} \cos \nu \kappa_z, \quad even \quad \mu,\nn\\
 &=& {2 \pi k_{00}k_{\mu \nu} J_\mu(\nu \kappa_F \tan \theta) \over \cos \theta} \cos \mu \phi_0 \times (-1)^{(\mu - 1) \over 2} \sin \nu \kappa_z, \quad odd \quad \mu,\nn\\
 \label{eq:Areamunu}
\eea
where $J_\mu$ is the $\mu$th Bessel function of the first kind. For a Fermi surface pocket that possesses such a complex structure that more than one warping parameter $k_{\mu \nu}$ are significant, then one must sum all the $\Delta A_{\mu \nu}$ and put them into the integral of  eq. \ref{eq:IntOsc} in order to calculate the magnetisation. This cannot be performed analytically but is done with a computer. Figure \ref{fig: QuarterPercentSim}, right, shows an example of a pattern produced by a simple warping parameter $k_{01}$ which has 0.25\% of the size of the in-plane Fermi wave vector $k_F$. 

Crystal symmetry restricts the possible values of the warping parameters. Contributions with $\mu = 0$ are allowed in all 2D crystal symmetries, since they are isotropic in the plane. But parameters with $\mu > 0$ must respect the in-plane symmetries since they provide the Fermi surface its in-plane shape, where for example a square involves parameters with $\mu = 4$. Consequently, in \TTS, for electron Fermi pockets in the centre of the zone, odd parameters or parameters that do not respect twofold rotation symmetry are forbidden. Effectively, $\mu = 1$ produces a surface that has no symmetry at all, and since the Brillouin zone must possess inversion symmetry, $E(\vec{\bf{k}}) = E(-\vec{\bf{k}})$, it follows that such a surface cannot be in the centre of the BZ. \footnote{If such a surface is not at the centre, it should appear at least twice, with inversion symmetry through the centre. Then, with dHvA and rotation, one should measure both at the same time, but with a difference in $\phi_0$ of 180$^{\circ}$, and due to the factor of $\cos\mu\phi_0$, it should cancel exactly, and cannot appear in the data.} Warping parameters of $\mu = 3, 6, 9 ...$ may only appear in hexagonal systems, hence not in any of the ruthenates, and higher odd numbers cannot be accommodated in any periodic system. Finally, the terms with $\nu = 0, \mu > 0$ are invisible to dHvA, where effectively, the in-plane shape cannot be determined. For \TTS, we will discuss only parameters with $\nu > 0, \mu = 0, 2, 4...$. 

Furthermore, while the warping situations where $\mu = 0$ are isotropic, for all warping within the class with $\nu > 0, \mu > 0$, some directions of rotation will produce no beat patterns at all, due to the factor $\cos\mu\phi_0$. For instance, with $k_{21}$, depending on the orientation of the Fermi surface pocket, one of the situations,  $\phi_0 = 45^{\circ}$ or $\phi_0 = 0^{\circ}$, should produce an area $\Delta A_{\mu \nu}$ that vanishes. This property may be used for the identification of such warping\footnote{For example, in the case of \TOF, the $\alpha$ pocket possesses a term in $k_{21}$, which has the symmetry of a screw, or as C. Bergemann calls it, the shape of a ``snake that swallowed a chain" \cite{bergemann}.}. Any difference in patterns between directions of rotation should be attributed to such parameters.

If only one warping parameter is relevant, as it is sometimes the case, then one can go further and calculate the integral of eq. \ref{eq:IntOsc}, with single warping $k_{\mu \nu}$, which is given in appendix~\ref{App:B}, and the result is 
\bea
M &\propto& \sin \bigg({\hbar X\over e}{ \pi k_{00}^2 \over \cos \theta}\bigg) {1\over \nu} J_0 \bigg({2\pi \hbar X\over e}{k_{00} k_{\mu \nu} \over \cos \theta} J_\mu(\nu k_{00} \tan \theta)\bigg)\nn\\
&=& \sin\bigg({2 \pi F_0 X \over \cos \theta} \bigg) {1\over \nu} J_0\bigg( {2 \pi \Delta F_{\mu \nu} X \over \cos \theta}J_\mu(\nu k_{00} \tan \theta)\bigg),
\label{eq:simplewarping}
\eea
where $F_0$ is the average dHvA frequency, and $\Delta F_{\mu \nu}$ the difference. The modulation of the dHvA oscillation is not a cosine but a Bessel function. When one uses small fields, though (when the argument of $J_0$ is larger than about 4), the modulation becomes very close to a cosine function of the form 
\beq
\sqrt{2\over \pi X}\cos\bigg({2\pi \Delta F_{\mu \nu} X \over \cos \theta} J_\mu(\nu k_{00} \tan \theta) - {\pi \over4}\bigg).
\label{eq:BesselCosApprox}
\eeq
There exists an angle at which the argument of the modulation factor becomes zero for all values of $X$, which arises when the angle $\theta$ is such that the function $J_\mu(\nu k_{00} \tan \theta)$ vanishes. In that case, there is no modulation and the oscillatory signal is maximal. This is called the Yamaji angle, and for one specific type of warping, it is only function of $k_{00}$, the average in-plane $k_F$. Figure \ref{fig: QuarterPercentSim}, right, shows an example of a beat pattern produced by a simple corrugation of type $k_{01}$ of 0.25\% of $k_F = 2.33 \times 10^9 m^{-1}$, corresponding to a frequency of 1.8~kT, in a Brillouin zone of height 606$\times 10^7 m^{-1}$, identical to that of \TTS. One can see the first Yamaji angle at 45$^{\circ}$, and the second one at 65$^{\circ}$.

In dHvA frequency space, peaks evolve with angle with the form
\beq
F(\theta) = {F_0 \over \cos \theta} \pm {\Delta F_{\mu \nu}\over \cos \theta} J_\mu(\nu k_{00} \tan \theta),\nn
\eeq
where, since $\Delta F_{\mu \nu}$ is small, one observes two peaks that oscillate around $F_0/\cos \theta$. At the Yamaji angles, the second term becomes zero and both peaks cross, the amplitude becoming large. Figure \ref{fig: QuarterPercentSim}, left, shows a simulated example of this phenomenon, where dHvA data was generated in the same way as for the beat pattern shown on the right, but using a larger warping of 2\%, for clarity of presentation.

Finally, we show one more calculation that will be used later, which aims to find curves of $B(\theta)$ where the amplitude of the modulation is either maximum or zero. Using the cosine approximation for the Bessel function of eq. \ref{eq:BesselCosApprox}, the condition for the modulation to vanish is 
\beq
{2\pi \Delta F_{\mu \nu} X \over \cos \theta} J_\mu(\nu \kappa_{00} \tan \theta) - {\pi \over 4} = {\pm n \pi \over 2}, \quad n = 1,2,3...\nn
\eeq
which translates as 
\beq
B(\theta) = {8 \Delta F_{\mu \nu} \over \cos \theta} {J_\mu(\nu \kappa_{00} \tan \theta) \over 1 \pm 4n},
\label{eq:BeatZeros}
\eeq
where the $\pm$ corresponds to positive for maxima, and negative for minima. Using $n = 1$, one obtains the first zero near the Yamaji angle, and with larger $n$, all the other zeros going to lower and lower field values as they intercept the $\theta = 0$ axis. Note that in the case of a single warping parameter, beat pattern nodes cannot cross one another.

\subsection{Spin splitting \label{sect:spinsplitting}}

\begin{figure}[t]
	\begin{minipage}[t]{7cm}
		\begin{center}
		\includegraphics[width=7cm]{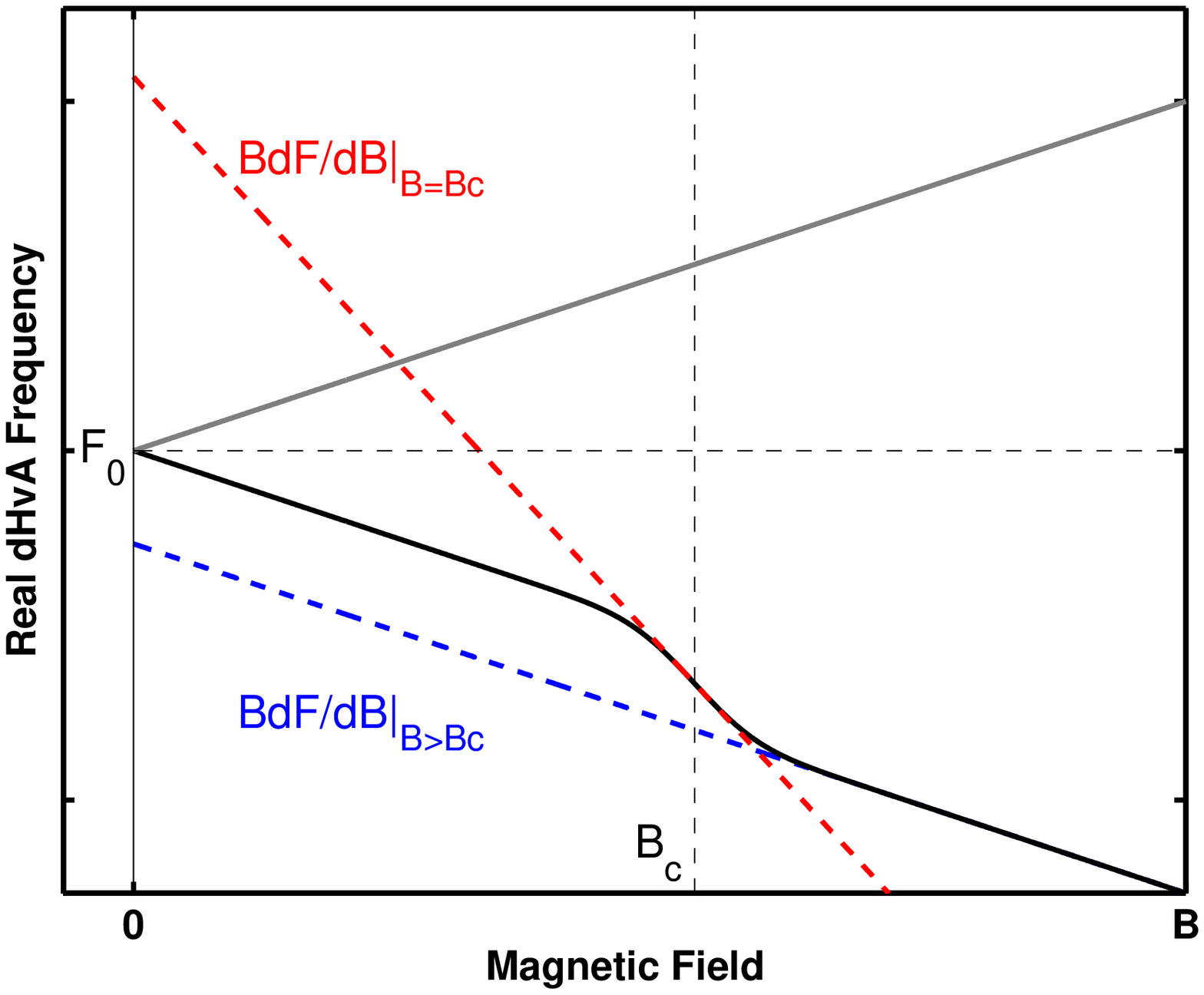}
		\end{center}
	\end{minipage}
	\hfill
	\begin{minipage}[t]{7cm}
		\begin{center}
		\includegraphics[width=7cm]{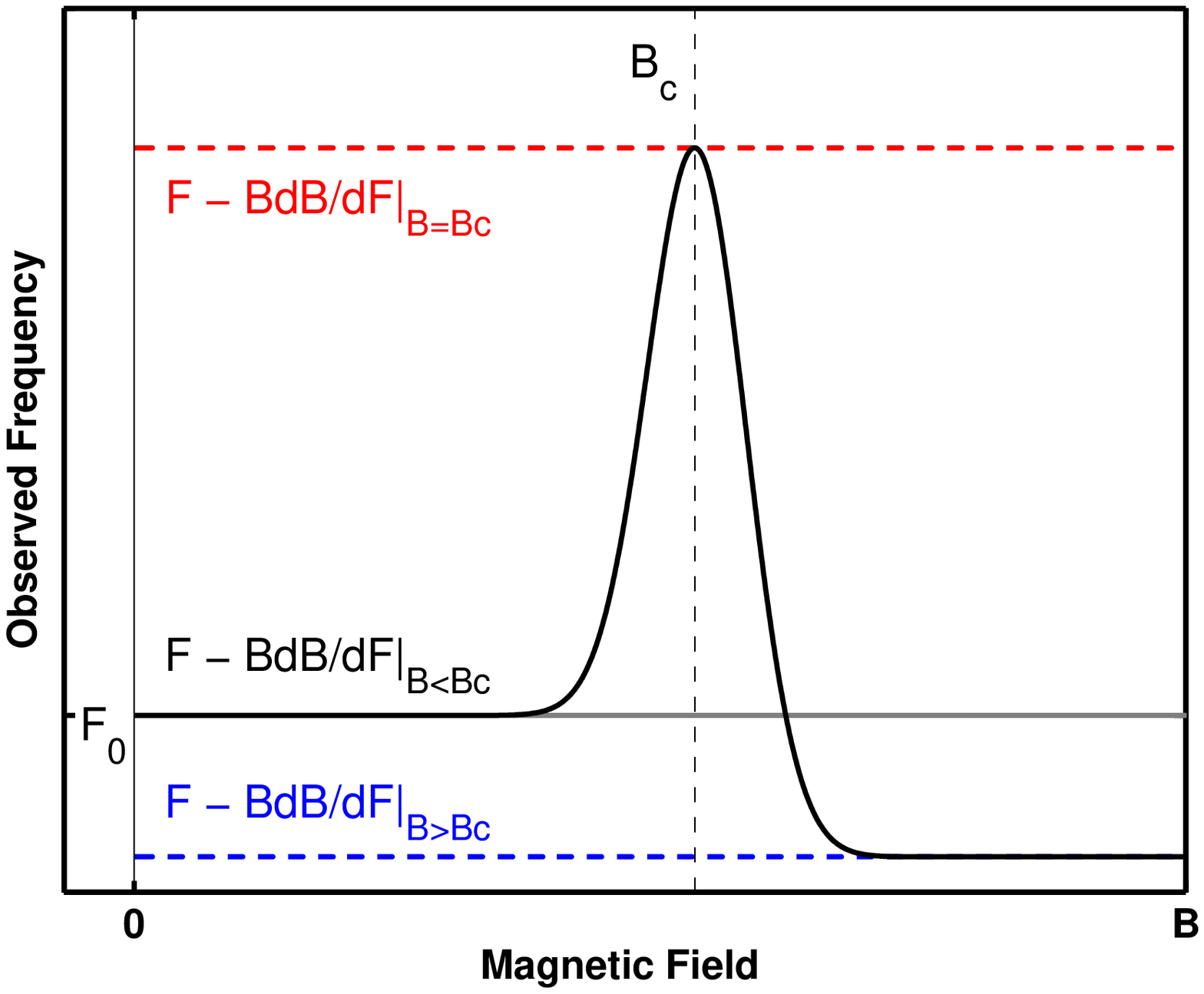}
		\end{center}
	\end{minipage}
	\caption[Illustration of spin splitting and back projection]{$Left$ Real dHvA frequency, proportional to the respective sizes of the Zeeman split FS. $Right$ Observed dHvA frequency, after the back projection.}
	\label{fig: SpinSplitting3}
\end{figure}

The previous sections have neglected the spin of the quasiparticle and Zeeman splitting of the FS leads to an additional amplitude reduction factor that requires attention. As one ramps the magnetic field, the FS splits into two, one for each spin flavour, one growing and the other shrinking in $\bf{k}$ space. In normal spin splitting \footnote{For details of anomalous spin splitting, see the review by Bergemann \cite{bergemann} pp. 658.}, when a paramagnetic metal possesses a constant magnetic susceptibility, the value of the average Fermi wave vector changes linearly in field,
\beq
k_F^{\uparrow \downarrow} = k_F \pm \chi_1 B,\nn
\eeq
where $\chi_1$, the rate at which $k_F$ changes with field, is proportional to the spin susceptibility. $\bf{k}$ space cross sectional areas increase with B, and so do the dHvA frequencies. In 2D systems, such areas are also be rescaled by the cosine of the angle $\theta$ between the sample and the field. Neglecting the second order term, for one cyclotron orbit, we obtain
\beq
\label{eq:Fupdown}
F^{\uparrow \downarrow}(B, \theta) = {F\over \cos \theta} \pm {\hbar \over e} {k_F \chi_1 \over \cos\theta}B = {F\over \cos\theta} \pm {g\over 2 \cos\theta}{m^*\over m_e}B.
\eeq
When put into the oscillatory term of dHvA, we obtain an additional phase $\phi =  {\pi g\over \cos\theta} {m^*\over m_e}$,
\beq
M^{\uparrow \downarrow}(B) \propto \sum_\pm \cos\bigg( {2\pi F \over B} \pm \phi \bigg) = 2\cos\bigg( {2\pi F \over B} \bigg)\cos \phi.
\label{eq:spinzero}
\eeq
This has no effect on the dHvA frequency spectrum, since only the phase changes. However, it affects the amplitude if the term $\cos \phi$ vanishes, which in 2D systems is called a spin zero, and the condition for that happening is
\beq
\phi = {\pi g\over \cos\theta} {m^*\over m_e} = {\pi \over 2} \quad \Rightarrow \cos\theta = 2g {m^*\over m_e},\nn
\eeq
One can thus find in the data a set of angles where field independent zeros appear, and obtain the $g$ factor in this way\footnote{This phase of the spin zero, proportional to $g m^*$, refers to non-interacting systems. In Fermi liquid systems, the spin susceptibility is enhanced by interactions, and one should use an effective $g^*$ factor, which is equal to ${g\over 1+ F_0^a}$, $F_0^a$ being one of the Landau Fermi liquid parameters. The phase then becomes $\phi =  {\pi g\over \cos\theta} {m^*/m_e\over 1+F_0^a}$, where $m^*$ remains the unrenormalised (band) effective mass.}. This additional amplitude modulation corresponds to interference of dHvA oscillations between the spin species.  

It is not the same situation when, in a certain field region, $B_c$, a metal undergoes a superlinear rise in magnetisation, as in a metamagnetic transition. In this case, $F^{\uparrow \downarrow}$ will increase faster than linearly\footnote{Where $\chi_2$ is much larger than the prefactor to the second order term that was dropped in eq. \ref{eq:Fupdown}.},
\beq
F^{\uparrow \downarrow}(\theta,B) = {F\over \cos \theta} \pm {\hbar \over e} {k_F\over \cos\theta}( \chi_1 B + \chi_2 B^2),\nn
\eeq
and one is left with terms periodic in $B$ (or higher orders of B),
\beq
\cos\bigg( {2\pi F \over B}\bigg)\cos\bigg(\phi+ 2\pi {\hbar \over e} {k_F \chi_2\over \cos\theta} B \bigg).\nn
\eeq
In such a case, the observed momentary frequency, measured from various Fourier transforms taken as a function of $1/B$, corresponds to the real frequency minus its linear part \cite{Ruitenbeek, JulianUPt3},
\beq
F_{obs} = F - B{dF\over dB}.
\label{eq:backprojection}
\eeq
For instance, if in a system, one of the spin FS sheets evolved with field faster than linearly, one would observe the behaviour described in figure \ref{fig: SpinSplitting3}, where the real frequency is depicted on the left and what is observed on the right. Below the critical field $B_c$, where simple paramagnetism is present, no difference in frequency is seen between both spins, but near $B_c$, one might detect splitting of the peak in the spectrum, which becomes very large at $B_c$ (shown in red), and decreases again at high fields (shown in blue), but never disappears. This means that in metamagnetic systems like CeRu$_2$Si$_2$ \cite{Aoki}, UPt$_3$ \cite{JulianUPt3}, \TTS\ \cite{borzi} and others, splitting is expected on the high field of the superlinear increase in magnetisation.

\begin{figure}[t]
	\begin{minipage}[t]{7cm}
		\begin{center}
		\includegraphics[width=7cm]{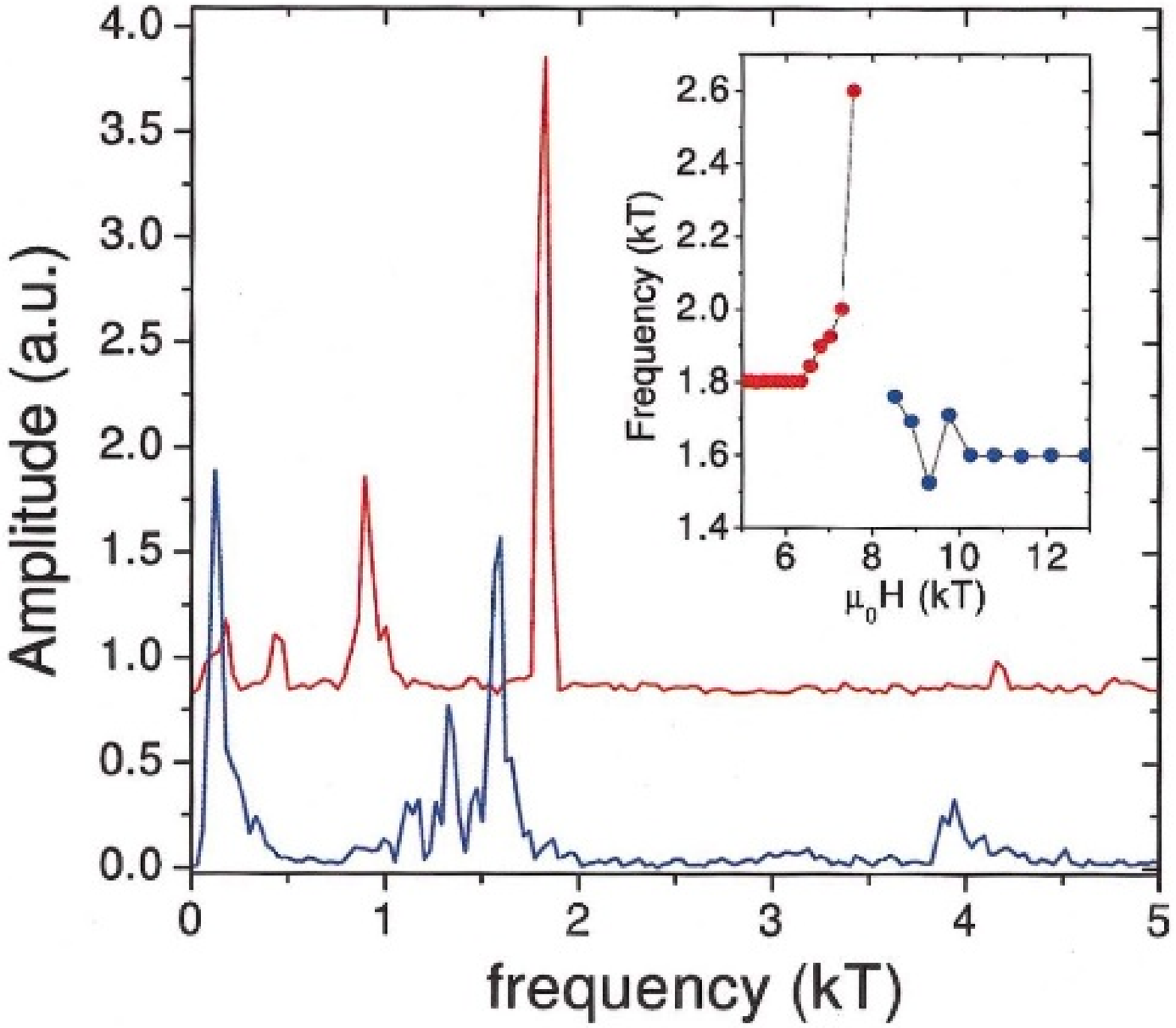}
		\end{center}
	\end{minipage}
	\hfill
	\begin{minipage}[t]{7cm}
		\begin{center}
		\includegraphics[width=7cm]{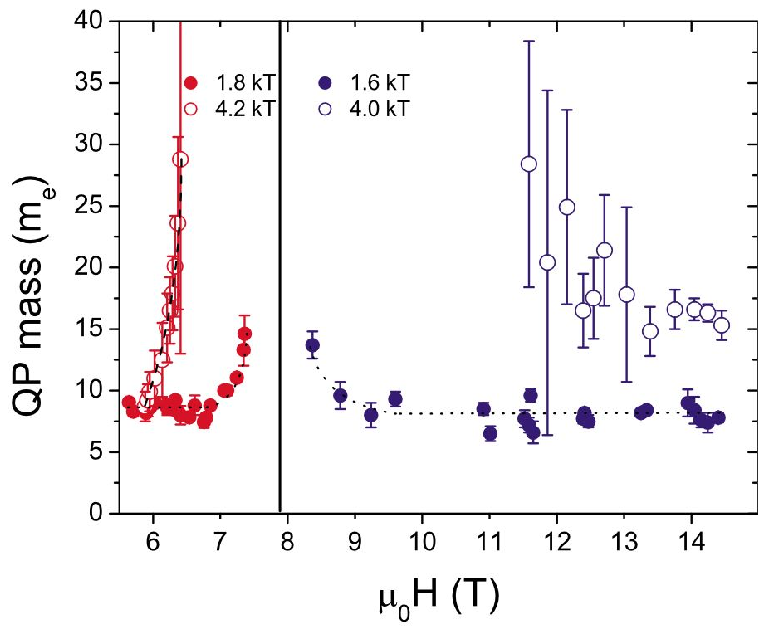}
		\end{center}
	\end{minipage}
	\caption[dHvA measurements by Borzi $et$ $al.$]{$Left$ dHvA spectra from the measurements by Borzi and co-workers \cite{borzi}. The red trace refers to the field range 5.5 to 6.5~T, and the blue one to 10 to 15~T. $Right$ Mass divergence of the 1.8 and 4.2~kT peaks, as reported by Borzi.}
	\label{fig: BorziSpectra}
\end{figure}

\subsection{First dHvA measurements in \TTS \label{sect:BorzidHvA}}

dHvA was measured in \TTS\ previously by Borzi and co-workers \cite{borzi} and the main properties of the FS have been explored. Quantum oscillations were obtained with the magnetic field parallel to the $c$-axis and five quasiparticle orbits were reported in the low field side of the metamagnetic transition, and four in the high field side. Figure \ref{fig: BorziSpectra}, left, shows both low and high field side spectra calculated from data in the ranges 5.5 to 6.5~T for the low field side, and 10 to 15~T for the high field side. Table \ref{tab:PRLBorzi} presents the value of the observed frequencies. Peak splitting was observed in the high field side, as one expects in a system with non-linear susceptibility (see section \ref{sect:spinsplitting}), but the picture that emerged was one that is similar in both field sides, with peaks located at similar frequency values, which possessed similar mass values\footnote{The mass values for the peaks of the high field side were not specifically quoted in the text.}. Moreover, closer to the metamagnetic transition on both field sides, some of the frequency peaks were reported to increase sharply, shown in the inset to figure \ref{fig: BorziSpectra}, left. The frequencies had no field dependence in the field ranges that were quoted as used for the FFTs of the plotted spectra.

Moreover, away from the metamagnetic transition, quasiparticle masses were calculated from the temperature dependence of the quantum oscillations, in the same field ranges used for the plotted spectra. Those were fairly high, in the range of 7 to 12$m_e$, and are presented in table \ref{tab:PRLBorzi}. Borzi also observed that they have a field dependence, and increase as one moves towards the QCEP from both field sides. Figure \ref{fig: BorziSpectra}, right, shows the field dependence of the mass of two of the frequencies, where the vertical line represents the metamagnetic transition, and masses close to 30$m_e$ were observed. This observation has attracted a fair amount of attention. It is an important measurement for the quantum critical picture in \TTS.

\begin{table}[t]
	\begin{center}
		\begin{tabular}[h]{|c|c|c|c|}
			\hline
			\multicolumn{2}{|c|}{5.5 to 6.5 T} & \multicolumn{2}{|c|}{10 to 15 T}\\
			\hline
			Peak (kT) & $m^*/m_e$ & Peak (kT) & $m^*/m_e$\\
			\hline
			0.2 & $6 \pm 3$ & $< 0.4$ & $-$\\
			0.5 & $13 \pm 4$ & &\\
			0.8 & $10 \pm 1$ & 1.1 - 1.5 & $-$\\
			1.8 & $8 \pm 1$ & 1.6 & $-$\\
			4.2 & $9\pm2$ & 4.0 & $-$\\
			\hline
		\end{tabular}
		\caption[Frequencies and masses observed by Borzi $et$ $al.$]{Table of the frequencies observed by Borzi, below and above the metamagnetic transition, with associated quasiparticle masses.}
		\label{tab:PRLBorzi}
	\end{center}
\end{table}

\chapter{Experimental methods and analysis procedures \label{chapter:experimental}}
\markright{Chapter~\ref{chapter:experimental}: Experimental methods and analysis procedures}

This chapter aims at describing in detail the experimental techniques and numerical methods that were used during this research project. These can be divided into two categories; those that relate to the de Haas van Alphen experiment and those that are part of the procedure for sample characterisation and sorting of single crystals of \TTS. 

Firstly, the dHvA experiment is a standard one, but some of the fine details are often overlooked. Thus, the author found it pertinent to present the relevant parts with a lot of detail. In particular, some aspects of the data analysis procedures are somewhat obscure in the standard textbooks, but also, special steps can be taken in the numerical methods to tackle our specific aims. We will emphasise particularly practices that may lead to systematic errors. 

Secondly, the most important part of this chapter may be the description of the main experiment, but it could not have been carried out successfully on average \TTS\ samples. The purity requirements of dHvA are extremely high, and they had not yet been achieved by material scientists in the early stages of this project. Indeed, the highest purity was reached by R. S. Perry in a collaboration with the author over several months, the information that was provided by the characterisation and sorting procedure of section \ref{sect:search} being used to optimise crystal growth conditions. Moreover, the best qualities were only reached in some spatial parts of single crystals, which required to be isolated. In consequence, an extensive string of measurements were applied to each sample cut individually in order to search for the best ones.  

This chapter first presents the dHvA experiment using the field modulation method. Next will be given the complete description of the various data analysis procedures and the rigorous checks that were made to ensure proper thermal equilibrium of the system. The second category of techniques will be introduced at this point, first describing the AC susceptibility probe that was built for the adiabatic demagnetisation refrigerator. The complete procedure for characterisation and sorting of samples will follow with the other already established techniques, along with data relating to the purest samples.

\section{AC susceptibility in a $^3$He/$^4$He dilution refrigerator}

We introduce in this section the main method that has been used in this experimental project, AC susceptibility. This technique can be used for several purposes, and in respect to this fact we begin with a general presentation of its operation through equations. We were, however, interested in measuring quantum oscillations, and carry on with a description of the case where the magnetisation is an oscillatory function of the magnetic field. Finally, we provide details of the specific experimental system that was used to obtain the data at the core of this thesis, along with quantities related to signal and noise levels.  

\subsection{Basics of AC susceptibility with equations}

Magnetic measurements using AC susceptibility is an established technique. The principle is to induce oscillations in the magnetisation of a sample and to measure the change in the magnetic flux using a coil. In order to improve sensitivity, one usually connects another empty identical coil to the first in opposition, in order to make the measurement differential. When measuring susceptibility as a function of magnetic field, one uses a small oscillating field $H_{AC}$ on top of a large slowly varying field $H_{DC}$:
\bea
H(t) &=& H_{DC}(t) + H_{AC}(t),\nn\\
&=& \alpha t + H_{AC}sin(\omega t).\nn
\eea
The magnetic induction in the first coil is
\beq
{B(t)\over \mu_0} = H(t) + M(t),\nn
\eeq
and the voltage on its terminals is
\beq
V_1 = {d\Phi \over dt} = A {dB\over dt} = A\mu_0 ({dH\over dt} + {dM\over dt}).\nn
\eeq
The voltage on the second, empty coil is equal to the first term only, and the substraction will leave a differential voltage proportional to the differential magnetic susceptibility
\beq
\Delta V = A \mu_0 {dM\over dH}{dH\over dt} = A \mu_0 \dot{H} \chi.
\label{eq:deltaVchi}
\eeq
The magnetisation as a function of magnetic field is usually a non-linear function and the excitation field $H_{AC}$ will produce harmonics, which can be detected with a phase sensitive detection device. Expanding the magnetisation as a function of $H_{AC}$, one obtains
\beq
M(H) = M(H_{DC}) + H_{AC} \bigg({dM \over dH}\bigg)_{H_{DC}} + {H_{AC}^2 \over 2} \bigg({d^2M \over dH^2}\bigg)_{H_{DC}} +  {H_{AC}^3 \over 6} \bigg({d^3M \over dH^3}\bigg)_{H_{DC}} + ...\nn
\eeq
One requires its derivative as a function of time. Assuming that the time derivative of the DC component is small compared to the oscillating terms, one finds
\bea
{dM \over dt} &=& H_{AC} \omega \cos(\omega t) \bigg({dM \over dH}\bigg)_{H_{DC}} + {H_{AC}^2\over 2} \omega (1-\sin(2\omega t)) \bigg({d^2M \over dH^2}\bigg)_{H_{DC}} \nn\\
&&+ {H_{AC}^3 \over 8}\omega(\cos(3\omega t) + \cos(\omega t)) \bigg({d^3M \over dH^3}\bigg)_{H_{DC}} + ...
\label{eq:Mexpansion}
\eea
One may observe two facts in this equation: first, the $n$th term includes the $n$th harmonic and those are multiplied by the oscillating field to the power $n$. Second, the phase changes by 90$^{\circ}$ at each harmonic.

\subsection{Oscillatory signal in the field modulation method \label{sect:oscsignal}}

\begin{figure}[t]
	\begin{minipage}[t]{7cm}
		\begin{center}
		\includegraphics[width=7cm]{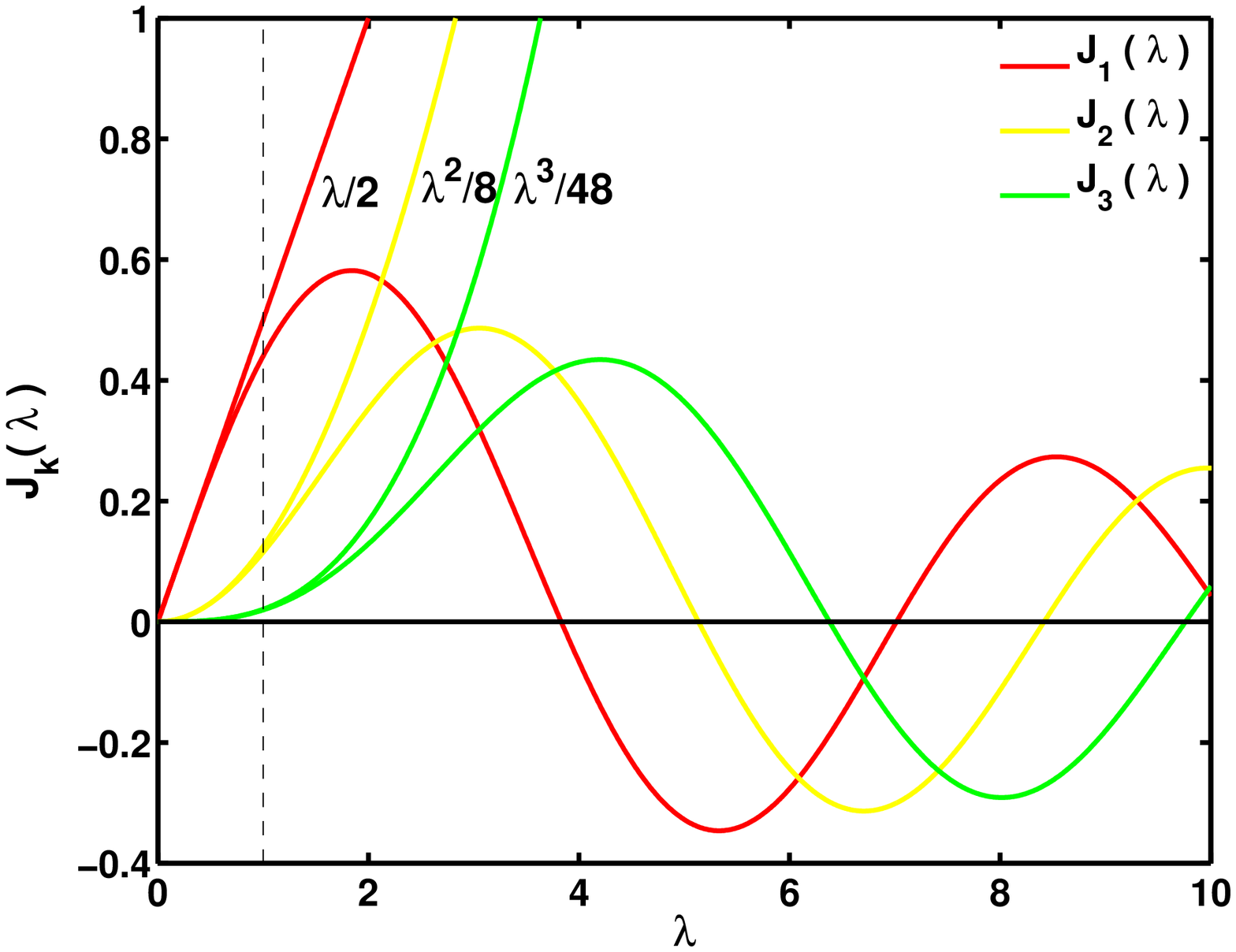}
		\end{center}
	\end{minipage}
	\hfill
	\begin{minipage}[t]{7cm}
		\begin{center}
		\includegraphics[width=7cm]{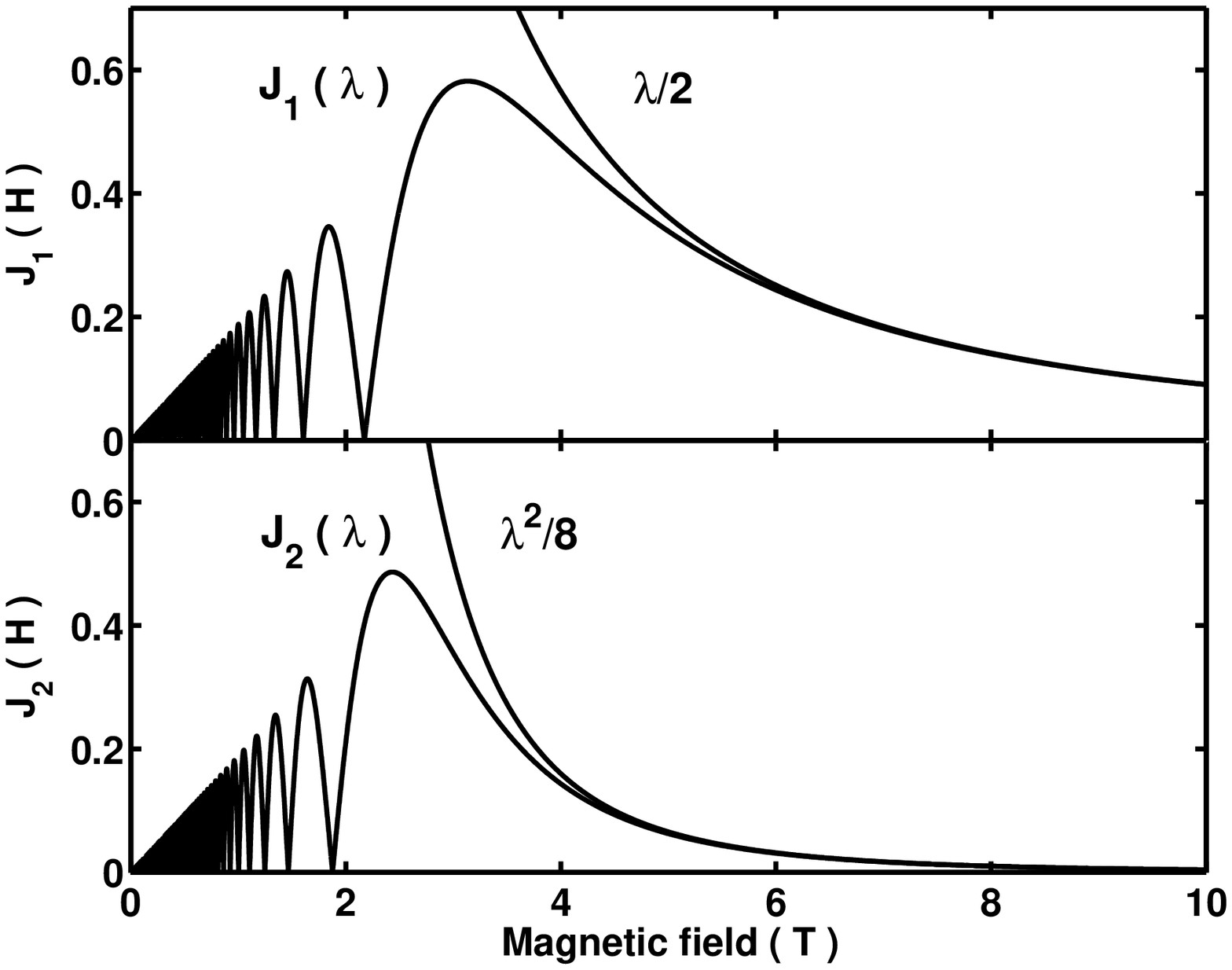}
		\end{center}
	\end{minipage}
	\caption[Bessel functions of the first kind]{($Left$) Bessel functions of the first kind. The case of $n=1$ is approximately linear at low values of $\lambda$, and the $n=2$ case is quadratic for small values of $\lambda$. (Right) Bessel functions of the first kind $J_k(\lambda)$ as a function of the DC magnetic field $H_{DC}$ for $k=1,2$, for a frequency $F = 1.8$~kT and an oscillating field value of $H_{AC} = 16$~G.}
	\label{fig:bessel}
\end{figure}

This section presents the calculation performed by Shoenberg (\cite{shoenberg}, pp 103-105), with emphasis on aspects that will be used later. During dHvA experiments, the magnetisation follows oscillations as a function of inverse magnetic field, of frequency $F$, which are assumed to be sinusoidal \footnote{When they are not, each oscillation harmonic, not to be confused with detection harmonic, can be treated independently.}:
\beq
M(H) = M_0(H) + A \sin\bigg({2\pi F \over \alpha t + H_{AC}\sin(\omega t)}\bigg),\nn
\eeq
where $M_0$ represents the magnetic background and is not oscillatory as a function of magnetic field.
This function is quite awkward to use, and a form that involves harmonics for phase sensitive detection is required. One uses the fact that the oscillating field amplitude is smaller than the magnitude of the DC field (usually by factor between 10$^3$ to 10$^5$), 
\beq
{1\over H_{DC}(t) + H_{AC}(t)} \simeq {1\over H_{DC}(t)} - {H_{AC}(t) \over H_{DC}^2(t)},\nn
\eeq
so that 
\bea
M(H) &\simeq& M_0(H) + A \sin\bigg({2\pi F \over H_{DC}}\bigg)\cos\big(\lambda \sin(\omega t)\big)\nn\\
&&-  A\cos\bigg({2\pi F \over H_{DC}}\bigg)\sin\big(\lambda \sin(\omega t)\big),\nn
\eea
with 
\beq
\lambda = {2\pi F H_{AC} \over H_{DC}^2}.
\label{eq:lambda}
\eeq
One then faces the complicated factors $\cos(\lambda \sin(\omega t))$ and $\sin(\lambda \sin(\omega t))$. These are periodic functions which can be expanded in Fourier series. This calculation is given in appendix~\ref{App:C} and the result involves Bessel functions of the first kind $J_k(\lambda)$, plotted in figure \ref{fig:bessel}:
\bea
M(H) &=& M_0(H) + A \sin\bigg({2\pi F \over H_{DC}}\bigg)\bigg( J_0(\lambda) + 2 \sum_{k=1}^{\infty}J_{2k}(\lambda) \cos(2k \omega t)\bigg)\nn\\
&&-  A\cos\bigg({2\pi F \over H_{DC}}\bigg)\bigg( 2 \sum_{k=1}^{\infty}J_{2k-1}(\lambda) \sin\big((2k-1) \omega t) \bigg).\nn
\eea
Each harmonic has as a prefactor, a Bessel function of order $k$ and the magnetic oscillations are in phase with $H_{AC}$ for the odd harmonics and ${\pi \over 2}$ out of phase for the even ones. 

In order to calculate the differential voltage, one requires the time derivative of $M(t)$. In real systems, $M_0$ usually features much slower field variations than the oscillations, and all higher derivatives are negligible, except, for instance, very close to a phase transition. Neglecting those, one obtains the harmonic expansion:
\bea
{dM \over dt} &=& {dM_0 \over dH} \big(\alpha + H_{AC}\omega \cos(\omega t)\big) +  2\omega A\sin \bigg({2\pi F \over H_{DC}}\bigg) \sum_{k=1}^{\infty} 2k J_{2k}(\lambda) \sin(2k\omega t)\nn\\
&&-  2\omega A\cos \bigg({2\pi F \over H_{DC}}\bigg)  \sum_{k=1}^{\infty} (2k-1)J_{2k-1}(\lambda) \cos\big((2k-1)\omega t\big).
\label{eq:dMdt}
\eea
One may conclude from this result that the slowly varying magnetic background is only present in the first harmonic, and it can be useful to use second or higher harmonic detection in order to remove it from the oscillatory signal.

It is an interesting fact that the dHvA amplitude is proportional to a function that oscillates with the AC field value and it is worth examining why. It corresponds to a smearing effect of the dHvA oscillations when the peak to peak value of the AC field is close to the DC field interval it takes for a specific dHvA frequency to complete a cycle. The condition for zero amplitude on this frequency is:
\beq
2 H_{AC} = \Delta H_{DC},\nn
\eeq
and the period corresponds to
\beq
{1 \over F} = {1\over H_2} - {1\over H_1} = {\Delta H_{DC} \over H_{DC}^2}.\nn
\eeq
The condition for zero amplitude is then
\beq
{2 H_{AC} F \over H_{DC}^2} = 1 =  {\lambda \over \pi}.\nn
\eeq
This is approximately true, since the bessel function $J_1(\lambda)$ has its first zero at $\lambda \simeq 3.81$. The reason is that the DC field continuously changes, and it takes a value for the oscillating field to be just a little over half the DC field interval. 

Finally, also shown in appendix~\ref{App:C}, when $\lambda < 1$, the time derivative of $M$ reduces to a simpler expression, since
\beq
J_k(\lambda) \simeq {\lambda^k \over 2^k k!}.
\label{eq:BesselApprox}
\eeq
indicating that each harmonic $k$ has an amplitude proportional to $H_{AC}^k / H_{DC}^{2k}$, a fact that will be used later. 

Figure \ref{fig:bessel} shows Bessel functions of the first kind $J_k(\lambda)$. From the left panel, one can observe that for low values of $\lambda$, the first harmonic has a much higher amplitude than the first, but above $\lambda \simeq 2.2$, the second harmonic possesses a higher intensity\footnote{Due to an additional factor of 4 in eq \ref{eq:dMdt} for $k=2$ compared to $k=1$.}. Thus, when the magnetic background is important, one may gain significantly in using a high modulation field and second harmonic, assuming that no eddy current heating is present (an effect that will be discussed in section \ref{sect:eddy}). The right panel of fig. \ref{fig:bessel} shows the behaviour of the $k=1,2$ bessel functions for fixed dHvA frequency and oscillating field, as a function of DC field. One can see that as $\lambda$ is function of the DC field, the system can pass through a zero of the bessel function. It is then important to set the oscillating field such that for all DC field and dHvA frequency values, the system does not cross a zero of the $n$th bessel function when using the $n$th harmonic.

\subsection{The dHvA apparatus \label{sect:Camprobe}}

\begin{figure}[p]
	\begin{minipage}[t]{9cm}
		\begin{center}
	\includegraphics[width=9cm]{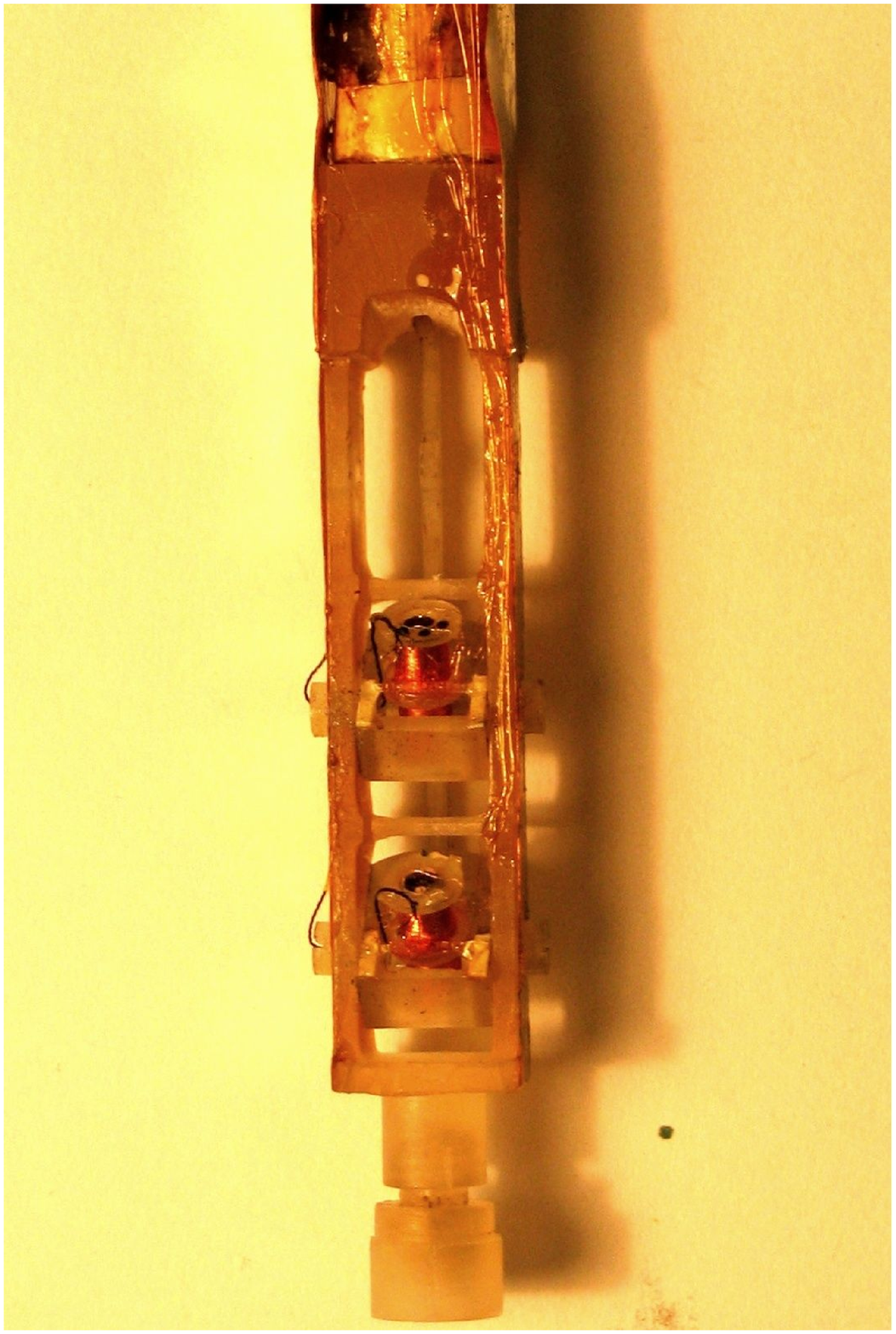}
		\end{center}
	\end{minipage}
	\hfill
	\begin{minipage}[t]{4.5cm}
		\begin{center}
		\includegraphics[width=4.5cm]{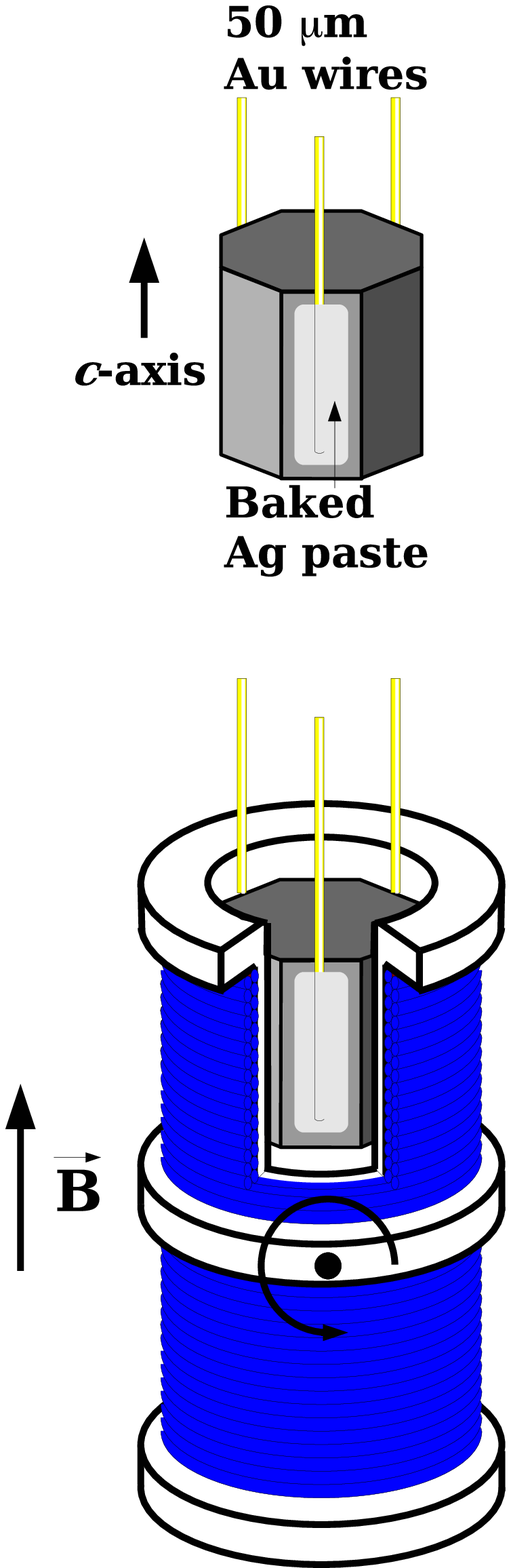}
		\end{center}
	\end{minipage}
	\caption[Experimental probe and schema of the coil pair]{$Left$ Photo of the plastic probe used in the Cambridge dilution refrigerator. $Top$ $Right$ Schematic representation of a heat sunk sample of \TTS. $Right$ Schematic representation of a compensated coil with a sample as was used in the susceptibility system. The rotation axis is shown with a curved arrow.}	
	\label{fig:Camprobe}
\end{figure}

The dHvA apparatus that was mainly used in this project is located in the Cavendish Laboratory of the University of Cambridge and operated by the Quantum Matter research group. The system is composed of a dilution refrigerator of base temperature of 6~mK\footnote{The base temperature is only reached in the best conditions, which was not the case in this project.}, which features an 18~T superconducting magnet \footnote{Details of the dilution refrigeration cycle can be found in the book by Pobell \cite{pobell}.}. The cryostat is located on a vibration isolation base, made of several tons of concrete placed on top of a rubber layer, contributing to very low vibrational noise levels.

The samples were placed inside coils situated on a rotation mechanism, inside the magnet bore. Figure \ref{fig:Camprobe}, left, shows a photo this system. It is composed of a frame, small bobbins onto which are fixed the compensated coil pairs, which can rotate around, and a rod that, by being pulled or pushed, controls the orientation of the bobbins, and spans around 90$^{\circ}$. The rotation was operated outside the cryostat by an electrical motor, the procedure and calibration of which is explained in appendix~\ref{App:E}. Figure \ref{fig:Camprobe}, left, presents a sketch of the sample mounting, heat sinking and coil system. The pick-up coils were wound in Cambridge by Swee~K. Goh. The pairs were well compensated, with coil resistances of 361~$\Omega$ and 361~$\Omega$ for the first pair used with sample C698I, and 350~$\Omega$ and 348~$\Omega$ for the second coil used with sample C698A, and all had approximately 1000 turns. The samples were heat sunk using three gold wires each, bonded to the material using sintered Dupont 6838 silver paste, 10~cm long, connected to a 1mm thick high purity silver wire tightly screwed against the mixing chamber of the dilution refrigerator. As we will see in section \ref{sect:ModFieldLK}, the thermalisation of the samples might not have been perfect at the lowest temperatures, but good above around 90~mK.

The modulation field was produced using a high quality power audio amplifier, to which was fed the oscillatory output voltage signal from a SR830 lock-in amplifier (LIA). A low modulation frequency of 7.9~Hz was used in order to obtain penetration depths longer than the sample size, but also, as we will show in section \ref{sect:eddy}, when using second harmonic, for constant sample heating by eddy currents, gains were made in signal amplitude by using as low a modulation frequency as was possible. A careful study of eddy current heating, presented in sections \ref{sect:eddy} and \ref{sect:CamEddyCurrents}, revealed that the optimum modulation field to use in this specific case was of around 20~G.

Each coil in a pair was connected in parallel with a variable resistor, using closely twisted wire triplets. Adjusting the resistors effectively changed the magnitude of the signal from each coil, and allowed improvements to the compensation of the coil pairs. The resulting signal was fed to low temperature transformers which were maintained at 1.5~K, and amplifed the signal by a factor 100. In order to damp noise at higher frequencies than the measurement one, 0.1 $\mu$F capacitors were added in parallel with the low temperature transformers. The resulting signal was fed to floating preamplifiers of amplification factor 1000, which contributed to matching the impedance between the low temperature transformers and the LIA, and an improvement in signal amplitude of approximately six was obtained in this way, normalised by the amplification factor.

\begin{figure}[t]
	\begin{minipage}[t]{6cm}
		\begin{center}
		\includegraphics[width=6cm]{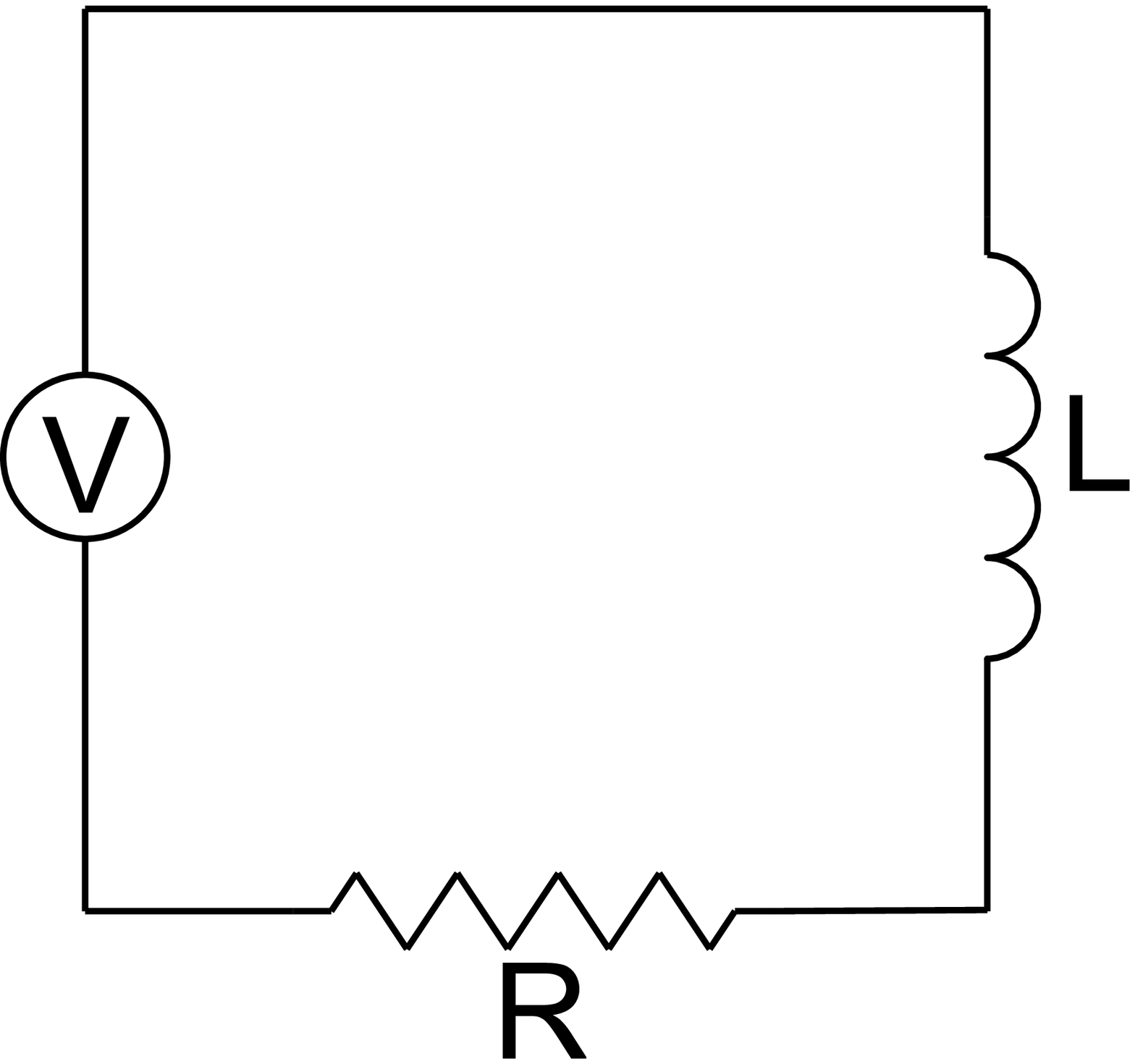}
		\end{center}
	\end{minipage}
	\hfill
	\begin{minipage}[t]{7cm}
		\begin{center}
		\includegraphics[width=7cm]{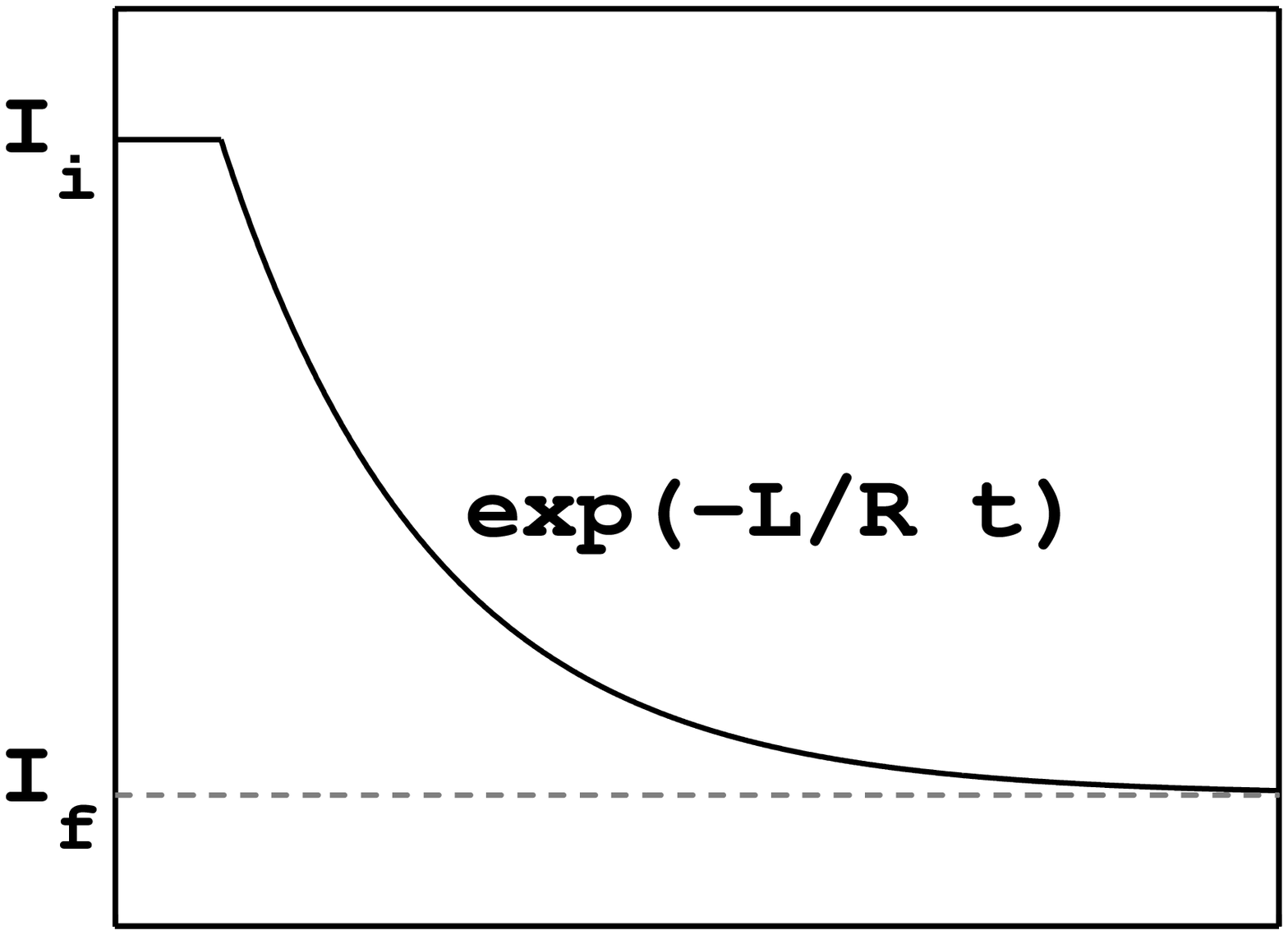}
		\end{center}
	\end{minipage}
	\includegraphics[width=1\columnwidth]{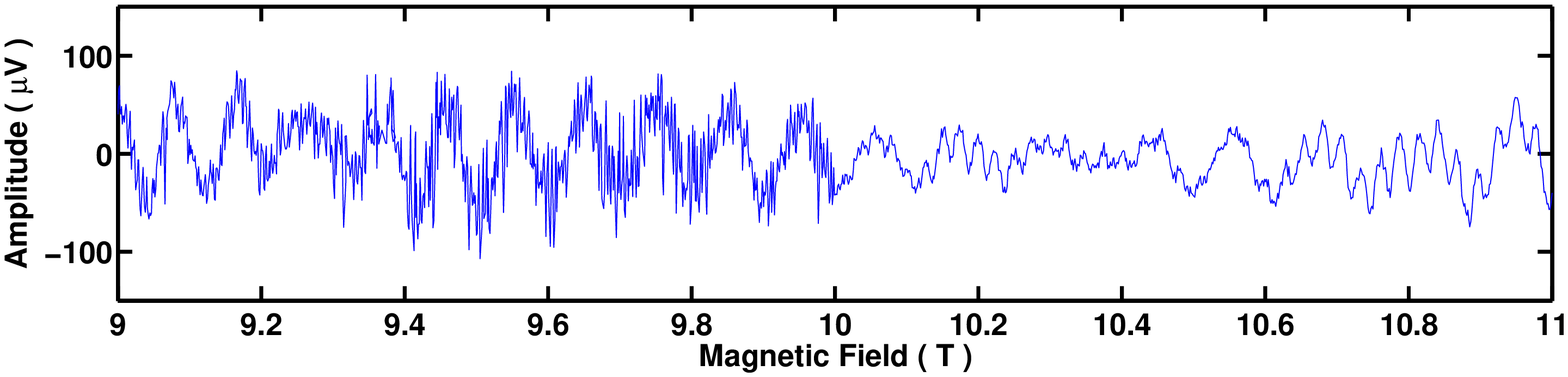}
	\caption[Voltage limited mode magnet operation]{$Top$ $left$. RL circuit that represents the magnet system. $Top$ $Right$. Time dependent profile for the magnet current when one abruptly changes the voltage on the leads of the magnet. $Bottom$. Effect of the voltage limited mode on the noise levels in a dHvA measurement. The system mode is changed at 10~T from the voltage limited to the linear one, when sweeping downwards. The change of noise levels, when normalised by the amplification factors, at 10~T is from 100 to 500~pV/$\sqrt{Hz}$.}	
	\label{fig:VLimitedMode}
\end{figure}

Throughout the measurements, an experimental procedure designed to reduce noise levels associated with the operation of the magnet power supply was used. By adding a large power resistor of small resistance value, the superconducting magnet system was transformed into an $RL$ circuit (see figure \ref{fig:VLimitedMode}, $Top$ $left$). When a current is present in such a system, along with a voltage at the power supply output, as one changes abruptly the voltage, the current adjusts exponentially to the new value, the field energy being absorbed by the power resistor. One may express this with a differential equation, where the current is $I$, the resistance $R$ and the inductance of the magnet $L$, with solution $I(t)$:
\bea
V = RI + L {dI \over dt}\nn\\
I(t) = (I_i - I_f) e^{-{L \over R} t} + I_f,\nn
\eea
with $I_i$ and $I_f = V_f / R$ the initial and final values for the current, $V_f$ being the new voltage value. Figure \ref{fig:VLimitedMode}, $Top$ $right$, sketches the profile of current when the voltage is abruptly changed from one value to another. With such an analog current sweep method, the magnetic field produced was more stable than with the normal mode and the noise levels picked up by the measurement coils became at least five times lower. It was not usable with the current system at low magnetic fields, since the sweep rates became too slow. Consequently, in the work presented in chapter~\ref{chapter:dHvA}, the long field sweeps from 18 to 2~T were performed in three parts, two at high fields, from 18 to 15~T and 15.1~T to 10~T, using the voltage limited mode at two different voltage values, resulting in sweep rates of around 0.04 T/min, and from 10.1~T to 2~T using the normal mode with a constant sweep rate of 0.02 T/min. Due to time constraints and the fact that the oscillations have longer periods in $field$ at high fields than at low fields, a higher sweep rate was used in the high field sweeps. A careful procedure was created in order to join the resulting data files using the regions of data overlap. Figure \ref{fig:VLimitedMode}, $Bottom$, shows a strip of data between 9 and 11~T, just near the region where the change from voltage limited to linear modes occurs, at 10~T, and a huge change in noise amplitude can be seen.

As a brief conclusion, we quote the noise levels that were obtained. When $not$ using the voltage limited sweep mode, the noise levels normalised by the total amplification factor of 100 000 were typically of 500~pV/$\sqrt{Hz}$ (peak to peak) and very stable. At high fields, when the voltage limited mode was used, the noise levels were of 100~pV/$\sqrt{Hz}$ (peak to peak). In the high field data, the noise was five times lower than in the quietest measurements performed previously with the current St Andrews system, but combined with the improvement to the signal due to a few more electronic components (the variable resistors, the capacitors and the preamplifiers), the total improvement in signal to noise, normalised by the square of the modulation field, was even higher.

\section{De Haas van Alphen data analysis \label{sect:dHvAanalysis}}

This section aims to present the details of the numerical analysis methods of dHvA data that were used in this project. This experiment usually produces huge amounts of data, which contains a lot of information, but complex numerical methods are required to be used in order to extract the parameters one is normally looking for. Moreover, since this experiment is at the core of this project, the author felt it relevant to describe also problems and pitfalls that exist in the analysis of quantum oscillations, which are not always well known by many condensed matter physicists. Such problems can lead, and have led in the past, to the production of systematic errors. A good understanding of this subject has been imperative throughout this project, and is hopefully well reviewed in this section.

We first review the analytical expressions to the various terms involved, which have been derived in the last chapter, in order to set a convention for the language used in this thesis, and sketch how these vary with temperature and field. We then describe how discrete Fourier transforms are performed using windowing. Since Fourier transforms presented in other work are not always appropriately normalised, but often feature arbitrary units, we felt that a description of how this was done in this project was not superfluous. One of the important pieces of information provided by dHvA is the quasiparticle mass, and we therefore carry on with a description of the non-linear fits involved in its extraction. The use of large field windows may lead in some case to systematic errors in the extraction of the quasiparticle mass, and we furthermore present how this can be avoided. Finally, the oscillatory amplitude contains valuable information which can be obtained by a method that we introduce. In particular, such data can provide knowledge of the mean free path of the quasiparticles through the Dingle factor.

\subsection{Analytical expression for the oscillations \label{sect:exprosc}}

\begin{figure}[p]
\begin{center}
	\includegraphics[width=.7\columnwidth]{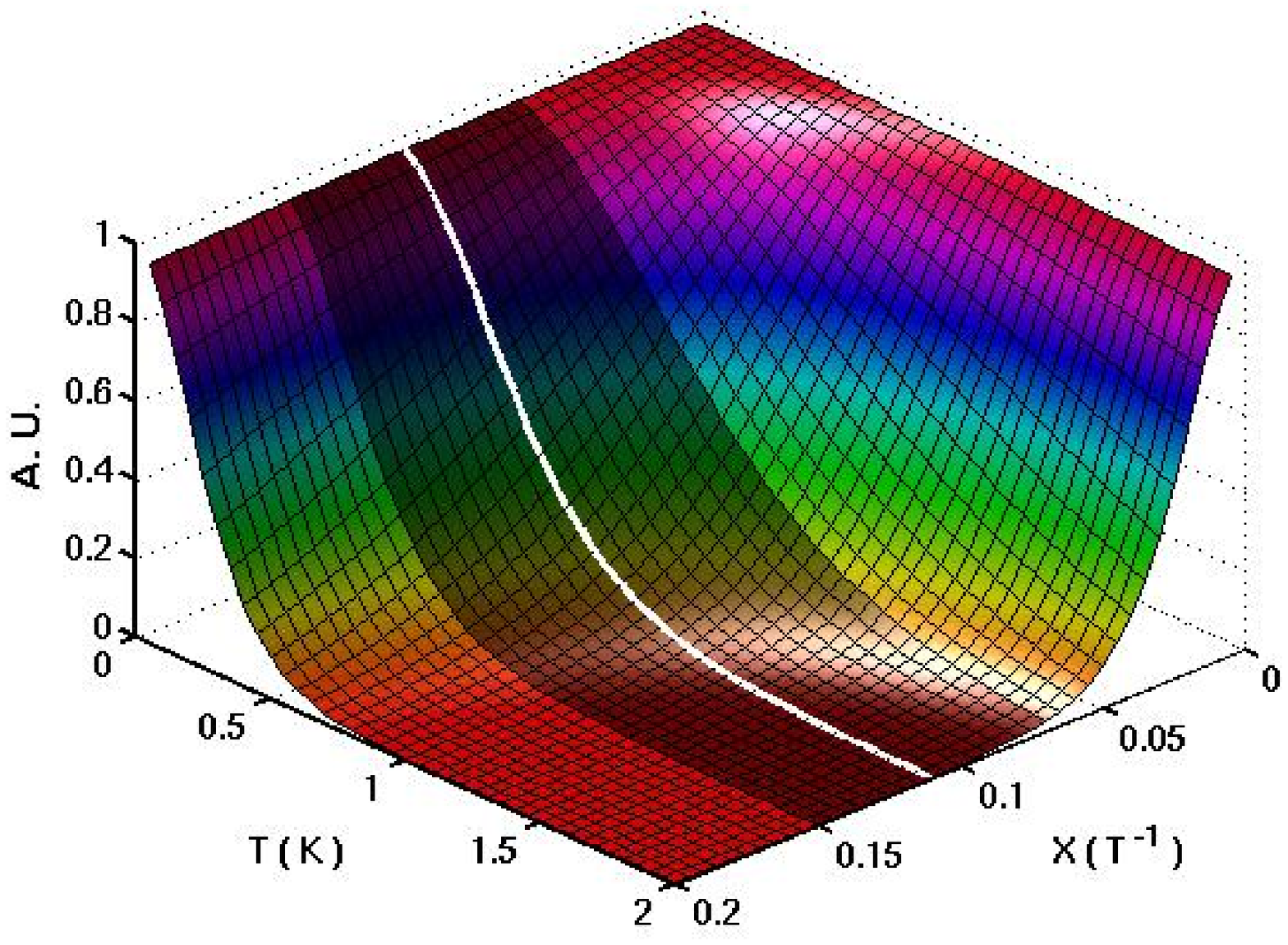}
	\caption[LK factor as a function of $T$ and $X$]{Three dimensional representation of the Lifshitz-Kosevitch factor as a function of temperature $T$ and inverse magnetic field $X$, for a constant mass $m^* =$~7~$m_e$. The shaded region illustrates a typical large $X$ measurement interval, where the magnetic field extends between 6 and 15~T. The white line indicates $X_0$, the average $X$ value for this interval.}
	\label{fig: LKsurf}
\end{center}

\begin{center}
	\includegraphics[width=.65\columnwidth]{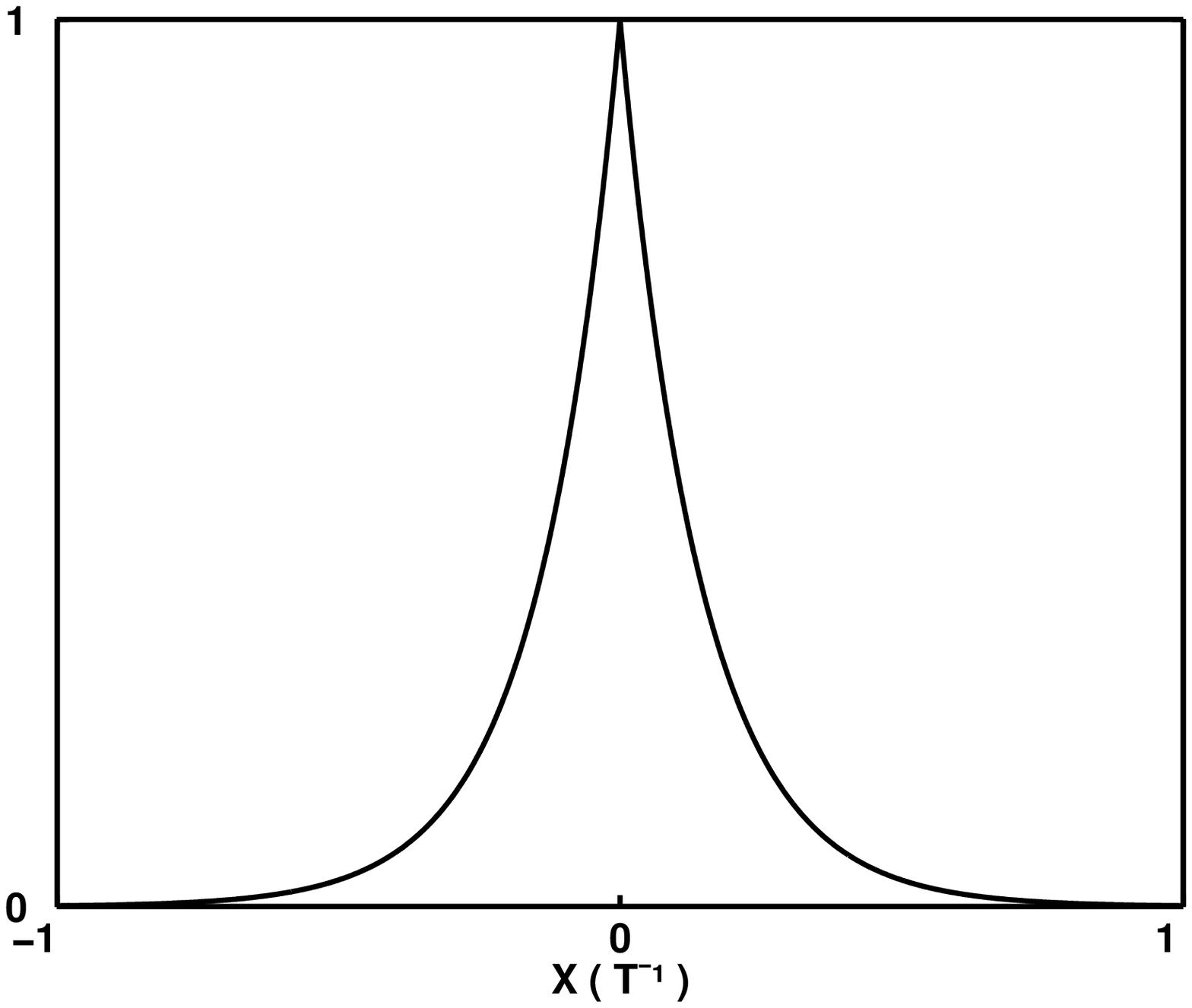}
	\caption[Dingle factor as a function of $X$]{Illustration of the Dingle factor as a function of $X$. It is symmetric around infinite magnetic field ($X=0$) and zero at zero field ($X \rightarrow \infty$).}
	\label{fig: D}
\end{center}
\end{figure}

We review in this section the analytical expressions for the different terms in the equation for the quantum oscillations, which were derived in the first chapter, and their dependence upon the different physical parameters that come into play in these experiments. This is done in a compact and consistent form that makes later descriptions of the numerical methods easier. 

We write the expression for the oscillations in the form of four functions, $J_k$, $LK$, $D$ and $E$, relating to the Bessel function modulation (eq. \ref{eq:dMdt}), the Lifshitz-Kosevitch (LK) factor (eq. \ref{eq:LK}), the Dingle factor (eq. \ref{eq:D}) and the finite size of a real data set. We express with the variables $X$ and $T$ the inverse field and the temperature, respectively, and with the integers $n$ and $k$, the various FS sheets and the detection harmonics. Moreover, we use $F_n$ as the reciprocal variable to $X$, the oscillatory frequency of FS sheet $n$, $X_0$ as the average inverse field and $m^*_n$ as the quasiparticle mass of the $n$th FS sheet. The complete expression for the oscillatory part of the susceptibility in a dHvA experiment is
\beq
\chi \propto \sum_{n,k = 1}^\infty J_k(X, F_n) LK_n(X,T,m^*_n) D_n(X) E(X-X_0) \cos(2 \pi F_n X).
\label{eq:Fullosc}
\eeq

As seen in section \ref{sect:oscsignal} (and appendix~\ref{App:C}), the oscillations are modulated by Bessel functions of the first kind
\beq
J_k(\lambda_n) = J_k\bigg({2\pi F_n H_{AC} \over H_{DC}^2}\bigg) = J_k(2\pi F_n H_{AC} X^2),\nn
\eeq
which when $\lambda_n < 1$ reduces approximately to
\beq
J_k(\lambda) \simeq {\lambda^k \over 2^k k!}  \propto {X^{2k} \over 2^k k!}.\nn
\eeq
This factor is present in dHvA but absent in Shubnikov-de Haas oscillations.

The second part of expression \ref{eq:Fullosc} is the $LK(X,T)$ function, which features the only temperature dependence of that equation,
\beq
LK_n(X,T) = {Cm^*_n X T \over \sinh Cm^*_n X T},\nn
\eeq
with variable $C$ containing various physical constants. When using $m^*_n$ as a quasiparticle mass enhancement without units,
\beq
C = {2 \pi^2 k_b m_e \over e \hbar } \simeq 14.7 \, \textrm{T}/\textrm{K}.\nn
\eeq

Figure \ref{fig: LKsurf} illustrates the dependence on $T$ and $X$ of the LK factor. It is identical for both variables, but values that can be reached in an experiment are very different.  Often, a small region in $X$ will be used, over which the LK factor varies very little, and $X$ may be safely replaced by its average $X_0$, shown as a white curve. In a case where a large $X$ interval is chosen, depicted with a shaded area, the dependence on $X$ has to be considered. The implications of this will be shown in the next two sections. 

The third factor modulating the oscillation amplitude is the Dingle factor $D(X)$, plotted in figure \ref{fig: D}, 
\beq
D(X) = e^{-\lambda_n |X|},\nn
\eeq
with the constant $\lambda_n$ proportional to the scattering rate of quasiparticles in FS sheet $n$.

The last coefficient in eq. \ref{eq:Fullosc} is the function $E(X)$. Its origin lies with the interval of $X$ over which data is available, where the data is assumed to vanish outside of that region. In order to improve the clarity of the Fourier spectra, this function is usually chosen such that is vanishes smoothly at each ends, a practice described in the next section. Without the function $E(X)$, the oscillations would extend to infinity and be symmetric around zero.

\subsection{Fourier transforms and windowing \label{sect:FFTwindowing}}

In order to study the Fermi surface structure of a material, one is required to separate the different contributions at different frequencies from its various FS sheets, using Fourier transforms. Unfortunately, it is not obvious what the analytical Fourier transform of equation \ref{eq:Fullosc} is. We may nevertheless show here why windowing is needed, which type we chose and its effect on the Fourier spectra.

The convolution theorem states that the transform of a product of functions is equal to their convolution in reciprocal space,
\beq
\mathcal{F}[f(x)g(x)] = \mathcal{F}[f(x)] \ast \mathcal{F}[g(x)] = \tilde{f}(x) \ast \tilde{g}(x).\nn
\eeq
Applied to eq. \ref{eq:Fullosc} expression, one obtains
\beq
\mathcal{F}[\chi(X,T)] \propto \sum_n \tilde{E}(F) \ast \bigg(\tilde{D_n}(F) \ast \big( \tilde{LK}(F,T) \ast (\delta(F-F_n) + \delta(F+F_n))\big)\bigg),\nn
\eeq
where the Fourier transform is taken as a function of $X$ and yields functions of the frequency $F$.

In a dHvA experiment involving short $X$ intervals, one assumes $D$ and $LK$ constant, and the Fourier transform is proportional to 
\beq
\sum_n \sum_\pm \tilde{E}(F \pm F_n),\nn
\eeq
the Fourier transform of the envelope $E$ centred at $\pm F_n$. If the data of a dHvA set extended from zero to infinite magnetic field, $\tilde{E}(F \pm F_n)$ would be a Dirac delta function centred at $\pm F_n$, but with finite size data,  $\tilde{E}(F \pm F_n)$ possesses a width and a shape defined by the nature of $E(X)$. If one takes $E(X)$ as a top hat function of width $2a$ centred at $X_0$, equal to one where dHvA data exists, and zero everywhere else, then $\tilde{E}(F \pm F_n)$ is
\beq
\tilde{E}(F) = {1\over \sqrt{2\pi}}\int_{X_0 - a}^{X_0 + a} e^{-iFX}dX = {2\over \sqrt{2\pi}}a e^{-iFX_0} \sinc(aF).\nn
\eeq
The $\sinc$ function oscillates around a central peak, and taking modulus of this result, the satellite peaks could be misinterpreted as additional dHvA frequencies. One generally uses another form of windowing for $E(X)$ that decreases smoothly at the edges of the data set. An article by Harris reviews most types of windowing techniques \cite{HarrisFFT}, but in general, it is the Hann or the Hamming windows that are the most often used. The first is defined as
\bea
E(X) &=& .5 + .5 \cos \bigg[{\pi \over a} (X - X_0)\bigg], \quad -a < X-X_0 < a \nn\\
&=& 0 \quad otherwise.\nn
\eea
Its Fourier transform is
\beq
\tilde{E}(F) = {1\over \sqrt{2\pi}}e^{iFX_0} \sinc(aF) {a \pi^2 \over \pi^2 - a^2F^2},\nn
\eeq
and decreases faster than the $\sinc$ function away from the main peak, having much smaller satellite peaks. The Hamming window, on the other hand, is tuned with a parameter $\alpha$ to suppress to zero the first satellite peak on each side of the main peak and is obtained by adding a very small top hat window with a Hann window:
\bea
E(X) &=& \alpha + (1-\alpha) \cos \bigg[{\pi \over a} (X - X_0)\bigg], \quad -a < X-X_0 < a \nn\\
&=& 0 \quad otherwise.\nn
\eea
The value of $\alpha \simeq 0.54$ gives the best results. Its Fourier transform is
\beq
\tilde{E}(F) = {2a \over \sqrt{2\pi}}e^{iFX_0} \sinc(aF) \bigg(\alpha + (1-\alpha){a^2F^2 \over \pi^2 - a^2F^2}\bigg),\nn
\eeq
The Hamming window was the one applied in this work.

Along with windowing, one generally also uses the technique called zero padding. It consists in artificially increasing the length of the windowing function $E(X)$ with values of zero in order to reduce the coarseness of the Fourier spectrum. It is equivalent to an appropriate interpolation of data points between existing ones, such that the spectrum is smoother and more continuous.

Finally, we may add a comment about the width of the Fourier transform of the windowing functions. The half width at half maximum of the $\sinc$ function is found by equating
\bea
{2a\over \sqrt{2\pi}} {\sin(a \Delta F) \over a \Delta F} &=& {a \over 2} \nn\\
\Rightarrow \sin(a \Delta F) &=& {a \Delta F \over 2},\nn
\eea
which yields numerically $a \Delta F = 0.3017$, where $\Delta F$ is measured from the centre of the function. The same can be done with the Hann and the Hamming windows, which give, respectively, 
\bea
\Delta F = 0.3017/a &&\quad Top\quad Hat \nn\\
\Delta F = 0.5 / a &&\quad Hann\nn\\
\Delta F = 0.4538 / a &&\quad Hamming\nn
\eea
These widths are function only of that of the window $E(X)$, regardless of, for instance, the number of cycles of oscillation within that set. For one to be able to compare spectral information at differing frequencies, the Fourier transforms have to be performed with equal inverse field widths, not the same number of cycles of oscillation. 

\subsection{Appropriate normalisation of Fourier transforms}

Throughout the data analysis of this thesis, it has been important to be able to compare the amplitude of various dHvA signals measured on different samples of \TTS, which has required a rigorous understanding of the normalisation procedure of dHvA spectra, with appropriate units. We present in this section the exact analytical method with which one should normalise and present dHvA spectra, and how to extract information from them.

As one measures voltages and takes Fourier transforms as a function of inverse field $X$ in T$^{-1}$, the frequency units are T and one expects that the units of the Fourier transforms should be $\textrm{V}/\sqrt{\textrm{T}}$. \footnote{In reference to time signal processing, where the spectral amplitude has units of V/$\sqrt{\textrm{Hz}}$.} The Fourier transform $\tilde{\Delta V}$ of the voltage measured $\Delta V$ is generally a complex number data set. The power spectrum $W(F)$ represents the power radiated per unit frequency, and is the square of the modulus divided by the interval (that we denote $2a$):
\beq
W(F) =  {1\over 2a}|\mathcal{F}[\Delta V(X)] |^2 = {1\over 2a} \tilde{\Delta V}^*\tilde{\Delta V},\nn
\eeq
which has units of V$^2/$T. In general, it is not the power spectrum that is plotted but its square root, with $\textrm{V}/\sqrt{\textrm{T}}$. In order to know the voltage amplitude of the signal for a certain frequency interval, the power spectrum is integrated within that interval, which yields units of V$^2$, and a square root of the result is taken. Due to spectral leakage, in order to know the voltage amplitude of a certain dHvA frequency, one is required in theory to integrate the power spectrum from minus infinity to infinity, since $\tilde{E}(F)$ is non-zero for all $F$. In practice, the presence of multiple frequencies makes this impossible and one may integrate a peak using its full width at half height as boundaries. This yields a fraction of the real amplitude, that depends on the integration limits but also on the windowing function used in the Fourier transform. 

The voltage measured on a coil may be expressed as
\beq
\Delta V(X) = \sum_n A_n \cos(F_n X), \nn
\eeq
with amplitude coefficients $A_n$ of units of V. The Fourier transform one performs is
\bea
\tilde{\Delta V}(F) &=& {1\over \sqrt{2\pi}} \int_{-\infty}^{\infty} E(X) \Delta V(X) e^{iFX} dX,\nn\\
&=& \sum_{n,\pm} A_n \tilde{E}(F\pm F_n),\nn
\eea
and
\beq 
W(F) = {1\over 2a} \bigg| \sum_{n,\pm} A_n \tilde{E}(F\pm F_n) \bigg|^2.\nn
\eeq
One usually requires the amplitudes $A_n$, assuming that the frequencies are well enough separated such that the various dHvA peaks $\tilde{E}(F-F_n)$ do not overlap. Since the coefficients $A_n$ are real, one requires the value of the integral of $|\tilde{E}(F)|^2$ around zero in order to find $A_n$. In the case of a top hat windowing function, the integral is
\beq
\int_{-\infty}^{\infty} |\tilde{E}(F)|^2 dF =  \int_{-\infty}^{\infty} {4a^2 \over 2 \pi} \sinc^2(aF) dF = 2a,\nn
\eeq
and the total power of the $n$th frequency is $W_n = A_n^2$. Since one generally integrates only the peak at positive frequency $F_n$, an additional factor of 2 is multiplied to the integrated power spectrum, and the formula to use when extracting $A_n$ from $\Delta V$ is
\beq
A_n = \sqrt{W_n} =  \sqrt{ 2 \int_{-\infty}^{\infty} {1\over 2a}|\tilde{\Delta V}(F)|^2 dF}.\nn
\eeq
In the case of a Hann window, the integral is 
\beq
\int_{-\infty}^{\infty} |\tilde{E}(F)|^2 dF =  \int_{-\infty}^{\infty} {a^2 \pi^3\over 2} {\sinc(aF) \over \pi^2 - a^2F^2} dF = {3a \over 4},\nn
\eeq
and the formula is
\beq
A_n =  \sqrt{ 2 {8 \over 3}\int_{-\infty}^{\infty} {1\over 2a}|\tilde{\Delta V}(F)|^2 dF}.\nn
\eeq
Finally, with a Hamming window, the integral is
\beq
\int_{-\infty}^{\infty} |\tilde{E}(F)|^2 dF = 2a (\alpha^2 + {1 \over 2}(1-\alpha)^2),\nn
\eeq
and with the usual parameter $\alpha = 0.54$, the result is ${1987 \over 2500}a = 0.7948a$. The formula to use is 
\beq
A_n =  \sqrt{ 2 {5000 \over 1987}\int_{-\infty}^{\infty} {1\over 2a}|\tilde{\Delta V}(F)|^2 dF}.\nn
\eeq

These formulae enables one to obtain average dHvA amplitude values over specific field ranges, and do not depend on the length of the data set if the amplitude remains constant. Thus, amplitudes from sets of differing lengths may be compared. In this thesis, quoted amplitudes were always obtained using this method, with a Hamming windowing scheme, and was proved correct by feeding simulated data to the calculation and obtaining the right result. The voltages thus calculated were directly comparable with voltage noise levels, measured in $\textrm{V}/\sqrt{\textrm{Hz}}$.

\subsection{Mass analysis \label{sect:mass}}

One of the parameters of primary interest in dHvA studies is the quasiparticle cyclotron mass. This value is extracted from the temperature dependence of the amplitude of the oscillations, and one such mass can be obtained from each frequency in the dHvA spectrum, and can consequently be assigned to a specific branch of the FS. In this section, we explain the numerical method used to extract the quasiparticle masses.

One assumes once again that the functions $LK(X,T)$ and $D(X)$ vary slowly enough and can be taken  out of the Fourier transform, by replacing $X$ by $X_0$. This is true most of the time and we will show in the next section the situations where it is $not$ possible to make this approximation, in other words, when the inverse field window is too large compared to $D$ and $LK$. One calculates the Fourier transform of eq. \ref{eq:Fullosc}:
\beq
\tilde{\chi}(F,T) \propto \sum_{n,\pm} LK_n(X_0,T) D_n(X_0) e^{iFX_0} \tilde{E}(F \pm F_n)\nn
\eeq
The temperature dependence is only present in the function $LK_n(X_0,T)$,
\beq
LK_n(X_0,T) = {C m_n X_0 T \over \sinh C m_n X_0 T}, \quad C \simeq 14.7 m^*_n\textrm{T}/\textrm{K},\nn
\eeq
where $m^*_n$ is the mass enhancement normalised by $m_e$. In order to extract $m^*_n$, one is required to perform a non-linear fit of the $LK$ function. This fit is carried out on values of the square root of integrals calculated over peaks in the power spectrum, as described in the last section, as a function of temperature. The fit possesses only two adjustable parameters, $A$ and $m$:
\beq
LK_{fit}(X_0,T) = A {14.7 m X_0 T \over \sinh 14.7 m X_0 T}.
\label{eq:LKfit}
\eeq
If in some cases a dHvA spectrum possesses a constant additive offset, identical for all temperatures, it is possible to use a three parameter non-linear fit, with parameters $A$, $m$ and $C$,
\beq
LK_{fit}(X_0,T) = A {14.7 m X_0 T \over \sinh 14.7 m X_0 T} + C.
\label{eq:3param}
\eeq
We will show in section \ref{sect:MassSystematic} that, in contrast to the two parameter LK fit, the use of this scheme for the extraction of the mass can lead to large systematic errors, and was possibly responsible for the mass enhancement measured by Borzi and co-workers \cite{borzi} which has not been reproduced by the author. 

\subsection{Fourier transforms using large field windows}

There are cases where the inverse field window one needs to use in order to calculate spectra with a desired resolution is too large compared to the width of the functions $LK(X)$ or $D(X)$. In this section, we demonstrate that in such a situation, the normal procedure of fitting $LK(X_0,T)$, the LK function at the average inverse field, always underestimates the real quasiparticle mass. This problem is generally overlooked, but the solution presented here has been critical for the analysis of dHvA data from one material, Sr$_2$RhO$_4$ \cite{PerryNJP}, where the total quasiparticle mass calculated with the standard method did not agree with the electronic specific heat. The analysis presented in this section was created by the author and contributed to finding of the ``missing mass" of Sr$_2$RhO$_4$.

When one works with well separated frequency peaks in the dHvA spectrum, one is not very limited in possible sizes of the inverse field window with which to perform a Fourier transform. Normally, noise is the limiting factor, as enlarging the inverse field width averages more data and reduces the noise floor in comparison to the signal. The situation where one is required to use large field windows is if dHvA peaks are too close to one another to be integrated separately when using suitably small field windows.

Parseval's theorem states that for an amplitude function and its Fourier transform, the area under the power spectrum is equal to the area under the square of the function. In the case of dHvA, the integrated power spectrum is the average of the square of the product of the amplitude functions,
\bea
W &=& 2a \big<|E(X-X_0) LK(X,T) D(X)|^2\big>\nn\\
&=& 2a \big<E^2(X-X_0) LK^2(X,T) D^2(X)\big>.\nn
\eea
Supposing for a moment that $D(X)$ is constant and that $E(X-X_0)$ is a top-hat window, one observes that the $T$ dependence of $\sqrt{W}$ is proportional to $2a \sqrt{\big<LK^2(X,T)\big>}$, which is not equal to $LK(X_0,T)$, since the width in temperature of $LK(X,T)$ varies with inverse field $X$ (depicted in figure \ref{fig: LKsurf}). Consequently, this averaging produces an $LK$-like function that cannot be adjusted adequately by a non-linear fit of $LK(X_0,T)$. Moreover, this average is always wider in $T$ than $LK(X_0,T)$, the $LK$ function $at$ the average field $X_0$. Consequently, the fit of $LK(X_0,T)$ always generates a mass that is smaller than the real value.

\begin{figure}[t]
\begin{center}
	\includegraphics[width=.6\columnwidth]{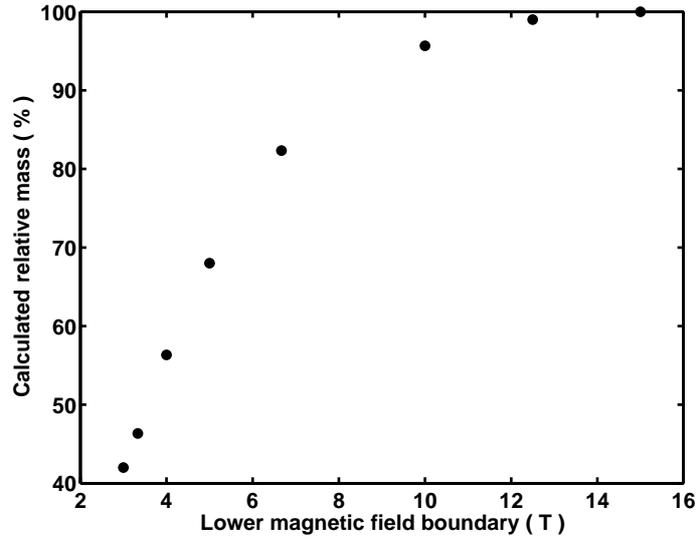}
	\caption[Error on the quasiparticle mass when using large field windows]{Relative value of the quasiparticle mass extracted with the normal fit of the the function $LK(X_0,T)$ applied to simulated dHvA data, to the real value. The relative mass is plotted as a function of the position in $B$ of the lower boundary of the field window, the higher boundary fixed at 15~T, where the oscillations are the highest. The frequency used was of 1~kT and the mass of 3 electron masses, parameters similar to those of Sr$_2$RhO$_4$. For a vanishingly small window, the calculation error is nil, and increases with the size of the window. The error can be higher than 50\%. }
	\label{fig:Mass_B}
	\end{center}
\end{figure}

Assuming that $D$ and $E$ are not constant worsens matters, since these functions influence the average of the LK function by weighing differently the amplitude at different values of $X$, $E$ towards the centre of the window and $D$ towards higher values. Nevertheless, as values at higher $X$ produce a higher contribution, the mass extracted by a fit of $LK(X_0,T)$ still underestimates the value.

Figure \ref{fig:Mass_B} shows $LK$ fits taken on simulated dHvA data, using the normal method of eq. \ref{eq:LKfit}, as a function of the position of the lower field boundary of the inverse field window, the higher one fixed. We observe that as the lower boundary moves towards the higher one, the mass approaches the real value.

Two methods of analysis may be used for such circumstances. The first is is to use the appropriate fit function, which is 
\beq
\sqrt{2a \big<E^2(X-X_0) LK^2(X,T) D^2(X)\big>}.\nn
\eeq
This requires knowledge of the Dingle temperature $T_D$ for the function $D(X)$, the method to extract will be shown in the next section. The second is to find the mass in a self-consistent way using a simulation. In this procedure, one uses data simulated with the real Dingle temperature and a chosen quasiparticle mass, performs a Fourier transform using the same large field window as with the experimental data. One then carries out an LK fit and adjusts the mass of the simulation until the LK fit yields the value obtained with the experimental data. The adjusted mass of the simulation corresponds to the real quasiparticle mass. This is the method that was used by Perry and co-workers \cite{PerryNJP}, as a result of the insight given in the present section.

\subsection{The envelope extraction method \label{sect:envelope}}

We describe here the method which will be used in section \ref{sect:CamEnvelopes} in order to extract the field dependence of the amplitude of the oscillations, taken from the review on Sr$_2$RuO$_4$ by Bergemann $et$ $al.$\cite{bergemann}, the envelope extraction. The envelope corresponds to the product of the functions $J_k$, $D$, $LK$ and some interference patterns, as discussed in section \ref{sect:BergemanAnalysis}. Using this technique, which consists of filtering, one can obtain continuous functions of the modulation amplitude for all individual dHvA frequencies. 

The dHvA oscillations take the form
\beq
\chi(X) = \sum_n A_n(X)\cos(F_nX),\nn
\eeq
The functions $A_n(X)$ are the envelopes one wishes to extract. Carrying out a Fourier transform yields
\bea
\tilde{\chi}(F)  &=& \sum_n{1 \over 2i \sqrt{2\pi}}\int_{-\infty}^\infty A_n(X) (e^{i(F_n-F)X} - e^{-i(F_n+F)X}) dX,\nn\\
&=& {1 \over 2i\sqrt{2\pi}}\sum_n[\tilde{A}_n(F - F_n) - \tilde{A}_n(F + F_n)]\nn
\eea
One requires to isolate one of the various $\tilde{A}_n(F)$, denoted $\tilde{A}_m(F)$, filtering the spectrum by using a frequency window that leaves only one peak and sets the rest to zero. If one also sets the negative frequencies, which appear symmetrically to the positive ones, to zero, one obtains,
\beq
\tilde{\chi}_{flt}(F) = {1 \over 2i\sqrt{2\pi}}\tilde{A}_m(F - F_n).\nn
\eeq
An inverse Fourier transform is then used to calculate the desired $A_m(X)$:
\beq
\chi_{flt}(X) = {1 \over 4i\pi}\int_{-\infty}^\infty \tilde{A}_m(F') e^{i(F' + F_n)X}dF' = {1 \over 4i\pi} A_m(X) e^{iF_nX}.\nn
\eeq
By taking the modulus, one is left with the envelope, without oscillations,
\beq
|\chi_{flt}(X)| = {1 \over 4\pi} |A_m(X)|.\nn
\eeq

It is possible to use this method even if the frequency peaks are split or vary with inverse field, where one includes everything in the filter window, where it is effectively the case with \TTS\ close to the metamagnetic transition. The shape of the filter window is moreover important for the same considerations as for the function $E$, where essentially the envelope extracted by this method consists of the real envelope convolved with the Fourier transform of the filter function. In particular, using a top-hat filter leads to oscillations added to the extracted envelopes. One can damp this effect by using a top-hat filter with ``rounded corners"\footnote{In exact terms, a top-hat function convolved with a gaussian of width $\sigma$.}:
\beq
\sqrt{2} \sigma \bigg[ \erf({F-a \over \sqrt{2}  \sigma}) - \erf({F+a \over \sqrt{2}  \sigma})\bigg],\nn
\eeq
where $\sigma$ is the roundness of the filter and $\erf$ is the error function. The rounder the filter, the fewer oscillations are present in the result.

The usefulness of this method is multiple. One can in principle fit the LK relation at each value of the inverse field $X$. This makes the calculated quasiparticle mass absolute for a value of $X$, instead of an average over the window $E$ used in a Fourier transform. If the mass is not a constant of the field, one can, in principle, calculate its field dependence using this method. The inconvenience is that it produces noisy temperature distributions and fits with large error bars. More importantly, the main interest in this method is in the analysis of modulating factors to the oscillations other than the LK function. As we will show in the next section, it is possible to extract the Dingle temperature with this method. But mainly, it will be used, section \ref{sect:CamEnvelopes}, in order to obtain interference patterns as a function of magnetic field and field angle.

\subsection{The Dingle analysis \label{sect:Dingle}}

When the temperature is strictly zero, the LK function is a constant of the inverse field. In an experiment, depending on the quasiparticle mass, it is often possible to find a temperature for which there will be an accessible region of inverse field where $LK(X,T)$ is very near or equal to one (High fields, low temperatures). When one works with a system without beat patterns or Bessel function coefficients (as in Shubnikov-de Haas measurements), using the envelope extraction method of the previous section, one can extract the Dingle temperature using a simple exponential fit of  the form
\beq
D(X) = e^{-\lambda |X|}, \quad \lambda = 14.7 m^*T_D,\nn
\eeq
and extract the Dingle temperature $T_D$.

In general, the task is slightly more complex, as it is not possible to remove the Bessel or beat pattern coefficients\footnote{These cross zero in certain field regions, and taking them out leads to divisions by noise.}. Thus one is compelled to fit a function which features all the contributions to the modulation of the oscillations. For dHvA data with a beat pattern due to simple warping (section \ref{sect:BergemanAnalysis}, eq. \ref{eq:BesselCosApprox}), this takes the form
\beq
A D(X) LK(X,T) J_k(X) \cos(\omega X + \phi),\nn
\eeq
\beq
=A e^{-\lambda |X|} {CXTm^* \over \sinh(CXTm^*)} \big|\cos(\omega X + \phi) J_k(2*pi*H_{AC}FX^2)\big|,\nn
\eeq
where the variables $A$ and $\alpha$ are free parameters. The constants $\omega$, $\phi$, $H_{AC}$, $F$, $m^*$, $C$ and $T$ are known values. The mean free path is extracted from $\lambda$ through the equation
\beq
\ell = {2\pi \over \lambda} \sqrt{{2 \hbar F \over e}}.
\label{eq:ell}
\eeq

\section{AC susceptibility in an adiabatic demagnetisation refrigerator}

The de Haas van Alphen experiment, as we know from section \ref{sect:Impurities}, depends on the impurity scattering rate of the electrons exponentially. As we explain later in section \ref{sect:search}, obtaining good samples in this respect was not trivial, but improved dHvA data immensely compared to using samples of average quality. Consequently, a sample sorting procedure was developed by the author. We will show that for high purity \TTS\ samples, mean free paths determined using the residual resistivity is not representative of bulk samples required for dHvA. The dHvA experiment itself may be used for that purpose, but the use of the apparatus described in section \ref{sect:Camprobe} is time consuming\footnote{Performed in a dilution refrigerator, it has a turnaround time of 2-3 days.}. This motivated the development in this project of an alternative AC susceptibility system for an adiabatic demagnetisation refrigerator (ADR), which could accommodate large unprepared samples. It was not intended for precision measurements but for ease of use and time of operation.

This system has served two different purposes, discussed in section \ref{sect:search}. The first was to determine the amplitude of the dHvA signal as a measure of the mean free path, and the second was to determine the superconducting volume fraction, which corresponds to that of \TOF\ inter-growths, by detecting zero DC field AC susceptibility as a function of temperature. This section describes the characteristics and testing of this new experimental probe, along with a short introduction to the ADR system.

\subsection{The adiabatic demagnetisation refrigerator}

\begin{figure}[t]
	\begin{minipage}[t]{7cm}
		\begin{center}
		\includegraphics[width=7cm]{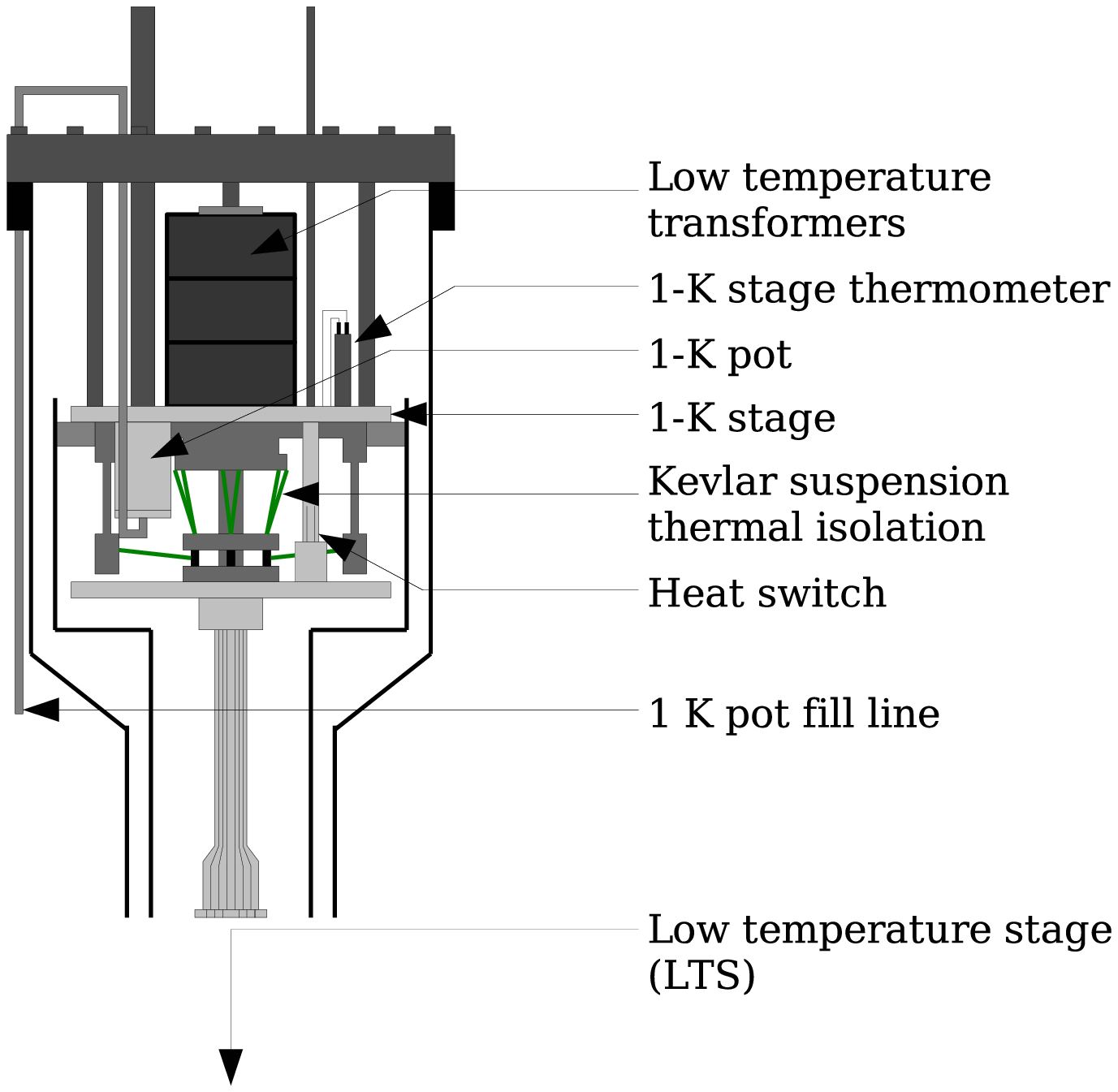}

		\end{center}
	\end{minipage}
	\hfill
	\begin{minipage}[t]{7cm}
		\begin{center}
		\includegraphics[width=7cm]{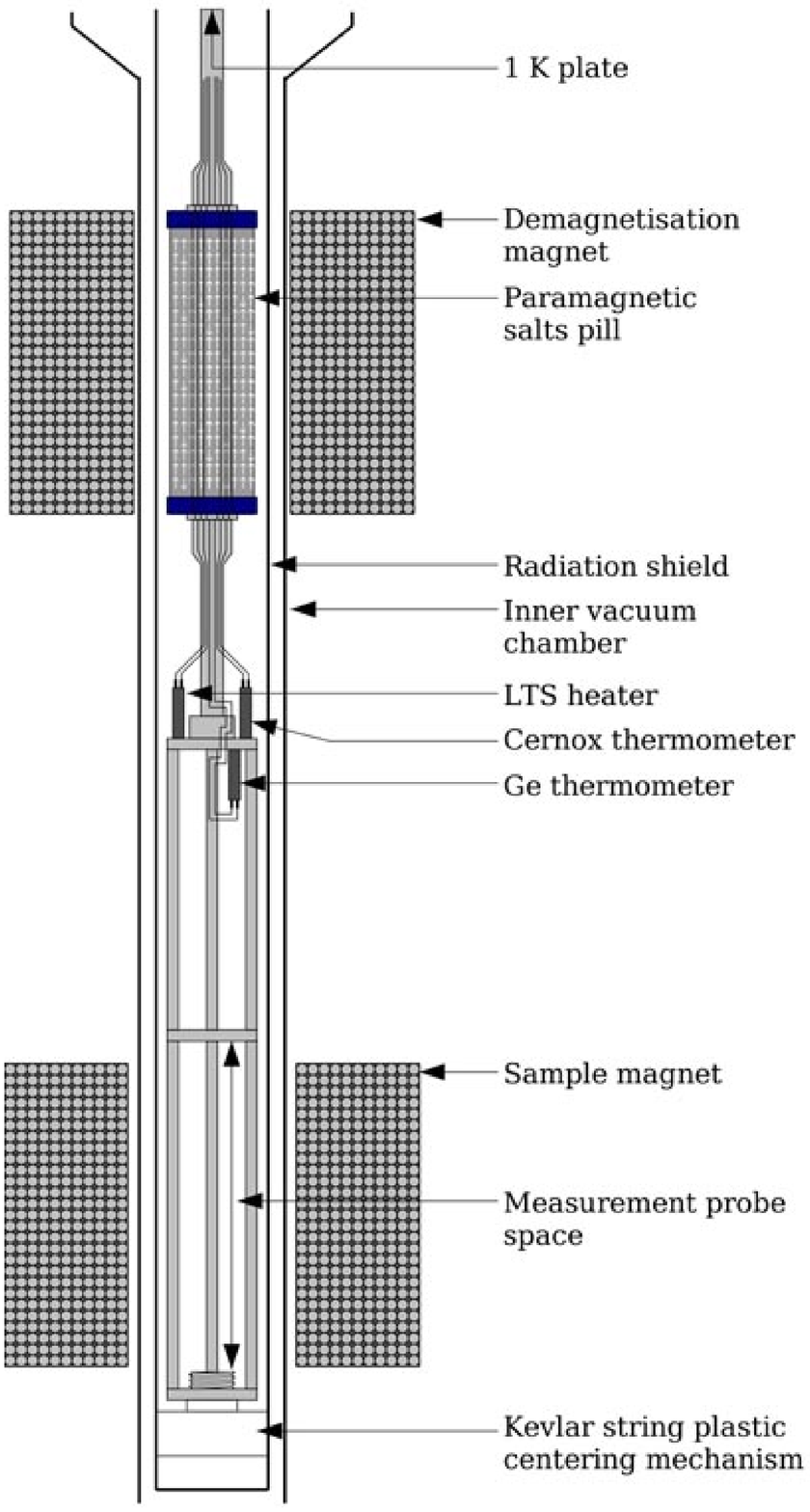}
		\end{center}
	\end{minipage}
	\caption[Sketch of the ADR components]{Sketch of the ADR components. The left side of the figure shows the top part, featuring the 1-K stage, and the right side the LTS.}
	\label{fig:ADFsketch}
\end{figure}

The adiabatic demagnetisation refrigerator\footnote{Cambridge Magnetic Refrigeration model mFridge mF-ADR/50.} (ADR) relies on the very high difference in entropy of a paramagnetic system between its states in high and low magnetic fields. The entropy is lowered by applying a high field, and when the system is put in thermal isolation, reducing its magnitude lowers the temperature by a transfer of entropy from the sample and measurement apparatus to the paramagnetic material.

Figure \ref{fig:ADFsketch} illustrates the main parts of the ADR. It is put into a dewar filled with liquid He$^4$. It features two independent stages that can be thermally connected or isolated, and possesses two cooling mechanisms, one on each part. The first is a standard 1-K pot, which refrigerates the apparatus by pumping $^4$He gas from its liquid surface, lowering its vapour pressure and temperature down to 1.5~K (left part of the figure). The second corresponds to a pill of paramagnetic salt situated in the bore of a 6~T magnet. 

Connecting both stages, by cooling down the system to 1.5~K using the 1-K pot while applying the maximum field of 6~T to the paramagnetic salt pill, the low temperature stage (right part of figure \ref{fig:ADFsketch}) is put in its state of low spin entropy and high temperature. Then, opening the thermal connection between both stages, the low temperature stage (LTS) may be stabilised at any temperature between 0.1 and 1.5~K by adjusting approximately linearly the value of the demagnetisation field. A heat leak is present, though, and the field is required to continuously decrease in order to compensate the heat input. Cooling is not possible anymore when the field reaches zero, and all temperature values will possess a different hold time, of zero at 0.1~K up to several hours near 0.5~K. Appendix~\ref{App:D} presents a calculation of the cooling power of an ideal ADR, where one finds that, effectively, the constant entropy is a function of $H/T$.

The ADR can moreover perform experiments with a sample magnetic field, independent of the demagnetisation field, produced by an 8~T superconducting magnet (lower right in the figure) at the sample region, at constant temperature. The LTS also features a compensated field region, where the thermometry is located, between the two magnets. The temperature is measured using two different thermometers, one for high and the other for low temperatures. For temperatures between 300 and 1~K,  a Cernox semiconductor resistive thermometer is used, and for temperatures between 1~K and 50~mK, the system switches to a Ge semiconductor resistive thermometer. Next to these is also situated a heater that can be used to obtain temperatures higher than the 1-K pot temperature, such that the system may be used continuously between 15 and 0.1~K.

\subsection{AC susceptibility apparatus for large crystals \label{sect:ADFprobe}}

\begin{figure}[p]
	\begin{center}
	\includegraphics[width=1\columnwidth]{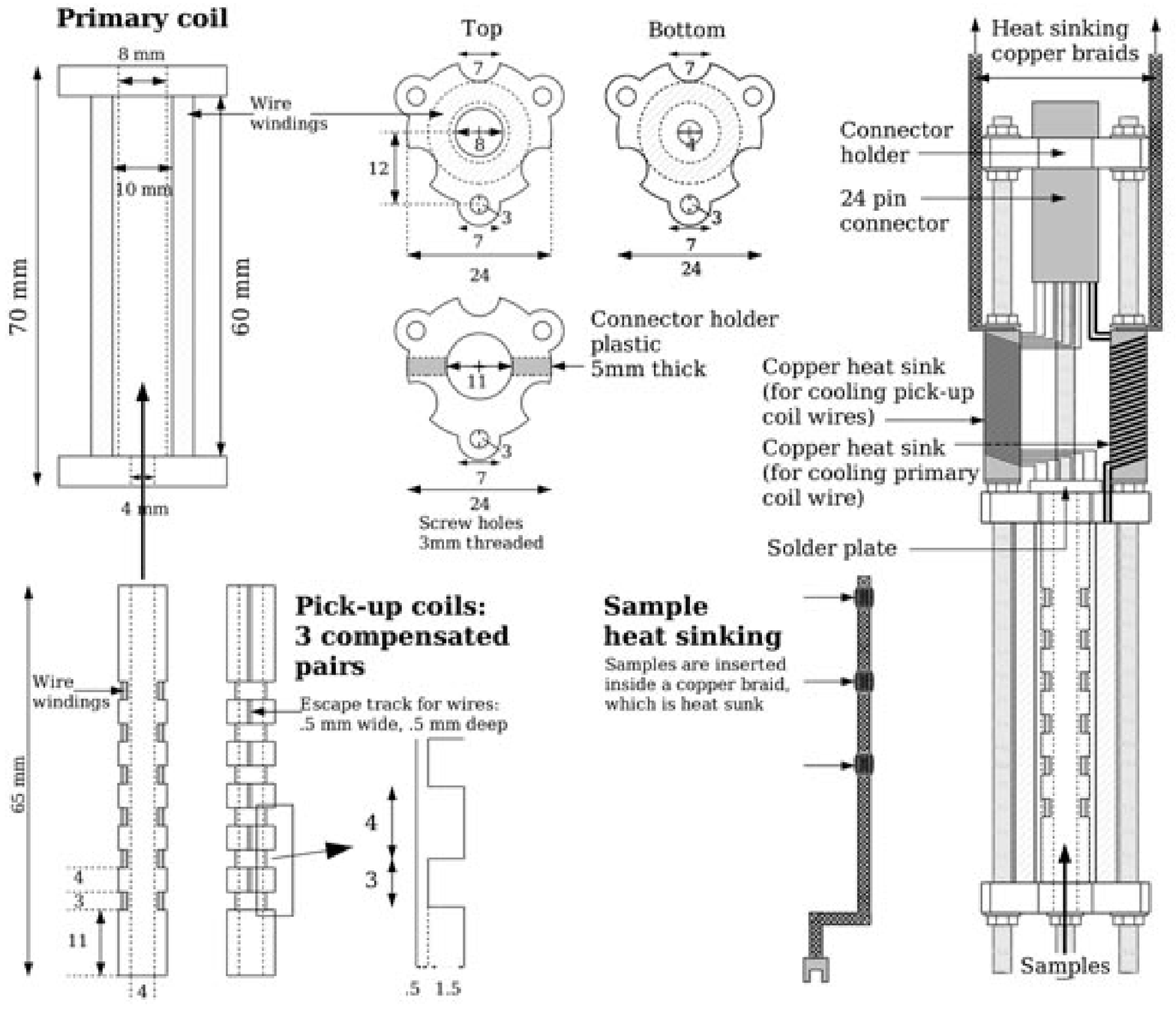}
		\begin{tabular}[h]{|c|c c c c c |}
			\hline
			&&&&&\\
			Coil& Turns & Resistance & Resistance & Sensitivity & Offset (paired)\\ 
			&& $\Omega$ & (paired) $\Omega$ & $\mu$V/GHz & $\mu$V/GHz\\
			\hline
			1 & 794$\pm$5 & 133 && 10.37 &\\
			2 & 607$\pm$15 & 104 &  \raisebox{1.5ex}[0pt]{237}& 7.97 &  \raisebox{1.5ex}[0pt]{2.366}\\
			&&&&&\\
			3 & 790$\pm$5 & 136 && 11.33 &\\
			4 & 800$\pm$2 & 138 &  \raisebox{1.5ex}[0pt]{274}& 11.68 &  \raisebox{1.5ex}[0pt]{.3254}\\
			&&&&&\\
			5 & 802$\pm$2 & 137 && 11.57 &\\
			6 & 839$\pm$1 & 138 &  \raisebox{1.5ex}[0pt]{275}& 10.74 &  \raisebox{1.5ex}[0pt]{0.8225}\\
			&&&&&\\
			Drive & 2555 & 42 &\multicolumn{3}{c|}{535 G/A}\\
			\hline
		\end{tabular}
		\caption[Schematics of the AC susceptibility probe for the ADR]{$top$ Schematics of the AC susceptibility probe for the adiabatic demagnetisation refrigerator. $Bottom$ Technical data for the various coils of this apparatus. }
	\label{fig:ADFCoils}
	\end{center}
\end{figure}

Figure \ref{fig:ADFCoils} shows the schematics of the AC susceptibility probe for the ADF, designed and built by the author in order to possess as many pick-up coils, along with a modulation coil, as the space inside the sample magnet on the LTS allowed.

All parts were made of tufnol\footnote{A hard plastic used for cryogenic applications.} or copper, except for supporting stainless steel threaded rods. The pick-up coils were made with a large bore diameter of 4~mm in order to accommodate large unprepared crystals, and the table in figure \ref{fig:ADFCoils} lists all their parameters, where resistance values were obtained at room temperature. They were wound using copper wire, of 125$\mu$m for the drive coil and 50$\mu$m for the six pick-up coils. The coils were numbered 1 to 6 from the top to the bottom of the probe, and the pairs 1,2; 3,4; 5,6 were connected in opposition. The sensitivity refers to the voltage on a coil in a field of 1~G at 1~Hz. When the probe is installed on the LTS, the pairs are normally connected to low temperature transformers located on the 1~K stage of the ADF, with an amplification factor of 100. This increases the signal to noise level of the measurement apparatus, and the noise levels range between 100~pV/$\sqrt{\textrm{Hz}}$ and 1~nV/$\sqrt{\textrm{Hz}}$ (peak to peak and normalised by the amplification factor).

Up to three samples may be placed inside the pick-up coils by fitting them inside a copper braid, which is inserted from the bottom of the apparatus, and subsequently thermally anchored to the low temperature stage. High vacuum grease is used to improve thermal contact with the braid. The thermal equilibrium of the samples with the system was verified, shown in the next section.

\subsection{Thermal equilibrium tests and calibration \label{sect:Tchecks}}

In order for the AC susceptibility measurement probe to be used reliably, we have characterised its performance. Two properties were investigated, the thermal equilibrium during experiments and the calibration of voltage to magnetic susceptibility. Both measurements were performed using the diamagnetic signal of superconductors. We present in this section the method and data which were used for these purposes.

In order to verify the thermal equilibrium of the samples during an experiment that uses temperature as a variable, we chose the measurement of the superconducting phase transition of different metals as an absolute thermometer. We used three metals of different critical temperatures, aluminium, zinc and \TOF, which possess critical temperatures ($T_c$) of 1.196, 0.875 \cite{ashcroft} and 1.430 K\footnote{Private communication from N. Kikugawa, who measured $T_c$ of the same \TOF\ samples in Kyoto.} respectively and critical fields $H_c$ of 105~G and 53~G for Al and Zn, and 750~G ($B$ $||$ $c$-axis) and 1.5~T ($B$ $||$ ab-plane) for \TOF\ \cite{kittel,mackenzieRMP}. AC susceptibility was measured on samples of these materials, which had shapes of rolled foil aligned with the oscillating field, for Al and Zn and 1~mm$^3$ cubes for \TOF, at zero sample field as a function of temperature. 

\begin{figure}[p]
	\begin{minipage}[t]{7cm}
		\begin{center}
			\includegraphics[width=7cm]{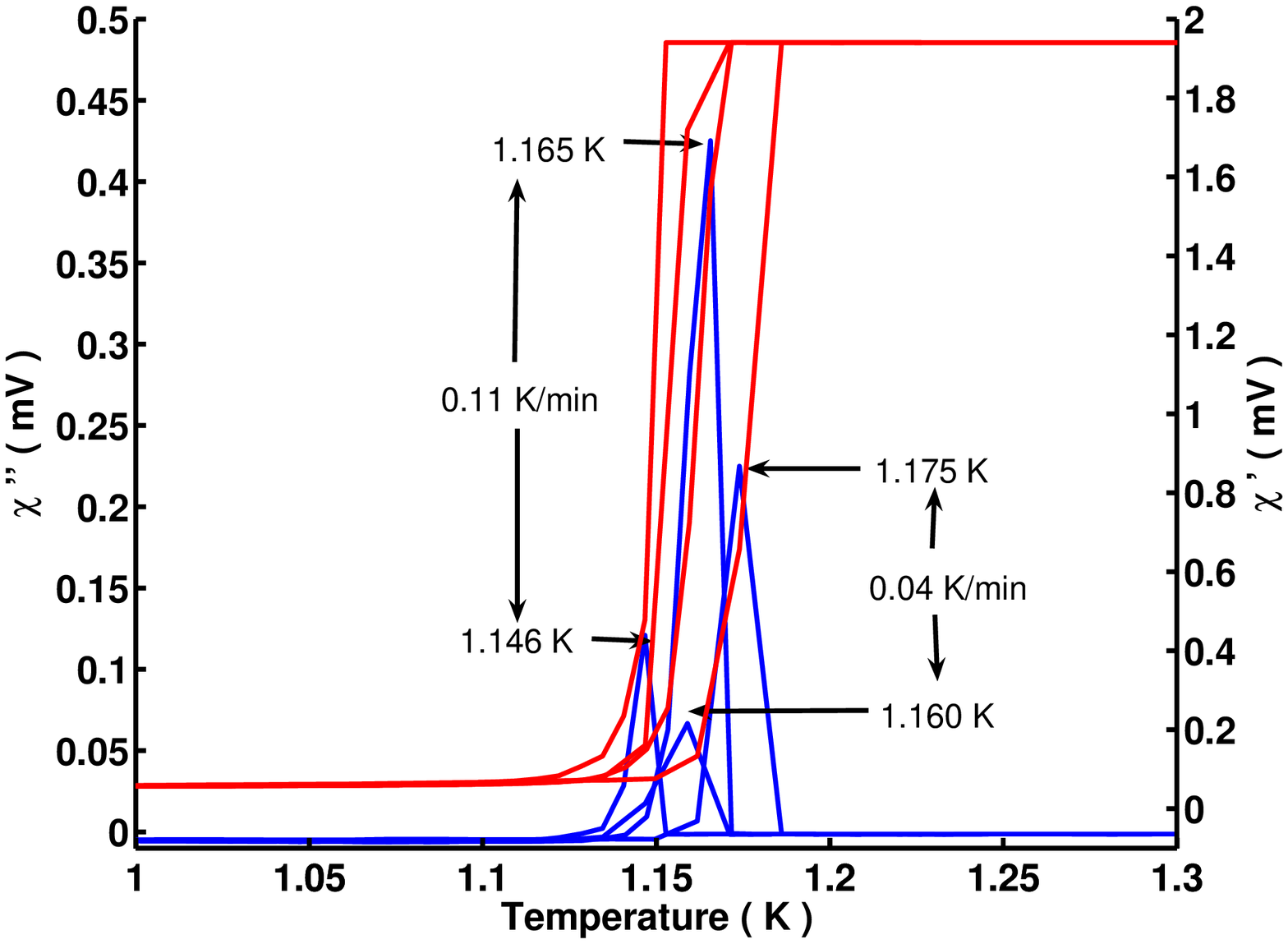}
			\includegraphics[width=7cm]{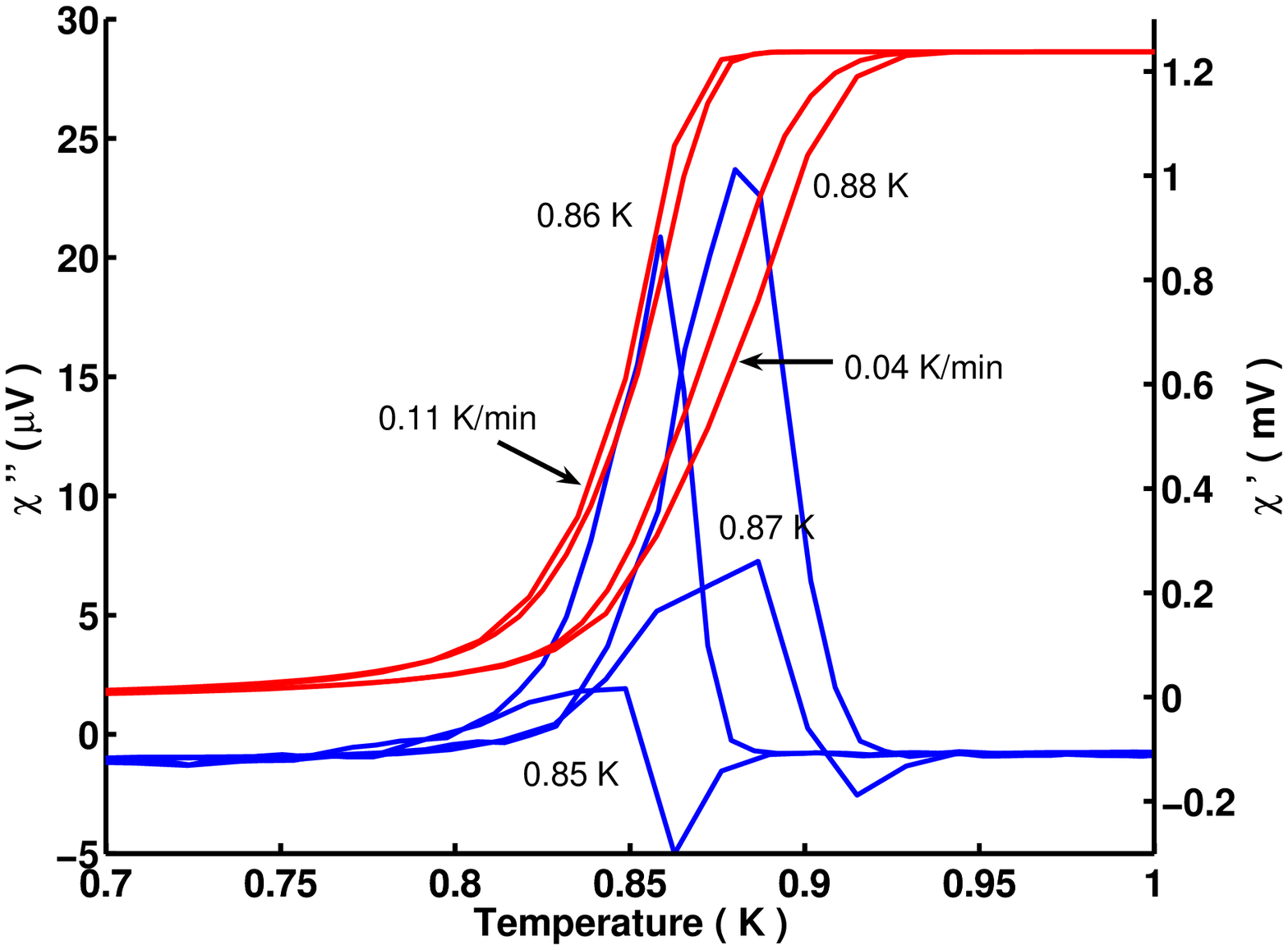}
			\includegraphics[width=7cm]{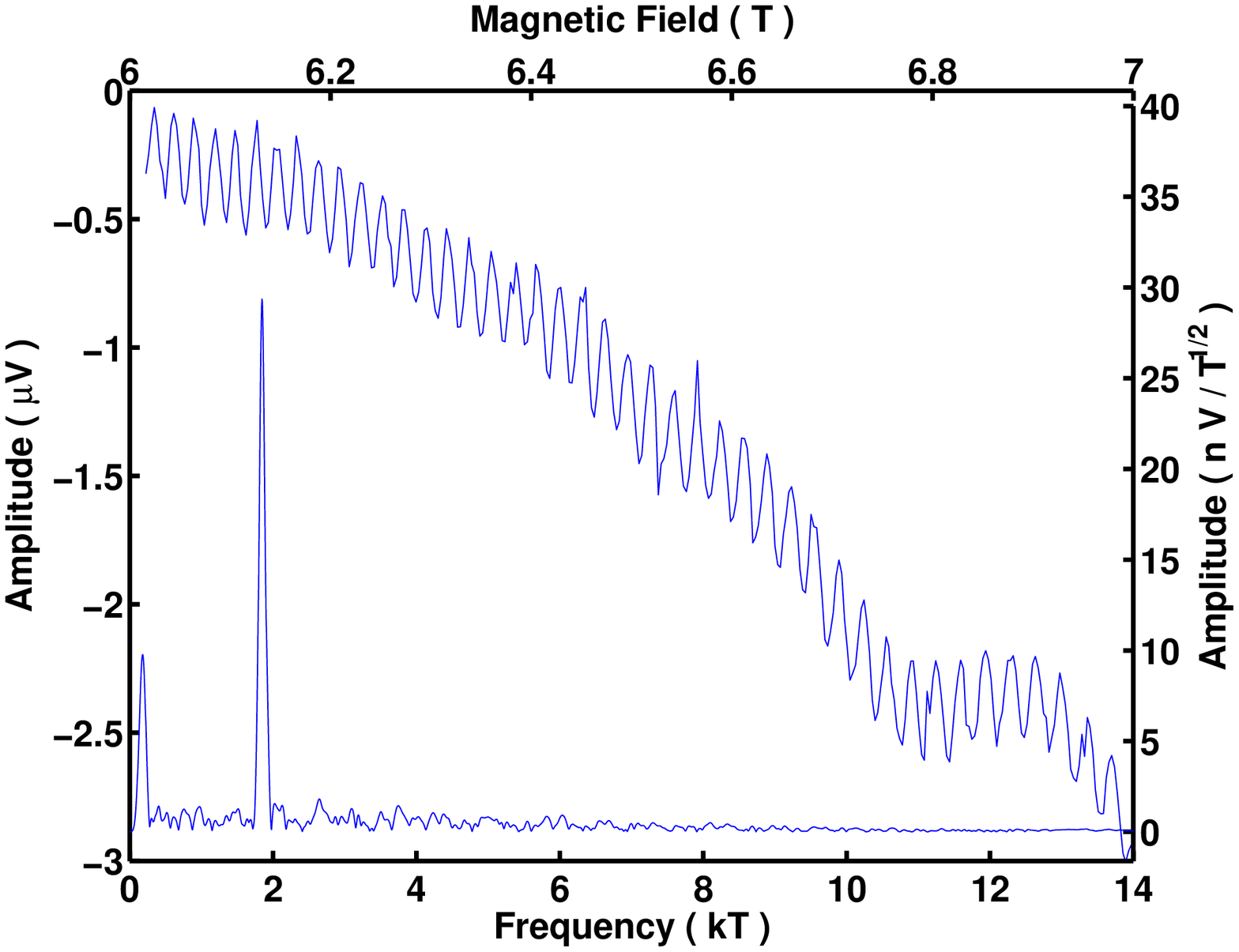}
		\end{center}
	\end{minipage}
	\hfill
	\begin{minipage}[t]{7cm}
		\begin{center}
			\includegraphics[width=7cm]{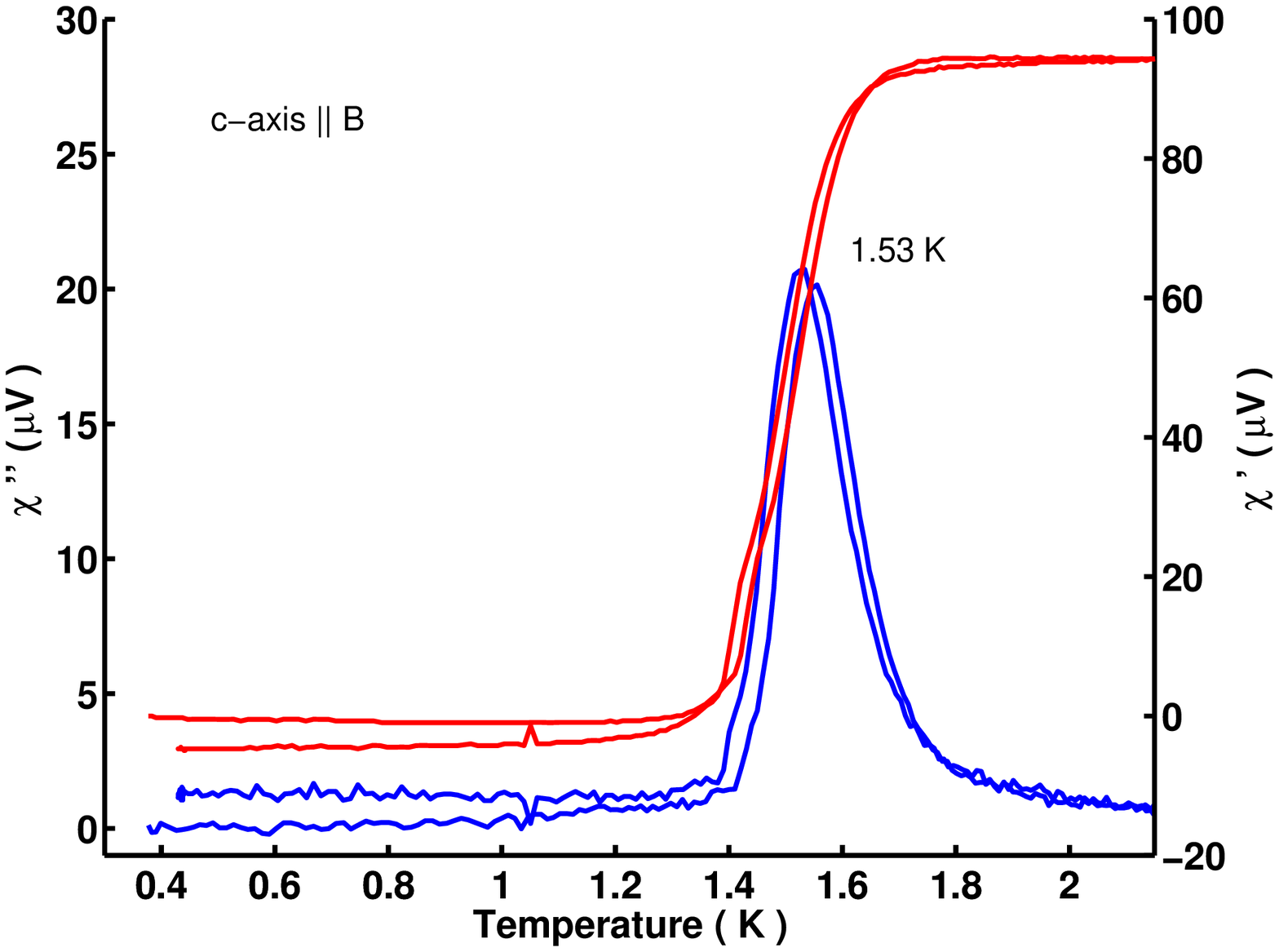}
			\includegraphics[width=7cm]{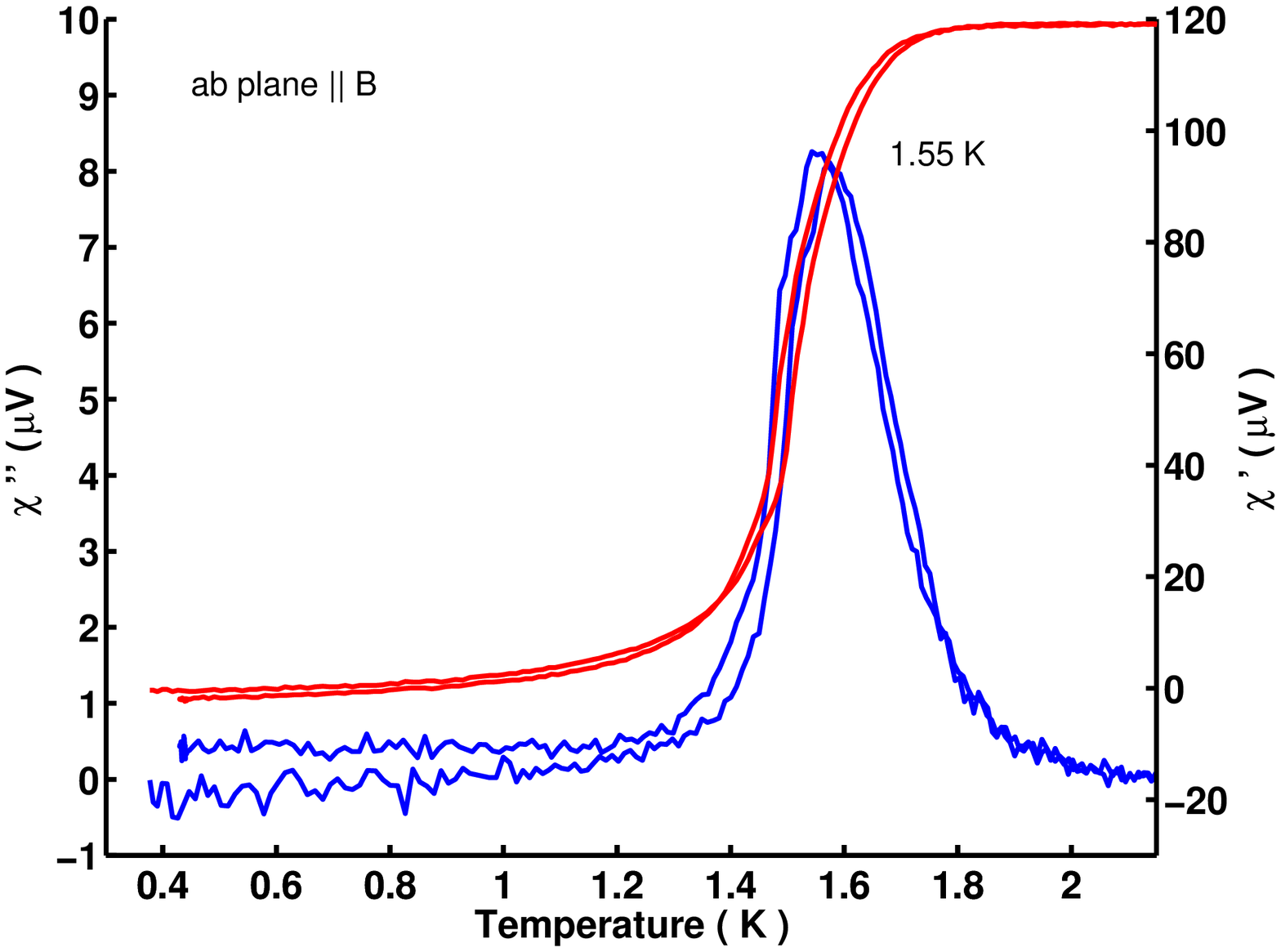}
			\includegraphics[width=7cm]{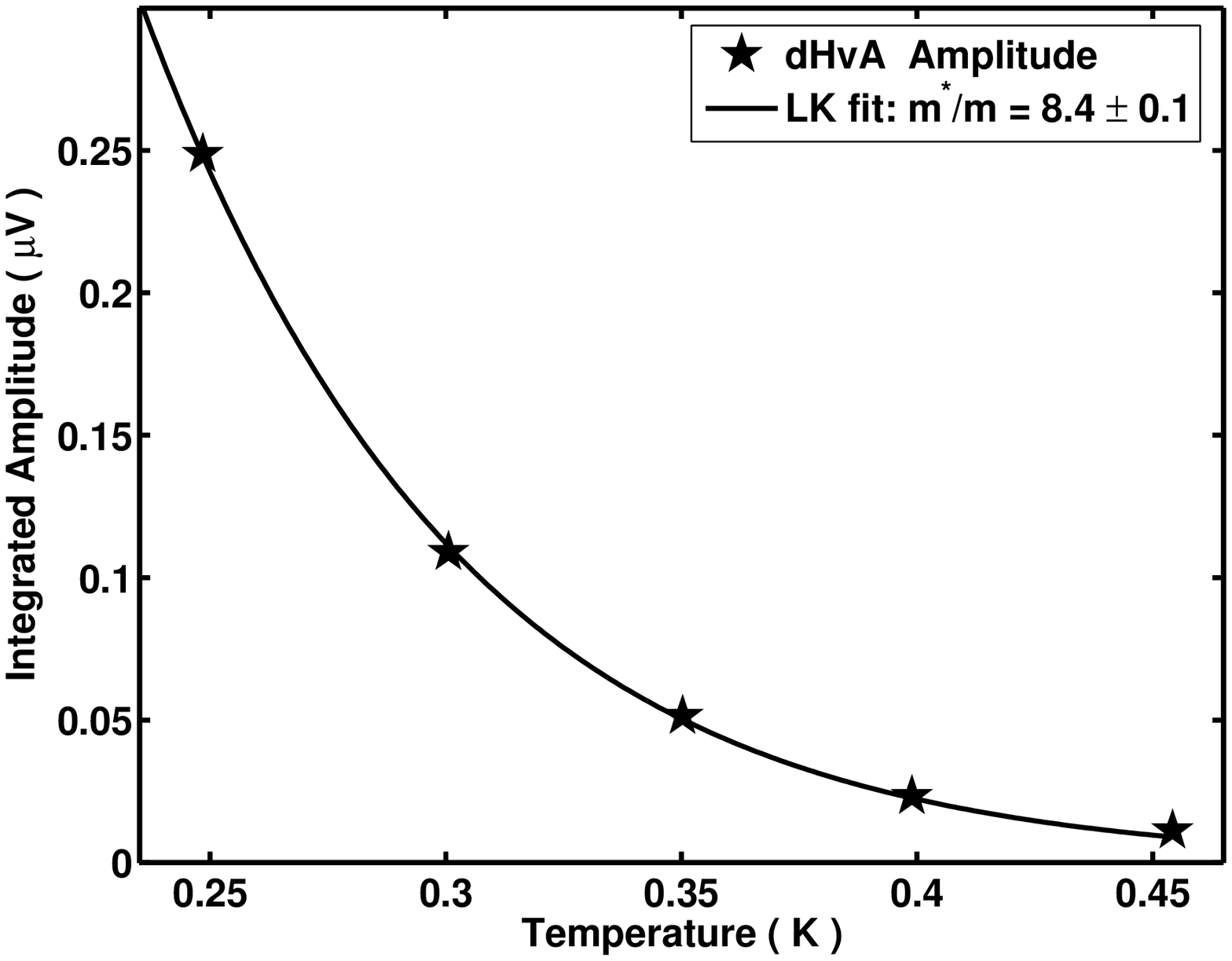}
		\end{center}
	\end{minipage}
	\caption[Thermal equilibrium tests]{ $Top$ $and$ $middle$ $left$ In-phase and out of phase susceptibility as a function of temperature for rolled foil shaped samples of high purity aluminium (top) and zinc (bottom) (99.999\%). The data was taken using two different sweeping rates, 0.11 K/min and 0.04 K/min, upwards and downwards in temperature. $Top$ $and$ $middle$ $Right$ Susceptibility as a function of temperature for samples of \TOF, one oriented with  H $||$ $c$-axis (top) and the other with  H $||$ $ab$-plane (bottom). $Bottom$ $left$ dHvA oscillations and spectrum in \TTS, sample C698A. $Bottom$ $right$ Amplitude of dHvA as a function of temperature, stars, and LK fit, solid line.}
	\label{fig: PureAlZnCoils56-06}
\end{figure}

Figure \ref{fig: PureAlZnCoils56-06} presents the in-phase (red curves) and out-of-phase (blue curves) data measured on these samples. With Al and Zn, susceptibility was measured with two different temperature sweep rates, and we obtained curves with hysteresis, indicating that a small residual magnetic field was present. We obtained critical temperatures of 1.175~K for Al and 0.880~K for Zn, within 20~mK of literature values\footnote{In a magnetic field, the superconducting transition is of first order. Hysteresis is present, and $T_c$ measured using both upwards and downwards field sweeps are lower than the zero field value (see, for instance, \cite{pobell}, pp 205-207). The value measured with an upwards field ramp is always closest to the zero field value.}, with hysteresis of 10-15~mK. In the case of \TOF, the critical temperature was of 1.53~K ( $B$ $||$ $c$-axis) and 1.55~K ($B$ $||$ ab-plane), with hysteresis of 25~mK, 120~mK higher than the value measured by N. Kikugawa in Kyoto\footnote{Later measurements on samples of \TOF\ grown in St Andrews reproduced these results.}.

From these values, we demonstrated that no significant temperature gradient was present, in the region from 0.8 to above 1.5~K, between the samples and the thermometer. Effectively, any gradient would result in the samples being warmer than the thermometer, since the thermometer is situated $between$ the cooling mechanism and the samples. The temperature at the thermometer position, from the superconducting critical temperatures, was either very near or higher than expected, suggesting a possible temperature calibration or thermometer self heating problem.

For further tests at lower temperatures, we measured the temperature dependence of dHvA in \TTS\ (figure \ref{fig: PureAlZnCoils56-06}, bottom two plots) and in \TOF\ (not shown), between 0.25 and 0.45~K (\TTS) or 0.8~K (\TOF). Using LK fits, we obtained quasiparticle mass values for the 1.8~kT peak of \TTS\ and the 3.0~kT peak of \TOF\ of $8.4 \pm 0.1$ and $3.1 \pm 0.1$ respectively. Again here, had the temperature of the samples been higher than that of the thermometers, the half-width of the LK function may have been larger, and the mass, proportional to the inverse of the width, would have been smaller than the real value. Since we know from the literature \cite{borzi, bergemann} that the masses are of $8\pm1$ and $3.3 \pm 0.3$, for \TOF\ and \TTS\ respectively, we conclude that there was no significant temperature gradient below 0.8~K.

Finally, from the diamagnetic signal of Al, we calibrated the voltages measured onto the coils into approximate real susceptibility values. We mainly used this information for calculating volume fractions of \TOF\ into \TTS\ samples (see section \ref{sect:superconductivity}), in the small fraction limit. From Figure \ref{fig: PureAlZnCoils56-06}, we obtained the amplitude of the diamagnetic signal of two samples with a demagnetisation factor close to zero\footnote{Effectively, their shape was of cylinders of rolled Al or Zn foil, which allows magnetic flux to penetrate between the layers.}. The voltage measured on a pair of compensated coils, with a sample placed into one of them, with an oscillating magnetic field $B$ is $\Delta V = A R \dot{B} \chi$, where $A$ is the total area within the turns of the coil containing the sample and $R$ the fraction of this area intersecting the sample, which is also equal to the volume fraction of the sample with respect to the coil, $R = v_{s}/v_{c}$, and $\chi$ is the susceptibility. 

The real coil volume is not an obvious quantity to measure, and for that we used the fact that when empty, the total coil area $A$ is equal to the measured sensitivity $S$ given in the table of figure \ref{fig:ADFCoils} \footnote{Since $\Delta V = \dot{\Phi} = A \dot{B_{AC}}$, we have $A = \Delta V/B_{AC}f$.} ,
\beq
\chi = {\Delta V v_{c}\over S f B_{AC} v_{s}}.
\label{eq:CoilCalibration}
\eeq
Using a voltage difference of 1.9~$mV$ for Al from figure \ref{fig: PureAlZnCoils56-06}, which includes an amplification factor of 100, an average sensitivity of 11.1~$\mu$VG$^{-1}$Hz$^{-1}$ (or, in area, 11.1$\times 10^{-2}$~m$^2$), a frequency of 71~Hz, a modulation field of 0.428~G, a mass of 52.6~mg and a density of 2.702~g/cm$^3$, we obtained a value for the volume of the coil of $v_{c} =$~0.346~cm$^3$. This was used to simplify our equation and we deduced an approximate calibration,
\beq
\chi = 3.12\times10^4 ({\textrm{cm}^3 \textrm{Hz} \textrm{G}  \over \textrm{V}}) {\Delta V \over B_{AC} f v_{s}}.\nn
\eeq
We conclude by noting that this was not used as an accurate calibration, but only as a measure the order of magnitude of volume fractions.


\section{Systematic procedure for the search for high purity samples \label{sect:search}}

\begin{figure}[ht]
\begin{center}
	\includegraphics[width=1\columnwidth]{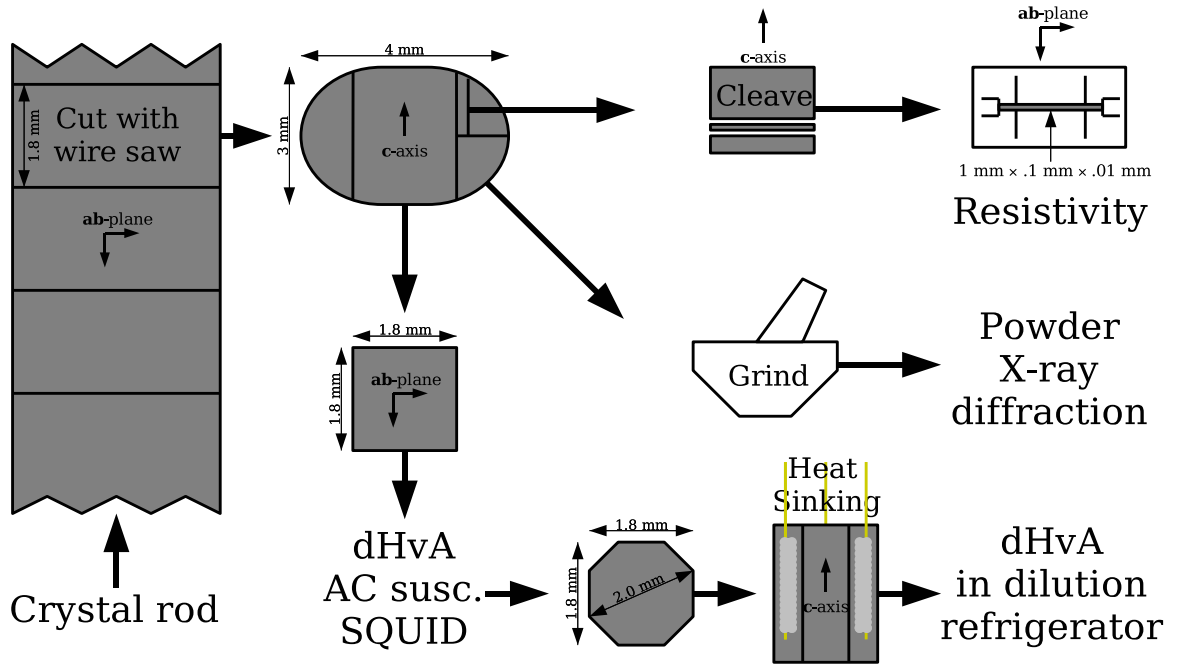}
	\caption[Recipe for cutting \TTS\ crystal rods]{Standard recipe for cutting crystal rods grown in the image furnace for the various experiments in the sorting procedure. The rods were cut into slices 1.8~mm thick. These were then further cut to obtain square prisms of dimensions $1.8\times1.8\times3$~mm. These parts were intended for the final dHvA experiment and were subjected to the susceptibility and magnetisation experiments. The off-cuts were used for measuring resistivity and the X-ray spectrum. The residual resistivity and the X-ray spectrum of the square prism were extrapolated from those of the off-cuts. The best two crystal prisms were finally polished into octagonal prisms, heat sunk and used in detailed dHvA experiments.}
	\label{fig: SampleRoad1}
	\end{center}
\end{figure}
 
We mentioned in the introduction for this chapter that the material purity requirements for dHvA experiments are very high. For this work, I was fortunate enough to interact with the materials scientist who was growing the crystals of \TTS, R.S. Perry, since this was done in the laboratory next door. The strategy we adopted was one of feedback style, where I carried out a chain of characterisation experiments on grown samples and provided the information back to R. S. Perry, in order to improve the growth parameters and restart the process. The sample sorting procedure was carried out using two types of characterisation methods. First, we will describe those used to detect different phases of the ruthenate family. Second, we will present the measurements of scattering from impurities and crystalline disorder.

In the first dHvA experiments of this project, we used an average quality sample of \TTS. We obtained relatively good quantum oscillation data, but used a very high modulation field in order to obtain a good signal to noise ratio, which led to eddy current sample heating. This sample also led us to erroneous interpretations when we discovered a dHvA frequency at 12.7~kT, which belongs to \TOF. It became clear that the sample was not purely of the bilayer type but had small inclusions of \TOF. This led us to suspect that it might also contain \FTT, \OOT, etc. Moreover, these inclusions of \TOF\ possessed a purity comparable to that of the \TTS, since we detected the oscillations well. Consequently, any good sample for dHvA should be thoroughly checked for all possible impurity phases before its oscillation spectrum can be interpreted. 

We used the magnetic properties of the different ruthenate phases to identify them quantitatively. The reason being that only \TTS\ is paramagnetic at all temperatures. The $n=\infty$, \OOT\ and $n = 3$, \FTT, members of the Ruddlesden-Popper series are ferromagnetic with Curie temperatures below room temperature, of 165 and 105~K respectively \cite{Cao,Crawford,Cao2, Kanbayasi}. We also expect the phases with $n > 3$ to be ferromagnetic, but no clear evidence of their presence was observed in this project. We used a Superconducting Quantum Interference Device (SQUID) magnetometer\footnote{Quantum Design MPMS model.} at small fields for detecting the spontaneous ferromagnetic moment of our samples. And finally, the $n=1$ phase, \TOF, is a superconductor with critical temperature of 1.48K or less \footnote{For complete information on \TOF, see \cite{mackenzieRMP}.}. We used AC susceptibility at zero field to detect the diamagnetic volume fraction of the samples. 

The next subsections will describe all the characterisations that were applied to the samples, in the same order that they were performed. The order was important for time saving. The first and quicker methods determined whether the samples were good enough to be submitted to the next. The first experiment performed on a newly grown single crystal rod was powder X-ray diffraction, which identified all the crystalline components of the material, but was also destructive, performed with a small piece ground to a fine powder, and the content of the crystal rod was extrapolated from that result, to first order. The second experiment was that of resistivity as a function of temperature, where one used a long and thin plate of crystal mounted in the way that will be described. When extrapolated to absolute zero, one obtains a value proportional to impurity and crystalline disorder scattering. Again, disorder in the larger crystal rod was extrapolated from the result. Crystal rods that had good properties were divided into as many samples as possible with the right dimensions for the dHvA apparatus of section \ref{sect:Camprobe}. Next, the samples that appeared promising were put into the ADR, for the measurement of AC susceptibility and dHvA, where we measured the \TOF\ volume fraction and the scattering, respectively. Finally, the ferromagnetic content of the samples was determined with the SQUID magnetometer.

This procedure was performed on more than 50 samples from around 10 crystal rods. 23 of them made it through the five experiments and were kept for this project. Figure \ref{fig: SampleRoad1} illustrates the way the crystal rods were divided for the various experiments. Data from the sample that we used in the early stages of this thesis project for dHvA experiments was kept as a standard with which to compare and against which to improve. The last subsection will present a summary of the data obtained with these samples. We will also demonstrate that the X-ray spectra and the residual resistivities agree only approximately with the other measurements, although they were still good starting points. Finally, the best two samples were retained for the dHvA experiments described in chapter~\ref{chapter:dHvA}, where unprecedented signal quality was achieved. Moreover, this sorting procedure turned out extremely useful to other experimentalists. In particular, five of the crystals of tables \ref{tab:327Samples1} and \ref{tab:327Samples2} were given to Angle Resolved Photoemission Spectroscopy (ARPES) experimentalists A. Tamai, F. Baumberger and co-workers, of St Andrews, who produced results that have been essential to this project through experiments performed at Stanford, presented in section \ref{sect:ZeroFieldFS}, and two were used for detailed specific heat and magnetocaloric measurements by A. Rost, in St Andrews. Others were distributed for various projects, and their characterisation data has proven a critical asset. 

\subsection{Powder X-ray diffraction}

\begin{figure}[h]
\begin{center}
	\includegraphics[width=1\columnwidth]{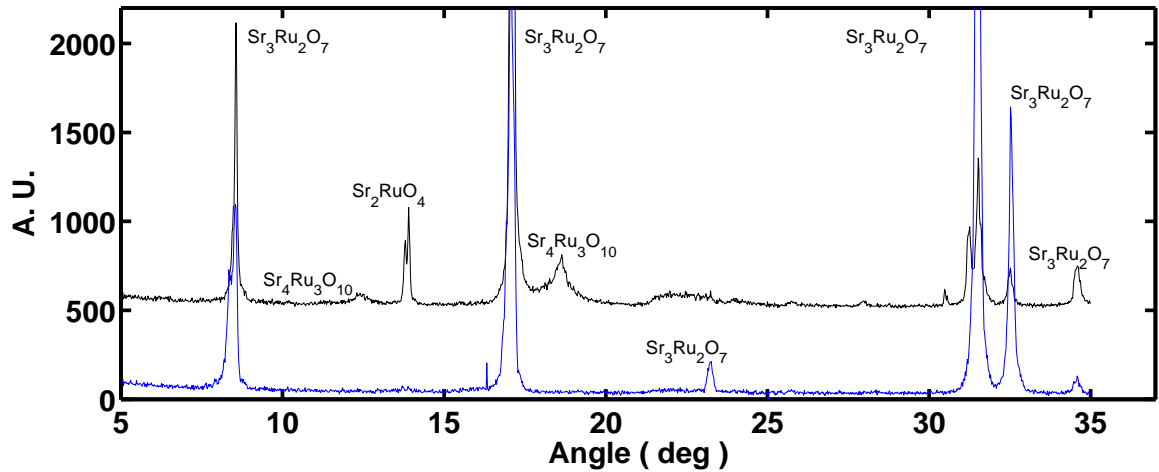}
	\caption[Powder X-ray diffraction measurements]{Two examples of powder X-ray diffraction measurements, offset for clarity. The top spectrum is of a crystal containing some detectable \TOF\ and \FTT\ intergrowths, while the bottom one was measured on a crystal where these phases were not detected.}
	\label{fig:Xrays}
	\end{center}
\end{figure}

Powder X-ray diffraction was measured first on all newly grown crystal batches. It took half an hour to perform and detected most phases of strontium ruthenate when present in large volume fractions. This method is of course destructive, so it was carried out on a small amounts of single crystal ground to powder. This measurement was always performed first in order to ascertain that the main phase produced during crystal growth was predominantly \TTS.

The measurement was produced with Philipps copper X-ray source and diffractometer, which emits at a wavelength of 1.54 \AA and detects the reflection intensity as a function of angle. The various atomic planes of the crystal reflect the X-rays at specific angles, which are tabulated in standard databases\footnote{From the International Centre for Diffraction Data.}. Figure \ref{fig:Xrays} shows two examples of spectra, offset for clarity, the first featuring only Bragg peaks associated with \TTS\ (in blue), and the other one exhibiting additional peaks, from \TOF\ and \FTT\ (in black). The chosen angle range was between 5$^{\circ}$ and 35$^{\circ}$. The reason behind this choice was that higher angles are populated with too many peaks from Bragg planes of \TTS\ and their identification is difficult. The main peaks that belong to \TTS\ between 5$^{\circ}$ and 35$^{\circ}$ were located at 8.6$^{\circ}$, 17.2$^{\circ}$, 23.3$^{\circ}$, 31.6$^{\circ}$ and 32.6$^{\circ}$. When present with \TTS, \TOF\ had only one visible peak which was easily distinguished from the others, at 13.9$^{\circ}$, and \FTT\ had two, at 12.4$^{\circ}$ and 18.6$^{\circ}$, the second of higher intensity. Unfortunately, \OOT\ featured peaks that were too close to those of \TTS\ or not intense enough to be used for its identification and none of the X-ray peaks of this phase were clearly observed. 

Since each X-ray spectrum featured an arbitrary intensity factor, only amplitude ratios between various peak intensities were meaningful. Consequently, when measuring the amount of \TOF\ or \FTT\ with this method, we took the ratio $R$, in the first case, of the peak at 8.6$^{\circ}$ of \TTS\ with that of the one at 13.9$^{\circ}$ of \TOF, and in the second case, the peak at 8.6$^{\circ}$ again with the peak at 18.6$^{\circ}$ of \FTT. Note that among the 23 samples that were kept for later experiments, only one featured a content of \FTT\ detectable with powder X-ray diffraction, C6102A, shown in figure \ref{fig:Xrays}. It thus meant that all the measured samples contained relatively small amounts of this type of intergrowths.

\subsection{Residual resistivity}

\begin{figure}[ht]
	\begin{minipage}[t]{9cm}
		\begin{center}
		\includegraphics[width=9cm]{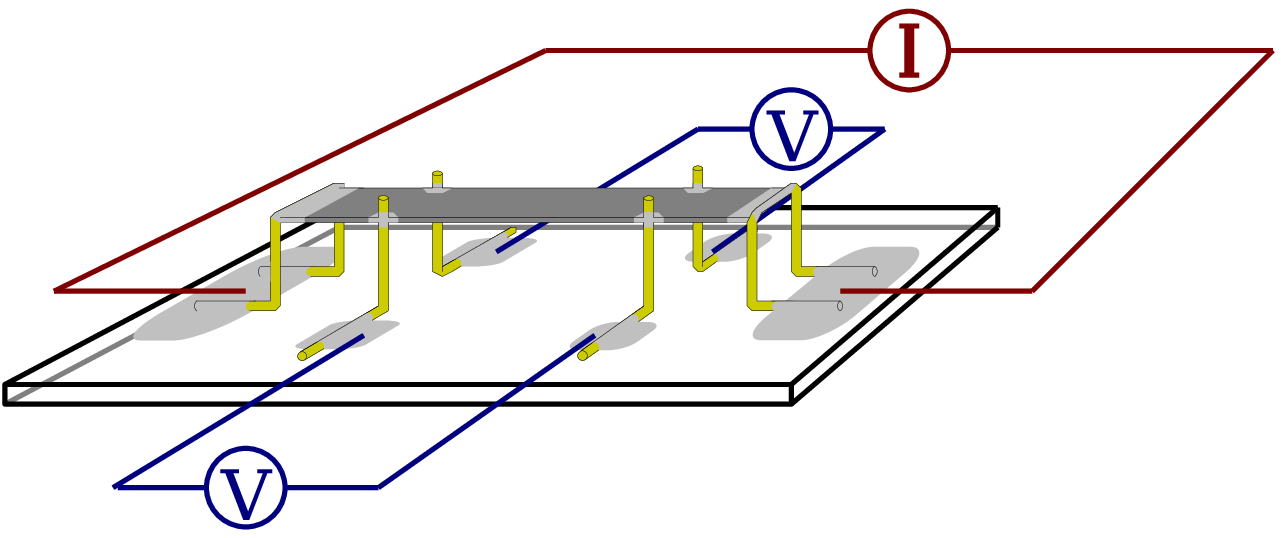}

		\end{center}
	\end{minipage}
	\hfill
	\begin{minipage}[t]{4.5cm}
		\begin{center}
		\includegraphics[width=4.5cm]{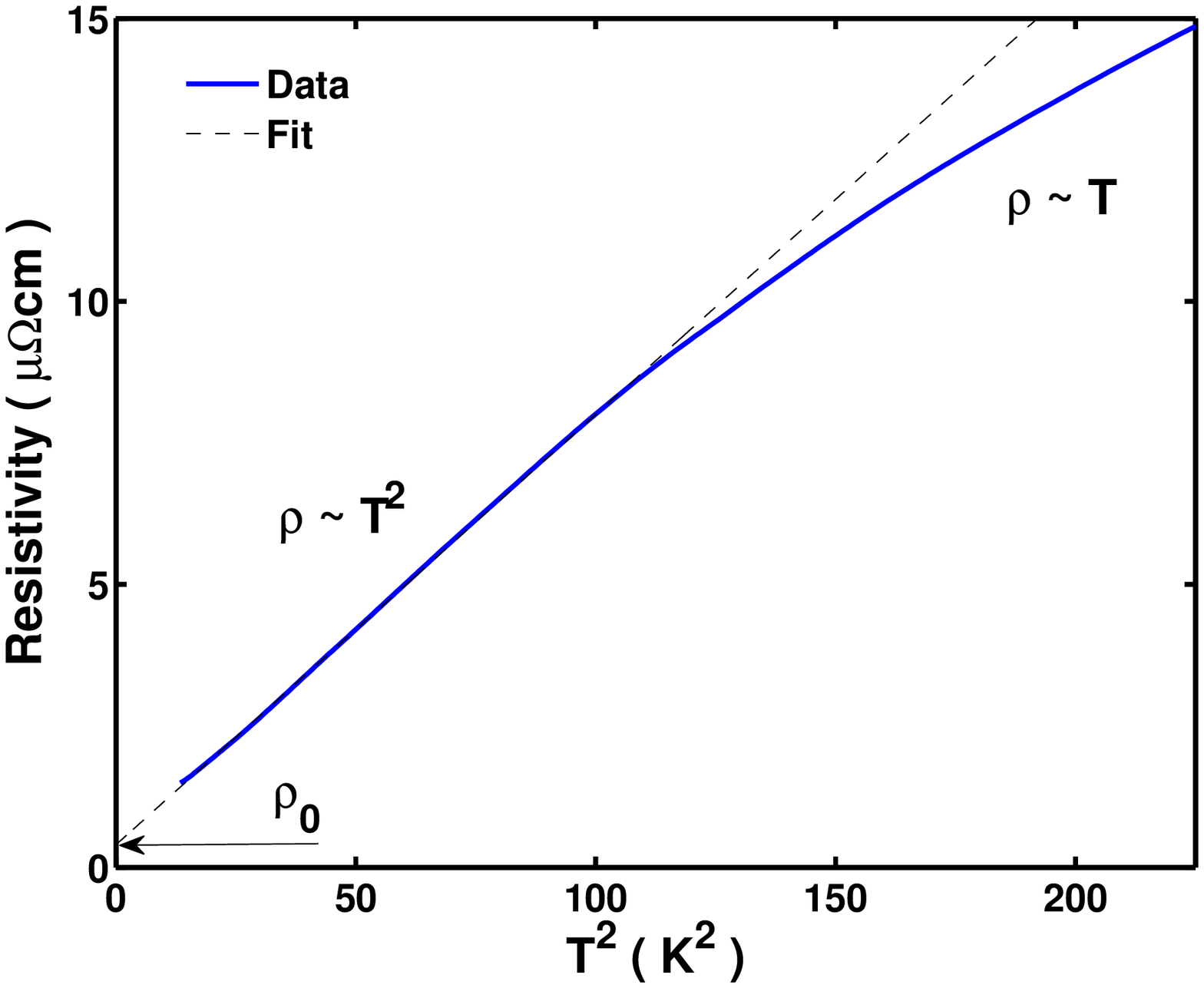}
		\end{center}
	\end{minipage}
	\caption[Residual resistivity measurement]{$Left$ Illustration of a mounted resistivity sample in a four point resistivity measurement configuration. The sample is oriented with the $ab$-plane in the horizontal direction. $Right$ Typical measurement of the resistivity as a function of temperature (blue), plot as a function of $T^2$, with a linear regression of expression \ref{eq:residualres} (broken line). Its extrapolation to zero temperature yields the residual resistivity $\rho_0$.}
	\label{fig:RRSample1}
\end{figure}

The standard method to determine disorder in a layered metallic sample is to measure its residual resistivity, in the $ab$-plane direction. This configuration was used since the $ab$-plane is the direction in which \TTS\ cleaves readily into thin plates, making it possible to obtain samples of very small cross-section\footnote{Plates of thickness 20-100 $\mu$m were obtained}. As was discussed in section \ref{sect:FermiLiquid}, in a highly correlated metal, resistivity as a function of temperature at low temperatures may be expressed as 
\beq
\rho(T) = \rho_0 + AT^2.
\label{eq:residualres}
\eeq
The $T^2$ term is fundamental to the Fermi liquid theory, the factor $A$ depending on parameters of the metallic state like the quasiparticle mass $m^{*2}$ and the Fermi wave vector $k_F$. In this case, we were interested in the constant $\rho_0$, which depends on static lattice disorder,
\beq
\rho_0 = {m^*\Gamma_0 \over ne} = {\hbar \bar{k_F} \over \ell n e},\nn
\eeq
with $n$ the density of carriers and $\Gamma_0$ the scattering rate of the electrons due to lattice imperfections and impurities, $\ell$ the associated mean free path and $\bar{k_F}$ the average Fermi vector. It is standard to measure disorder in units of $\mu\Omega$~cm, and one simply needs to adjust the low temperature data with a linear regression of $T^2$, extrapolating the data to absolute zero. Previous crystal growth experience led us to believe that only \TTS\ samples that featured a $\rho_0$ of $1 \mu\Omega$~cm or less were likely to exhibit good quantum oscillations. The best known \TTS\ samples had $\rho_0 \simeq 0.25 \mu\Omega$~cm.

An illustration of a sample mounted in a four point resistivity measurement in the $ab$-plane configuration is given in the left side of figure \ref{fig:RRSample1}. Very thin single crystal were cleaved from bulk samples, and then mounted on quartz substrates such that they were held in mid-air by thin gold wires. 50 $\mu m$ thick gold wire was used for current and voltage contacts, fixed with high temperature cured silver epoxy, shown in light grey\footnote{Type Dupont 6838.}. This system allows for the relaxation of thermal stress during cooldown through the flexibility of the gold wires. In order to obtain a good signal to noise ratio, the sample geometry needed a very high aspect ratio. Typical dimensions were of 2~mm $\times$ 0.2~mm $\times$ 0.02~mm. Six contacts were used and two resistivity measurement were made per sample, one on each side of the crystal. The quality of the electrical contacts was determined by comparing the measurement performed on both sides, which normally yielded the same $\rho_0$. The most common problem that arose was a $c$-axis contribution to the resistivity, where bad quality electrical contacts at the current leads force the current to travel in the $c$-axis direction, for which the resistivity tensor component is around 300 times higher at 0.3~K \cite{Ikeda2000}, and leads to erroneous residual resistivities. The way by which this was avoided was to make sure all contacts covered the thickness of the sample.

The measurements were performed in a simple He flow cryostat, which can cool down from room temperature to 3.5~K. For \TTS, this base temperature was low enough to obtain a reasonable amount of data in the $T^2$ dependence region of $\rho(T)$ that enabled us to extrapolate it to absolute zero. The measured resistance was converted to resistivity using the room temperature value of 232 $\mu \Omega$~cm. The right side of figure \ref{fig:RRSample1} shows a typical resistivity measurement and $T^2$ extrapolation. 

\subsection{Zero field AC susceptibility \label{sect:superconductivity}}

\begin{figure}[h]
\begin{center}
	\includegraphics[width=0.8\columnwidth]{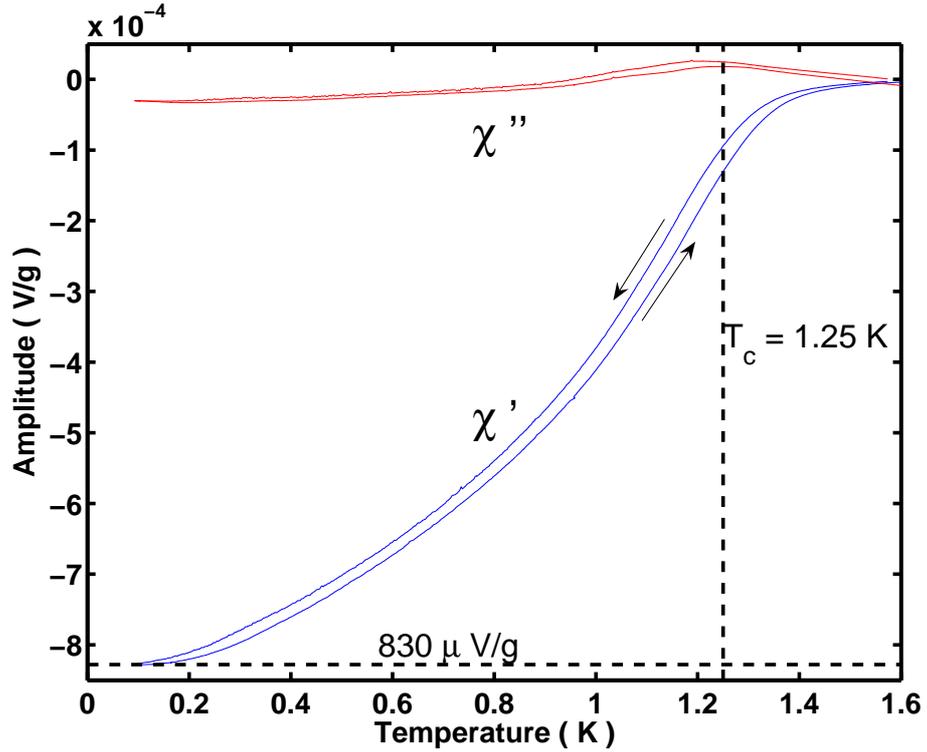}
	\caption[Zero field AC susceptibility measurement]{Typical zero field AC susceptibility measurement. The top curve (red) is proportional to the out-of-phase susceptibility $\chi ''$, while the lower one (blue) is proportional to the in-phase susceptibility $\chi '$. The data was divided by the mass of the sample. The peak in $\chi ''$ yields the critical temperature of the superconducting transition, while the amplitude of the diamagnetic signal, the value at which $\chi '$ saturates towards low temperature, is, to first order, proportional to the superconducting volume fraction of the sample. The arrows indicate the direction of the temperature variation: it was first lowered from 1.6~K to .1~K, and then raised again to 1.6~K.}
	\label{fig: ACsuscC698I}
	\end{center}
\end{figure}

The most sensitive method to determine the presence of the superconducting phase \TOF\ in samples that we intend to use for dHvA experiments is to measure AC susceptibility at zero magnetic field, as a function of temperature below 2~K. The transition from the metallic state to superconductivity is observed by a jump in the in-phase AC susceptibility ($\chi '$) and a peak in the out-of-phase AC susceptibility ($\chi ''$) (see figure \ref{fig: PureAlZnCoils56-06} for AC susceptibility data of pure \TOF). In the limit where the demagnetisation factor of these inclusions is near one, the amplitude of the jump in $\chi '$ corresponds to the volume fraction of superconducting material. In our case here, since we were searching for crystals without inclusions of \TOF, we were only looking for the order of magnitude of the volume fraction.

The measurements were done with the opportunity that it was made in the same apparatus as for the dHvA experiments of the next section, and consequently, with the same cooldown of the ADR. It also implied that the samples were positioned with their $c$-axis parallel to the axis of the coils, as is required for dHvA. Due to time constraints, we were not able to measure each sample with $B$ parallel to the $ab$-plane. We expected, from polarised light optical microscopy experiments done by Fittipaldi $et$ $al.$ \cite{fittipaldi}, \TOF\ inclusions to have a geometry of flat islands perpendicular to the $c$-axis, and the $c$-axis of \TOF\ to be in the same direction as for the \TTS. Due to demagnetisation effects, this geometry made our calculation of the superconducting fraction difficult, but in order to obtain the order of magnitude of the content of \TOF, we assumed that the diamagnetic signal was proportional to the volume fraction to first order. 

The parameters of these experiments were the following. The magnetic field was set to zero, which meant that only a remanent field of less than 50~G was present\footnote{The remanent field of the magnet was due to the use of field for the dHvA experiments that were also performed at each cooldown.}. The temperature was swept from the 1-K pot temperature to the base temperature and back to the 1-K pot temperature of the ADR at a rate of 25~mK/min. Finally, the samples were excited with a modulation field of 0.428~G at a frequency of 71~Hz.

In order to estimate an order of magnitude of the volume fraction from the amplitude of the diamagnetic signal measured in the $c$-axis direction, some assumption had to be made about the demagnetisation factor $n$. We supposed the following: the average demagnetisation factor in the $c$-axis direction is to first order roughly the same in all samples of interest or, in other words, that the shape of the inclusions had a distribution that was statistically the same. This is reasonable for samples of small \TOF\ content, according to the polarised light experiments. The second assumption we used was that the demagnetisation factor in the $ab$-plane direction is $n=0$, such that in that direction we measured the real susceptibility, and hence the real volume fraction. Then, measuring the AC susceptibility of only one sample in both directions should provide us with an approximate proportionality factor for all the other samples, and thus with an upper boundary value for the volume fraction of superconducting material. 

\begin{figure}[h]
\begin{center}
	\includegraphics[width=0.8\columnwidth]{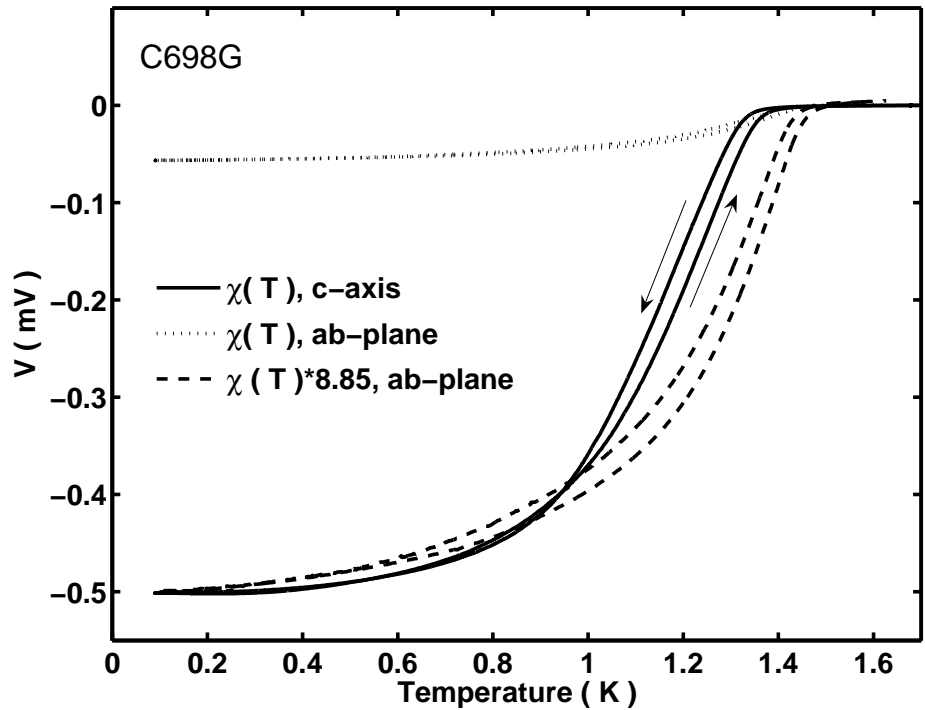}
	\caption[Superconducting volume fraction measurement]{Susceptibility measurements as a function of temperature performed on sample C698G, of \TTS, with the excitation field aligned with the $c$-axis and the $ab$-plane of the crystal. The demagnetisation effect produced a larger difference in voltage when the field was aligned with the $c$-axis by a factor 8.85.}
	\label{fig: ACsuscDemagFactor}
	\end{center}
\end{figure}

We measured the AC susceptibility in both directions of sample C698G (figure \ref{fig: ACsuscDemagFactor}), which yielded a ratio at base temperature of $n+1$ = 8.85 for the $c$-axis value with respect to the $ab$-plane value. Taking the normalisation factor of section \ref{sect:Tchecks}, eq. \ref{eq:CoilCalibration}, the values of AC field and frequency, and transforming the volume into a mass of \TTS\ using the density of 6.117g/cm$^3$, we obtained
\beq
\chi = 62.7 {\Delta V\over (n+1) m} = 7.1 {\Delta V \over m},\nn
\eeq
which ranges between 0 and -1. This corresponds effectively to a fraction of the real susceptibility of a sample entirely superconducting, and consequently to the volume fraction of \TOF\ into \TTS. Since a large error is associated with our assumptions regarding the demagnetisation factor, these could be wrong by up to a factor 2 or more, but we considered reasonable their order of magnitude.

\subsection{The de Haas van Alphen effect as a measure of disorder\label{sect:dHvAADF}}

\begin{figure}[ht]
\begin{center}
	\includegraphics[width=1\columnwidth]{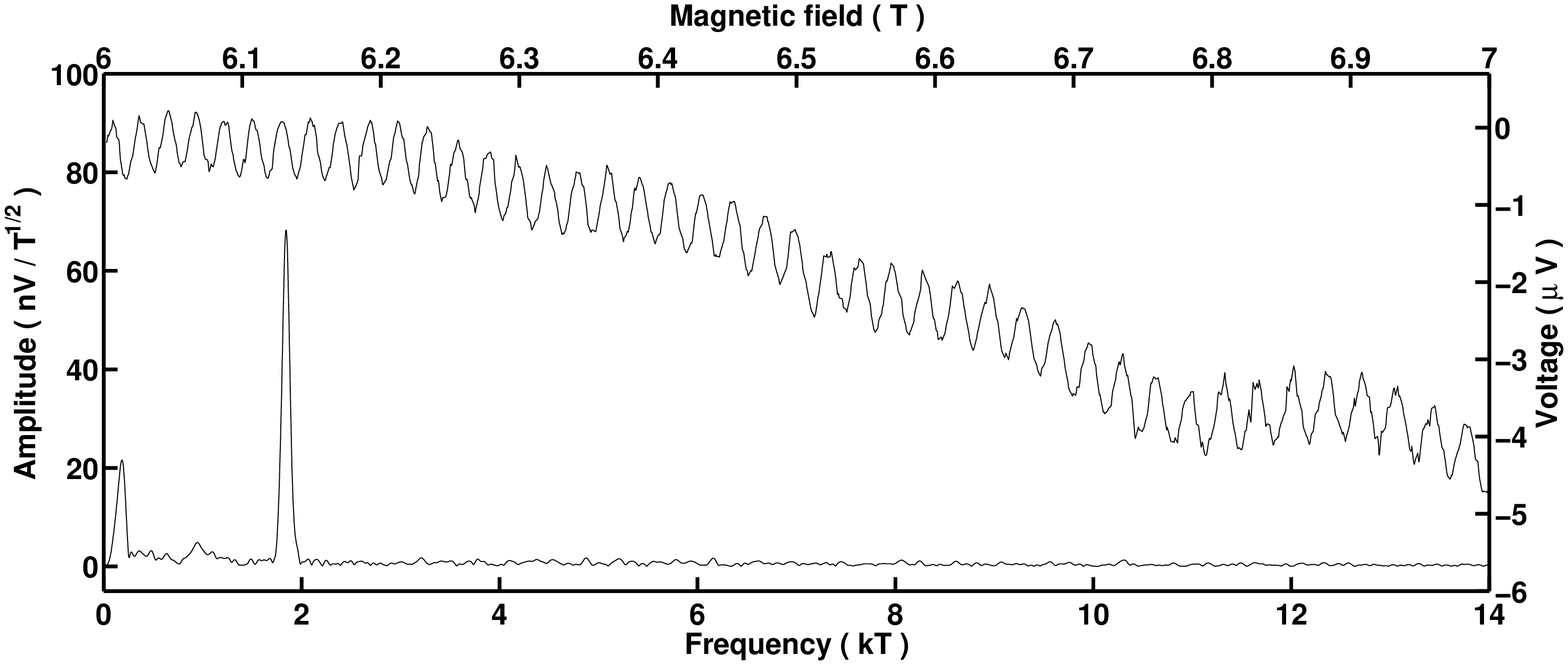}
	\caption[Disorder measurement using dHvA]{Typical dHvA measurement performed in the ADR. The top curves, with axes at the right and top, feature the oscillations in the measured voltage, which is proportional to the in-phase susceptibility. The lower curve is the absolute value of the Fourier spectrum, with axes at the left and bottom. The integral of the peak at 1.8~kT in the spectrum divided by the sample masses yields an intensity for the oscillations of 100~nV/g in this case, the mass being of 80~mg.}
	\label{fig:dHvA+FFT_C698I}
	\end{center}
\end{figure}

The best way to determine in a comparative way the disorder in large crystals of high quality is to perform a dHvA experiment on them. Using the AC susceptibility system described in section \ref{sect:ADFprobe}, we systematically measured dHvA on samples using the same parameters such that the amplitude of dHvA was comparable between runs. In order to be able to compare the quality of the various samples, we used the square root of the integral of the power spectrum, divided by the mass, as a measure of the amplitude, which according to eq. \ref{eq:D2}, depends exponentially on the inverse of the mean free path. We obtained values between zero and 100~nV/g. Note that these values were always measured with the $c$-axis parallel to the magnetic field.

Intensity values of 10~nV/g were usually close to the noise levels of our susceptibility system\footnote{See section \ref{sect:ADFprobe} for details}. Effectively, as we always used a time constant for detection of one second, transformers with amplification factors of 100 and sample masses of about 50~mg, this corresponded to intensities between .5 and 10~nV/g. When only noise was observed we quoted the intensity value with a $less$ $than$ symbol but still retained the integrated intensity in the usual frequency region. The parameters used for measurement were the following. We applied a modulation field of 0.428~G at a frequency of 71~Hz. We used the field region between 6 and 7~T, where, for \TTS, the oscillations have the highest amplitude, and varied the DC field at a rate of 20 mT/min. We verified carefully that these parameters did not cause eddy current heating, as was described in section \ref{sect:Tchecks} (see, in particular, figure \ref{fig: PureAlZnCoils56-06}).

\subsection{Magnetisation \label{sect:SQUID}}

\begin{figure}[p]
\begin{center}
	\includegraphics[width=1\columnwidth]{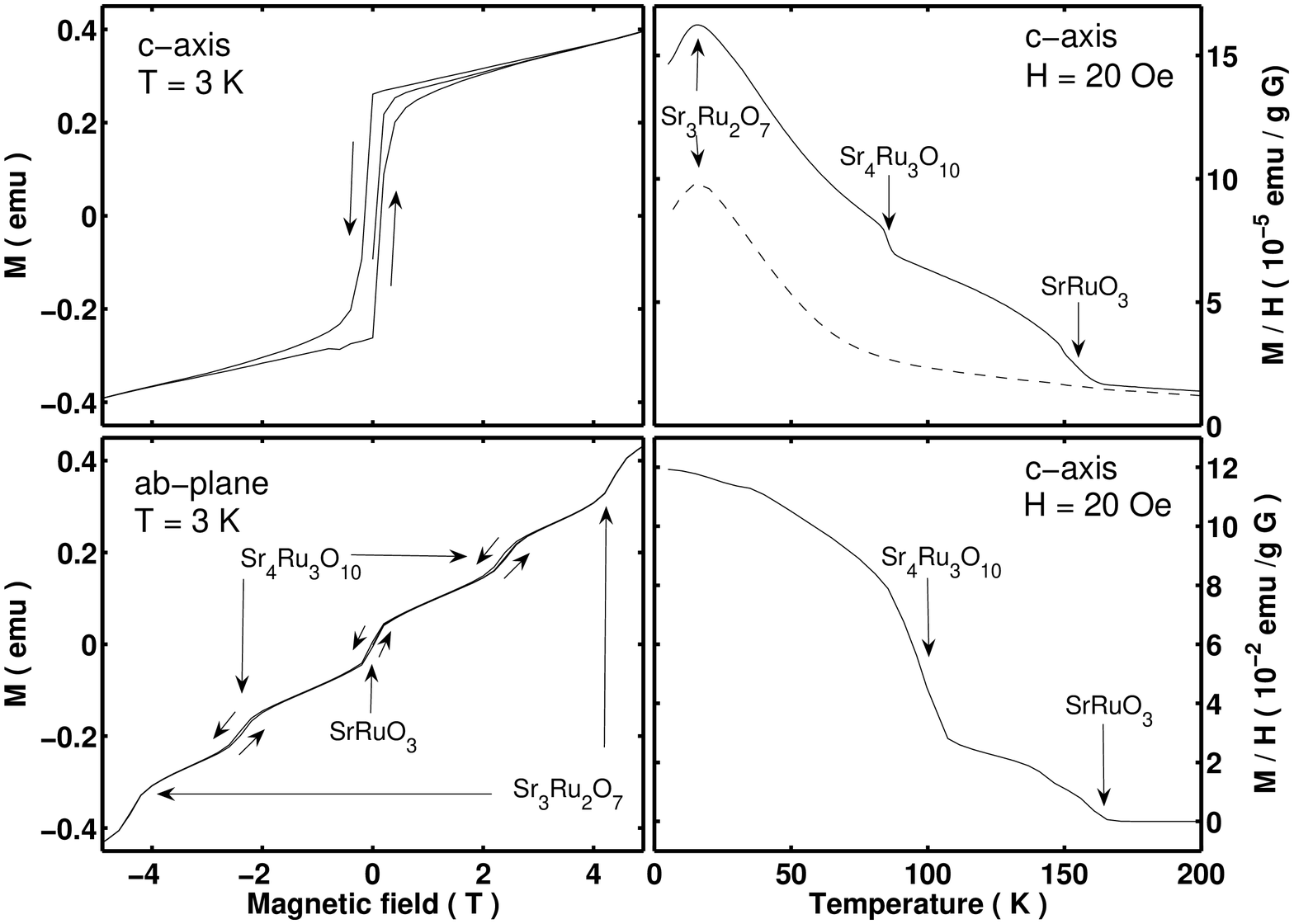}
	\caption[Magnetisation measurement of \OOT\ and \FTT]{Magnetisation measurements performed with the SQUID magnetometer. $Left$ Hysteresis loops measured on sample C6102A, which featured large amounts of \OOT\ and \FTT. The top graph is for a measurement with the magnetic field along the $c$-axis, and the bottom one for the $ab$-plane. Arrows indicate regions of hysteresis. $Top$ $right$ Typical SQUID mass susceptibility measurements as a function of temperature, with the sample field cooled at a magnetic field of 20~G. The dashed curve is for a very pure sample of \TTS, while the solid one was measured on a sample featuring small amounts of \OOT\ and \FTT. $Bottom$ $right$ Field cooled mass susceptibility as a function of temperature for sample C6102A, with a magnetic field of 20~G.}
	\label{fig:SquidExample2}
	\end{center}
\end{figure}

\begin{figure}[t]
\begin{center}
	\includegraphics[width=.8\columnwidth]{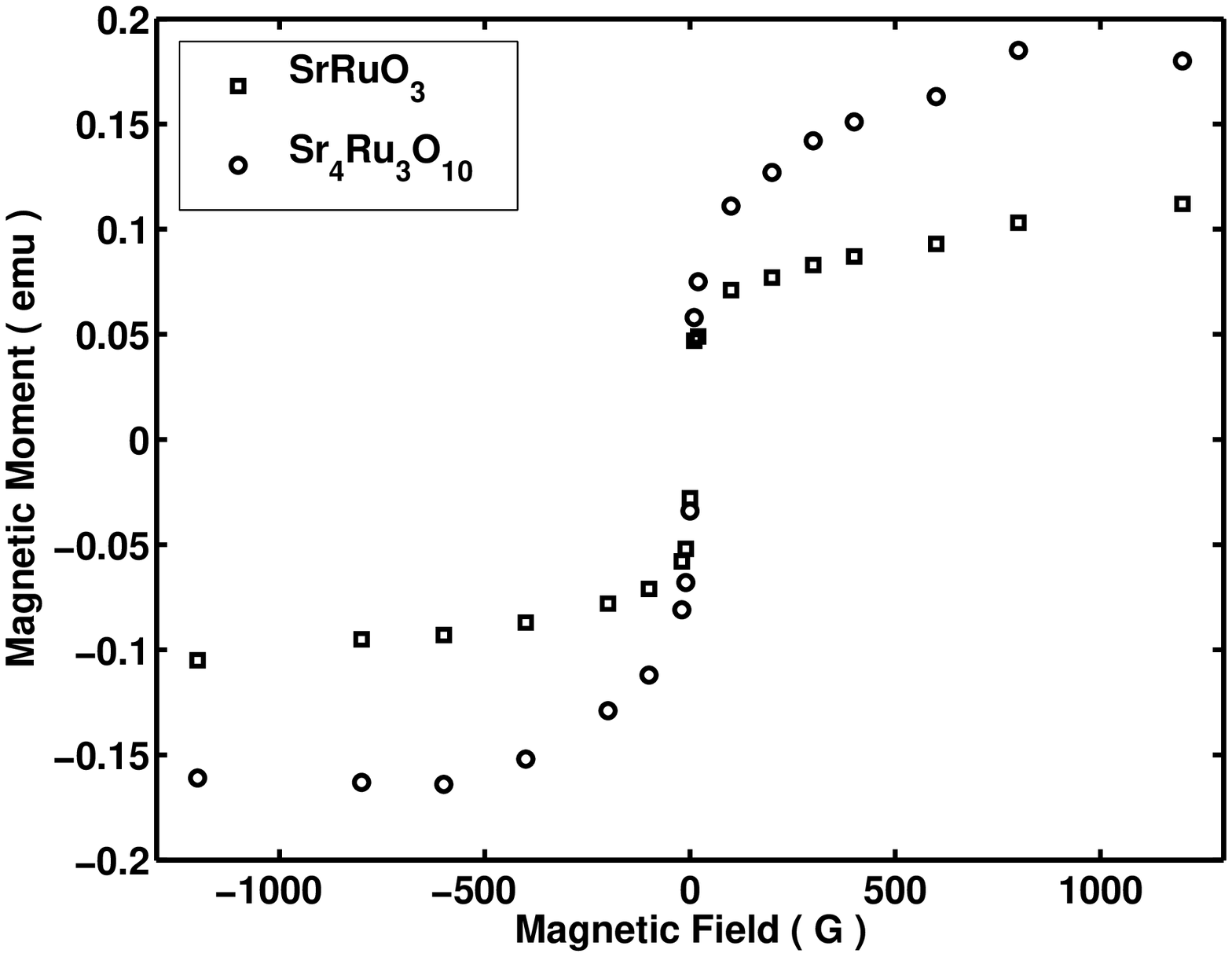}
	\caption[Evaluation of Ferromagnetic volume fractions]{Values of the extrapolation to 0~K of the field cooled spontaneous magnetisation performed at various magnetic fields on sample C6102A. The squares are for \OOT\ and the circles for \FTT.}
	\label{fig:FerroSignal0K}
	\end{center}
\end{figure}

Measuring the magnetisation offered the easiest way, in our laboratory, to detect ferromagnetic signals in samples with extremely high sensitivity. We had the opportunity to use a Quantum Design MPMS Superconducting Quantum Interference Device magnetometer (SQUID). This system can reach temperatures between room temperature and 2~K, and fields of up to 5~T. It allows, among other things, to measure the DC magnetisation as a function of temperature or magnetic field.

When measuring the magnetisation of our samples as a function of temperature, starting at room temperature and decreasing to 5~K, at small fields, we generally observed two super-linear rises of the total magnetic moment, at around 105~K and 165~K (see figure \ref{fig:SquidExample2}, top right, solid line). The first is the Curie point of \FTT\ \cite{Cao} and the second, that of \OOT\ \cite{Cao2}. These were superimposed on the paramagnetic signal of \TTS, which features a peak at 16~K \cite{Ikeda2000} (see figure \ref{fig:SquidExample2}, top right, dashed line). This moment being smallest at low fields, it often had lower magnitude than the spontaneous ferromagnetic moment of small inclusions \OOT\ and \FTT. It was thus relatively easy in this way to detect their presence quantitatively, and for that we only needed to calibrate the amplitude of the two ferromagnetic signals into volume fractions of samples. Figure \ref{fig:SquidExample2}, bottom right, shows the field cooled spontaneous magnetisation of sample C6102A, the one that exhibited the ferromagnetic signals of highest amplitude, and which was used for the calibration of the volume fractions.

To further demonstrate the presence of ferromagnetic materials in our samples, we performed hysteresis loops. The left side of figure \ref{fig:SquidExample2} shows the one performed on sample C6102A, for magnetic fields aligned with the $c$-axis (top) and the $ab$-plane (bottom). In the first case, the ferromagnetic signal for both \TTS\ and \OOT\ were indistinguishable and saturated below 1T, indicating that the easy axis of \OOT\ was probably not in the same direction as that of \FTT, but lay in the direction of the bilayer of \TTS. In the second case, it was \FTT, as we know from the work of Cao and co-workers \cite{Cao}, that featured hysteresis loops near 2.5~T, making its identification easy. We saw three sharp rises in magnetic moment, at 0, 2.5 and above 4~T. They correspond, in order, to \OOT\ \cite{Cao2}, \FTT\ and \TTS, \cite{perry1}. The first two produced hysteresis, but not the third one, as one expects for two ferromagnetic and a paramagnetic materials. 

In order to calibrate the measurement of the field cooled spontaneous magnetisation as a function of temperature for all samples, we used sample C6102A, for which we could neglect the signal of \TTS. We used a simple extrapolation to zero temperature of the total moment of each ferromagnetic phase. Figure \ref{fig:FerroSignal0K} shows how these moments evolved with the magnetic field at which the samples were cooled, for both \OOT\ and \FTT. One sees that the data are symmetric, indicating that the remanent field of the SQUID magnet is very small, and both data sets saturated above 0.1~T. Using saturated moments of 0.112~emu (\OOT) and 0.180~emu (\FTT), and those given by Cao (\cite{Cao,Cao2}) of respectively approximately 1.0 $\mu_B$/Ru and 1.1 $\mu_B$/Ru, which correspond to 23.6~emu/g and 22.7~emu/g, we obtained total masses of 4.7~mg of \OOT\ and 7.9~mg of \FTT. These corresponded to 9\% and 16\% of the total sample mass.

From the lower right graph of figure \ref{fig:SquidExample2}, we extrapolated the moments to 0K, and obtained 0.049~emu for \OOT\ and 0.075~emu for \FTT. With the masses previously calculated, we obtained the values of magnetic moment at 20~G of 12.9~emu/g and 10.3~emu/g. These are the numbers we used to calibrate the volume fractions of \OOT\ and \FTT\ in all the other samples.

\subsection{Complete quantitative results of the search \label{sect:samplelist}}

\begin{table}[p]
\begin{center}
	\includegraphics[width=1\columnwidth]{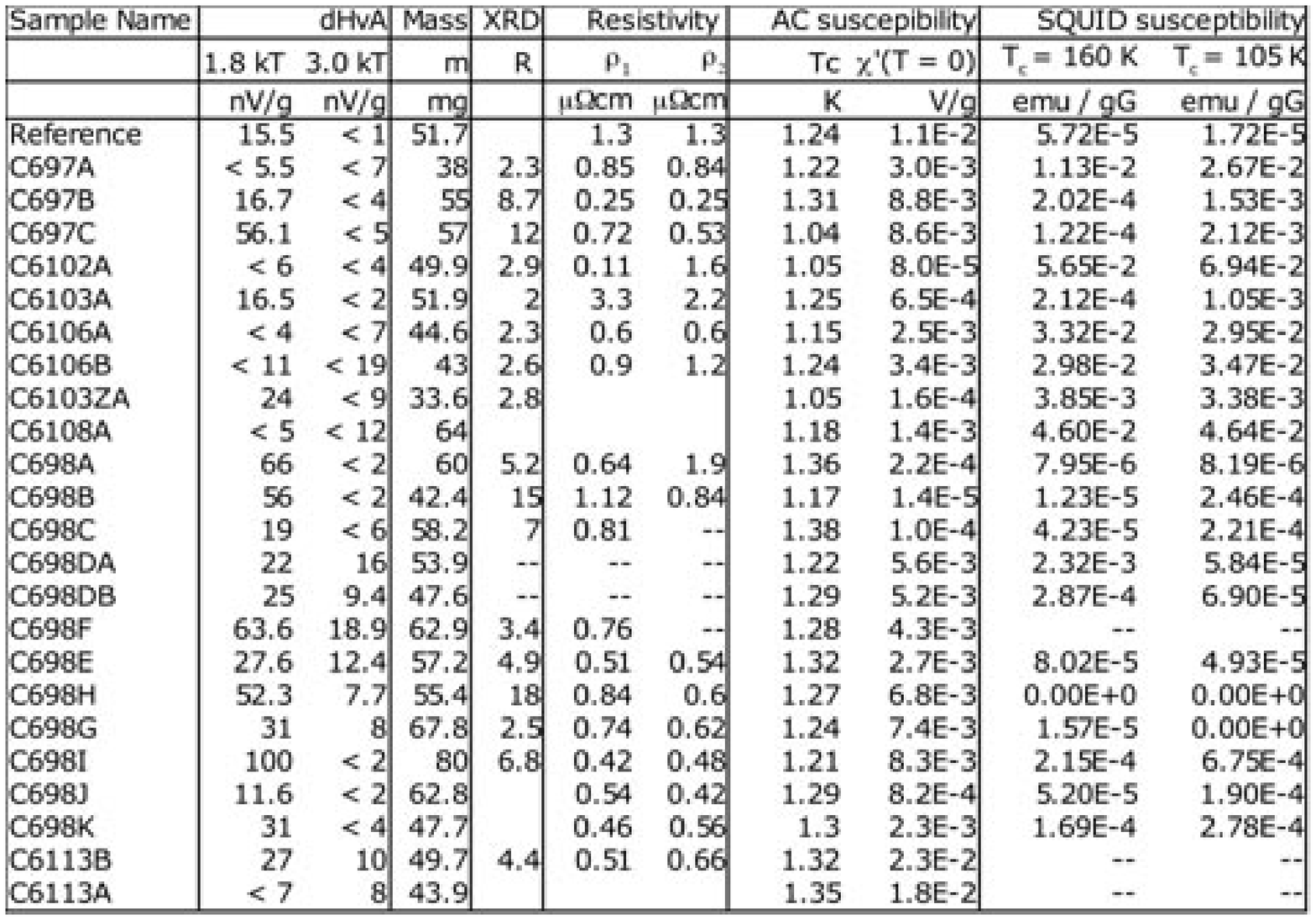}
	\caption[Quantitative results of sample characterisation]{Complete quantative results of all characterisation methods for all the samples that were analysed. The second and third columns give the integrated dHvA amplitude for the 1.8~kT peak, related to \TTS, and the 3.0~kT peak, related to \TOF, normalised by the mass of the samples. The fifth column is the ratio R of the powder diffraction peaks at 8.6$^{\circ}$, associated with \TTS, and 13.9$^{\circ}$, of \TOF. The next two show the residual resistivities measured on each side of the resistivity crystals. Next are shown the columns with data from AC susceptibility, the superconducting critical temperatures and the amplitude of the diamagnetic signal. Finally, in the last two columns we present the amplitude of the ferromagnetic signals of \OOT\ and \FTT\ measured with the SQUID, extrapolated to absolute zero. The first row of data belongs to the sample that was used in preliminary experiments, and was the standard with which to compare the new samples. The empty elements of the table indicate measurements that were impossible or failed for various technical reasons.}
	\label{tab:327Samples1}
	\end{center}
\end{table}

\begin{table}[p]
\begin{center}
	\includegraphics[width=1\columnwidth]{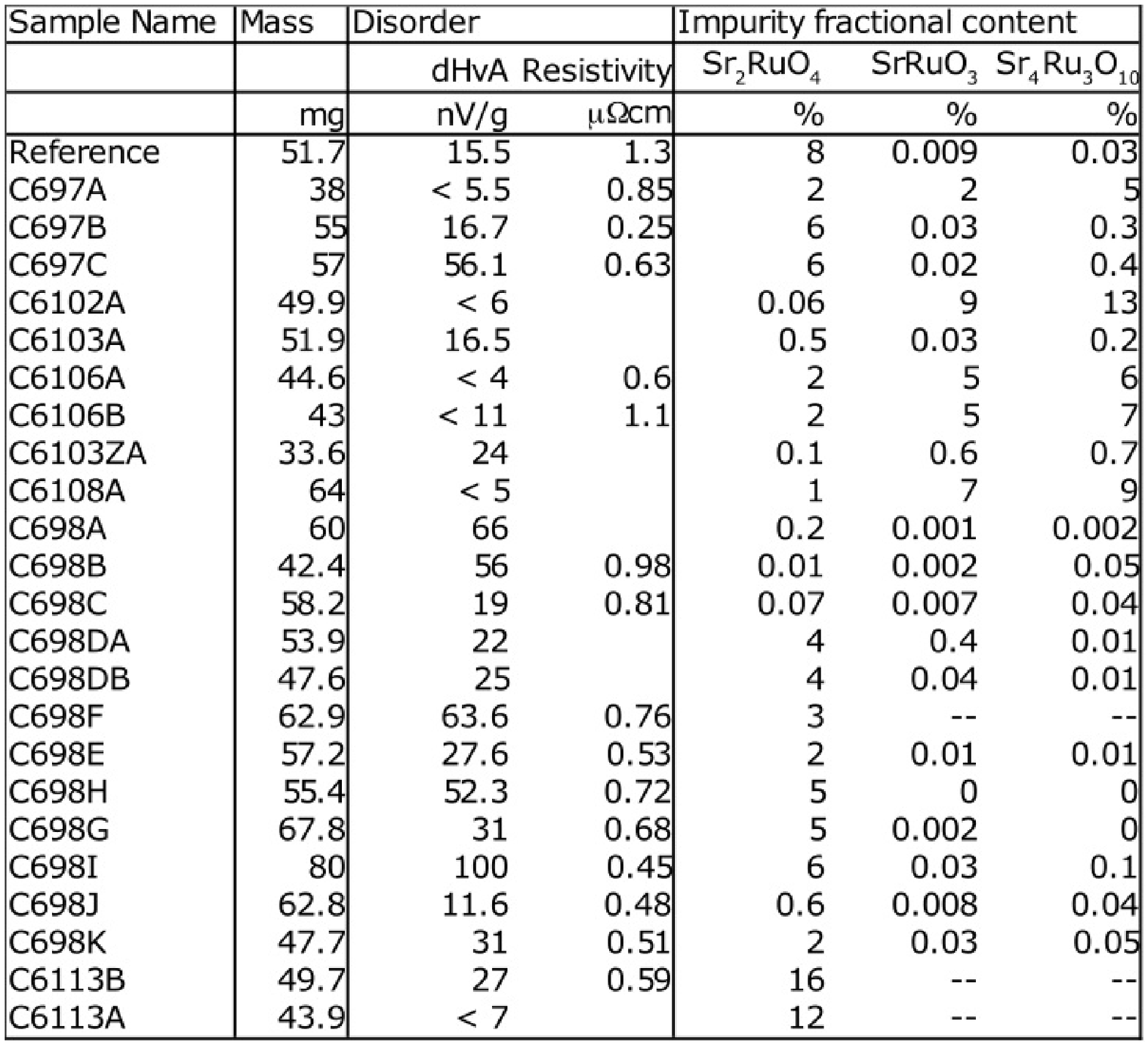}
	\caption[Quantative fractional mass of impurity phases]{Fractional mass of the impurity phases \TOF, \OOT\ and \FTT\ in ppm, for all the measured samples. These are quoted along with the amplitude values of dHvA and the average residual resistivity, for convenience. Five samples feature remarkable dHvA properties. Among them, one possesses by far a much higher dHvA signal, C698I, while one contains negligible amounts of the impurity phases, C698A.}
	\label{tab:327Samples2}
	\end{center}
\end{table}

In this section, we will present the complete results of the sorting procedure for 23 new samples of \TTS. Table \ref{tab:327Samples1} presents all quantities measured using the characterisation methods described in this section: powder X-ray diffraction, residual resistivity, dHvA, zero field AC susceptibility and low field magnetisation. The same measurements were performed on the sample that was used in preliminary experiments\footnote{The results of these experiments are not shown in this thesis, since equivalent higher quality measurements are presented in chapter~\ref{chapter:dHvA}.}, shown here in the first row of data named ``Reference", for comparison purposes.

Using the various calculations described in the previous sections, we estimated volume fractions of \TOF, \OOT\ and \FTT\ for each sample. Table \ref{tab:327Samples2} summarises these values, along with the same amplitude values for dHvA as the previous table, and the average residual resistivities. Reading carefully this table reveals the strategy to employ when choosing samples for extensive dHvA experiments, which we describe here.

The very first column of interest is that of the amplitude of the quantum oscillations at 1.8~kT. The reference sample provided us with a value for comparison: all samples with a value higher than 15.5~nV/g were worth using in dHvA experiments, but six of them had exceptional values, which were C697C, C698A, C698B, C698F, C698H and C698I. As we know from experience with crystal growth, it was hardly surprising that most of them originated from the same crystal growth experiment, of batch name C698. According to these data, sample C698I featured by far the highest amplitude, 100~nV/g, over 6.5 times that of the reference sample.

The next column of interest is that of the mass fraction of \TOF\ calculated from the amplitude of the superconducting diamagnetic signal. We observed that even within a single crystal batch, for instance C698, the amount of \TOF\ varied by orders of magnitude. Among our exceptional samples, it was sample C698A that featured the smallest amount, 50 times less than the reference sample. It was also interesting to find that the highest value belonged to the reference sample, meaning that in this respect, all the new samples were cleaner. Moreover, the sample which was best for dHvA, C698I, featured a lot more \TOF\ than C698A, by a factor 38.

The third set of values to examine was the mass fraction of \OOT\ and \FTT. The two values, for each sample, seemed correlated. This is not surprising, since we intuitively know that these inclusions correspond to stacking faults. The $n > 2$ members of the Ruddlesden-Popper series possess physical properties that resemble more those of the $n = \infty$ member as $n$ increases, and it is probable that slight deviations in growth parameters can produce a whole variety of them, and members with very large $n$ are essentially indistinguishable from $n = \infty$. However, the fractional values for \OOT\ and \FTT\ were visibly not correlated with those for \TOF. The sample with the smallest amounts of \OOT\ and \FTT\ was C698H, for which the ferromagnetic signals were not at all observed. In this regard, the reference sample contained very small amounts of these, while for sample C698I the fractions are 4 (\OOT) and 39 (\FTT) times larger. Sample C698A also contained very small amounts of ferromagnetic inclusions, being one of the samples with the best phase purity.

Moreover, the residual resistivities were surprisingly not correlated with dHvA amplitude values. This effect can only be explained by the fact that the resistivity crystals were not representative of large dHvA samples, much less of whole crystal rods, at these levels of purity. One should only rely on residual resistivity to infer disorder for bulk crystals only in a larger scale of impurity content, where quantum oscillation experiments are not possible, or to a very small spatial scale within a crystal. 

Before we finish this section, we make a few more observations. Examining the data from powder X-ray diffraction, in table \ref{tab:327Samples1}, one finds that it also possessed no apparent correlation with the amplitude of the diamagnetic signal produced by \TOF. These data were not representative of large crystals, except in the limit of large volume fractions. A signal was almost always observed at 13.9$^{\circ}$, even in cases where very low amounts were present in the neighbouring dHvA sample. Also, in some instances, the quantum oscillation signal of the 3.0~kT orbit of \TOF\ was observed, and its amplitude was not correlated with that of the diamagnetic signal. Since in such cases, the quality of the inclusions was very good, the disorder in \TOF\ intergrowths was not correlated with the fractional amount. However, the correlation could probably be made between the amplitude of the signal of the 3.0~kT dHvA frequency and the superconducting critical temperature, since these measurements were performed on the same crystal and, as was demonstrated by Mackenzie $et$ $al.$ \cite{PRLMackenzie}, a correlation exists between disorder and the superconducting $T_c$.

In conclusion, we summarise by describing the strategy we have used for choosing crystals for the extensive dHvA experiments on \TTS\ that is at the core of this thesis project. In order to maximise the dHvA signal, we chose the sample with highest dHvA amplitude at 1.8~kT, C698I. But we know, from preliminary experiments and our sorting procedure, that it is possible that with higher resolution, dHvA peaks produced by \TOF, \OOT\ or \FTT\ are detected using this sample since the volume fractions of these impurity phases are not negligible. Consequently, we chose to use as well a sample with a slightly higher disorder level but lower amounts of impurity phases, C698A, in order to make rule out interpretation errors due dHvA frequencies of other phases than \TTS.

\section{Laue orientation of the samples \label{sect:Laue}}

\begin{figure}[!p]
	\begin{minipage}[t]{7cm}
		\begin{center}
		\includegraphics[width=7cm]{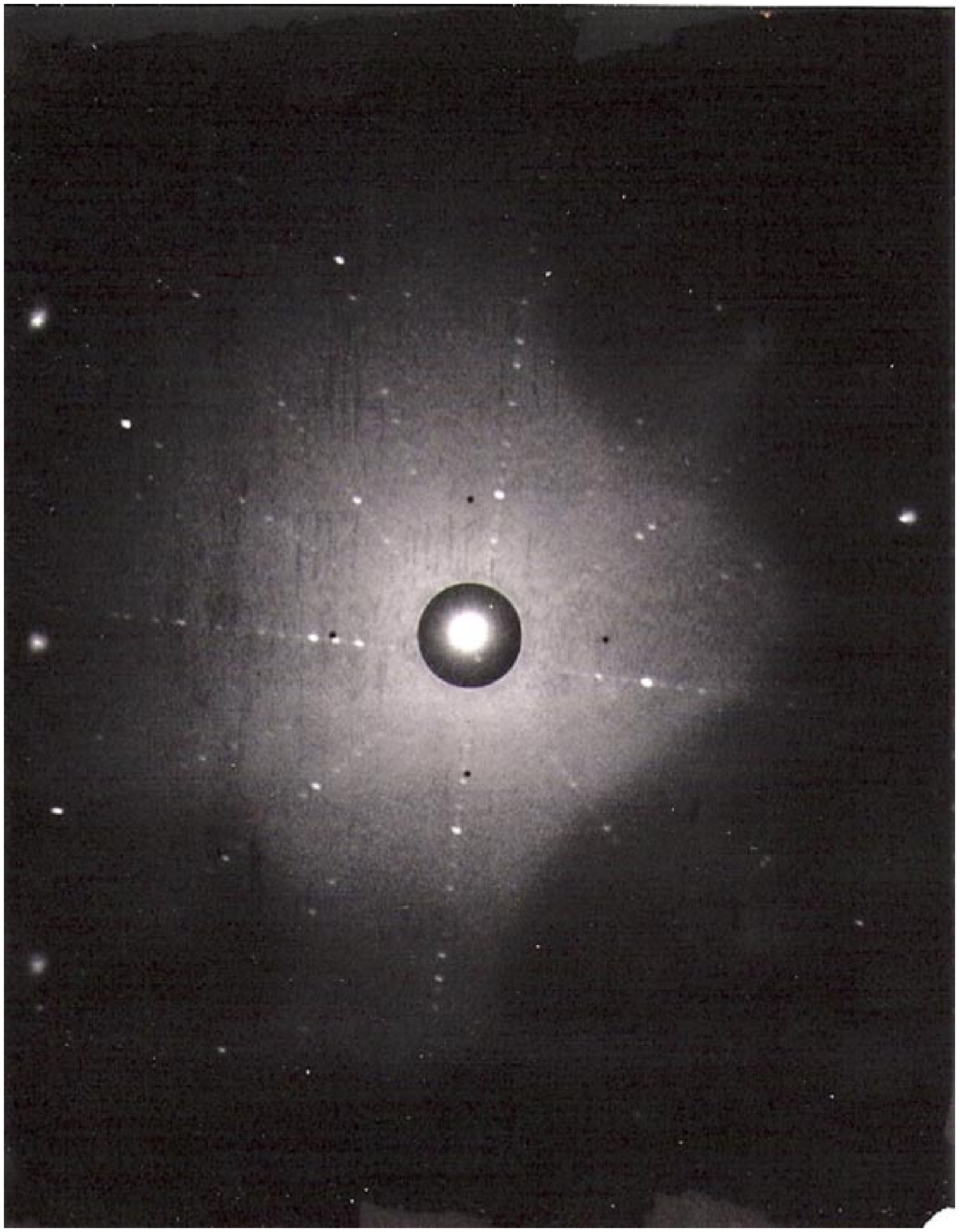}
		\includegraphics[width=7cm]{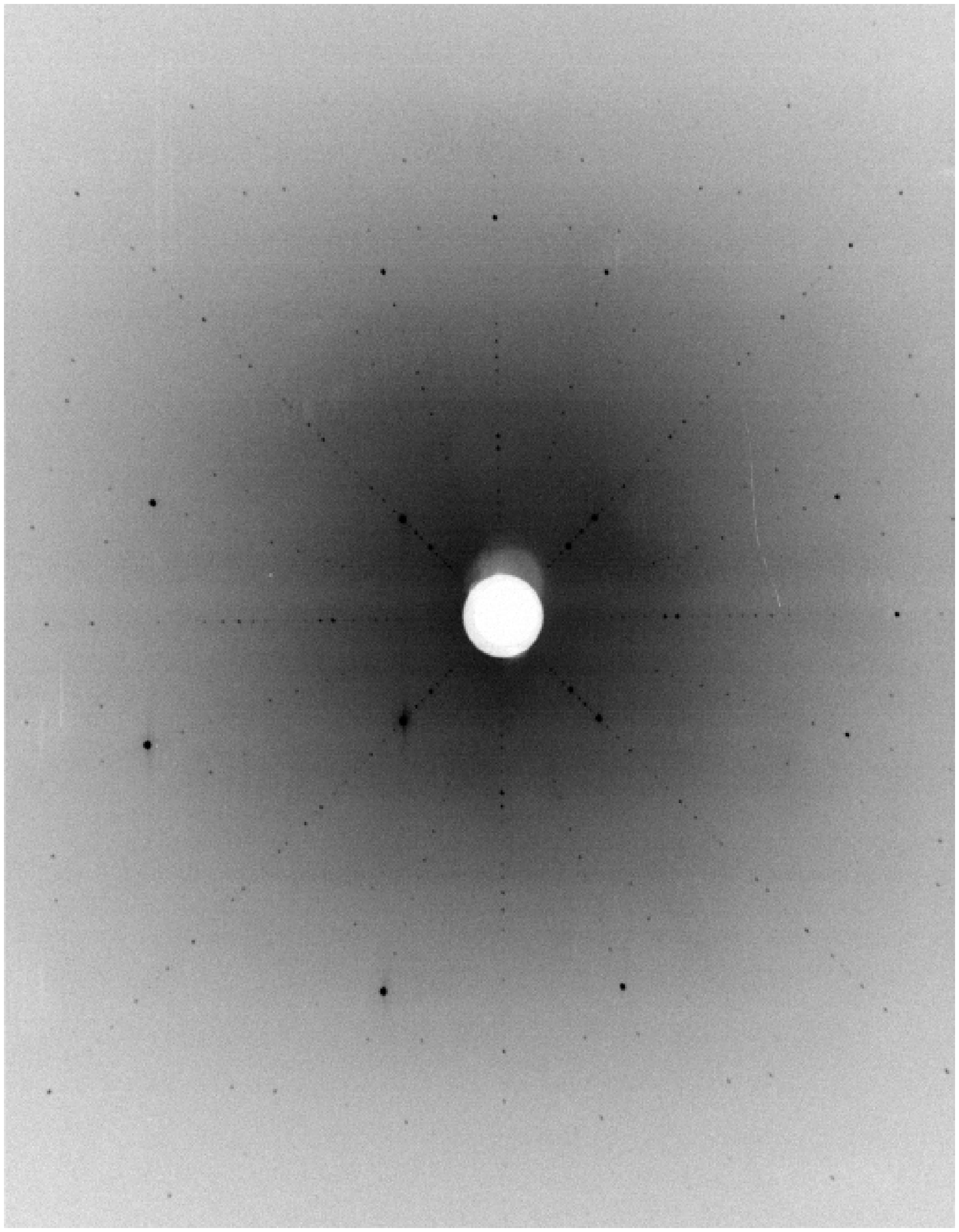}
		\end{center}
	\end{minipage}
	\hfill
	\begin{minipage}[t]{7cm}
		\begin{center}
		\includegraphics[width=7cm]{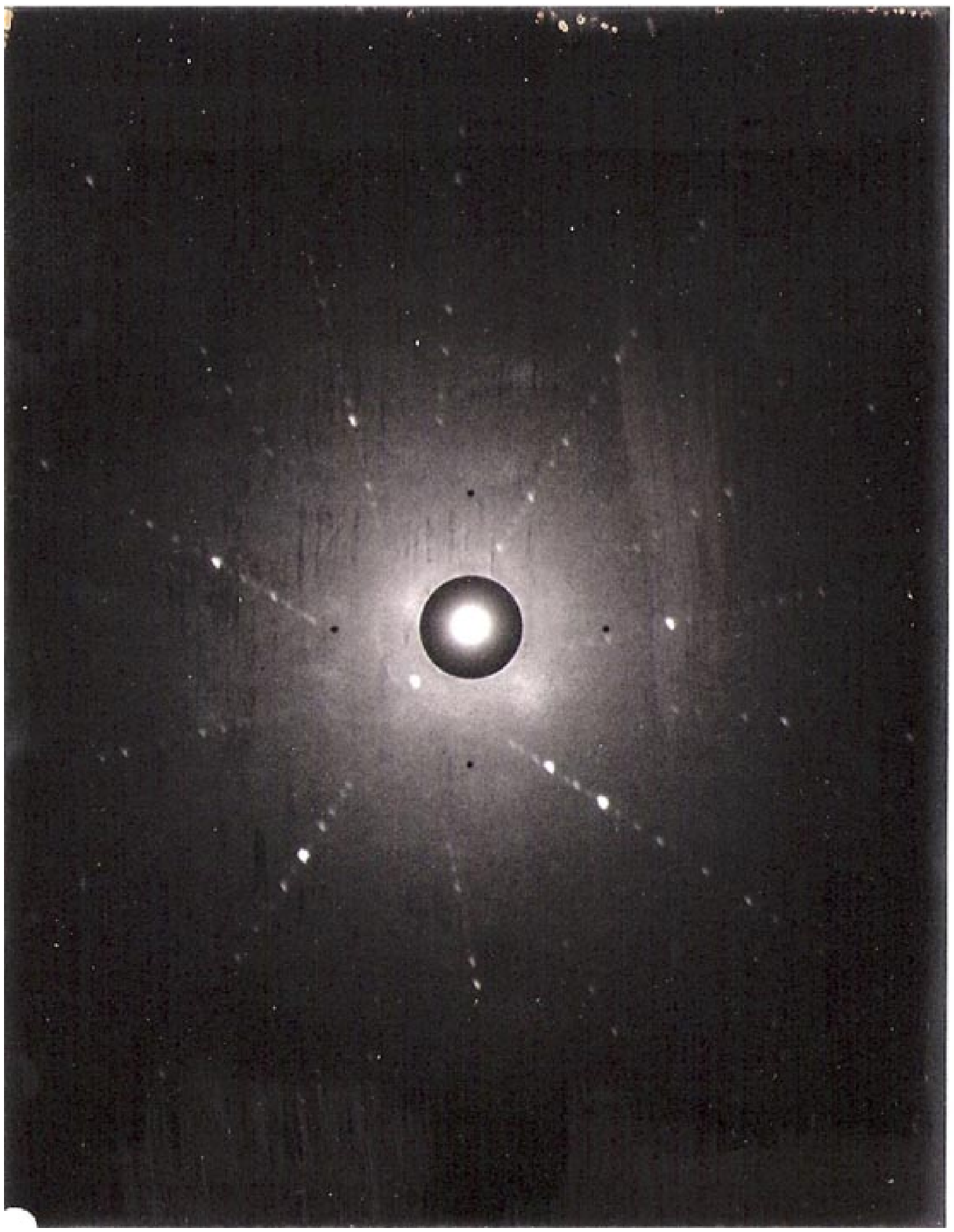}
		\includegraphics[width=7cm]{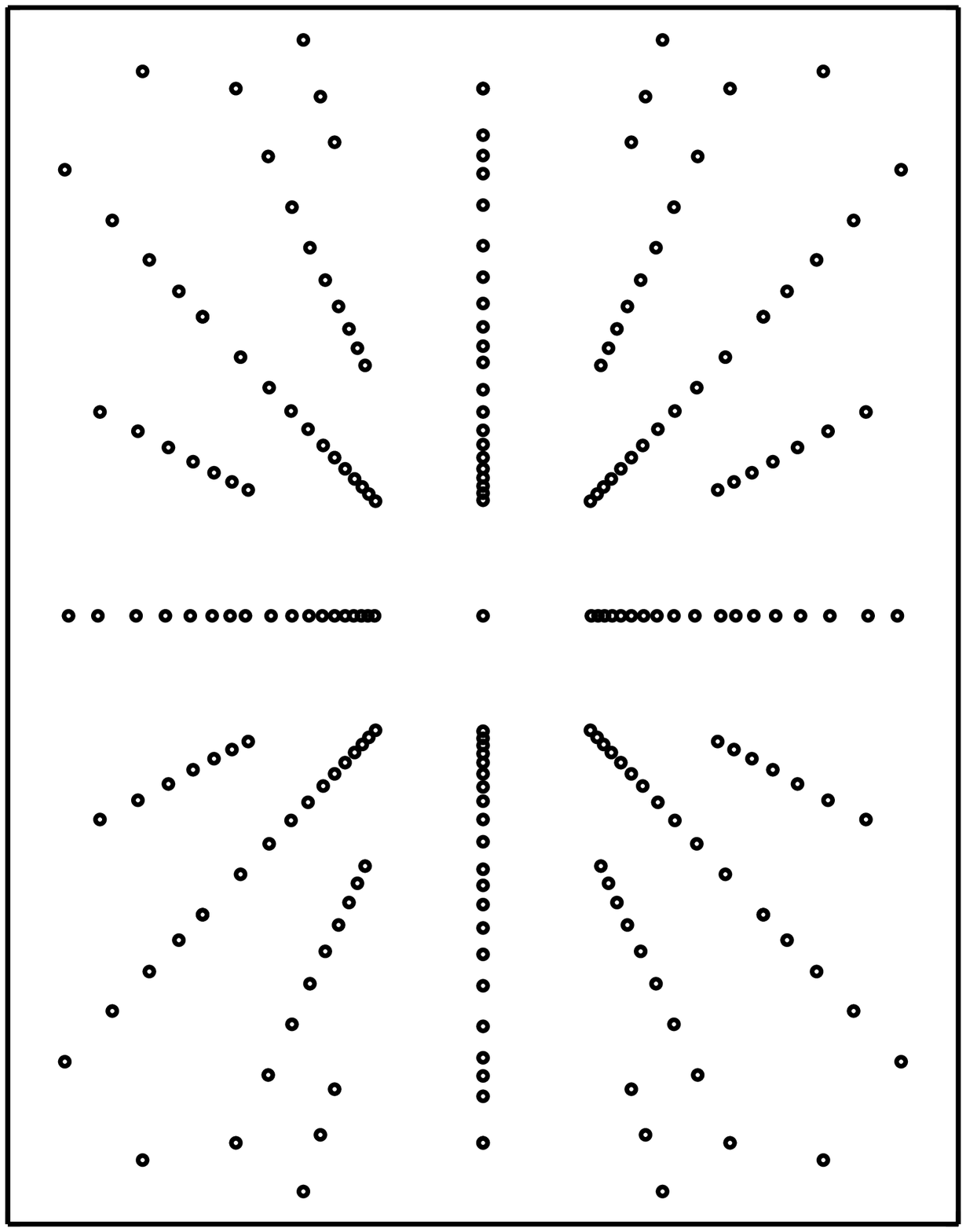}
		\end{center}
	\end{minipage}
	\caption[Laue orientation of samples]{$Top$ $left$ Scanned Laue photograph taken on sample C698I. $Top$ $right$ Scanned Laue X-ray photograph taken on sample C698A. $Bottom$ $left$ Laue photograph taken on sample C698K by A. Rost with the system at the University of Edinburgh. $Bottom$ $right$ Simulation of the Laue pattern using the software Orient Express, where the $a$ and $b$ directions are along the figure edges, and equivalent.}
	\label{fig:LauePhotos}
\end{figure}

Before placing samples into pick-up coils of a dHvA system, in order to measure and interpret the angular dependence of the quantum oscillations, one is required to find the direction of the crystallographic axes using Laue X-ray diffraction. For \TTS, the $c$-axis is easy to determine, as the material readily cleaves along the ab-plane, but finding the orientation of the $a$ and $b$ directions is impossible by eye. One is required to use the standard procedure with Laue X-ray diffraction, which is presented here, with the resulting photos.

We used the Laue X-ray source and camera situated in Cambridge, a model that employs polaroid photographs. We placed the samples such that their $c$-axis was parallel to the X-ray beam, and the photographic plate was put between the source and the sample, the sensitive side towards the sample. The beam then passed through the centre of the plate and was reflected by the sample back to the photographic plate. We obtained complex dot patterns along straight lines, shown in figure \ref{fig:LauePhotos}, top left and right, for samples C698I and C698A, respectively. 

In order to find out which of those lines of spots corresponded to which crystallographic direction, we simulated such patterns with a computer program, Orient Express, using the space group of $I4mmm$\footnote{The rotation of the RuO octahedra does not affect the Laue pattern in a significant way, and it looks qualitatively the same as that of \TOF.}, with the right lattice parameters of $a = b = 3.88 \AA$ and $c = 20.7 \AA$. We obtained the plot on the bottom right of figure \ref{fig:LauePhotos}. In this simulation, one recognises the [100] from the [110] directions by using the sets of spots between the vertical or horizontal lines and the diagonal ones, which lie slightly closer in angle to the diagonals than to the vertical or horizontal lines. We chose to place sample C698I in a pick-up coil of the system described in section \ref{sect:Camprobe} such that the magnetic field rotates from [001] towards [110], and towards [100] for sample C698A.

The orientation procedure of samples C698I and C698A was verified independently by A. Rost, with a different Laue system located at the University of Edinburgh, using sample C698K, in preparation for specific heat experiments, shown in the lower left panel of figure \ref{fig:LauePhotos}. Sample C698K was a single crystal neighbour of C698I, and consequently possessed the same orientation. The sample was rotated until the Laue picture presented crystallographic directions aligned with the photo edges, and yielded the same rotation angle.

\section{Avoiding eddy current heating in dHvA experiments\label{sect:eddy}}

The effect of eddy current heating is to increase the sample temperature compared to that of the thermometer, which in a dilution refrigerator, is located outside of the magnetic field region, at the mixing chamber. In dHvA experiments, even for small eddy current heating, the systematic error on the thermometry produces dramatic effects on the calculation of the quasiparticle effective masses during LK fits. This can be explained by considering that the effective mass is inversely proportional to the width at half maximum of the LK function (see figure \ref{fig: RLK}). A positive systematic error on the measured sample temperature always leads the non-linear fit of the LK function to underestimate the quasiparticle mass. Moreover, for a same heat input, this effect is the more pronounced the higher the mass is. Overall, this may have the effect of suppressing, partly or completely, an enhancement of the quasiparticle mass as a function of magnetic field, and serious measures had to be taken in this project to ensure that such heating was not significant during dHvA experiments. 

The source of heat in this phenomenon is produced proportionally to the rate of change of magnetic field, and the modulation field normally possesses the largest time derivative. For a conducting cylinder of volume $V$ and resistivity $\rho$ in a varying field, the power dissipation is \footnote{see the book by Pobell, \cite{pobell}, p. 177}
\beq
P_{eddy} \propto V {\dot{B}^2 \over \rho}.
\label{eq:eddy}
\eeq
Therefore, one is required to use the lowest possible excitation field, which also reduces the amplitude of the signal, and a compromise has to be found where the temperature difference between the thermometer and the sample is negligible. The quasiparticle mass of one dHvA frequency extracted with a non-linear LK fit is a good measure of sample heating. The mass saturates to a maximum value as the excitation field is decreased, and for high eddy current heating, significant deviations from LK are measured.

The guidelines we used stem from the following considerations. We showed in section \ref{sect:oscsignal} (see eq. \ref{eq:eddy} and \ref{eq:BesselApprox}) that, in the limit where $\lambda < 2$, 
\begin{enumerate}
\item Eddy current heating is proportional to $\dot{H}^2 \propto \omega^2 H_{AC}^2$,
\item The signal in the first harmonic is proportional to $\omega H_{AC}$,
\item The signal in the second harmonic is proportional to $\omega H_{AC}^2$.
\end{enumerate}
We conclude that increasing by the same factor the AC field or its frequency produces the same increase in eddy current heating, as well as the same increase in amplitude of the first harmonic, and therefore increasing one of these parameters while keeping the product $\omega H_{AC}$ constant results in constant eddy currents and first harmonic amplitude. However, it is not the case with the second harmonic, where reducing the frequency and increasing the modulation field increases the signal while leaving the heating constant. Thus, when using the second harmonic, one should use the lowest frequency possible.

\begin{figure}[t]
	\begin{minipage}[t]{7cm}
		\begin{center}
		\includegraphics[width=7cm]{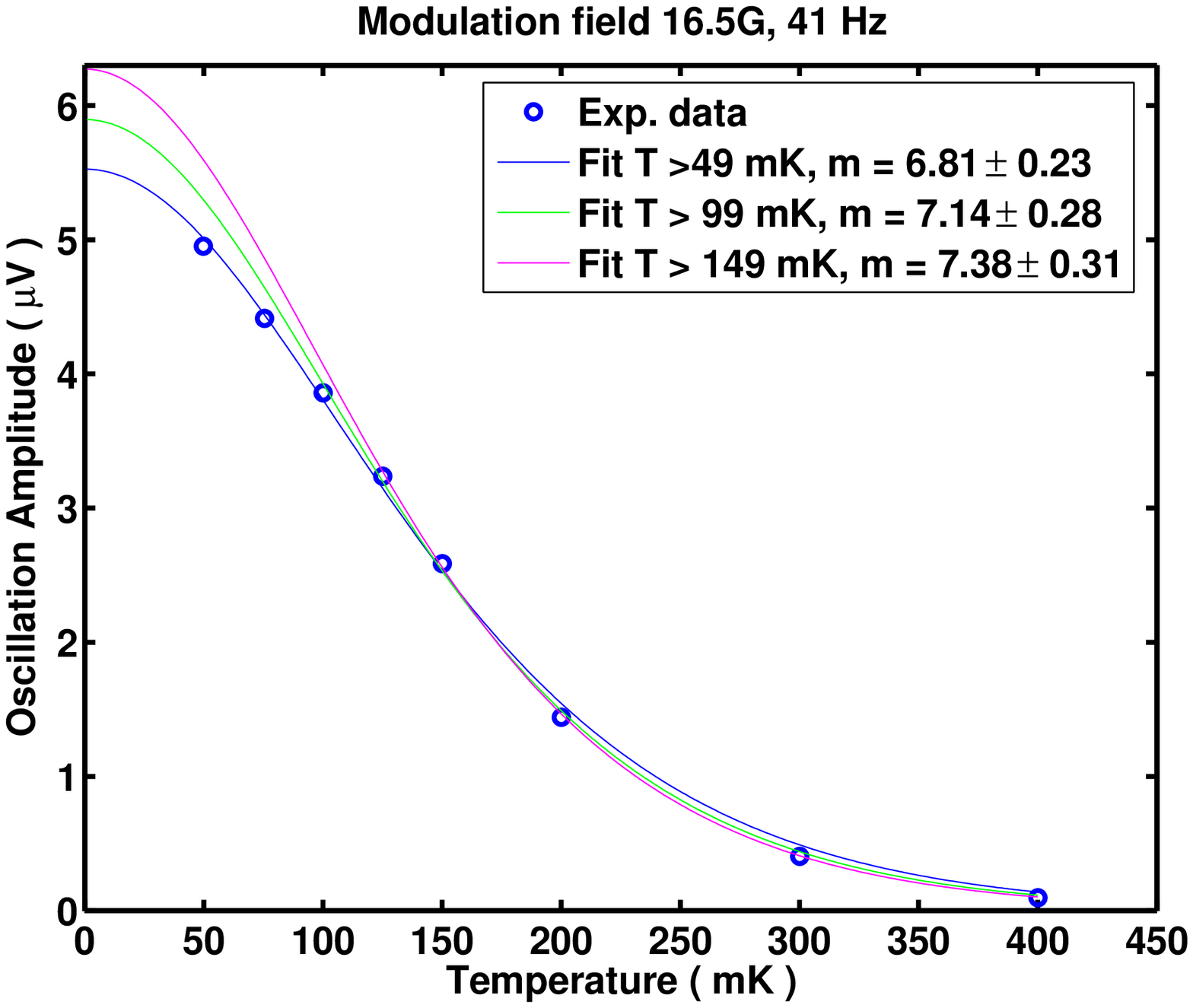}

		\end{center}
	\end{minipage}
	\hfill
	\begin{minipage}[t]{7cm}
		\begin{center}
		\includegraphics[width=7cm]{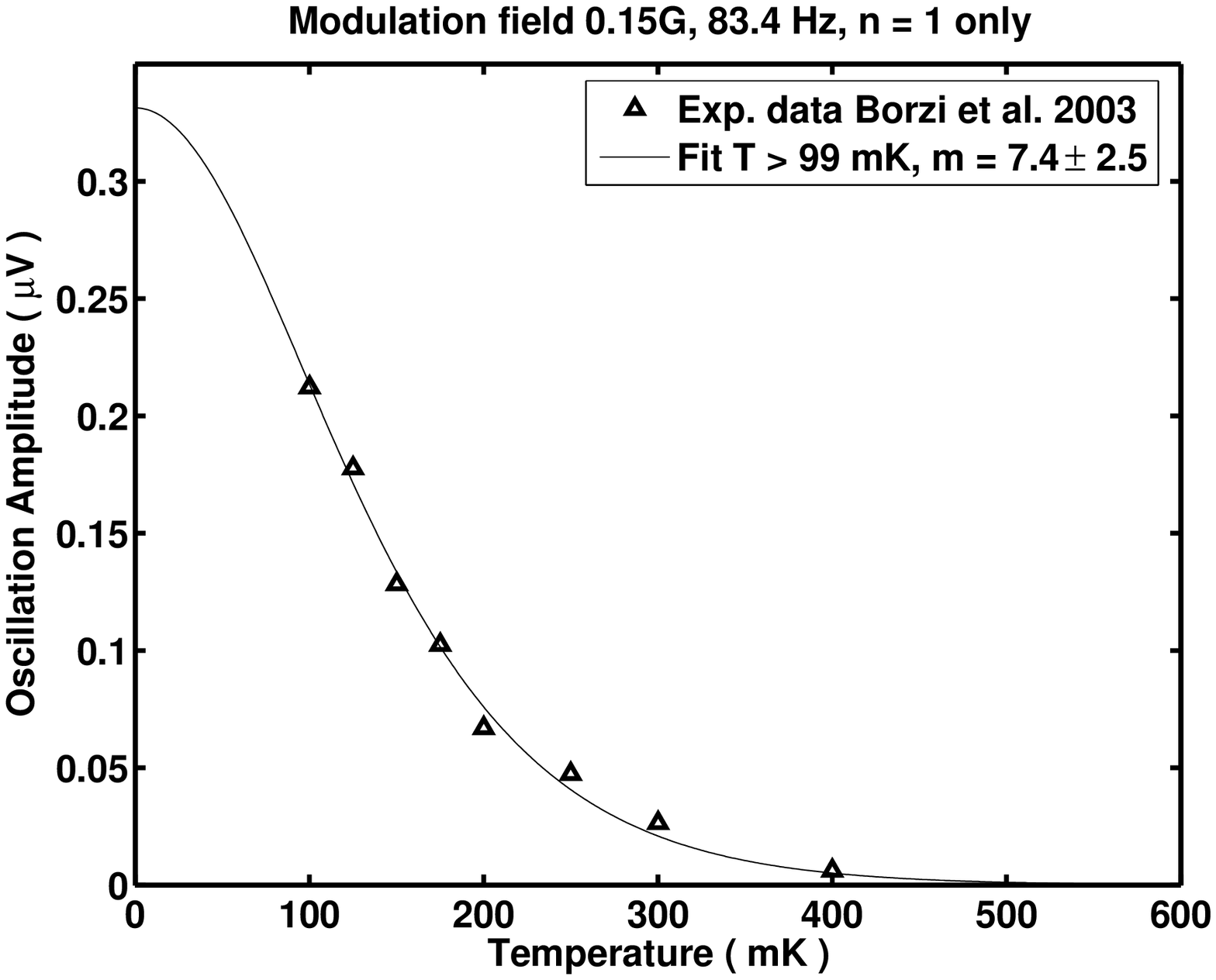}
		\end{center}
	\end{minipage}
	\caption[Eddy current heating]{Data for the temperature dependence of the oscillation amplitude of the 1.8~kT peak between 6 and 7~T for two values of the product of modulation field and frequency: $\omega H_{AC} = 676.5$ G$\cdot$Hz ($left$) and 12.5 G$\cdot$Hz ($right$). The data was measured with different but equivalent experimental apparatuses, so the absolute values of amplitude cannot be compared directly. Solid lines correspond to LK fits and the results are given in the legends.}
	\label{fig:MassRun16G41Hz}
\end{figure}

Figure \ref{fig:MassRun16G41Hz} presents two examples of the temperature dependence of the dHvA amplitude and LK fits at two extremes, the first with a product of modulation field times frequency high enough that eddy current heating is significant, left, and the second with such a product 54 times smaller, producing eddy current heating nearly 3000 times smaller. We found in the first case that we could not properly fit the LK relation to the data. The adjusted curve overestimates intermediate temperature points. Since, with eddy current heating present, the error on temperature values decreases as temperature increases, deviations to LK decrease at high temperature points. By excluding low temperature points, we found that better fits could be obtained, and that using only data above 149~mK produced the same LK fit result as in the low modulation field case. The quasiparticle mass obtained increased as we removed data points saturated to the value measured with low heating, and in doing so the fit quality improved at intermediate points. Therefore, for extended dHvA experiments in a material that respect LK temperature dependence, one should use the highest modulation field at which the LK distribution is not significantly disturbed.

\chapter{The de Haas van Alphen experiment on \TTS\ \label{chapter:dHvA}}
\markright{Chapter~\ref{chapter:dHvA}: The de Haas van Alphen experiment on \TTS\ }

This chapter presents the complete experimental data obtained from the de Haas van Alphen experiments that we performed on \TTS. These measurements were carried out in three sets, performed in 2005, 2007 and 2008, the first two in St Andrews and the third in Cambridge. The St Andrews experiments did not yield a complete set of information for the completion of this project, but mainly served to generate the motivation for a collaboration with the dHvA group of Michael Sutherland in Cambridge. The combination of the high quality samples that we provided and the performance of the dHvA apparatus in Cambridge, described in section \ref{sect:Camprobe}, has been the key to the success of this project.

dHvA has been measured in the Cavendish Laboratory in Cambridge for decades, mainly by the late professor David Shoenberg, who was also the author of the main reference on the subject, still used today \cite{shoenberg}, and responsible for the determination of the Fermi surfaces of a large number of elements, notably the noble metals. The current Cambridge dHvA system was developed mainly by Stephen Julian, Gilbert Lonzarich and Christoph Bergemann, and complete Fermi surfaces of many complex systems were determined with it, notably those of \TOF, CaVO$_3$, URu$_2$Si$_2$, CeRu$_2$Si$_2$, UPt$_3$ and recently Ag$_5$Pb$_2$O$_6$. Charge of this apparatus was recently taken by Michael Sutherland, under the direction of whom our experiments have taken place. The work was performed by myself in close collaboration with Cambridge postgraduate students Swee K. Goh and Eoin C. T. O'Farrell. All subsequent data processing and analysis were performed by myself only.

All the samples that were used were grown in St Andrews by Robin Perry and thoroughly characterised by myself. The 2005 series of experiments were performed using the sample labelled ``Reference", in comparison to which we strove to improve the purity of crystals\footnote{See section \ref{sect:search} for details.}. The 2007 and 2008 series of experiments were carried out using simultaneously two samples of the highest quality recorded by our sorting procedure, labelled C698I and C698A, which featured very low disorder and high phase purity. Finally, since most measurements performed in St Andrews were repeated in Cambridge, where the signal quality was superior, only data from the 2008 experiment series in Cambridge are presented, except where stated otherwise.

The first section compares detection harmonics, followed by the lengthy way in which we determined the modulation field to use in order to avoid eddy current heating. This involved measuring the temperature dependence of the dHvA signal in a certain field region for many different modulation fields. The next section shows the general description of the dHvA data, spectra and quasiparticle masses away from the metamagnetic transition. We then present the largest data set, the angular dependence of dHvA. Next, we reinvestigate the temperature dependence of dHvA in \TTS\ and the field dependence of the quasiparticle masses, in which we contradict one of the conclusions of the previous work by Borzi \cite{borzi}. This is finally followed by another very successful experiment, the measurement of dHvA inside the nematic phase. 

\section{Determination of experimental parameters \label{sect:CamEddyCurrents}}

This section is dedicated to the method used to select the optimal parameters with which we performed the dHvA measurements. There were many ways in which we could perform this experiment, and we desired to tailor it to our specific needs. In principle, one can use various values of modulation field and different detection harmonics, which enhance different aspects of the data, as was seen in section \ref{sect:oscsignal}. However, the modulation field can lead to significant heating through the process of eddy currents (see section \ref{sect:eddy}), and this aspect also needed attention. 

The method with which we determined the amount of heat generated by eddy current heating relied on the principle that a systematic difference between the measured amplitude of dHvA for one orbit and that predicted by theory appears when strong heating occurs at the sample such that there is a temperature difference between the sample and the thermometer. It stems from the fact that, by the LK relation (eq. \ref{eq:LK}, p. \pageref{sect:LK}), the dHvA amplitude decreases monotonically as temperature increases, and when heating occurs, the measured value is $lower$ than the prediction. There are two ways by which such a difference can be measured. The first consists in measuring dHvA as a function of modulation field and compare the measured amplitude to the Bessel function it is expected to follow (eq. \ref{eq:dMdt}, p. \pageref{sect:oscsignal}). The second is to look at the temperature dependence of dHvA for many modulation fields and compare with the LK relation.

We first show why second harmonic detection is appropriate for measuring dHvA in \TTS, through comparison of the first and second harmonics. Then, we present the modulation field dependence of the second harmonic amplitude, which we also used to calibrate accurately the magnitude of the modulation field. Finally, we show which modulation field should be used in order to avoid eddy current heating.

\subsection{Comparison of detection harmonics \label{sect:dataharmonics}}

\begin{figure}[t]
  \begin{center}
	\includegraphics[width=1\columnwidth]{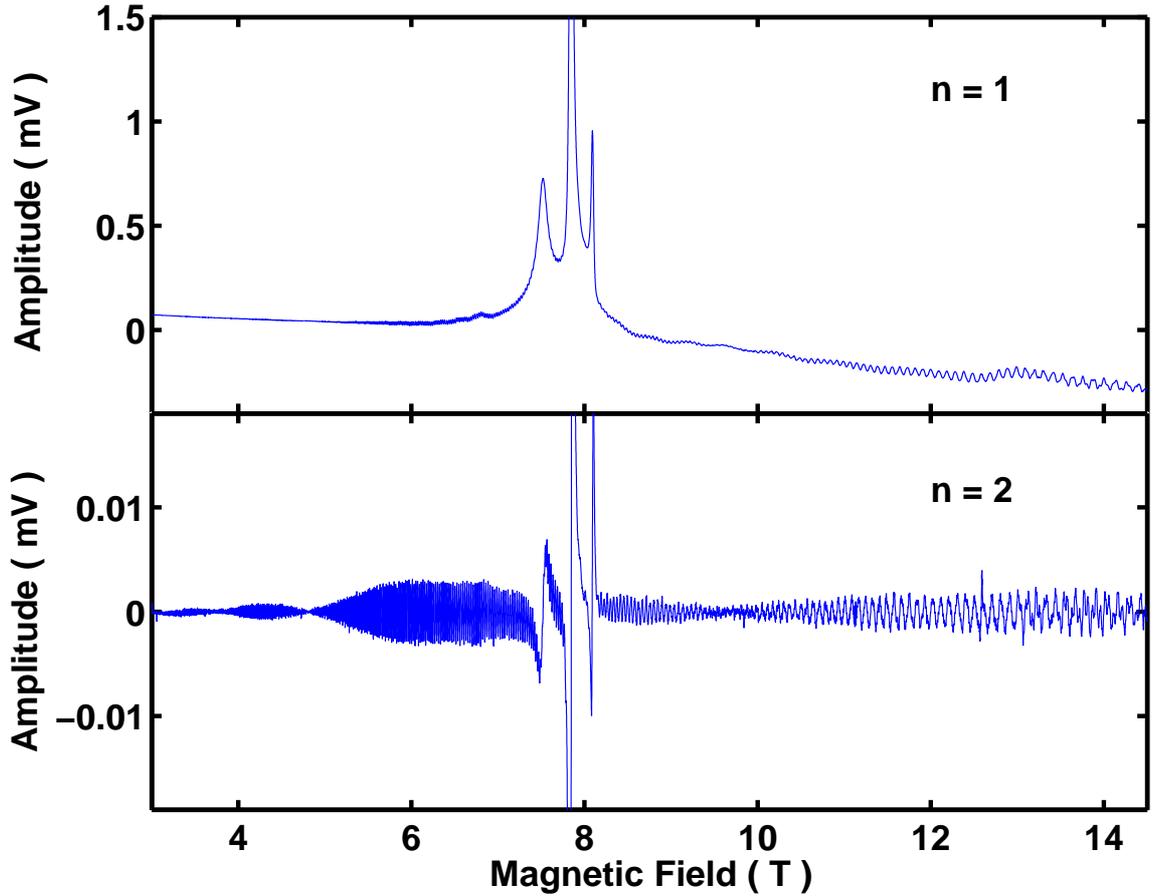}
	\caption[First and second harmonic dHvA in \TTS]{dHvA measured using first ($top$) and second ($bottom$) detection harmonics. The data were measured on the sample labelled ``Reference" in section \ref{sect:search}, using a modulation field of 16~G at a frequency of 41~Hz. The total amplification factor was of 100, and the plots were not normalised.}
	\label{fig:1nHarmOsc}
	\end{center}
\end{figure}

We showed in section \ref{sect:oscsignal} that an important reduction of non-oscillatory magnetic background in AC susceptibility measurements can be obtained by using second harmonic detection. It is particularly the case in \TTS, where the magnetic signal produced by the metamagnetic transition is of several orders of magnitude larger than the oscillations. These measures led to significant improvements in sensitivity compared to the experiments of Borzi \cite{borzi}. 

Figure \ref{fig:1nHarmOsc} presents a comparison of first and second detection harmonics when used with \TTS, where the top plot refers to the first and the bottom the second. The data was measured using the St Andrews system, with sample labelled ``Reference" in section \ref{sect:search}, during preliminary experiments. The first harmonic corresponds to the in-phase component of the AC-susceptibility, the first field derivative of the magnetisation of \TTS\ and features three peaks. Those correspond to a crossover at 7.5~T, and two first order metamagnetic transitions at 7.9 and 8.1~T, and are the same as those in the data published by Grigera \cite{science2}. With such a signal, setting the LIA sensitivity high enough that the oscillations away from the peaks are measured without digitisation results in the saturation of the LIA input in the peak region. 

The bottom plot of figure \ref{fig:1nHarmOsc} shows the second harmonic, which corresponds to the out-of-phase component of the second derivative of the magnetisation. One can observe that the magnetic background is reduced to very sharp asymmetric peaks at 7.5, 7.9 and 8.1~T of smaller magnitude compared to the oscillations, and full sensitivity can be achieved on the LIA using a simple offset. With such data sets, dHvA can be analysed through the complete field range available. We conclude from this that in order to analyse properly dHvA near the metamagnetic transition, second harmonic detection was the appropriate method to use.

\subsection{Modulation field dependence of dHvA}

\begin{figure}[t]
         \begin{center}
	\includegraphics[width=0.8\columnwidth]{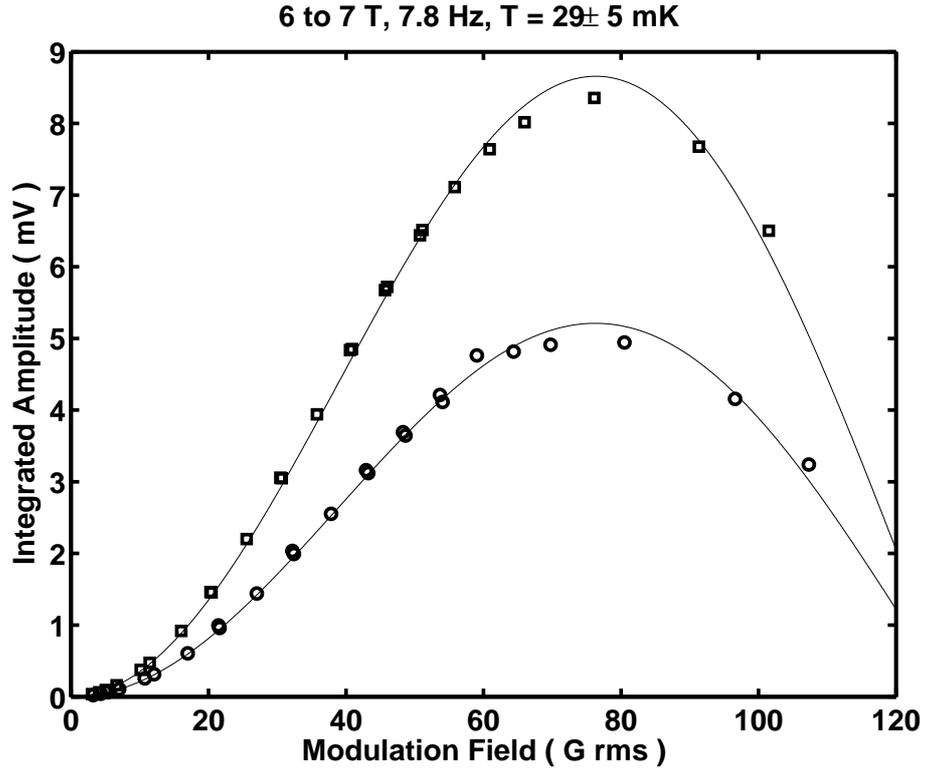}
	\caption[Measurement of the Bessel function dependence of dHvA in \TTS]{Amplitude of the dHvA signal for the 1.8~kT frequency between 6 and 7~T, as a function of the $rms$ amplitude of the modulation field at 7.9~Hz, compared with the predicted Bessel functions. The squares correspond to sample C698I, the circles to sample C698A, and the solid lines to fits of Bessel functions.}
	\label{fig:BesselModulation2kT}
	\end{center}
\end{figure}

Figure \ref{fig:BesselModulation2kT} shows the modulation field dependence of the amplitude of the 1.8~kT frequency between 6 and 7~T, measured on both samples C698I (squares) and C698A (circles), using a modulation frequency of 7.9~Hz. The solid lines correspond to non-linear fits of Bessel functions. As we will see in section \ref{sect:FirstRotation}, the alignment of the crystallographic $c$-axis of these was not identical, and a difference in field angle was present between them. The measurement was performed with sample C698I positioned with its $c$-axis parallel to the magnetic field, and simultaneously with C698A at an angle of 10$^{\circ}$. 

Bessel functions were fitted to the data with scaling parameters $a$ and $b$, using the form 
\beq
f(\lambda) = a J_k ( b \lambda ), \quad k = 2, \quad \lambda = {2\pi F \sqrt{2} H_{AC} \over H_{DC}^2},
\label{eq:BesselScaling}
\eeq
where $F$ is the dHvA frequency and $H_{AC}$ the $rms$ modulation field\footnote{Which requires a factor $\sqrt{2}$ in order to be consistent with the calculation of section \ref{sect:oscsignal} (eq. \ref{eq:lambda}), in which $H_{AC}$ refers to half the peak-to-peak amplitude.}. When using the value of the voltage over a resistance of 1.7 $\Omega$ in series with the modulation coil as a variable, we obtained the values for $b$ of 57.5 and 60.7 G/V using the coils containing C698I and C698A, respectively, corresponding to approximately 98 and 103 G/A for the modulation coil. Assuming it is possible that each pick-up coil pair experienced a slightly different modulation field amplitude due to field inhomogeneity in space, this allowed us to scale properly the abscissa of figure \ref{fig:BesselModulation2kT} for both data sets. 

It is not obvious, however whether there are deviations in the data due to eddy current heating. The fit reproduces extremely well the data below 60~G. There is a difference between the fit and the data near the peak of the function, where the fit overestimates the data. As we move to higher modulation, the fit joins the data again, suggesting that this is not the result of eddy current heating. Had it been the case, we would have expected the difference between the fit and the data to increase indefinitely with modulation field. Since there were no clear signs of eddy current heating in the modulation field dependence of the amplitude of the oscillations, we next looked at the temperature dependence and the fits of the LK relation.

\subsection{Analysis of eddy current heating \label{sect:ModFieldLK}}
\begin{figure}[!p]
	\begin{minipage}[t]{7cm}
		\begin{center}
		\includegraphics[width=7cm]{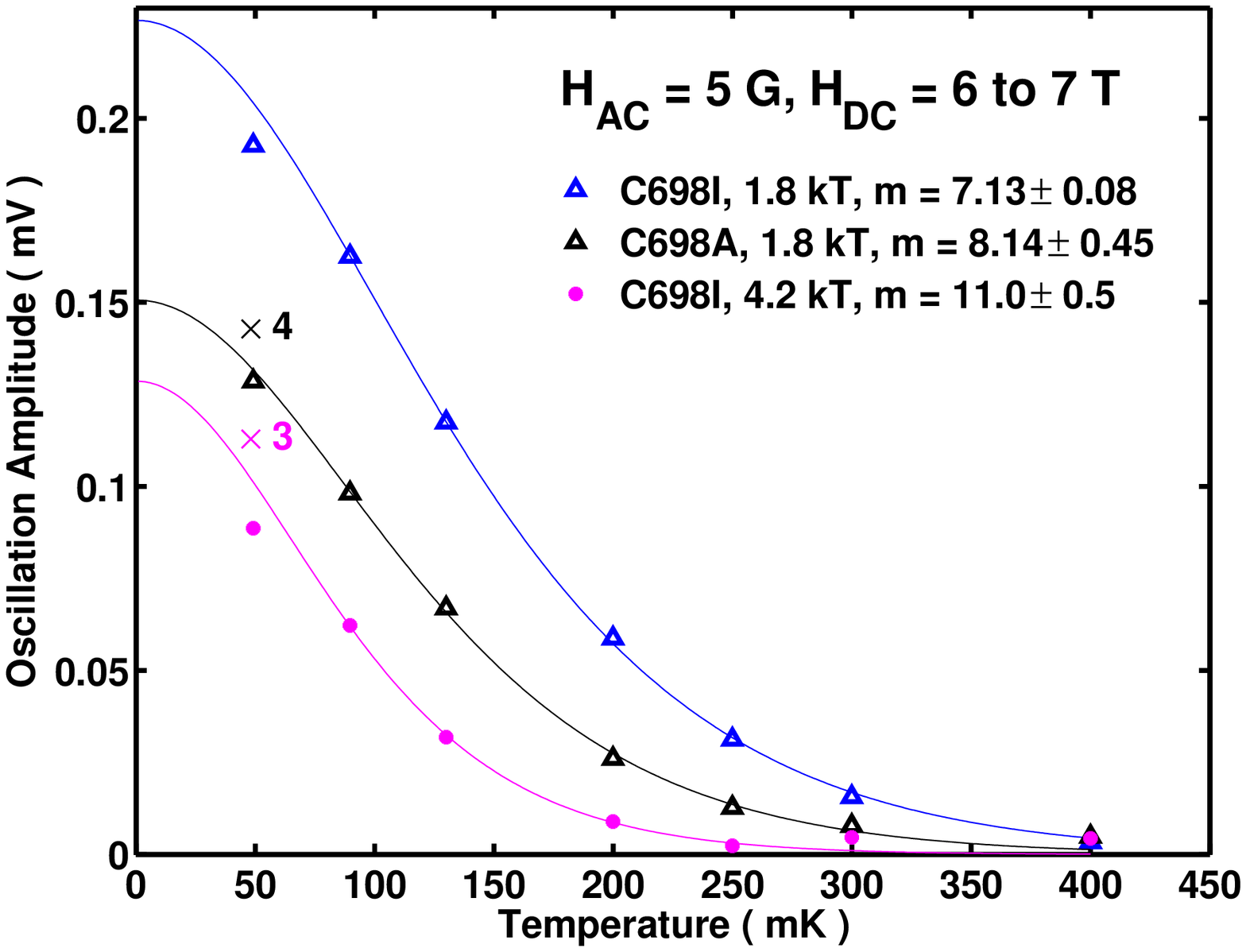}
		\includegraphics[width=7cm]{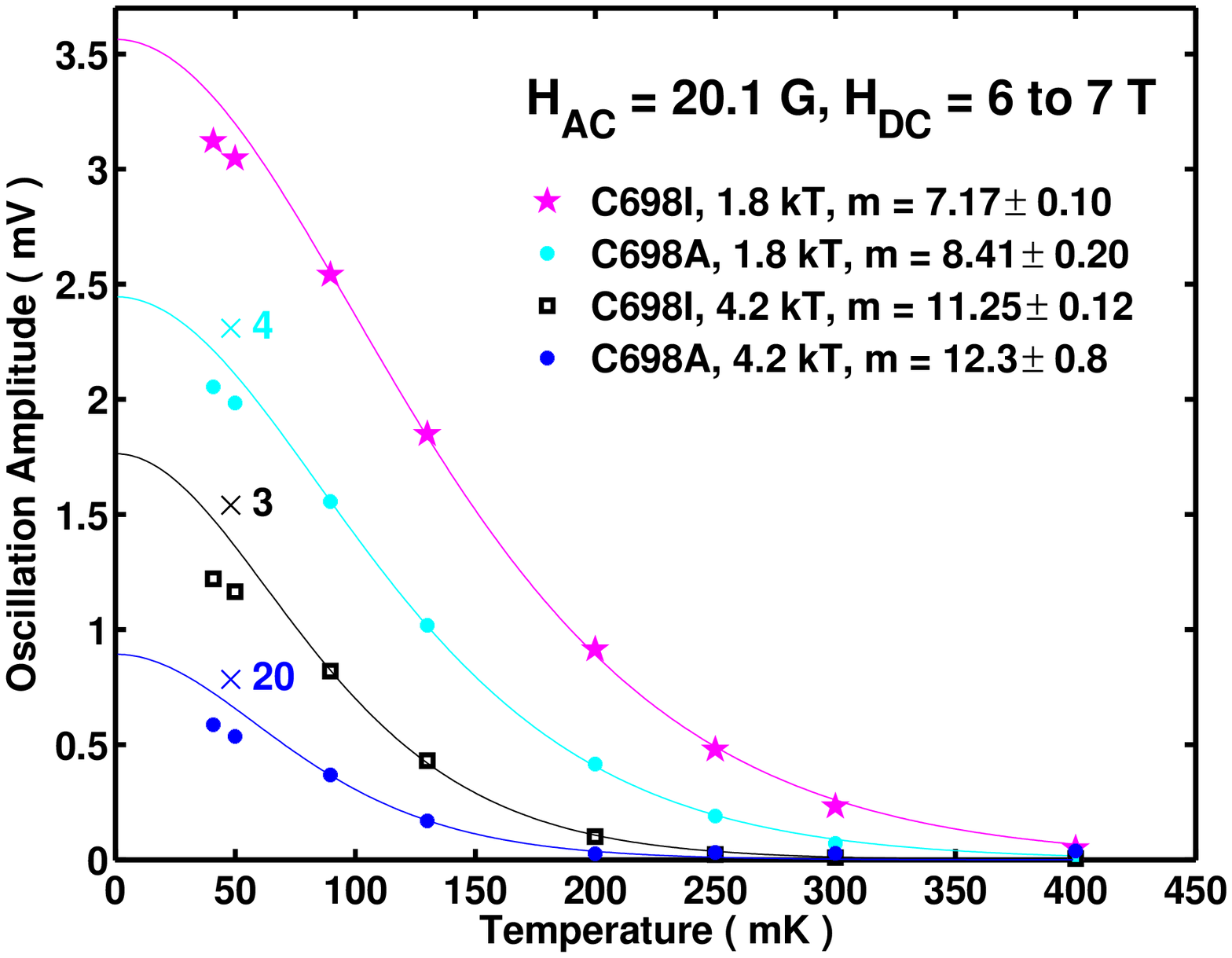}
		\includegraphics[width=7cm]{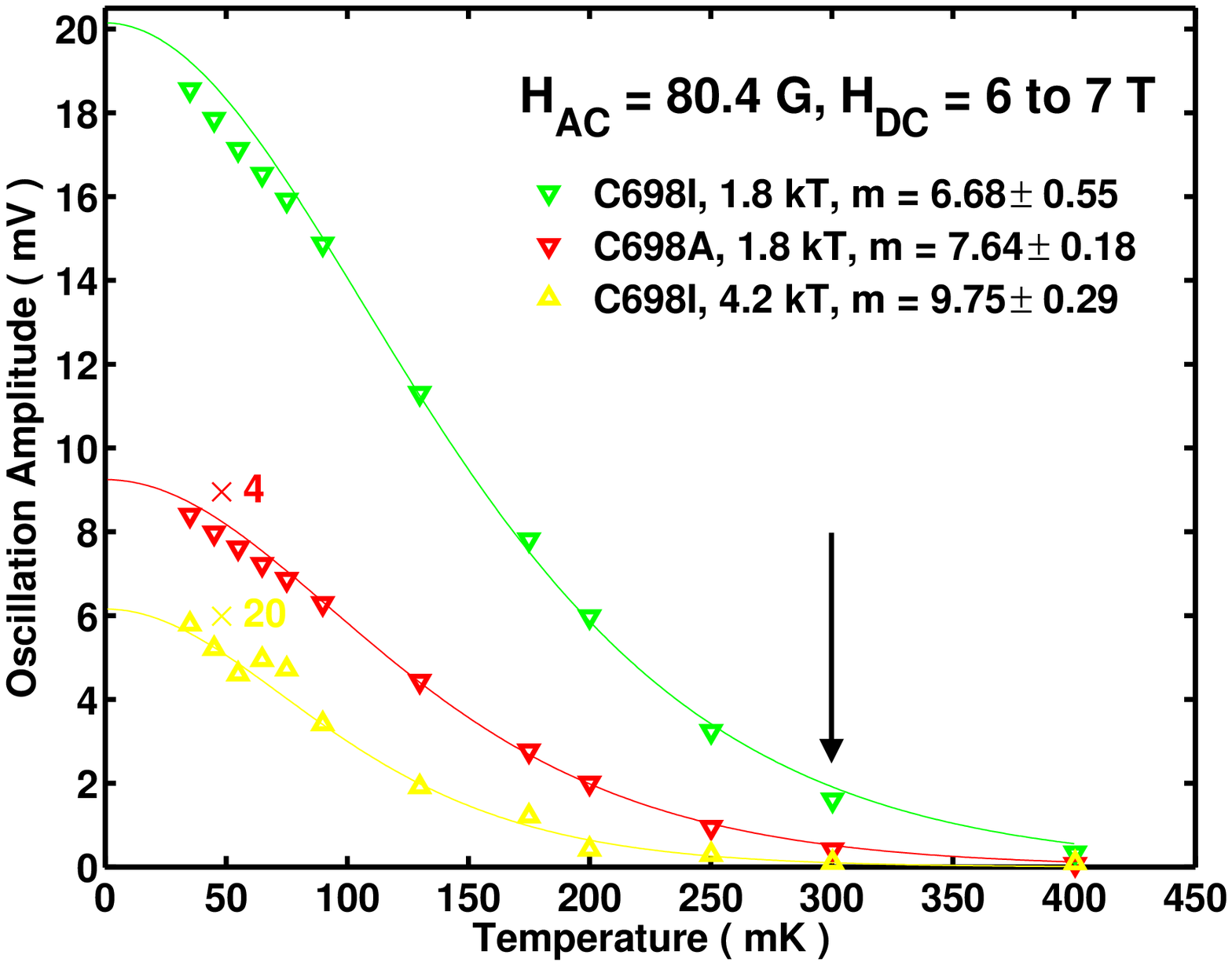}
		\end{center}
	\end{minipage}
	\hfill
	\begin{minipage}[t]{7cm}
		\begin{center}
		\includegraphics[width=7cm]{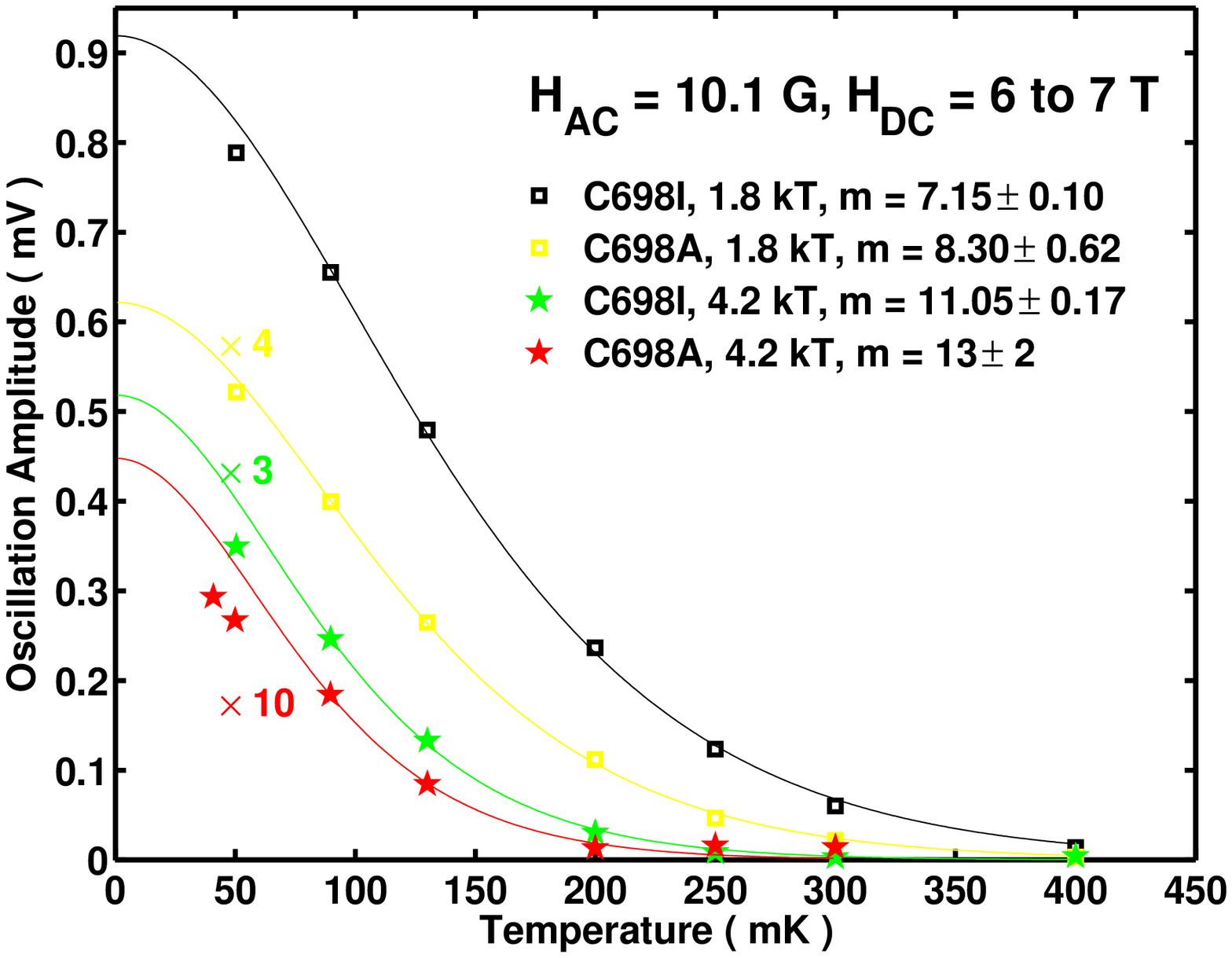}
		\includegraphics[width=7cm]{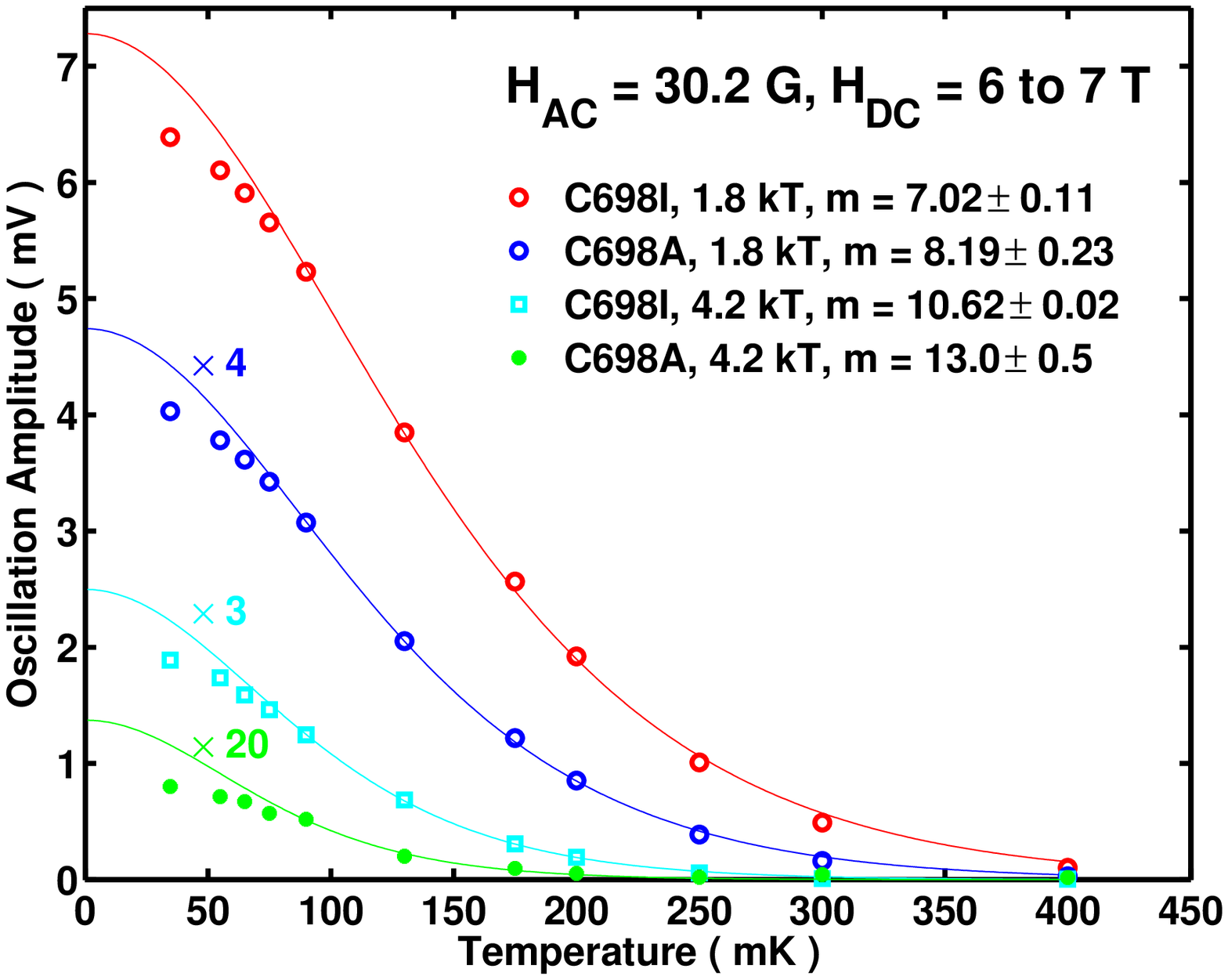}
		\includegraphics[width=7cm]{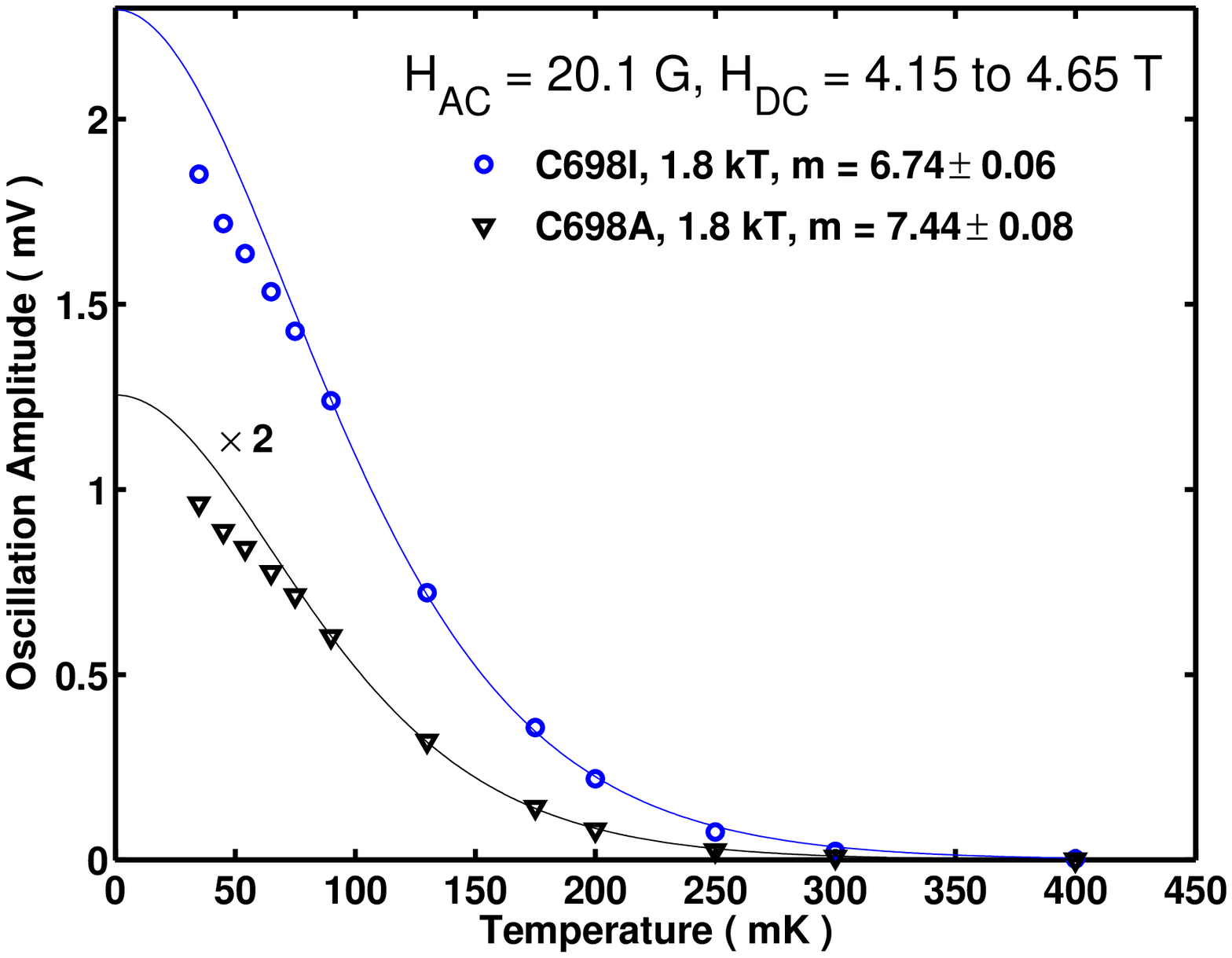}
		\end{center}
	\end{minipage}
	\caption[Modulation field study of eddy current heating]{Modulation field study of eddy current heating. The temperature dependence of the amplitude of the 1.8 and 4.2~kT peaks were measured for different modulation fields, from left to right, top to bottom: 5, 10.1, 20.1, 30.2, 80.4~G and 20.1 again, at 7.9~Hz.  The data were measured on samples C698I and C698A, where C698I was aligned with its $c$-axis parallel to the magnetic field, and C698A lied with an angle of 10$^{\circ}$. All Fourier transforms were performed between 6 and 7~T, except for the bottom right graph, which was measured between 4.15 and 4.65~T. Solid lines correspond to LK fits, and the resulting masses are quoted in the legend. Error values are those given by the fit procedure. }
	\label{fig:QMR_05G}
\end{figure}

The ultimate test for eddy current heating lies with the non-linear fit of the LK function. We expect the temperature dependence of the amplitude of dHvA in \TTS, away from the metamagnetic transition, to follow the LK function. Since the specific heat of most materials reduces with decreasing temperature, a constant heat input at the sample produces an error in temperature that becomes larger the lower the temperature. We conclude that any temperature gradient between the sample and the thermometer situated at the cooling apparatus will result in the low temperature part of the data saturating to a value lower than it would normally reach in normal conditions.  This subtle change can make the data appear obviously non-LK, but it could also appear like an LK function with a lower quasiparticle mass than reality. 

We measured the temperature dependence of dHvA between 6 and 7~T, at seven different modulation fields between 5 and 80.4~G at 7.9~Hz, for both the 1.8 and the 4.2~kT peaks. Figure \ref{fig:QMR_05G} shows the data for both frequencies, both samples for the modulation values of 5, 10.1, 20.1, 30.2 and 80.4~G. We observed a systematic deviation at low temperatures, which made the LK fits impossible when using all temperature points. By looking at the modulation field dependence of this deviation, we found there was none and that the deviation was identical at all modulation field values, which span over more than an order of magnitude. We concluded that this deviation was not related to the modulation field. By excluding temperature points below 90~mK, we found that the LK function fits the data well up to 80.4~G, where signs of eddy currents start to appear\footnote{Note that sample C698A lies with an angle of 10$^{\circ}$ between the field and its $c$-axis, leading to slightly higher masses and frequencies since, as seen from eqns \ref{eq:dAde} and \ref{eq:cosinelaw}, the extremal area and the mass, in 2D systems, scale with the inverse of $\cos\theta$}. The fits are shown in solid lines in figure \ref{fig:QMR_05G} and the results are presented in the legends. Slight deviations in the low temperature points generally force the LK fit to a lower mass, and one usually observes a deviation between fit and data at intermediate temperatures (as indicated by a black arrow in the lower left plot, figure \ref{fig:QMR_05G}). However, the value of the extracted mass was the best indication of heating: we observed a deviation from the normal on-axis value of $7.15\pm0.10$m$_e$ for the 1.8~kT peak, and $11.0 \pm 0.2$m$_e$ for the 4.2~kT peak, at 30~G and above \footnote{The same can be said of the values measured on C698A, slightly off-axis, which produced higher mass values, but the same deviation.}. We concluded that eddy current heating is only significant at such a modulation field or higher, and that the appropriate value to use safely lies near 20.1~G, at 7.9~Hz.

We furthermore verified whether the deviation had its origin in the proximity to the metamagnetic transition, where effects like spin-dependent masses or non-LK behaviour could potentially be observed. We measured the temperature dependence of dHvA in another field region of similar inverse width, 4.15 to 4.65~T, much further away from the metamagnetic transition (see lower right graph of figure \ref{fig:QMR_05G}). The same unusual behaviour was observed at low temperatures, excluding this possibility. Moreover, the calibration of the thermometer was later verified by comparison with another new RuO thermometer that was installed on the mixing chamber of the refrigerator, which revealed deviations too small to explain this problem. We conclude that if a temperature gradient was present in our system, it was not due to eddy current heating, and it is still not resolved at this point whether the deviation is part of the physics taking place in \TTS\ or if it was an experimental problem. As a result, all LK fits performed in this thesis were done using data from 90~mK and above.

\section{dHvA oscillations and spectra \label{sect:CamOscSpectra}}

We present in this section information that was extracted from the dHvA data for the $c$-axis direction. We first show raw data of second harmonic dHvA for both samples. We then reveal spectra that were calculated over large field ranges both in the low and high field sides of the metamagnetic transition, and provide a table of the extracted frequencies and quasiparticle masses. These are average values, mainly of interest for their comparison with, for instance, ARPES data (introduced in section \ref{sect:ZeroFieldFS}). Finally, we show plots of the field dependence of the dHvA frequencies, which will be used later in section \ref{sect:electrontransfer} in order to demonstrate that quasiparticle transfers occur in \TTS\ at the metamagnetic transition.

\subsection{dHvA oscillations}
\begin{figure}[t]
         \begin{center}
	\includegraphics[width=1\columnwidth]{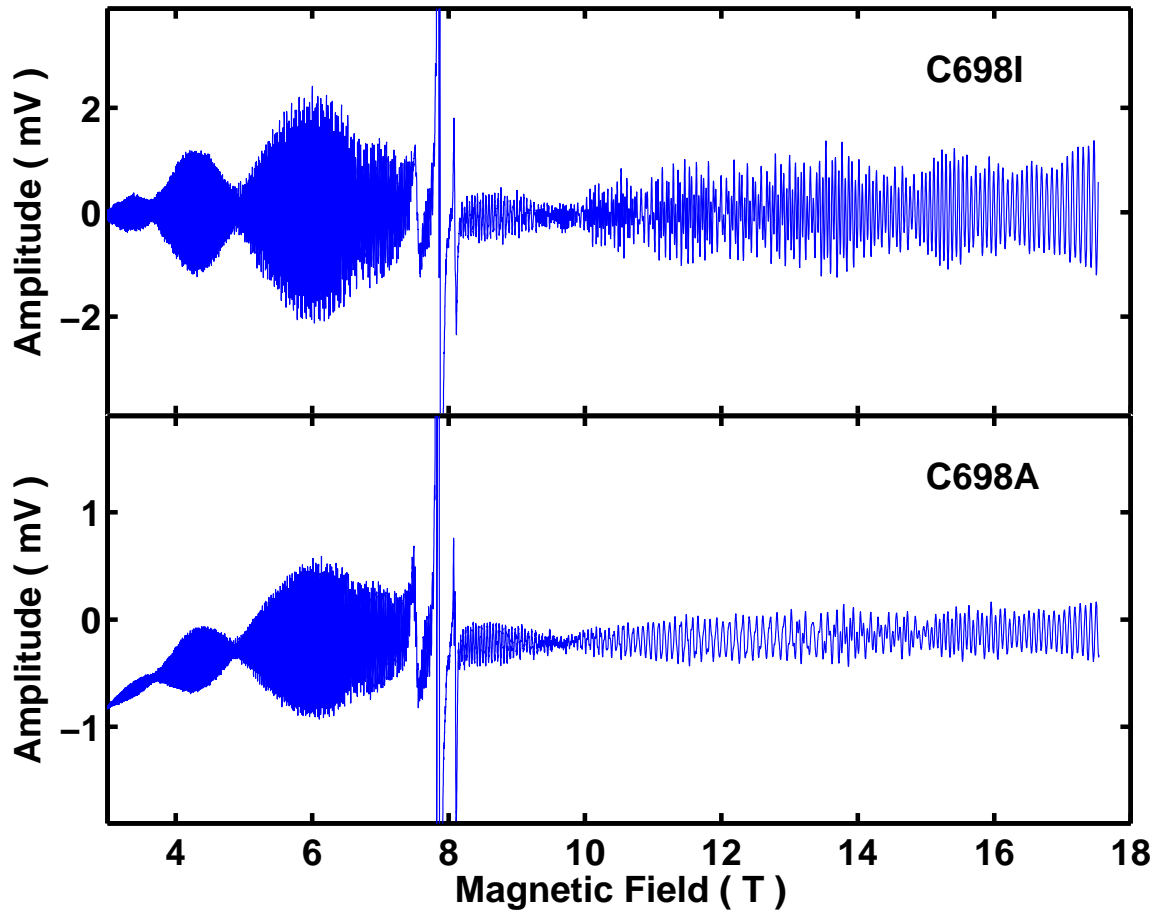}
	\caption[Raw second harmonic data at 25~mK in \TTS]{Raw second harmonic data measured at 25~mK, using a modulation of 20.1~G at 7.9~Hz, for both samples in the $c$-axis direction. The total amplification factor is of 10$^{5}$.}
	\label{fig: Raw_c-axis_25mK}
	\end{center}
\end{figure}

The oscillations measured in Cambridge were of exceptional quality compared to any of our previous experiments. The samples, C698I and C698A, were measured simultaneously with second harmonic detection only, using the AC susceptibility probe described in section \ref{sect:Camprobe}, with a modulation field of 20.1~G at 7.9~Hz, and a sweep rate of approximately 0.04 T/min above 10~T, and 0.02 T/min below, as described in section \ref{sect:Camprobe}. 

Temperatures of 25~mK or above were reached, and angles between -10$^{\circ}$ and 54$^{\circ}$ (see appendix~\ref{App:E} for details) were probed. We found a discrepancy between the DC field positions of the metamagnetic transitions as measured in these experiments compared to previous ones performed in St Andrews, of approximately 3\%. Since the values measured in St Andrews were previously confirmed by two other laboratories elsewhere in the world\footnote{At the Max Planck Institute in Dresden, and at the University of Kyoto}, we corrected all Cambridge data sets by the appropriate scaling factor\footnote{The conversion number used was of 0.9743, the field values of Cambridge being too high.}. 

Figure \ref{fig: Raw_c-axis_25mK} shows typical dHvA data at 25~mK for both samples, where the amplitude is approximately twice higher with C698I. These were not measured simultaneously, but at two different rotator positions, such that the two crystals were aligned with their $c$-axis parallel to the magnetic field. Since the rotation study was performed before the mass analysis, we knew precisely which rotator positions aligned each sample.

\subsection{dHvA spectra and quasiparticle masses at $c$-axis \label{sect:CamSpectra}}

\begin{figure}[p]
         \begin{center}
	\includegraphics[width=1\columnwidth]{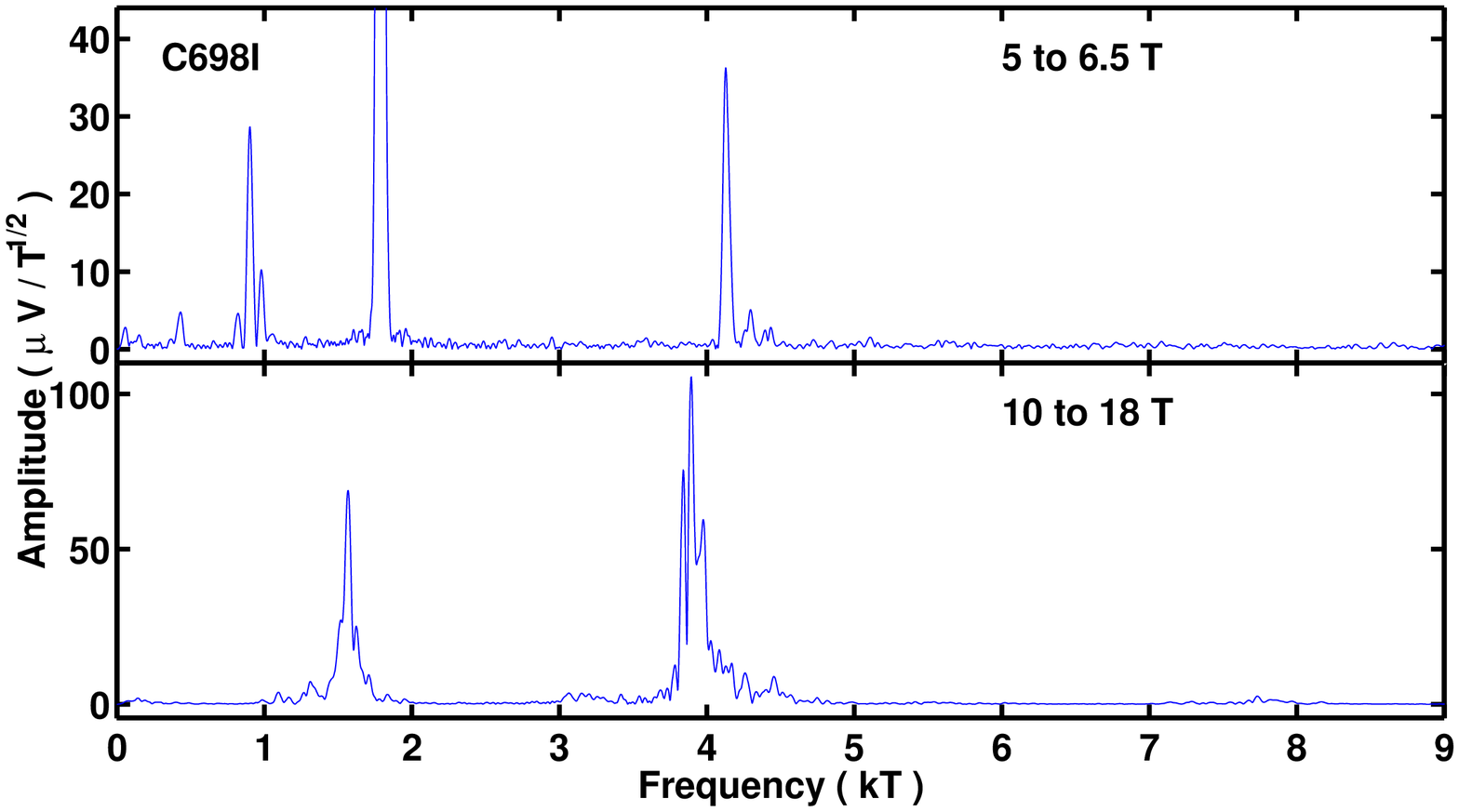}
	\includegraphics[width=1\columnwidth]{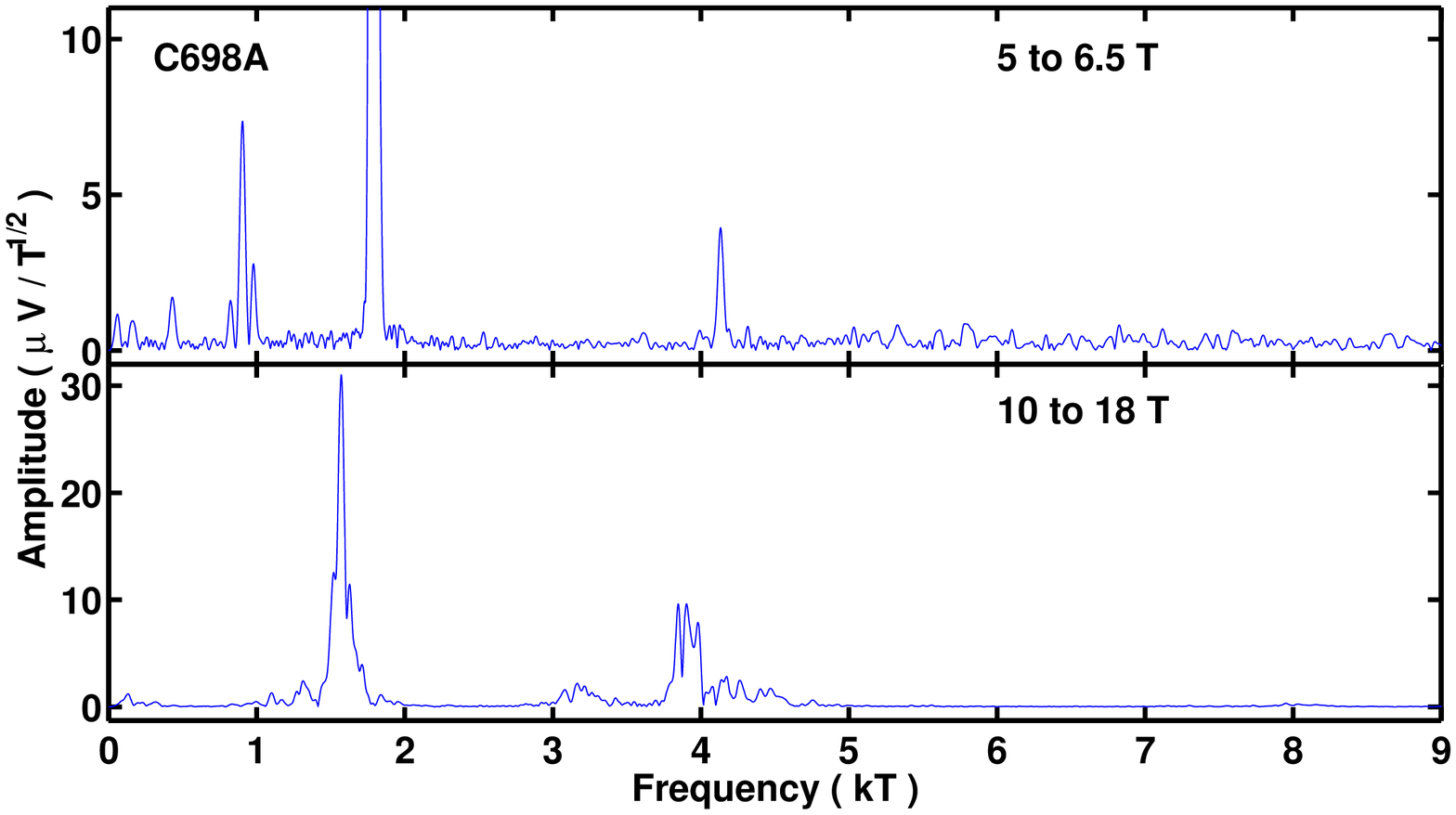}
	\caption[Fourier spectra between 5 and 6.5 T]{Fourier transforms of the dHvA data for both samples C698I and C698A, in the low field side (top parts of the graphes), using data between 5 and 6.5~T, and the high field side (bottom parts of the graphs), using data between 10 and 18~T. The peak at 1.8~kT was cut off for clarity at 350 $\mu$V/T and 120 $\mu$V/T for C698I and C698A respectively.}
	\label{fig: FFT_C698I}
	\end{center}
\end{figure}

\begin{table}[p]
	\begin{center}
		\begin{tabular}[h]{|c|c|c|c|}
			\hline
			\multicolumn{2}{|c|}{5 to 6.5 T} & \multicolumn{2}{|c|}{10 to 18 T}\\
			\hline
			Peak (kT) & $m^*/m$ & Peak (kT) & $m^*/m$\\
			\hline
			0.15 & $5.6 \pm 0.7$ &0.14&$4.9 \pm 0.1$\\	
			0.43 & $8.4 \pm 0.3$ &1.01&$10.9 \pm 0.2$\\
			0.82 & $7.7 \pm 0.7$ &1.12&$10.57 \pm 0.6$\\			
			0.91 & $7.7 \pm 0.8$ &1.20&$11.6 \pm 0.2$\\
			0.98 & $8.0 \pm 0.3$ &1.30&$9.3 \pm 0.3$\\
			1.78 & $6.6 \pm 0.4$ &1.35&$10.6 \pm 0.8$\\						
			4.13 & $10.1 \pm 0.1$ &1.45 to 1.73&$7.3 \pm 0.2$\\
			4.30 & $10.1 \pm 0.1$ &1.75&$8.4 \pm 0.5$\\
			4.40 & $ -- $ &1.88&$12.2 \pm 0.1$\\
			&&2.00&$14.4 \pm 0.1$\\
			&&3.14&$13.3 \pm 0.2$\\
			&&3.26&$11.7 \pm 1.5$\\
			&&3.36&$12.5 \pm 0.1$\\
			&&3.51&$14.9 \pm 0.1$\\
			&&3.65&$12.6 \pm 0.1$\\
			&&3.74 to 4.16&$11.3 \pm 0.3$\\
			&&4.16 to 4.32&$13.9 \pm 0.2$\\
			&&4.37&$13.0 \pm 0.1$\\
			&&4.57&$15.4 \pm 0.2$\\
			&&7.94&$17.2 \pm 0.3$\\
			\hline
		\end{tabular}
		\caption[Observed frequencies and associated quasiparticle masses]{Table of the observed frequencies in the Fourier transforms of the dHvA data from sample C698I, along with associated quasiparticle masses. The Fourier transforms were taken from 5 to 6.5~T, for the low field side, and from 10 to 18~T for the high field side. Some peaks in the high field side were difficult to integrate individually, where in such cases larger frequency intervals were used in the integrals for the non-linear LK fits. Such intervals are indicated.}
	\label{tab:freqmassesCambridge}
	\end{center}
\end{table}

\begin{figure}[p]
         \begin{center}
	\includegraphics[width=1\columnwidth]{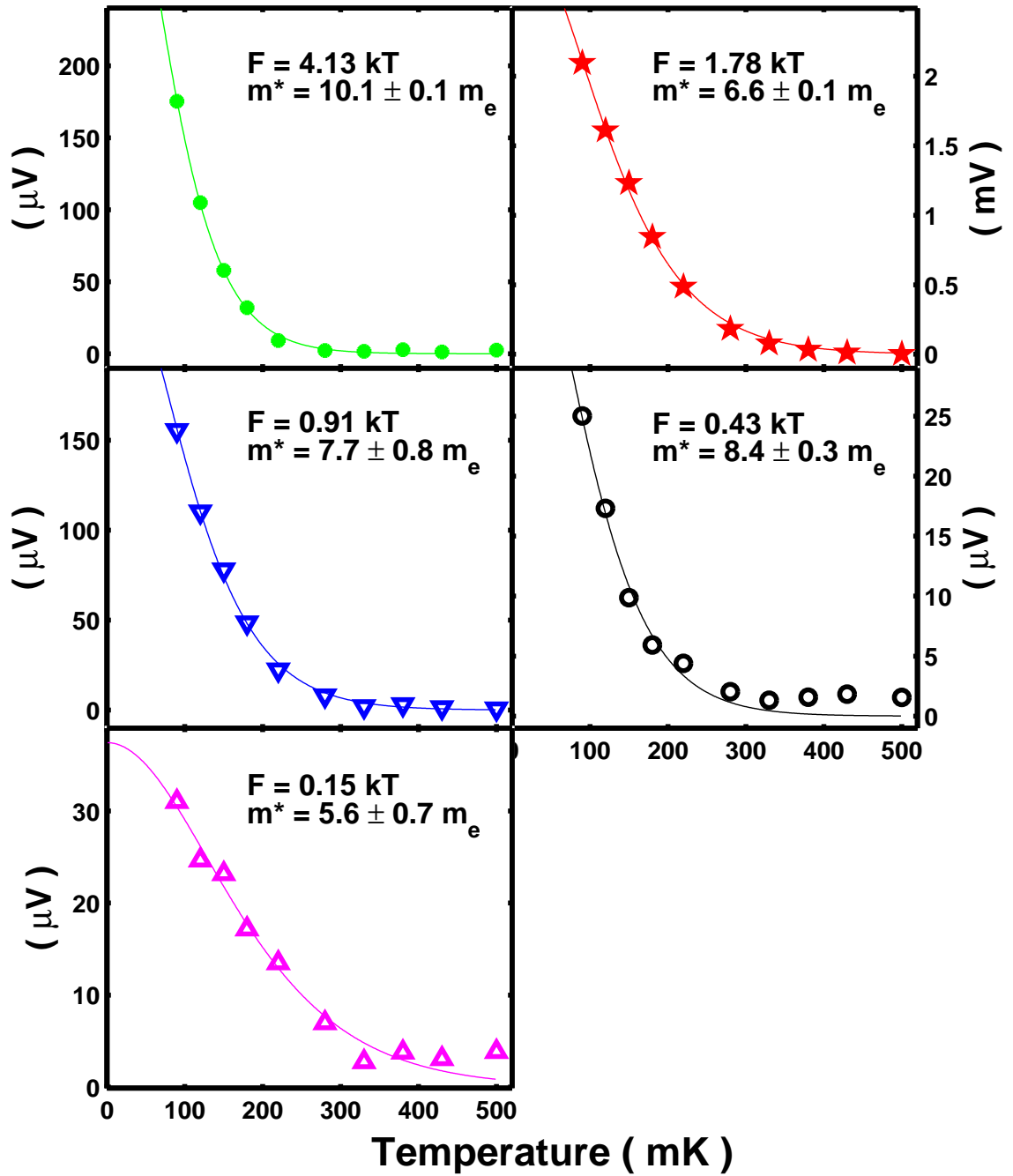}
	\caption[LK fits for the five frequencies observed in the low field side]{Temperature dependence of the amplitude of the oscillations for the five main frequencies present in the low field side, along with LK fits, shown in solid lines. Fourier transforms were taken from 5 to 6.5~T in all cases except for $F = 0.15$ kT, of which the Fourier transform was taken from 6 to 7~T.}
	\label{fig: LowFieldMasses4}
	\end{center}
\end{figure}

\begin{figure}[p]
         \begin{center}
	\includegraphics[width=0.8\columnwidth]{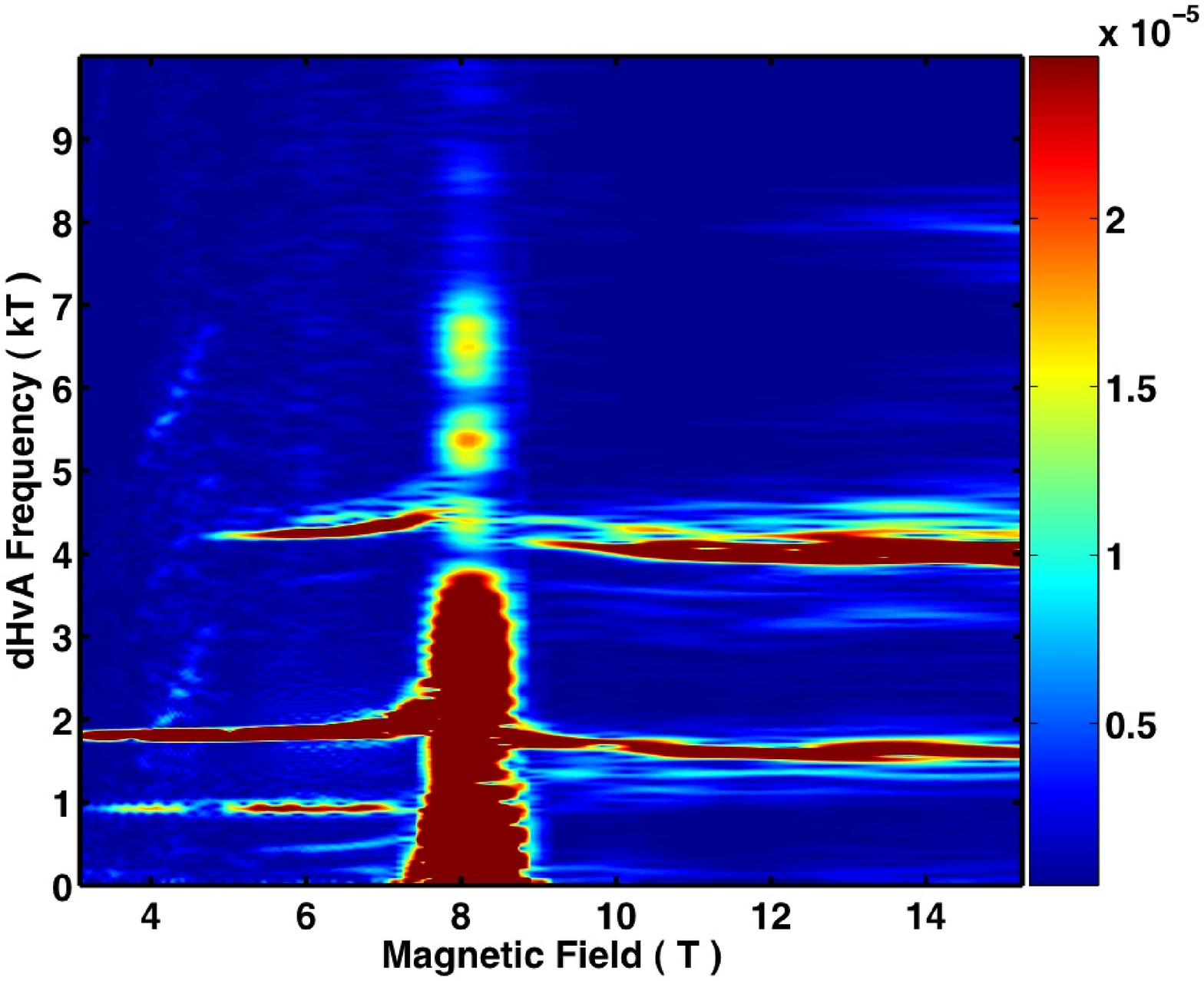}
	\includegraphics[width=0.8\columnwidth]{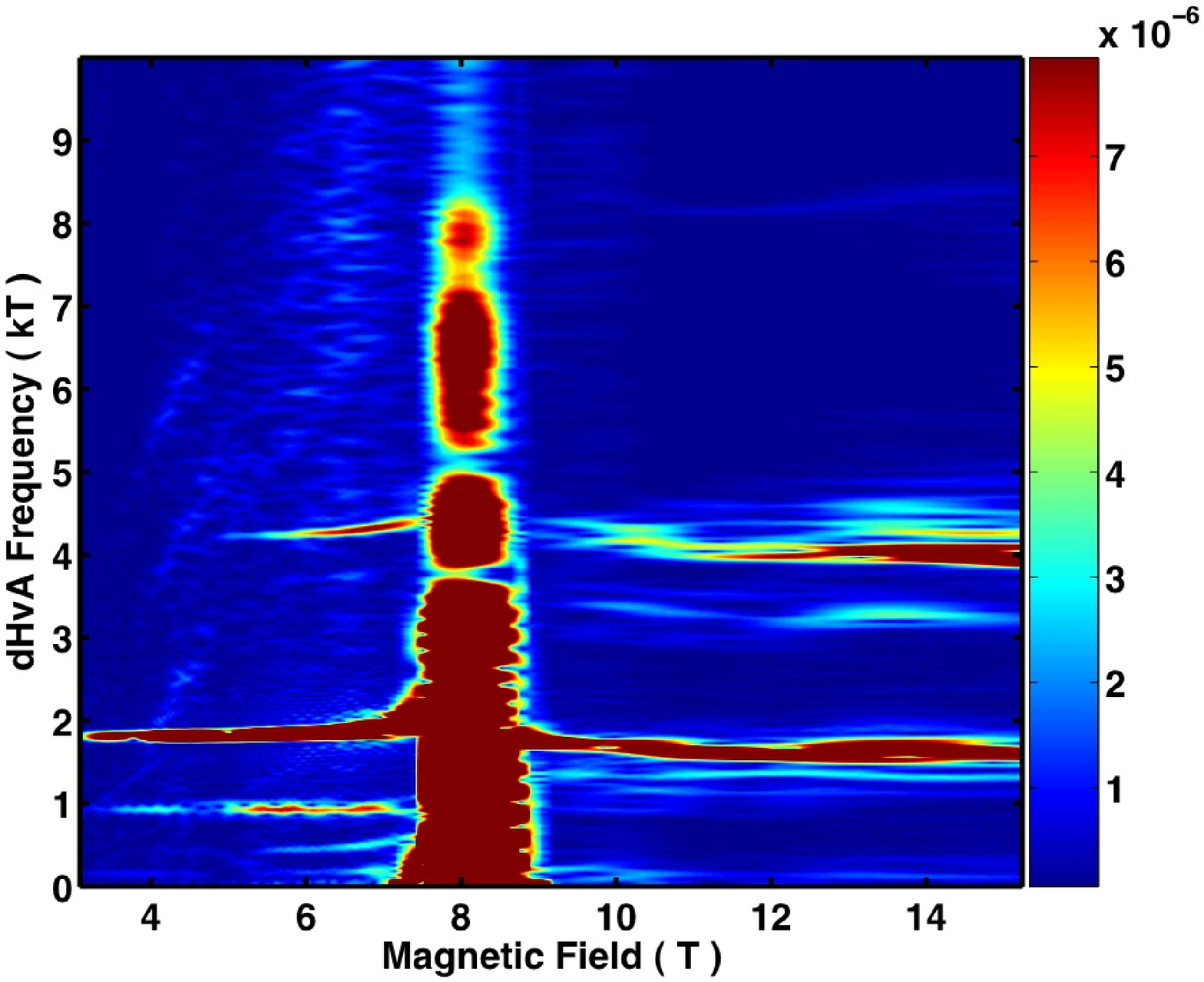}
	\caption[Field dependence of the dHvA spectra]{Field dependence of the dHvA spectra for samples C698I ($top$) and C698A ($bottom$), at 25~mK. The intensity of the data was cut off at 25 ($top$) and 8 $\mu$V/$\sqrt{T}$ ($bottom$) in order to produce contrast for the low amplitude details.}
	\label{fig:FvsBCam}
	\end{center}
\end{figure}

The dHvA spectra away from the metamagnetic transition were obtained by using Fourier transforms applied to the data within the intervals of 5 to 6.5 and 10 to 18~T, which correspond to similar inverse field intervals. Since, as we know from the work of Borzi \cite{borzi}, the dHvA frequencies vary close to 8~T, we could not include that region in the calculation. Moreover, field ranges below 5~T do not improve the clarity of the spectra.

Figure \ref{fig: FFT_C698I} shows Fourier spectra for both samples calculated using these field regions. In the low field side, we obtained five peaks or groups of peaks, positionned at 0.15, 0.43, 0.9, 1.8 and 4.1~kT. These are essentially the same as those measured by Borzi, but we observed that the peak at 4.2~kT is split into three and the one near 0.9~kT into four\footnote{Only three are visible in figure \ref{fig: FFT_C698I}, with a larger field range, the central one splits into two. Larger field ranges reduce the clarity of the overall spectrum.}. Moreover, both samples produced the same results, with a resolution slightly lower with C698A. In the high field side, the improved data quality revealed more complexity than in the work of Borzi. We observed four groups of peaks, each of which appears to correspond to a superposition of a great number of frequencies and lie near 1.6, 3.2, 4.0 and 7.9~kT.  

Table \ref{tab:freqmassesCambridge} presents two lists of frequencies and corresponding masses for both field sides of the metamagnetic transition in sample C698I. An identical set of values was also calculated for sample C698A, which turned out very similar, and is not shown. The table moreover lists the value of the quasiparticle mass associated with each of these frequency peaks calculated with non-linear LK fits applied to the temperature dependence of their amplitude, as described in section \ref{sect:mass}. For low field side data, figure \ref{fig: LowFieldMasses4} presents these data and LK fits for the five measured frequencies, where data from 90~mK and above were used, for the reason described in section \ref{sect:ModFieldLK}. All masses calculated are more precise and accurate than in the previous work by Borzi, as error bars are smaller and only two parameter fits were used as opposed to three\footnote{We will show in section \ref{sect:MassSystematic} how three parameter fits induce systematic errors to LK fits.}.

The complexity of the high field side does not simply arise due to the additional intrinsic inner magnetic field produced by the sample itself, producing a complex dependence of the oscillations on $B$. Using the work by Perry $et$ $al.$ \cite{perry1}, we calculated its magnitude, and obtained around 0.015~T, which, when added to the measured field values, does not make any visible difference to the spectra; much higher offsets are required to produce broadening. Moreover, the multiple peaks of the high field side do not originate from simple spin splitting as one would normally expect in a metamagnetic system\footnote{see section \ref{sect:spinsplitting} for details}. In such a scenario, each peak of the low field side splits into only two as the system crosses the metamagnetic transition. Part of the complexity of the high field side spectrum may originate from spin splitting, but many more peaks are found. Finally, the quasiparticle masses associated with these peaks are not all consistent within each group; instead, we find different values ranging between 7 and 17~$m_e$. 

Since the Fermi surface undergoes an evolution across the metamagnetic transition, it is interesting to look at the field dependence of the spectrum shown in figure \ref{fig:FvsBCam}. This calculation was performed by taking Fourier transforms over 150 small overlapping inverse field intervals of width $\Delta X = $0.02~T$^{-1}$, and plotting the intensity of its power spectrum at each average field position. Resulting horizontal lines of intensity correspond to quasiparticle orbits, and the vertical feature near 8~T  is produced from the Fourier transform of the metamagnetic peaks. 

One can see, in the low field side, that the dHvA frequencies at 0.45 1.8 and 4.2~kT undergo a dramatic increase as they come towards 8~T. The peaks at 1.8 and 4.2~kT decrease again in the high field side towards 0.16 and 4.0, respectively, while the peaks at 0.45 and 0.9~kT disappear. Finally, one observes that the complexity of the high field side gradually increases with field.

\subsection{Dingle analysis of $c$-axis dHvA \label{sect:CamDingle}}

\begin{figure}[t]
         \begin{center}
	\includegraphics[width=0.8\columnwidth]{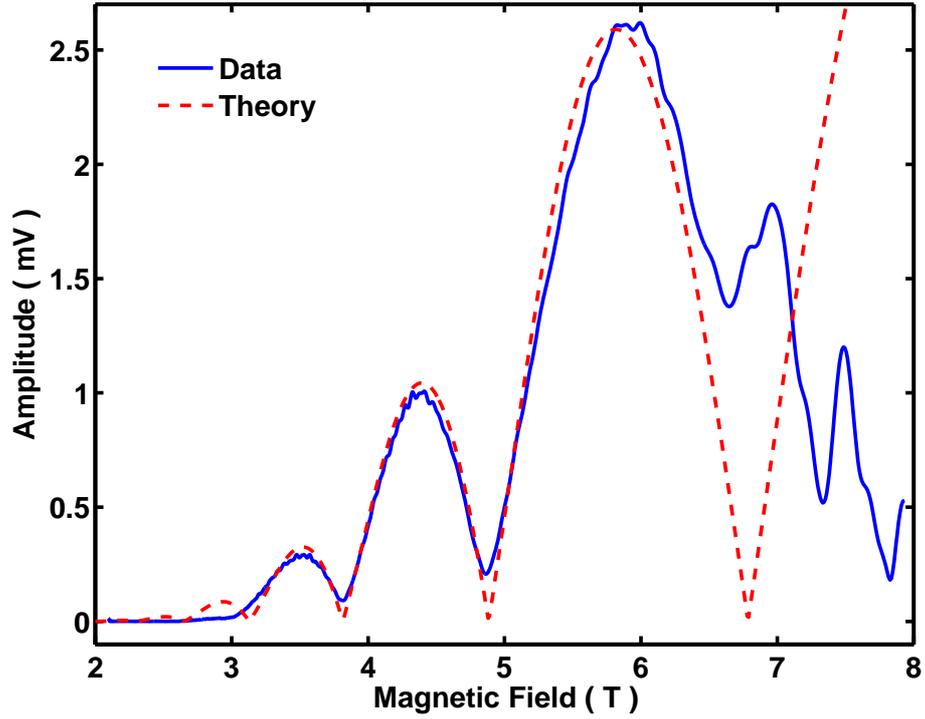}
	\caption[Dingle analysis of the amplitude of the 1.8~kT frequency.]{Dingle analysis of the amplitude of the 1.8~kT frequency in the low field side with the field in the $c$-axis direction. The experimental amplitude function is shown with a solid blue line, and the broken red curve corresponds to eq. \ref{eq:DingleFit}. }
	\label{fig: DingleAnalysis}
	\end{center}
\end{figure}

The low field side data at $c$-axis, 25~mK, may be used in order to calculate the mean free path of the quasiparticles, using the method described in section \ref{sect:Dingle}. Only the 1.8~kT peak possesses a modulation that is simple enough to model, and thus will be the only one used. We employed the envelope extraction method, section \ref{sect:envelope}, and adjusted a function of the type
\beq
A e^{-\lambda |X|} {CXTm^* \over \sinh(CXTm^*)} \big|\cos(\omega X + \phi) B_J(2*pi*H_{AC}FX^2)\big|,
\label{eq:DingleFit}
\eeq
with free variables $A$ and $\lambda$, assuming all the other parameters known. 

Figure \ref{fig: DingleAnalysis} presents the analysis, with the envelope of the 1.8~kT peak plotted in blue and eq. \ref{eq:DingleFit} overlaid in red. For most of the low field range, reasonable agreement was found. The data differs from the model near the metamagnetic transition, as a result of dephasing due to spin splitting, which will be discussed in section \ref{sect:LowFieldSideModel}. We extracted the value $\lambda = 36$~T, which, using eq. \ref{eq:ell}, corresponds to a mean free path of around $\ell = 270~$nm. 

\section{Angular dependence of dHvA \label{sect:FirstRotation}}

The angular dependence of dHvA in \TTS\ was measured in Cambridge only during this project. Traditionally, the angular study of dHvA is used to obtain information about the three dimensional shape of a Fermi surface. In two dimensional layered materials, as we described extensively in section \ref{sect:BergemanAnalysis}, the angular study of dHvA allows one to determine in detail the corrugation of the various Fermi surface pockets. It can also help one to determine their symmetry and position inside the BZ. This type of analysis was not performed completely in the past in \TTS, and we present here the data resulting from the first attempt at describing the complete three dimensional shape of the Fermi surface in \TTS.

The samples were cooled down and kept at a constant temperature of 30$\pm$5~mK. These were oriented along crystallographic directions, described in section \ref{sect:Laue}. The rotation of the samples was performed using the system described in section \ref{sect:Camprobe} and appendix~\ref{App:E}, such that sample C698I was rotated from [001] towards [110] and C698A towards [100]. We found that errors in the installation of one coil and in the cutting of sample C698A resulted in a difference of alignment between the samples, which lies in the plane of rotation \footnote{see appendix~\ref{App:E} for details}. Over 35 angles at steps of 1.7$^{\circ}$, second harmonic AC susceptibility was recorded along the full field range between 18 and 2~T, using the method described in section \ref{sect:Camprobe}. The resulting data possessed an unexpected complexity which, as we will present in section \ref{sect:CamEnvelopes}, is not explained by the approach of Bergemann \cite{bergemann}, and a model will be presented in section \ref{sect:LowFieldSideModel} suggesting an origin to this phenomenon. Appendix~\ref{App:E} presents plots of the raw data at all angles, for the interest of the keen reader, where the complexity of the beat patterns can be appreciated, as well as the behaviour of the metamagnetic transition signal.

\TTS\ is close to tetragonal; although the $a$ and $b$ axes are not equivalent from a symmetry point of view, these lattice parameters are equal in length and make those directions indistinguishable from a Laue experiment. Effectively, \TTS\ is orthorhombic from the crystal structure, which lacks 90$^{\circ}$ rotational symmetry, and hence the Fermi surface lacks that symmetry as well. Using the work of Mazin and Singh \cite{singh} as a guiding line, we nevertheless assumed that the $a$ and $b$ were $almost$ equivalent, in that the Fermi surface is $almost$ symmetrical under rotation by 90$^{\circ}$. We found later that, from the work of A. Tamai $et$ $al.$ \cite{tamai}(described in section \ref{sect:ZeroFieldFS}), this is actually the case.

We discuss in the first section an aspect of the data independent of dHvA but of importance in \TTS, that of the phase diagram and the mapping of the metamagnetic transition in the field-angle plane. We will then present the angular dependence of the dHvA spectra, followed by a discussion of its meaning. Finally, we will describe beat patterns, along with an attempted simulation. 

\subsection{Mapping of the metamagnetic transition \label{sect:MMTpositions}}

\begin{figure}[t]
         \begin{center}
	\includegraphics[width=0.8\columnwidth]{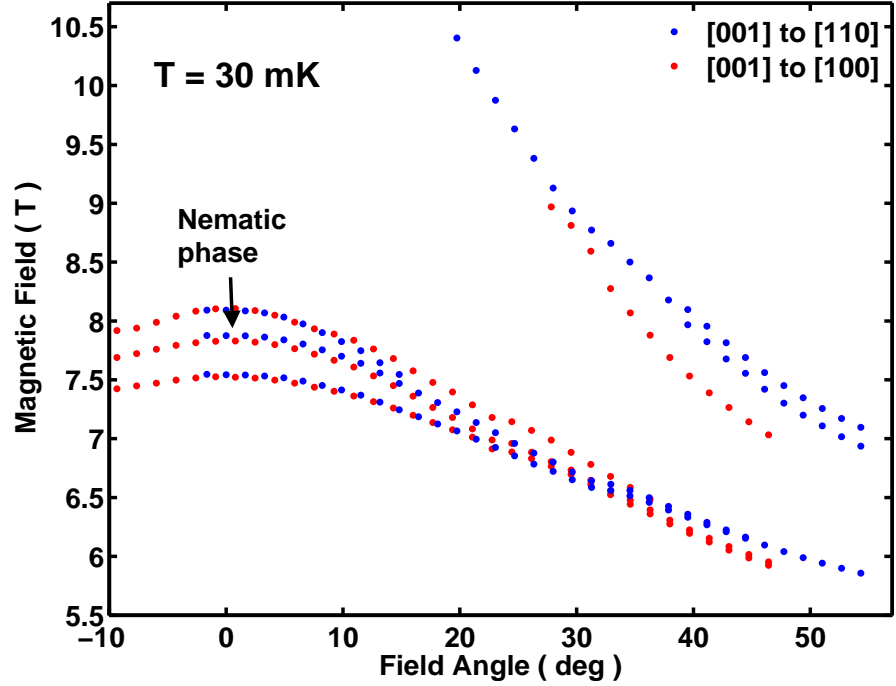}
	\caption[Phase diagram of \TTS\ from AC susceptibility]{Phase diagram of \TTS. Mapping of the positions of the features in the magnetic susceptibility, taken from the second harmonic data of the rotation study, for both samples. Sample C698I was rotating towards [110] and sample C698A towards [100], where the zero of the angle is at [001]. Three features seen in the $c$-axis data are visible at zero angle, and join to a single one at higher angle. The nematic phase region lies between the features at 7.9 and 8.1~T near zero angle.}
	\label{fig:MMTpositions}
	\end{center}
\end{figure}

The field derivative of the AC susceptibility provides information about the metamagnetic transition and the phase diagram of \TTS. The field values at which the metamagnetic features appear depend strongly on field angle \cite{grigeraPRB}, and it was possible to map out these in the present data set. Figure \ref{fig:MMTpositions} shows the positions in field as a function of field angle for two directions of rotation, from [001] towards [110] and [100]. It is interesting to note that the two directions are not equivalent in terms of metamagnetic behaviour. At zero angle, one can see three features, the nematic phase lying between the top two. Increasing the angle, we observe that the top transition line joins the middle one at an angle of about 15$^{\circ}$, for the [110] direction, and at about 36$^{\circ}$ for the [100] direction. Moreover, the middle transition moves towards 6~T at high angle in a different manner for both directions. Also, there is a feature arising at higher fields near 20$^{\circ}$ which moves fast to lower fields and splits into two, and also behaves differently in the two directions.

These remarks about the direction of rotation were first observed with this data. It is not absolutely surprising that the metamagnetic behaviour should be different in these directions, as they are not equivalent\footnote{see section \ref{sect:Cstructure} for details about the crystal symmetry}. In other words, the magnetism of \TTS\ is not isotropic in the plane. It has not been observed before, and suggests that a new phase diagram needs to be measured with the highest purity samples. It was also recently found, with specific heat experiment by A. Rost \footnote{Unpublished work which will form part of  A. Rost's thesis \cite{rost}.}, that there are significant differences in the thermal properties of \TTS\ with rotation direction, such that they are not equivalent when the field is in the plane, whether it points towards [100] or [110].

\subsection{Angular dependence of dHvA spectra}

\begin{figure}[p]
         \begin{center}
	\includegraphics[width=0.8\columnwidth]{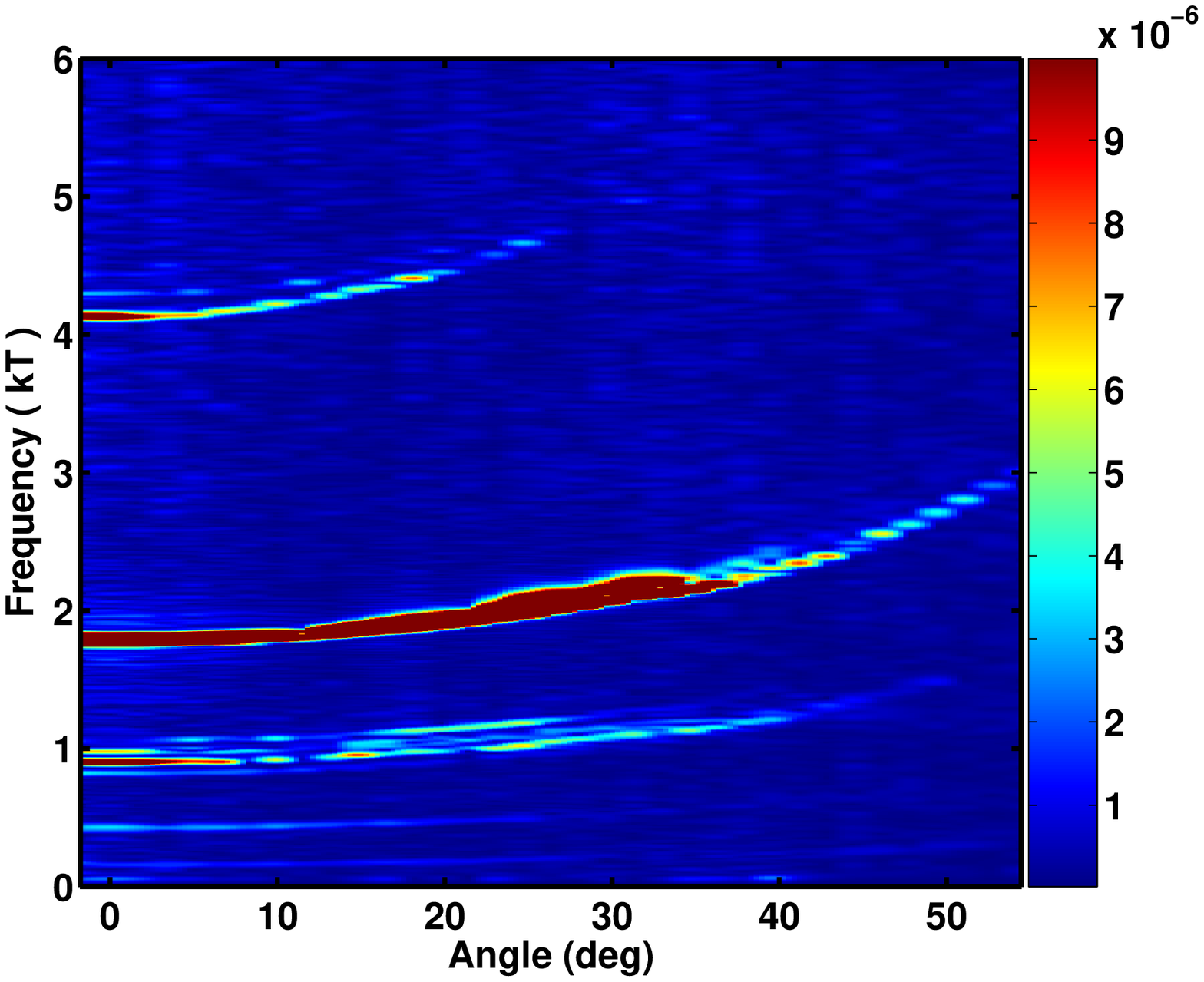}
	\includegraphics[width=0.8\columnwidth]{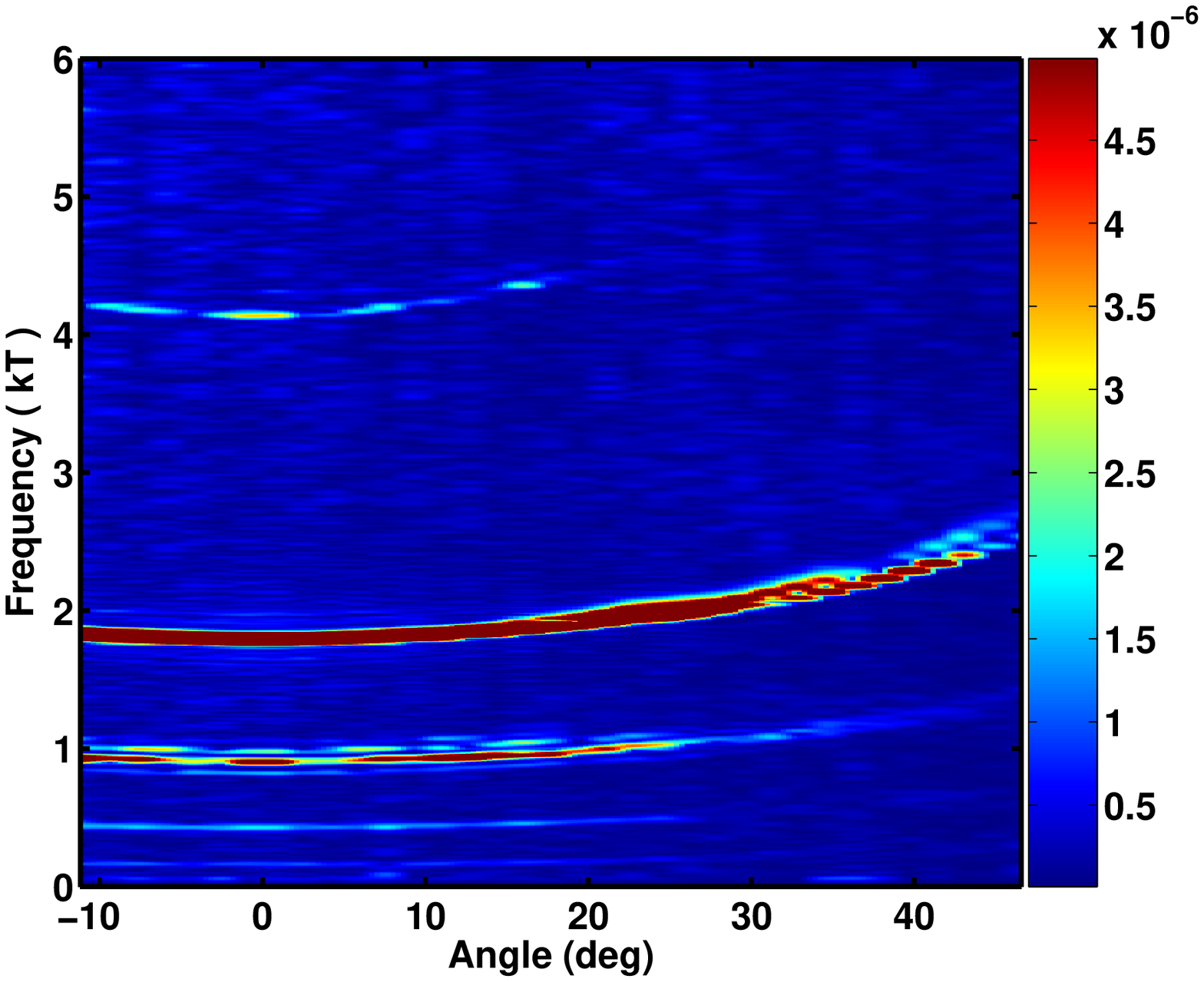}
	\caption[Angular dependence of dHvA spectra in the low field side]{Angular dependence of the dHvA spectra in the low field side for sample C698I, $top$, which was rotated from [001] towards [110], and sample C698A, $botttom$, which was rotated from [001] towards [100]. }
	\label{fig: FFT5to6p5T_A}
	\end{center}
\end{figure}

\begin{figure}[p]
         \begin{center}
	\includegraphics[width=0.8\columnwidth]{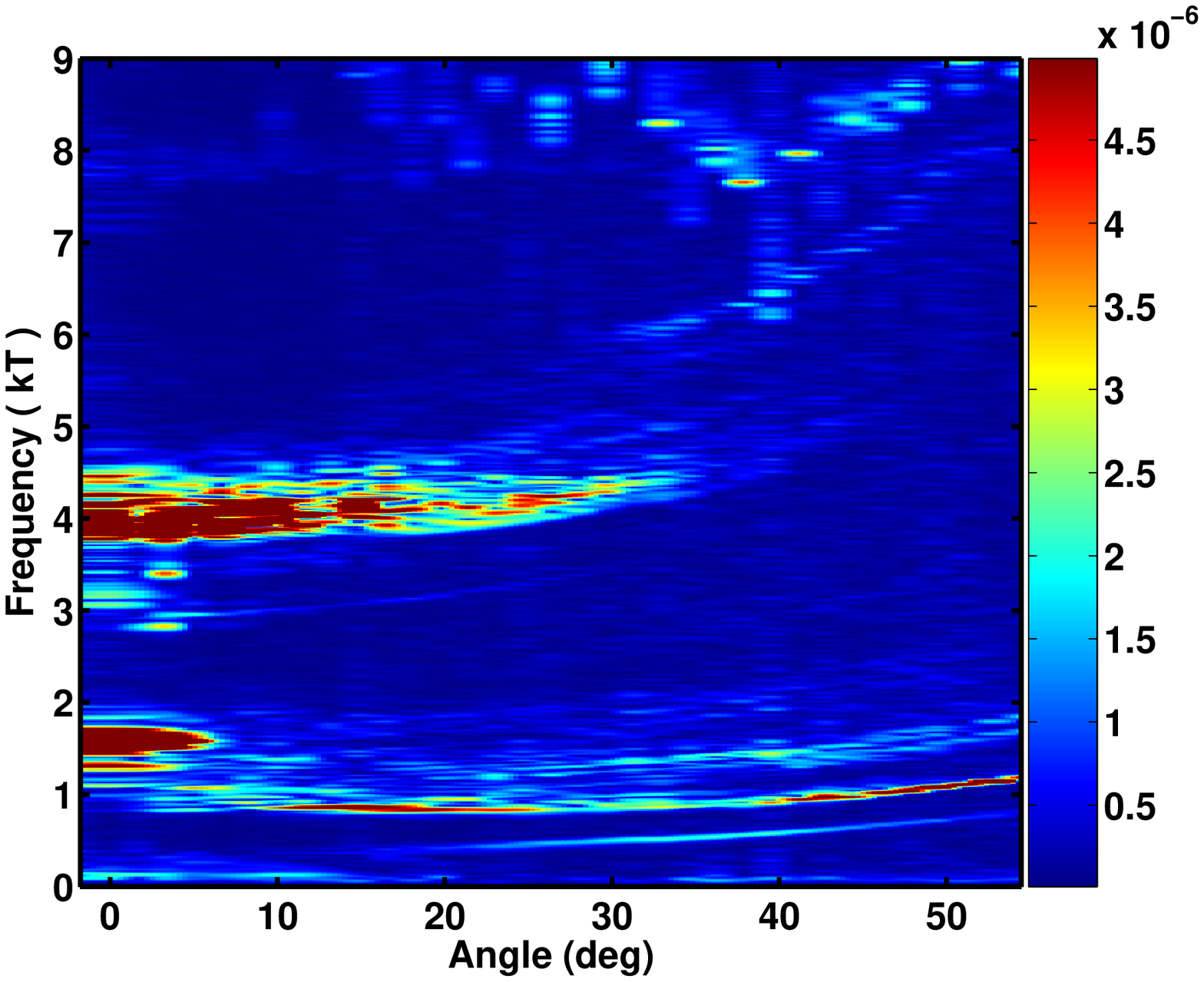}
	\includegraphics[width=0.8\columnwidth]{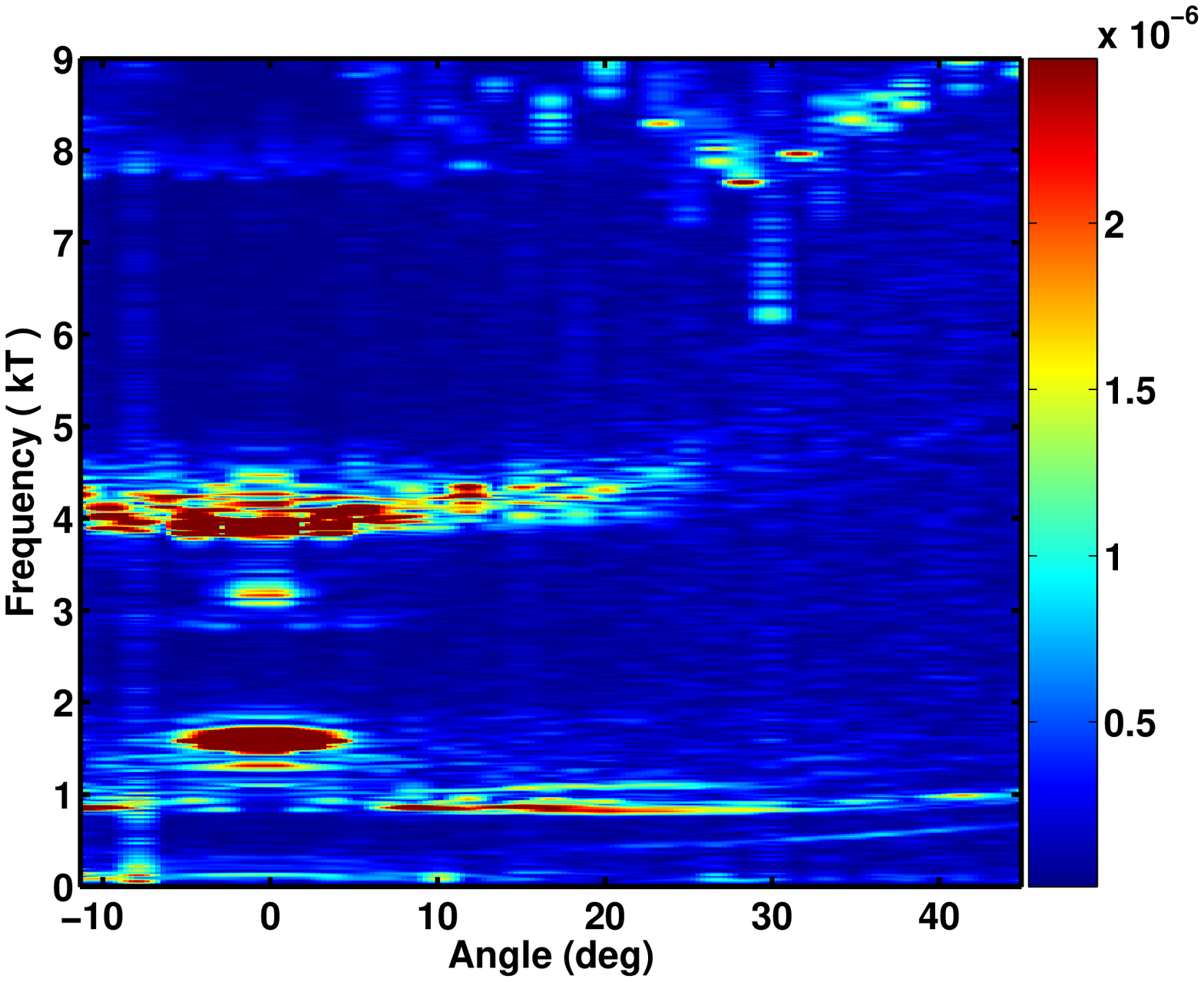}
	\caption[Angular dependence of dHvA spectra in the high field side]{Angular dependence of the dHvA spectra in the high field side for each sample, as described in figure \ref{fig: FFT5to6p5T_A}.}
	\label{fig: FFT10to18T_A}
	\end{center}
\end{figure}

Figure \ref{fig: FFT5to6p5T_A} presents the angular dependence of the dHvA spectra for both samples in the low field side, one per sample, or direction of rotation. The colour in the plot corresponds to the square root of the amplitude of Fourier transform power spectra for each field angle centred at the angle value. The Fourier transforms were taken from 5 to 6.5~T, except at high angles, where the interval was taken from 5 up to the position of the metamagnetic transition \footnote{As can be seen in figure \ref{fig:MMTpositions}, the metamagnetic transition enters the region from 5 to 6.5~T near 30$^{\circ}$. The Fourier transform of the metamagnetic signal produces large intensities near zero frequency, which may interfere with interesting features.}. The intensity of the data was furthermore cut off in order to provide contrast for low amplitude features. 

\begin{figure}[p]
	\begin{center}
		\includegraphics[width=0.8\columnwidth]{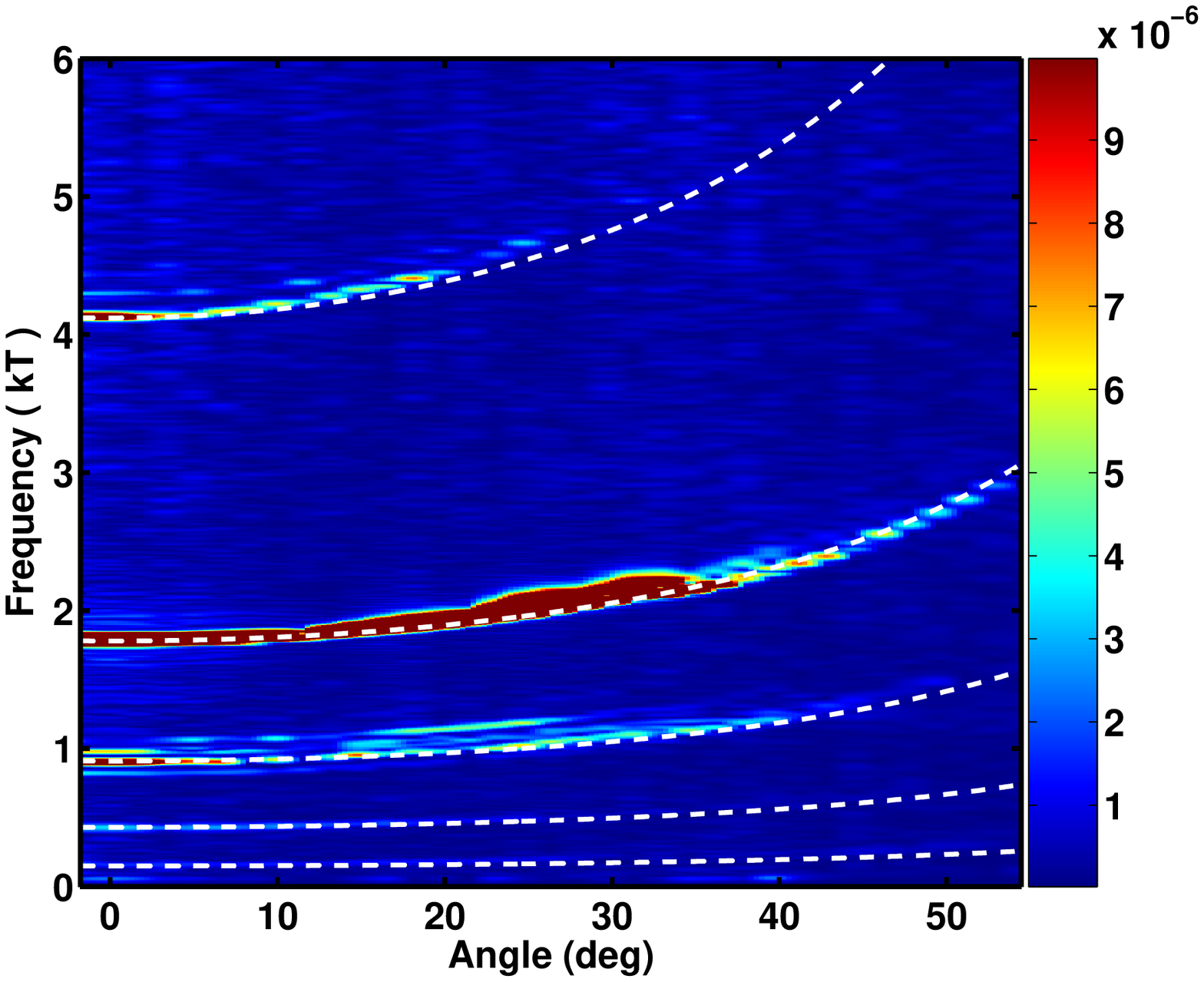}
		\includegraphics[width=0.8\columnwidth]{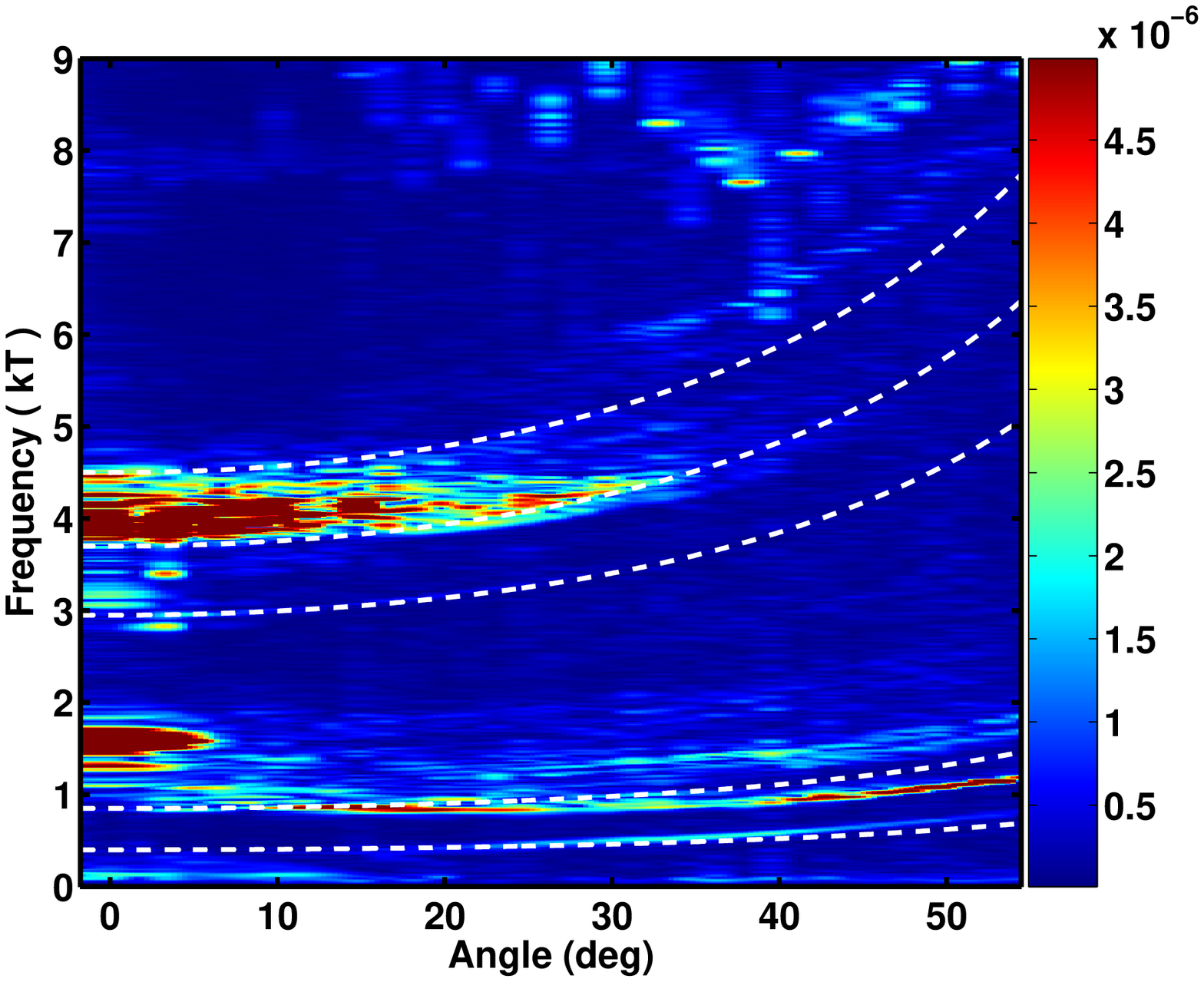}
	\end{center}
	\caption[Angular dependence of dHvA with expected curves]{Angular dependence of dHvA for sample C698I (rotation towards [110]), in the low field side $(top)$ and the high field side $(bottom)$, with expected frequency curves overlaid on top, in white dashed lines. The dashed lines correspond to the function $F = F_0/\cos(\theta)$.}
	\label{fig: FFT5to6p5T_A_cos}
\end{figure}

We observed that the frequencies possess the normal upwards curvature expected from two dimensional materials (described in section \ref{sect:BergemanAnalysis}). Figure \ref{fig: FFT5to6p5T_A_cos}, top graph, shows a comparison between this data and the function 
\beq
F = {F_0\over \cos(\theta)},\nn
\eeq
with $F_0$, the $c$-axis direction frequency values, from table \ref{tab:freqmassesCambridge}, of 0.15, 0.43, 0.91, 1.78 and 4.13~kT~\footnote{Note that these curves are not fits.}. We conclude that \TTS\ is quasi two-dimensional in the low field side, and that the slight deviations from the function $F_0/\cos(\theta)$ originate from $k_z$ dispersion. Finally, we note that a similar result was obtained in both directions of rotation.


Turning our attention to the high field side, shown in figure \ref{fig: FFT10to18T_A}, we observed that the behaviour is completely different from the low field side, and that it does not simply follow the function $F_0/\cos(\theta)$. We mentioned in section \ref{sect:CamSpectra} that at $c$-axis, the peaks were multiple, a feature that we found is present at all angles. Secondly, we observed that the amplitude also possesses a complex angular dependence, such that some frequencies $appear$ while some others $disappear$ at specific angles. In particular, the 1.6~kT frequency vanishes between 5$^{\circ}$ and 10$^{\circ}$, while a 0.9~kT peak appears near the same angle, and a peak near 0.5~kT appears between 15$^{\circ}$ and 20$^{\circ}$. The large structure near 4~kT is also interesting. It does not have the same behaviour as that of the 1.6~kT peak, but still vanishes at a lower angle than other peaks and is broken down into more than 25 peaks. It follows an upwards curve with increasing $\theta$, though, and features very complex beat patterns. Finally, the group of peaks near 7.9~kT is very faint and it is difficult to observe its angular dependence. Note that the spots in the upper right corner correspond to low frequency time-periodic noise, and contains no dHvA~\footnote{It is due to the voltage limited mode (see section \ref{sect:Camprobe}), which makes the time and $1/B$ frequency domains similar. When the field is swept exponentially as a function of time, noise at low time frequencies appear periodic in $-\log (B/B_0)$, which is closer to $1/B$ than in a linear field sweep.}.

Figure \ref{fig: FFT5to6p5T_A_cos}, bottom panel, presents the high field side data overlaid with curves of $F_0/\cos(\theta)$, using for $F_0$ the values 0.40, 0.85, 2.95, 3.7 and 4.5~kT. We observed that the the data did not follow this trend, but curved either significantly more or less than $F_0/\cos(\theta)$. These observations suggest that the system in the high field side has more three dimensional character than in the low field side. And finally, no significant differences were seen between the two directions of rotation.

In conclusion, we stress that the electronic structure of \TTS\ changes possibly more radically than was previously thought across the metamagnetic transition. The correspondence between the low and high field sides is not as obvious as previously thought. Moreover, the phenomenon of dHvA peaks appearing and disappearing with angle is difficult to explain, as is the change in the angular dependence of the frequency values across the transition. 

\subsection{Interference patterns in the angular dependence \label{sect:CamEnvelopes}}

\begin{figure}[p]
         \begin{center}
	\includegraphics[width=0.8\columnwidth]{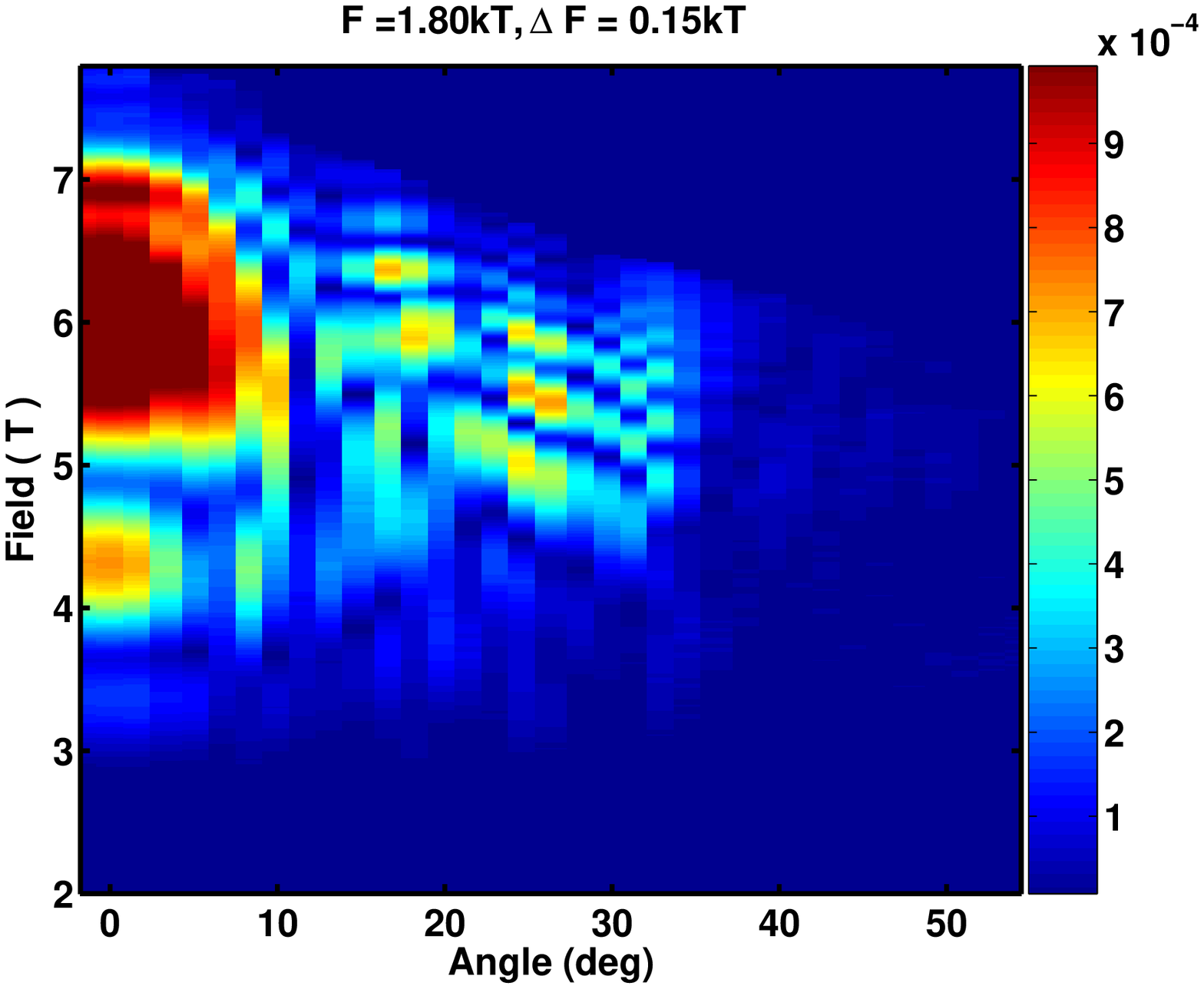}
	\includegraphics[width=0.8\columnwidth]{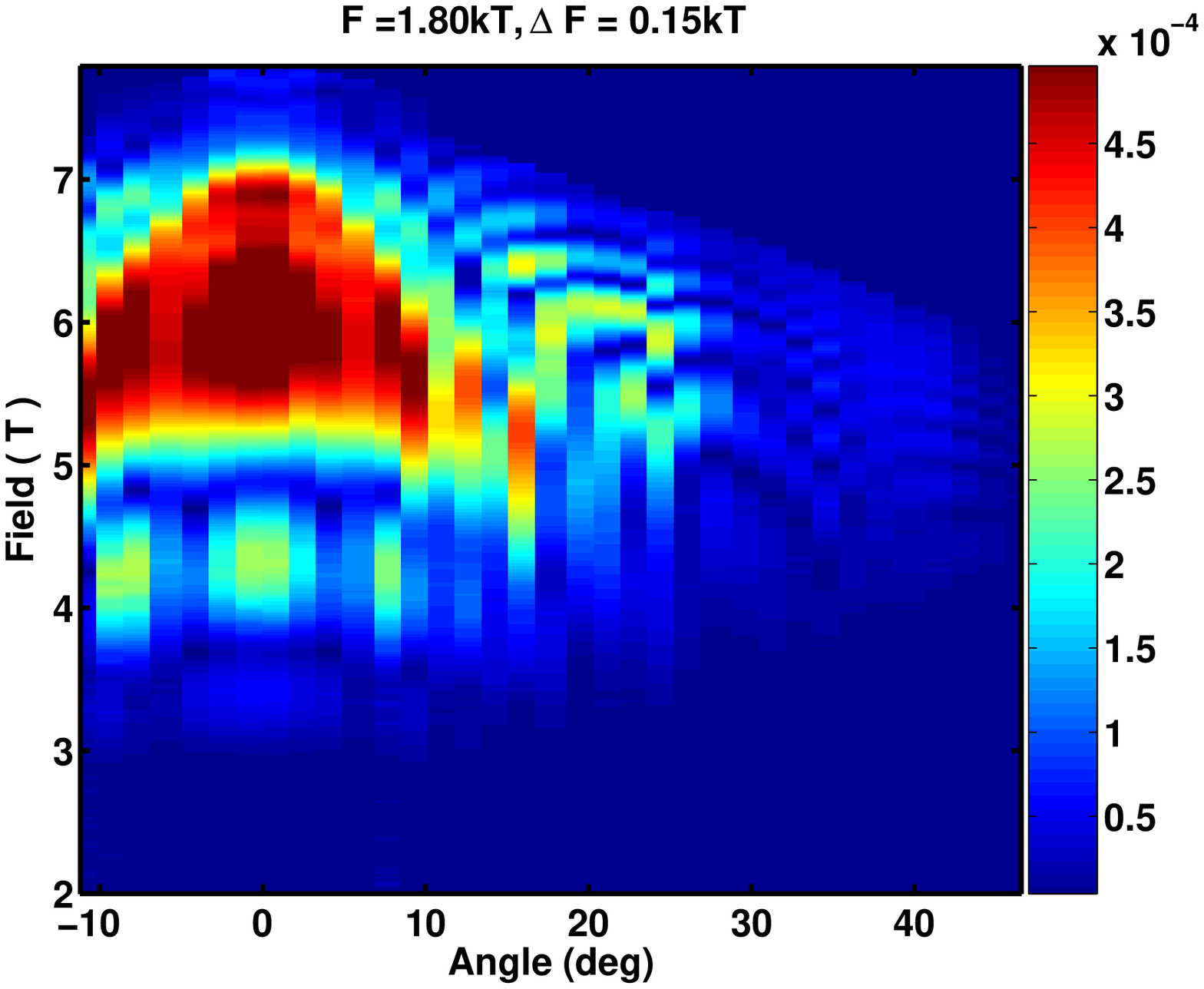}
	\caption[Amplitude of the 1.8~kT frequency as a function of field and angle in the low field side]{Amplitude of the oscillations at 1.8~kT as a function of field and angle in the low field side, for sample C698I, which was rotated from [001] towards [110], $top$, and C698A, which was rotated towards [100], $bottom$.}
	\label{fig: LF1p8kT-A}
	\end{center}
\end{figure}

\begin{figure}[t]
	\begin{minipage}[t]{7cm}
		\begin{center}
		\includegraphics[width=1\columnwidth]{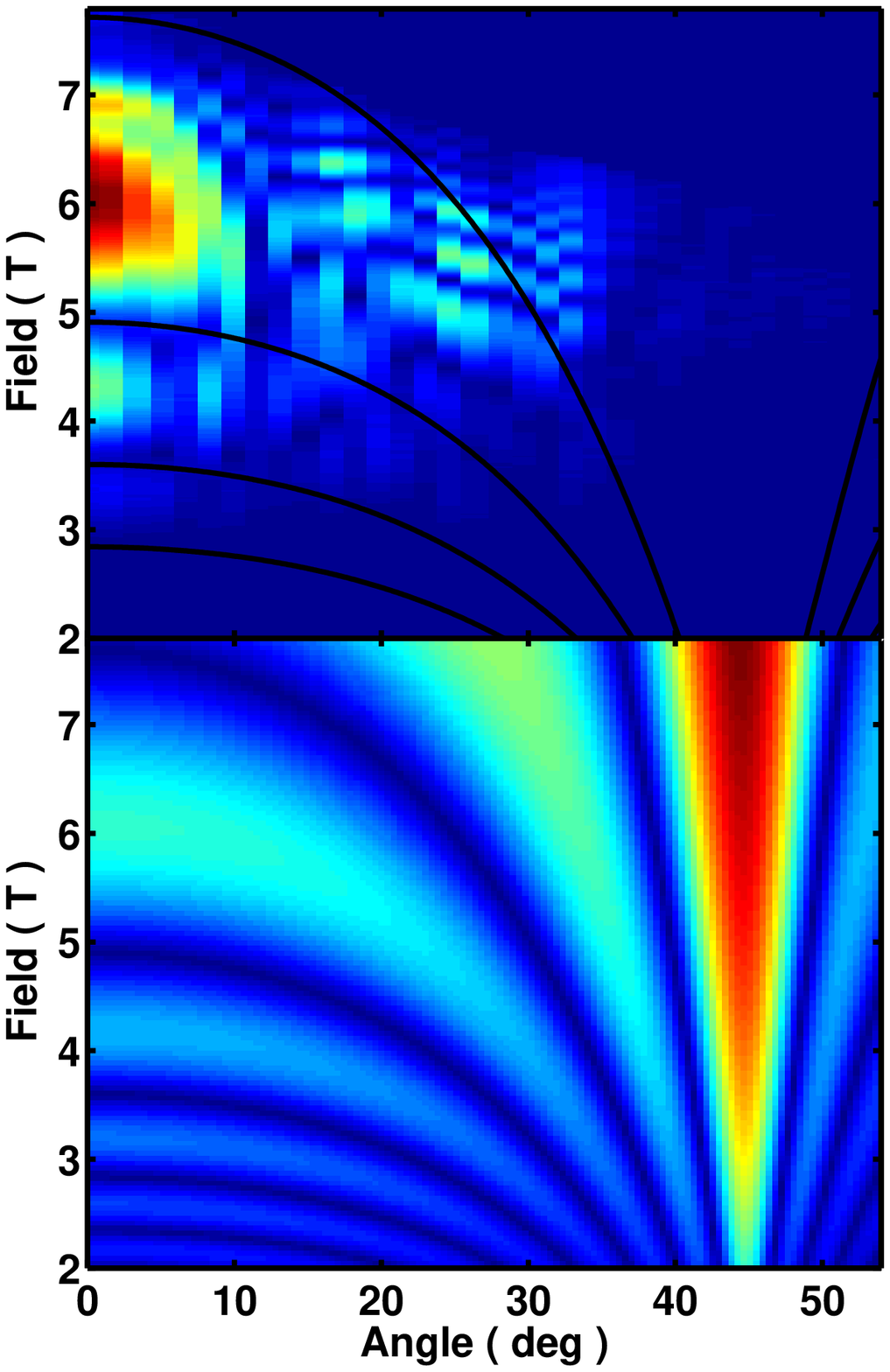}
		\end{center}
	\end{minipage}
	\hfill
	\begin{minipage}[t]{7cm}
		\begin{center}
		\includegraphics[width=1\columnwidth]{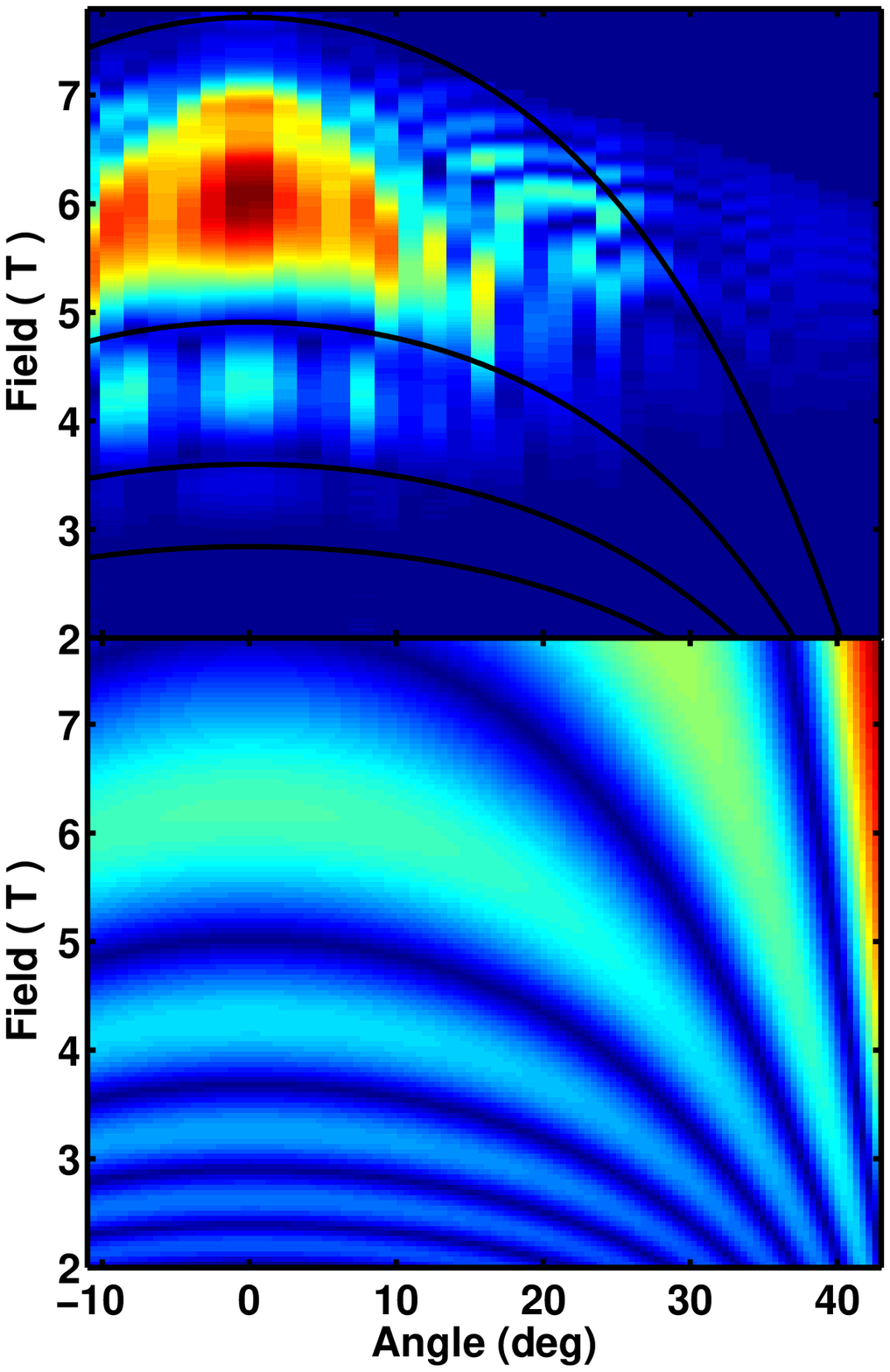}
		\end{center}
	\end{minipage}
	\caption[Comparison of data and simulation of beat patterns for simple warping]{Comparison of the data and simulations of beat patterns as a function of field and angle, using a single parameter in the cylindrical expansion of $k_{01} = 4.4 \times 10^6$~m$^{-1}$. The plots on the top presents the same data as in the previous figure for the rotation towards [110], $left$, and [100], $right$, overlaid with black curves corresponding to the nodes of the beat pattern. The plots on the bottom present the simulation using the single warping parameter, using the same angular span as the data. }
	\label{fig: LF1p8kT_Sim2_A2}
\end{figure}


dHvA data contains more information than that which is immediately visible in its Fourier transform. As was described in section \ref{sect:BergemanAnalysis}, closely spaced frequencies normally appear as weakly split peaks in the transform, but may have more dramatic consequences in the overall appearance of the raw data, as beat patterns affect their amplitude. Hence, in that view, it is sometimes the case that examining the amplitude of modulation of frequencies is more effective than analysing its Fourier transform, as was introduced by Bergemann and co-workers \cite{bergemann}. In simple 2D materials, the beat patterns are unique fingerprints of the specific type of corrugation of the Fermi cylinders forming the FS, but the interference may originate from other phenomena as non-linear magnetisation and spin splitting (introduced in section \ref{sect:spinsplitting}). 

In \TTS, it is well known that non-linear magnetisation occurs near 8~T, accompanied by changes in the dHvA frequencies in the low field side of the metamagnetic transition. Consequently, we anticipated that this analysis would become more complex than for \TOF. Effectively, performing the extraction of the beat patterns of all the frequencies present in the low field side revealed patterns that were too complex for us to model except in the case of the 1.8~kT frequency. This Fermi cylinder has consequently become the centre of our attention in this section.

Figure \ref{fig: LF1p8kT-A} shows the amplitude of the frequency at 1.8~kT in the low field side as a function of field and angle, for both rotation directions. This was calculated using the method described in section \ref{sect:envelope}, using a $rounded$ $top$-$hat$ window of width 0.3 kT, centred at (1.8 kT)$/\cos \theta$, which was chosen to include all the data even where the frequency changes\footnote{Since the frequencies evolve with magnetic field.}. The regions in the top right corner, top plot, and both top corners, bottom plot, lie above the metamagnetic transition, and were excluded from the computation. These calculations revealed that the beat patterns for both directions of rotation are similar, and the corrugation of this part of the Fermi surface appears isotropic apart from the in-plane shape (of warping parameters $k_{\mu 0}$), which is not detectable with dHvA\footnote{This information was provided to us by ARPES, section \ref{sect:ZeroFieldFS}.}. We deduced from this fact that all warping parameters with $\mu > 0, \nu > 0$ are absent or very small (for example, $k_{21}$, $k_{41}$, $k_{22}$, $k_{42}$, etc.), which leaves an isotropic warping of the type $\mu = 0, \nu > 0$, most probably $k_{01}$ or $k_{02}$. Our second observation was that the $c$-axis region appears normal, with a simple beat pattern that may be used to extract the value of $k_{0\nu}$, where effectively, we remarked that the beat nodes, or zeros, curve downwards with angle as expected (see, for example, figure \ref{fig: QuarterPercentSim}, section \ref{sect:BergemanAnalysis}). 

However, we found that the high angle data appears anomalous, where effectively, we expected to find a Yamaji angle. Instead, we observed rapid beating, of which the period becomes shorter as field increases towards the metamagnetic transition, in opposition to the normal situation, where beat nodes appear at equal intervals of inverse field. We will show in section \ref{sect:LowFieldSideModel} that this can be partly explained by non-linear magnetisation and spin splitting resulting from a peak in the DOS. We note that in the region where the beat pattern is very rapid, the angle step of about 1.6$^{\circ}$ is not small enough and the pattern is under-sampled, and warn the reader that the appearance of the data can be misleading: the beat pattern in a data set where the beating period in angle is comparable or smaller than the sampling rate may appear in some cases as a checkerboard pattern, in others as horizontal nodes, where in reality the nodes may cross the vertical with a sharp angle. 

\begin{figure}[p]
         \begin{center}
	\includegraphics[width=0.8\columnwidth]{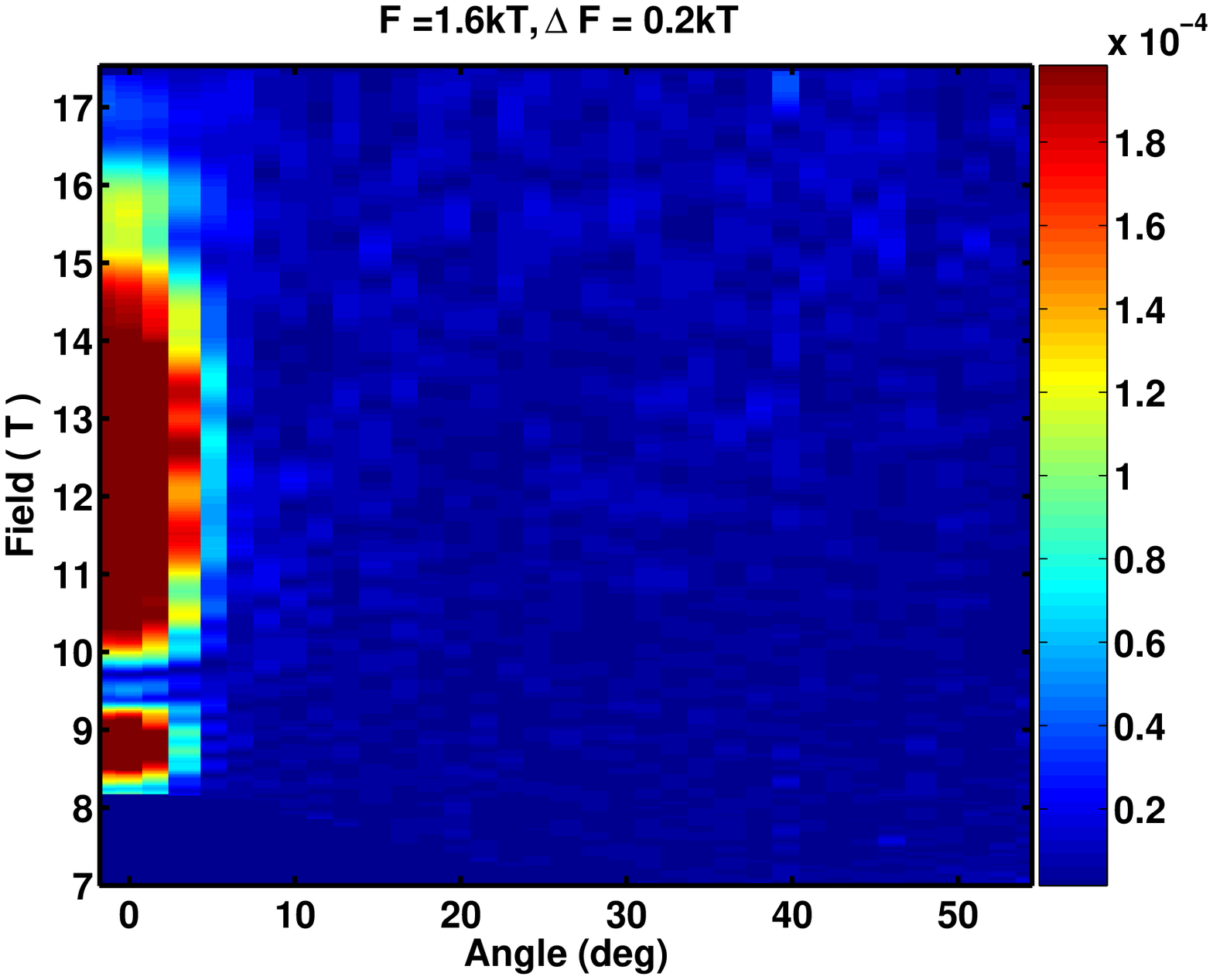}
	\includegraphics[width=0.8\columnwidth]{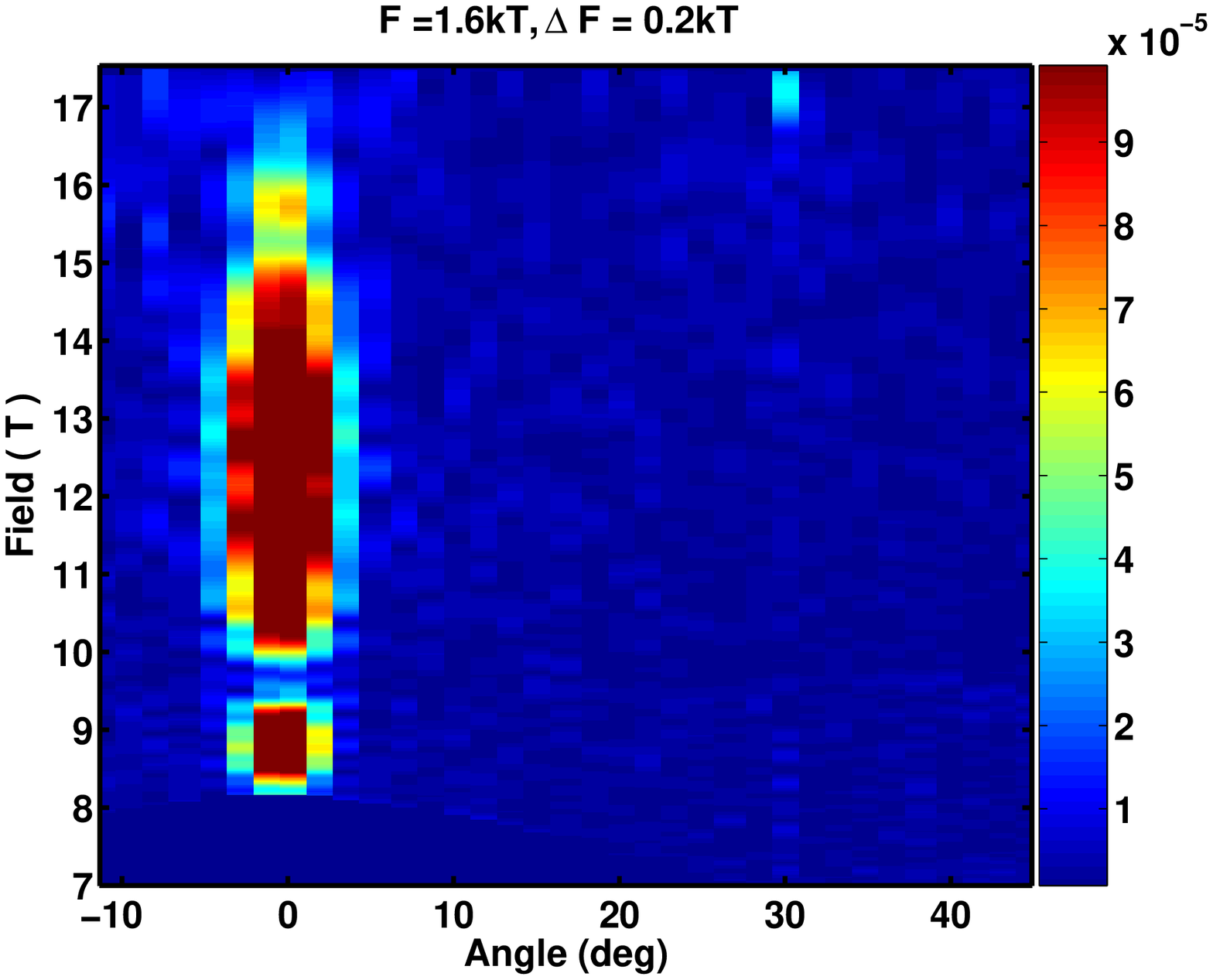}
	\caption[Amplitude of the 1.6~kT frequency as a function of field and angle in the high field side]{Amplitude of the frequency at 1.6~kT in the high field side, as a function of field and angle, for sample C698I, which was rotated from [001] towards [110], $top$, and C698A, which was rotated towards [100], $bottom$.}
	\label{fig: HF1p6kT-A}
	\end{center}
\end{figure}

In order to find whether the warping is of type $k_{01}$, $k_{02}$, simulations were produced in order to compare with the data. We also used equation \ref{eq:BeatZeros} for plotting the positions of the zeros of the beat patterns. Figure \ref{fig: LF1p8kT_Sim2_A2} presents such an analysis, where data is shown on the top overlaid with beat pattern nodes in black, and simulations on the bottom. Although the agreement is far from perfect, the best estimate we can make is that $k_{01} = 4.4 \times 10^6$~m$^{-1}$ is the dominant warping parameter, which reproduces the low angle part of the data. With $k_{0\nu} = 4.4 \times 10^6$~m$^{-1}$, $\nu > 1$, we also reproduced properly the $c$-axis direction beat pattern, but the nodes fell towards zero field too sharply in angle compared to the data. Using a combination of more than one parameter of the type $k_{0\nu}$ did not reproduce the data, but instead changed the beat pattern at $c$-axis. Since parameters with $\mu > 0$, $\nu > 0$ were already excluded, we conclude that no warping parameter of significant amplitude produces the additional modulation to the pattern from simple warping $k_{01}$. We therefore associated the remaining modulation to non-linear magnetisation and spin splitting, as will be discussed in section \ref{sect:LowFieldSideModel}.

The envelope extraction was also performed with the high field data, where even more surprising results were found, in line with the observations of figure \ref{fig: FFT10to18T_A}. The calculation was performed using a window of width 0.4 kT, centred at (1.6 kT)$/\cos \theta$, over the region between 7 and 18~T, excluding the region near $c$-axis which lies in the low field side of the transition. Figure \ref{fig: HF1p6kT-A} shows the result, the shape of which reminds the author of a simple exclamation mark. We observed that the oscillations around 1.6~kT are completely suppressed from an angle of about 5$^{\circ}$ from $c$-axis, a result that is very difficult to explain. We stress that only a very complex combination of many warping parameters could reproduce such an angular dependence of the oscillations, which is not probable. Moreover, even though non-linear magnetisation and spin splitting are likely to produce complex beat patterns in the high field side, they cannot suppress the oscillations altogether. Consequently, we found no explanation for this puzzling behaviour.

We conclude this section by stating that the amplitude modulation of the 1.8~kT frequency does not behave like the theory predicts at all fields and angles except near $c$-axis in the low field side. In the low field side, we found a crossover from normal to abnormal behaviour at an angle of about 15$^{\circ}$, above which we observed fast beat patterns that are not periodic in inverse field and assume that this unusual modulation is the result of non-linear magnetisation and spin splitting, which will be modelled in section \ref{sect:LowFieldSideModel}. In contrast, we discovered that the high field side of the transition exhibits a behaviour that is not explained either by warping or non-linear magnetisation and spin splitting and remains mysterious.

\section{Field dependent quasiparticle masses \label{sect:MassNonDiv}}

This section presents a reinversigation of the enhancement of the quasiparticle masses near the quantum critical end point in \TTS. Since the beginning of this project, discrepancies have been found between the measurements of Borzi $et$ $al.$ \cite{borzi} (see section \ref{sect:BorzidHvA}, figure \ref{fig: BorziSpectra}) and ours. Several problems with eddy current heating and high noise levels prevented us producing a definitive answer to the problem until we performed measurements with the system in Cambridge. These prove that the conclusion in the work of Borzi that there exists an enhancement of the quasiparticle mass of the 1.8-1.6~kT and 4.2 - 4.0~kT peaks was erroneous. It furthermore demonstrates that none of the masses of the other measured peaks exhibit any increase either. 

We saw in section \ref{sect:CamEddyCurrents} that we have performed extremely careful checks on the thermometry during the Cambridge experiments, and showed that no eddy currents were present. In such a situation, we have eliminated the main source of systematic errors on the thermometry. Moreover, in the Cambridge data the signal to noise was high enough for us to be able to remove the third parameter in the LK fit \footnote{see section \ref{sect:mass} for details LK fits and the number of parameters}, which made our analysis much less prone to systematic errors, and thus more reliable than in previous work. We came to the conclusion that a combination of field varying low signal to noise ratio and three parameter LK fits may have led to the spurious continuous enhancement of the quasiparticle mass in the work of Borzi. This section presents the quasiparticle masses calculated with two parameter fits, for all the major dHvA frequencies of the low field side. We will explain in the next chapter, section \ref{sect:MassSystematic}, how Borzi $et$ $al.$ probably obtained enhancements of those masses through systematic errors.

\subsection{All masses, at $c$-axis and 10$^{\circ}$}

\begin{figure}[!t]
	\begin{minipage}[t]{.5\columnwidth}
		\begin{center}
		\includegraphics[width=\columnwidth]{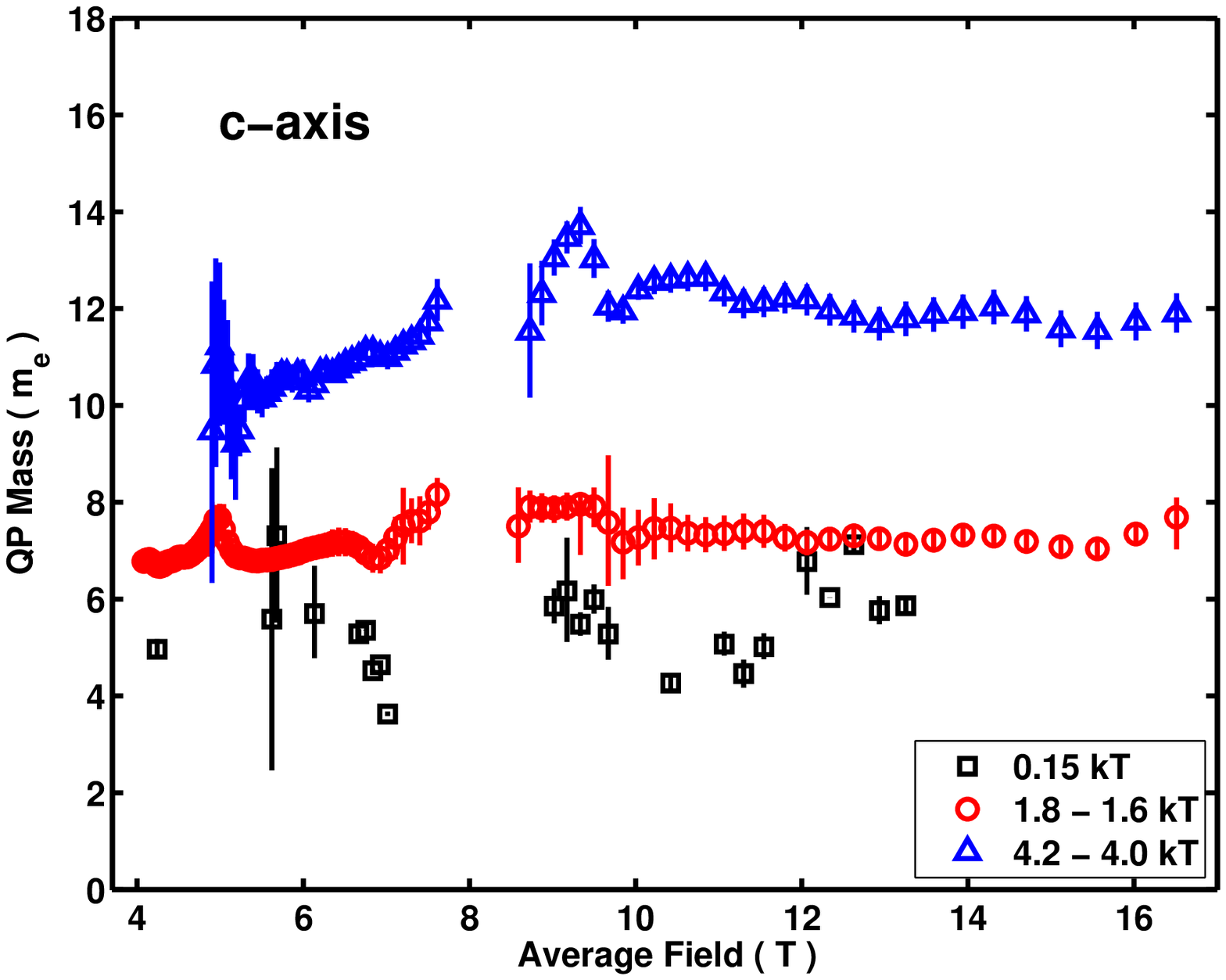}
		\includegraphics[width=\columnwidth]{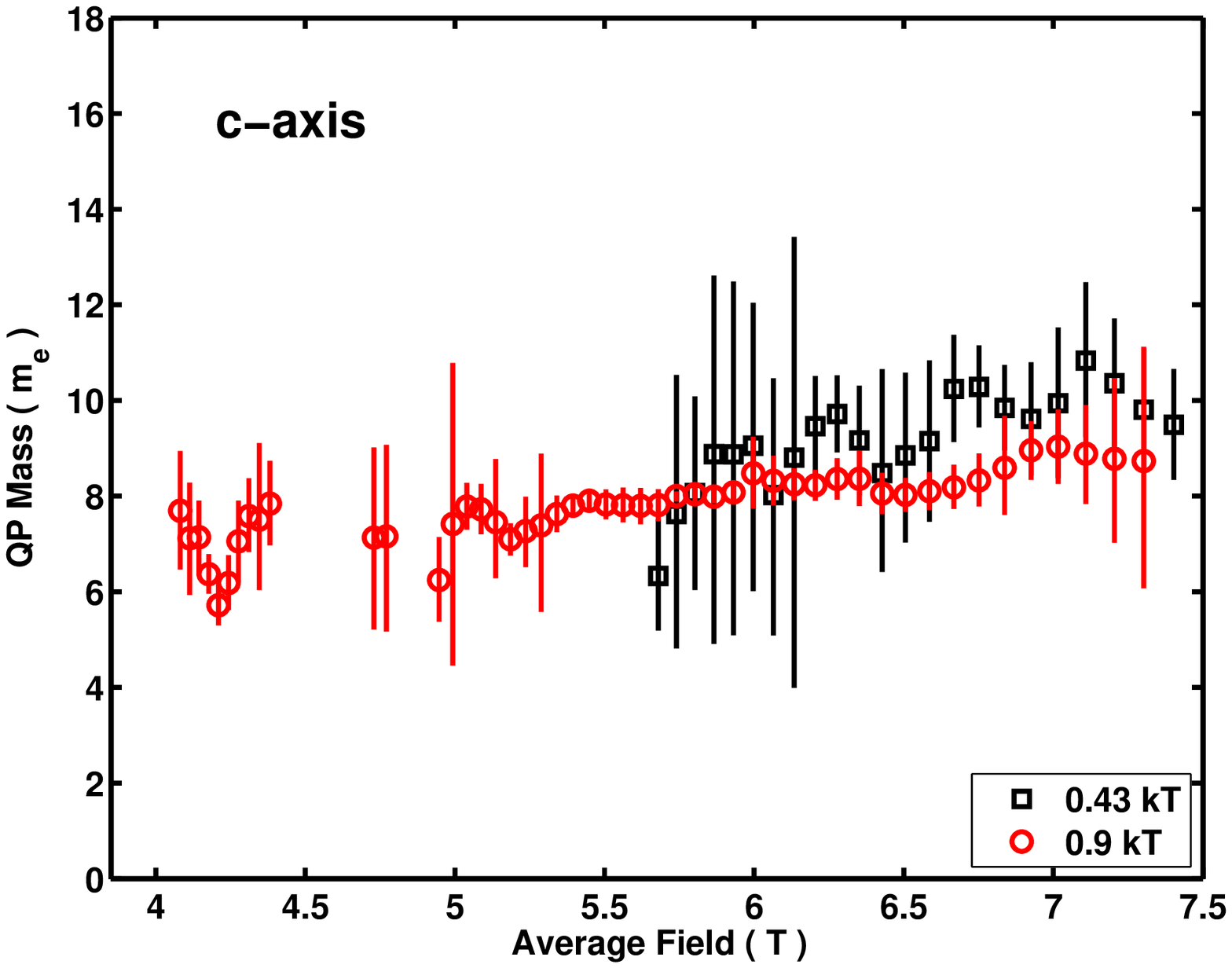}
		\end{center}
	\end{minipage}
	\hfill
	\begin{minipage}[t]{.5\columnwidth}
		\begin{center}
		\includegraphics[width=\columnwidth]{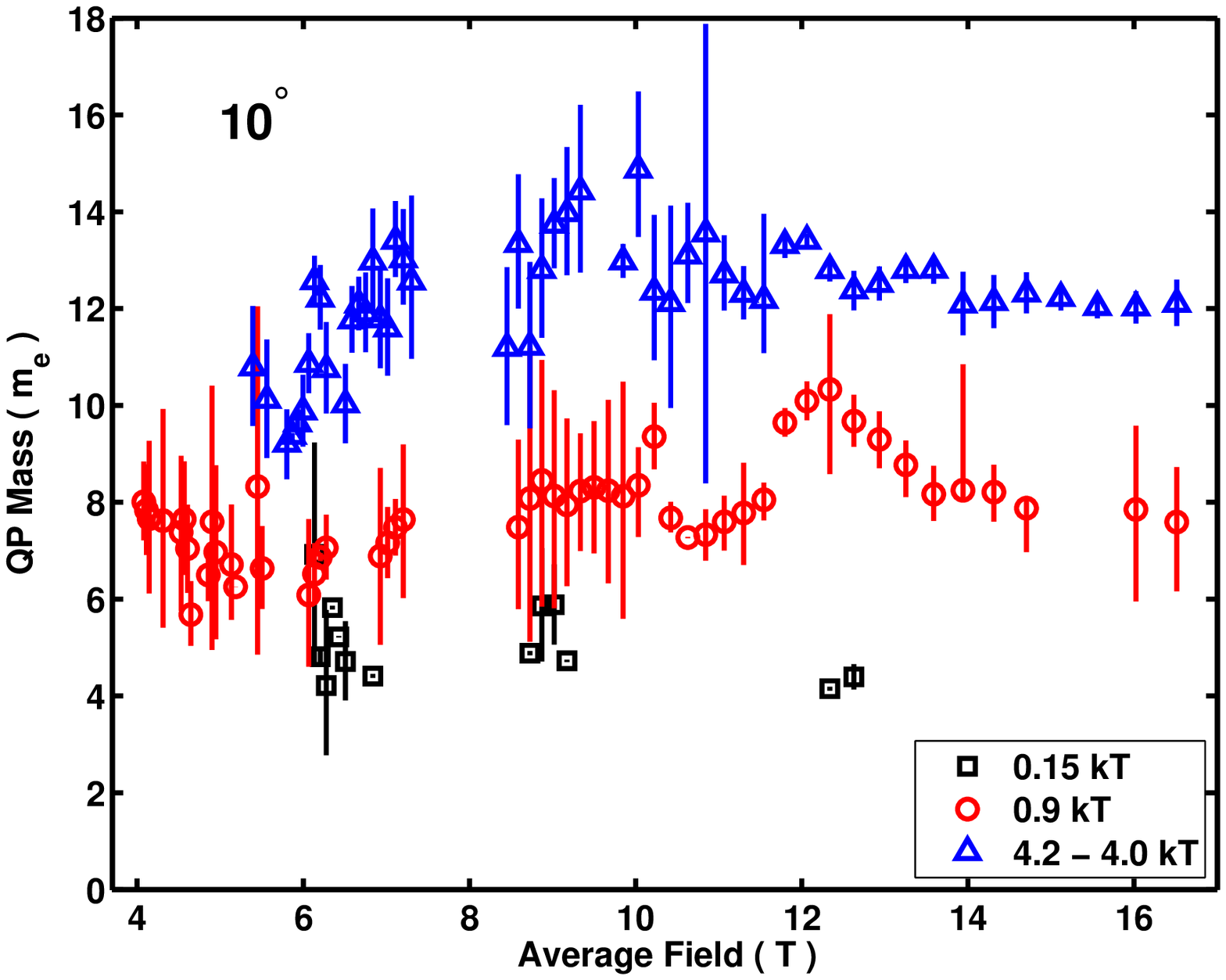}
		\includegraphics[width=\columnwidth]{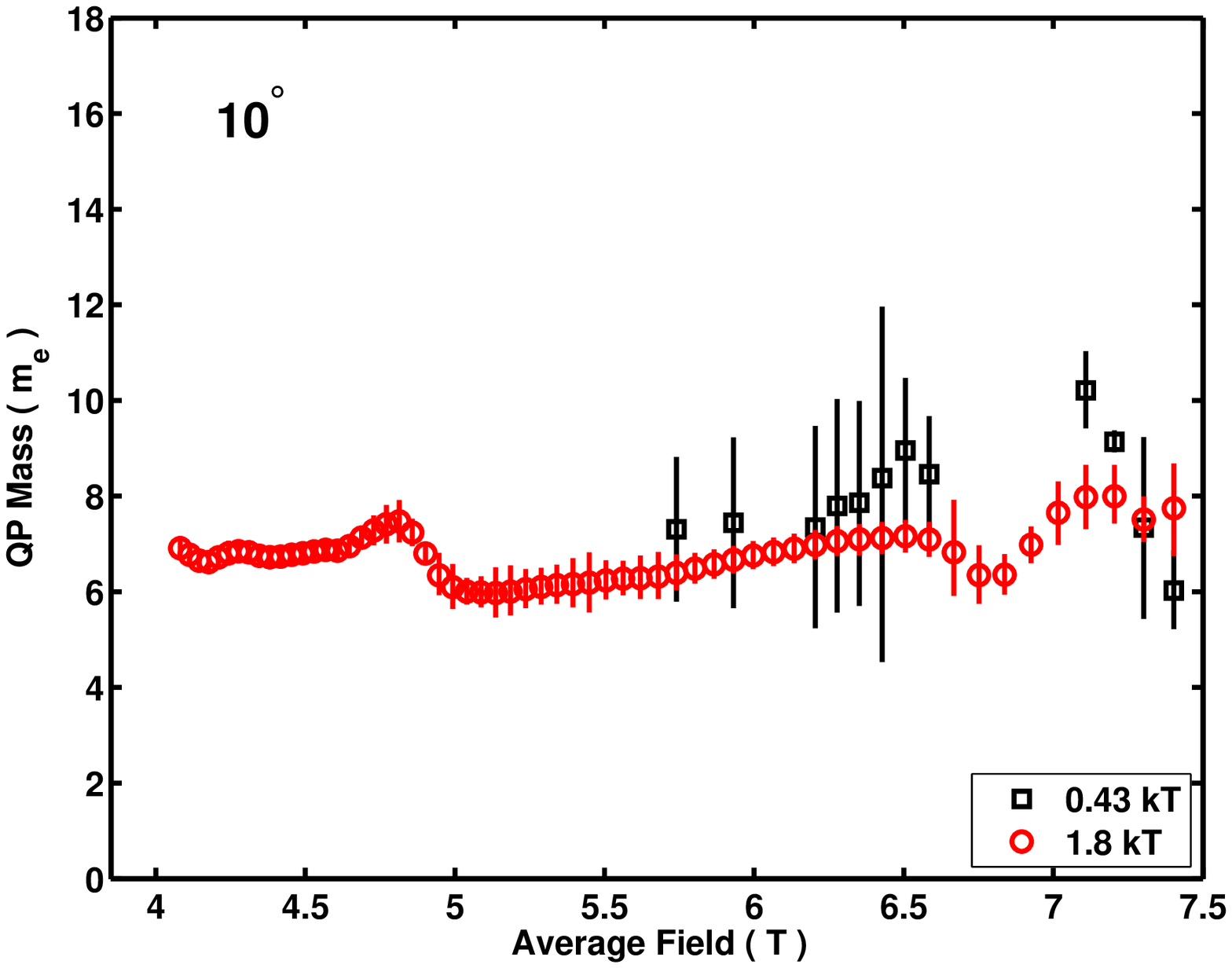}
		\end{center}
	\end{minipage}
	\caption[Field dependence of the quasiparticle masses for sample C698I]{Field dependence of the quasiparticle masses in sample C698I, measured with the magnetic field aligned in the $c$-axis direction, $left$, and at 10$^{\circ}$ towards [110], $right$. The frequencies used were of 0.15, 0.43, 0.9, 1.8 and 4.2~kT. The inverse field window of 0.01 T$^{-1}$ is approximately 5.5 data points wide.}
	\label{fig: p9p45kT_c-axis_a}
\end{figure}
\begin{figure}[t]
	\begin{minipage}[t]{.5\columnwidth}
		\begin{center}
		\includegraphics[width=\columnwidth]{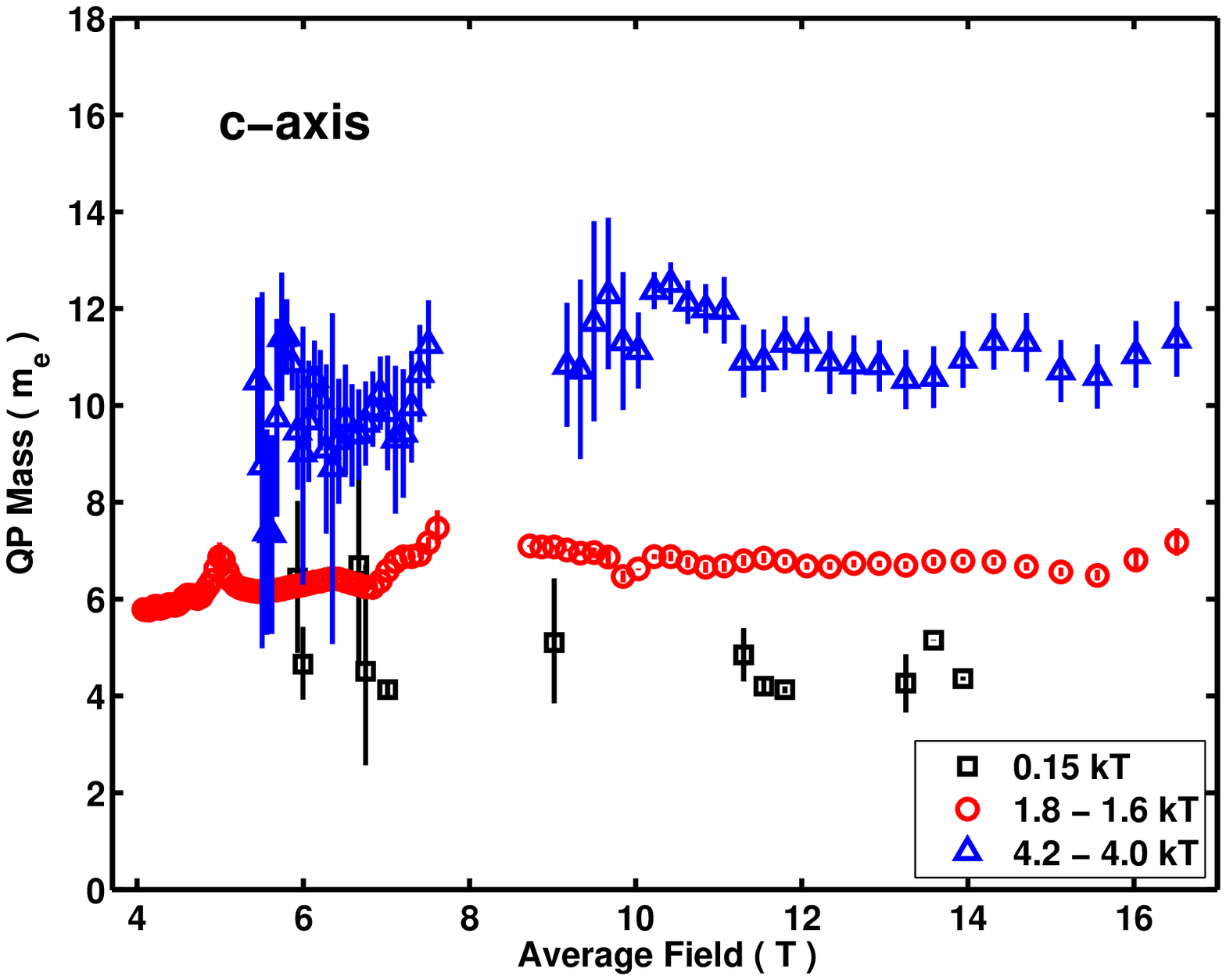}
		\includegraphics[width=\columnwidth]{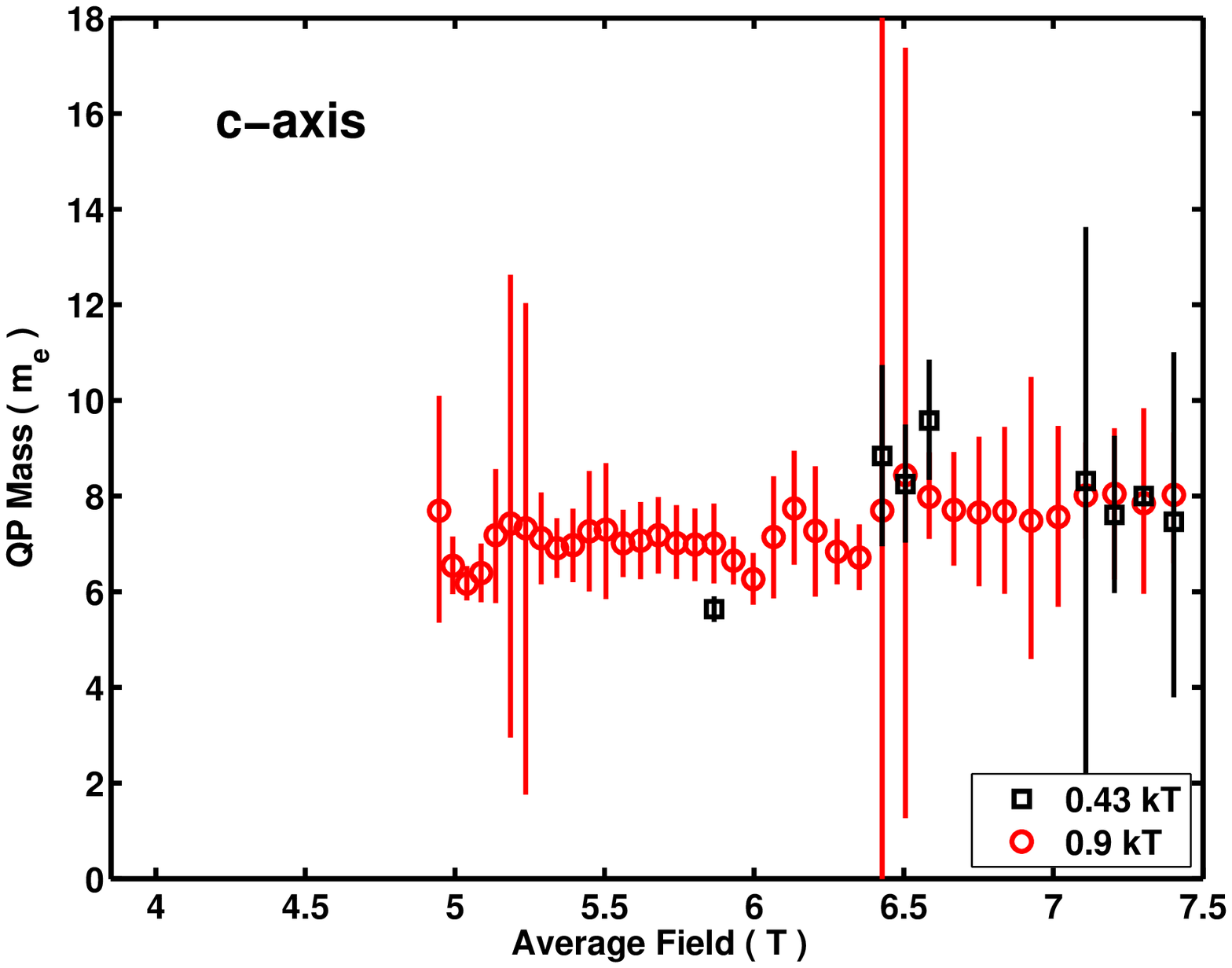}
		\end{center}
	\end{minipage}
	\hfill
	\begin{minipage}[t]{.5\columnwidth}
		\begin{center}
		\includegraphics[width=\columnwidth]{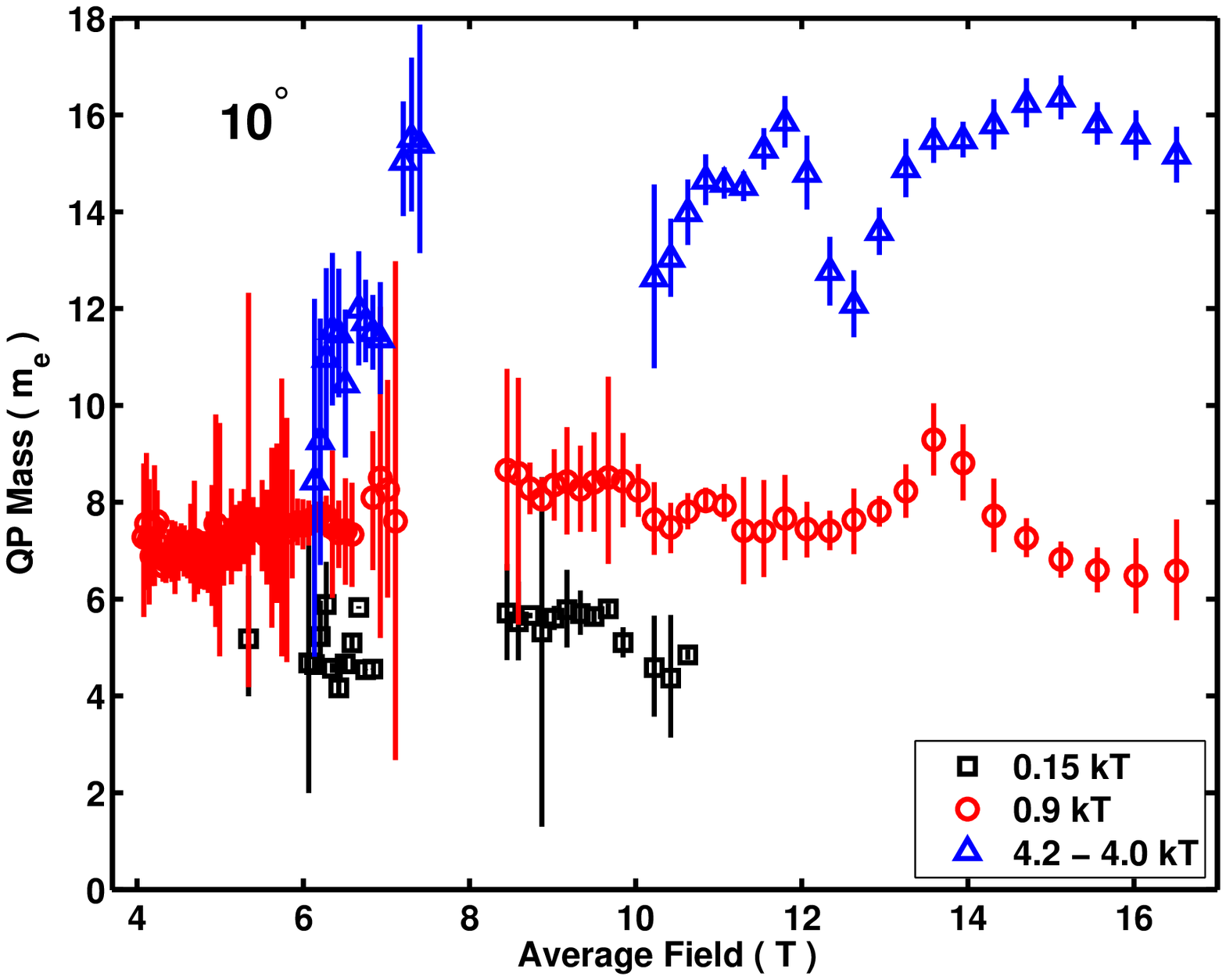}
		\includegraphics[width=\columnwidth]{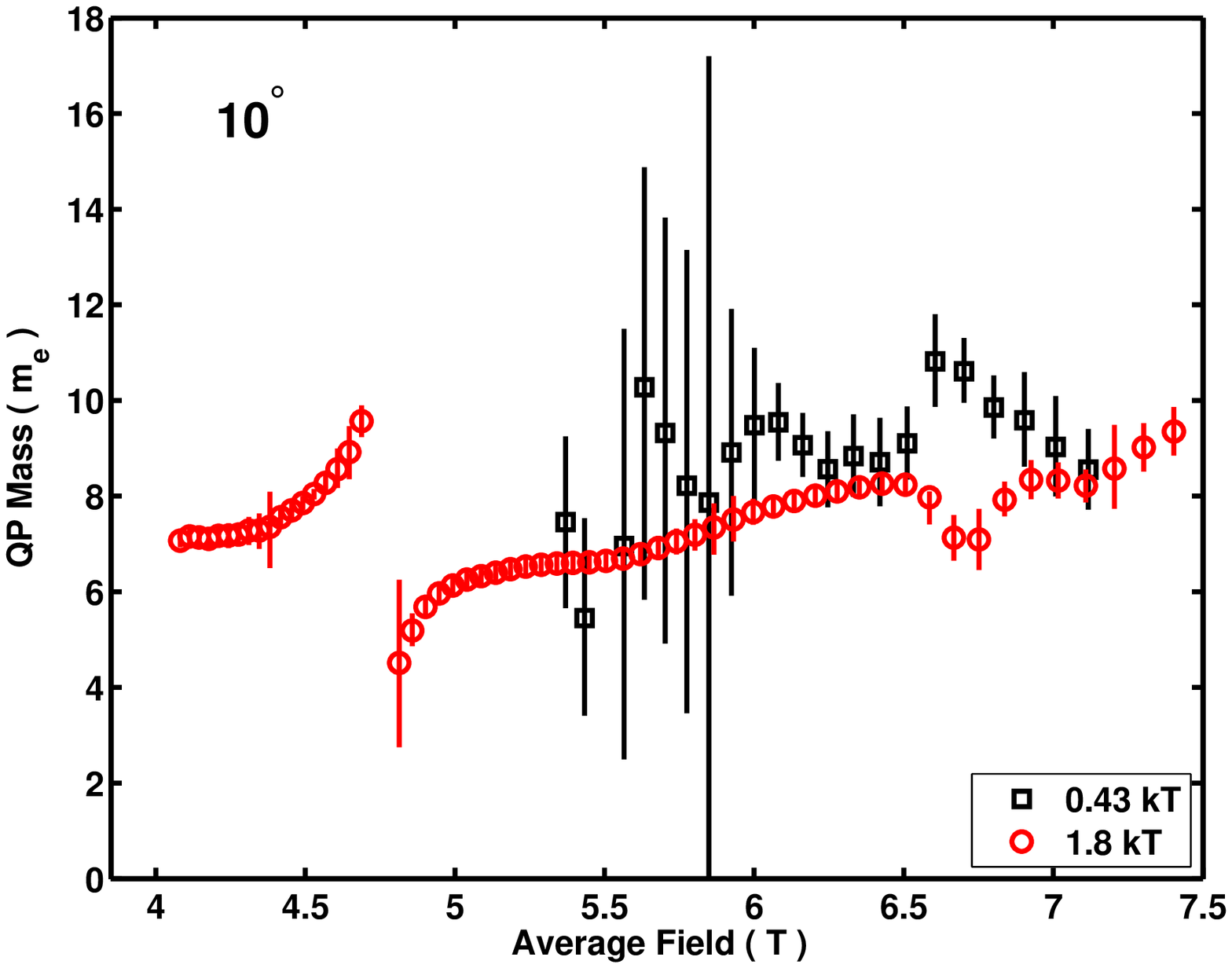}
		\end{center}
	\end{minipage}
	\caption[Field dependence of the quasiparticle masses for sample C698A]{Field dependence of the quasiparticle masses in sample C698A, measured with the magnetic field aligned in the $c$-axis direction, $left$, and at 10$^{\circ}$ towards [100], $right$.}
	\label{fig: p9p45kT_c-axis_b}
\end{figure}

We have measured during the Cambridge series of measurements dHvA as a function of temperature using sample C698I with its $c$-axis aligned with the magnetic field, and simultaneously sample C698A aligned with its $c$-axis 10$^{\circ}$ away from the field towards [100], at 10 different temperatures between 90 and 500~mK. Subsequently, we performed a similar measurement using C698A aligned with the field and C698I at 10$^{\circ}$ away towards [110], at 8 different temperatures within the same boundaries. Measurements performed at lower temperatures were not used for this section due to possible thermal equilibrium issues, as explained in section in section \ref{sect:ModFieldLK}.

We chose to use the frequencies present in the low field side for the calculation of the field dependence of the quasiparticle masses, those at 0.15, 0.43, 0.9, 1.8 and 4.2~kT (see table \ref{tab:freqmassesCambridge}). The reason for this choice lies with the simplicity of the low field side and the overwhelming complexity of the high field side. In the $c$-axis direction, the low field side frequencies at 0.15, 1.8 and 4.2~kT were in the high field side as well, at values of 0.14, 1.6 and 4.0 kT, while those at 0.43 and 0.9 were present only in the low field side (see figure \ref{fig:FvsBCam}). In contrast, at 10$^{\circ}$, the frequency at 1.8~kT was not present in the high field side, but the one at 0.9~kT was (see figures \ref{fig: FFT10to18T_A} and \ref{fig: FFT5to6p5T_A}). The ones among these that extended to the high field side in either directions were used over the whole field range, but the others were used only in the low field side. 

Figures \ref{fig: p9p45kT_c-axis_a}, for C698I, and \ref{fig: p9p45kT_c-axis_b}, for C698A present the field dependence of the mass of these frequencies, at $c$-axis in the left side graphs, and at 10$^{\circ}$ on the right. The calculations were performed using an inverse field window of 0.01 T$^{-1}$, and a non-linear LK fit of the type (see section \ref{sect:mass})
\beq	
LK(X_0,T) = P_1 { P_2 14.7 X_0~T \over \sinh( P_2 14.7 X_0~T )},\nn
\eeq
where $P_1$ and $P_2$ are the only two fit parameters. One hundred such fits were performed for each frequency, between 18 and 4~T, and the quality of each single one was evaluated by eye, along with the temperature dependence of the Fourier spectrum at each field location, in order to verify the nature of the data that was being used, and all fits based on inappropriate data sets were rejected. The error bars represent the fit error, as defined by the least squares method. For most of the fits, the third parameter, if it had been introduced, would have been small compared to the amplitude $P_1$ of the LK function. Nevertheless, three parameter fits were also performed using only the 1.8 - 1.6~kT and 4.2 - 4.0~kT frequencies, and no difference was observed in the resulting field dependent masses, and are not shown here. 

We observed no significant enhancement of the quasiparticle mass in any of the low field side frequencies as a function of field, either at $c$-axis or at 10$^{\circ}$. The only variations that were seen always corresponded to regions where the signal to noise vanishes, for example at nodes of beat patterns\footnote{See, for instance, the 1.8~kT frequency near 5 T}, where not only enhancements but also decreases in mass were observed, and all of those were of no more than 4 electron masses. Moreover, in contrast to the first harmonic data of the previous work by Borzi $et$ $al.$, in these data sets the signal to noise does not vanish near the metamagnetic transition, and the metamagnetic background only becomes significant at field values much closer to the transition (see \cite{borzi}, figure 1, compared to \ref{fig: Raw_c-axis_25mK}). In that region, only windows covering the metamagnetic background signal were discarded\footnote{The Fourier transform of the metamagnetic background leads, as in figure \ref{fig:FvsBCam}, to a large peak at zero frequency, the tail of which extends to the regions of interest, and could lead to systematic errors.}. Thus, we conclude that there is no enhancement of the quasiparticle mass of any of the  Fermi surface sheets observed with dHvA near the quantum critical end point in \TTS.

\section{dHvA in the nematic phase \label{sect:nematicCambridge}}

The discovery of a new phase of the electron liquid in \TTS\ in 2005 \cite{science2} has raised many new ideas, but also many questions about its nature \cite{PRLgreen,PRLBinz}. The findings about its nematic nature in 2006 \cite{science3} attracted significant interest in the theoretical community \cite{PRLdoh,yamase}. One of the issues that have been raised regards the behaviour of the FS in this nematic phase, and the most appropriate probe for gaining this type of information is dHvA, the only other experiment that can measure FS sheet specific information being ARPES, which cannot be performed in a magnetic field. This section presents the first successful observation of dHvA inside the nematic phase, which was carried out in Cambridge.

The oscillations were first discovered in 2007 in St Andrews but exhibited a very faint amplitude. The experiment was repeated in Cambridge using second harmonic and the voltage limited mode described in section \ref{sect:Camprobe}, and very low noise data were obtained in and around the nematic region. The oscillations  were studied on axis as well as at 10$^{\circ}$ as a function of temperature. We also discovered that they possessed a strong angular dependence, which motivated a detailed analysis as a function of angle. The first part to this section presents the raw oscillations as a function of temperature, for both samples, and their detailed analysis. Then, the second part shows how they evolve with field angle.

\subsection{Temperature dependence of the oscillations}

\begin{figure}[!t]
	\begin{minipage}[t]{.5\columnwidth}
		\begin{center}
		\includegraphics[width=\columnwidth]{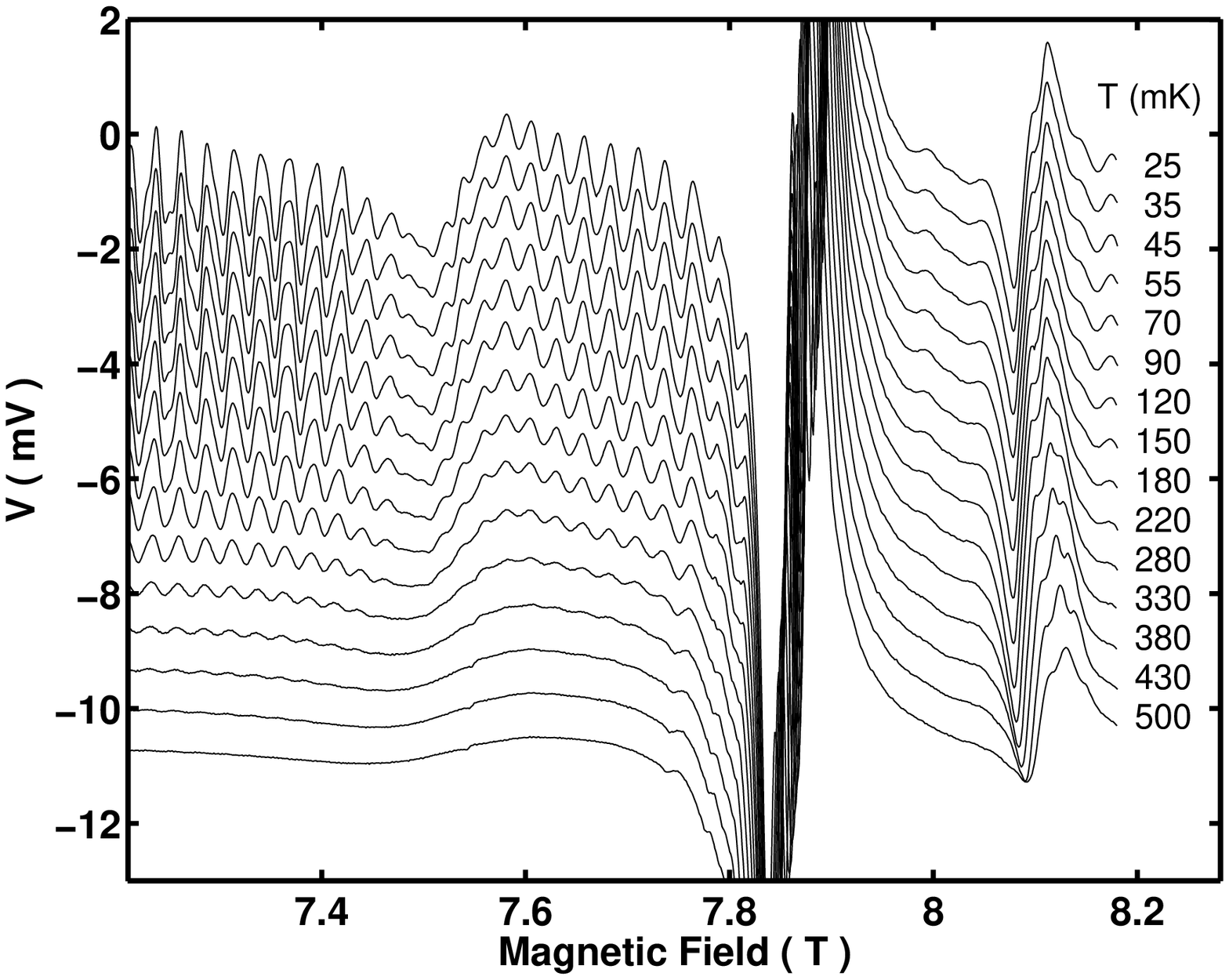}
		\includegraphics[width=\columnwidth]{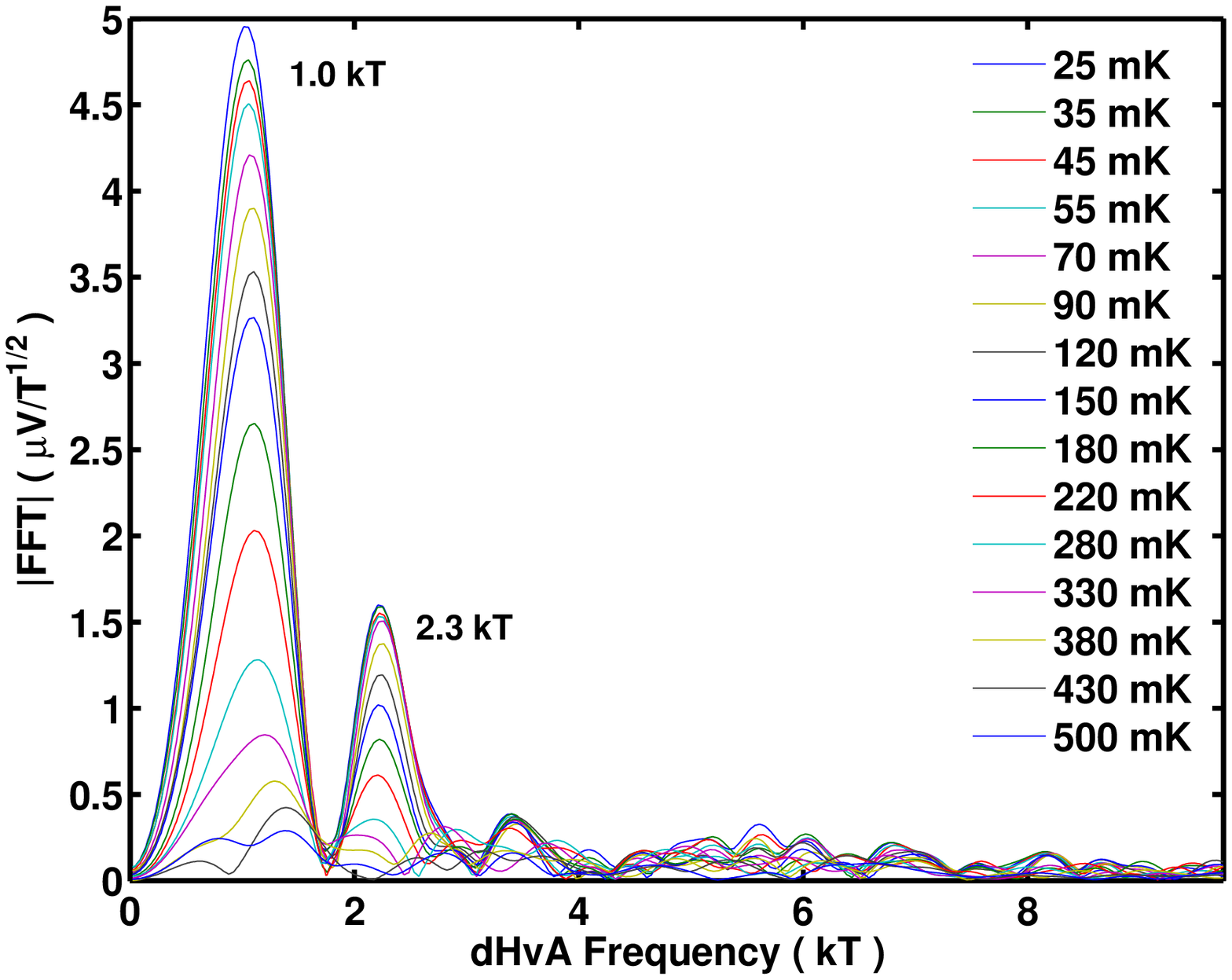}
		\end{center}
	\end{minipage}
	\hfill
	\begin{minipage}[t]{.5\columnwidth}
		\begin{center}
		\includegraphics[width=\columnwidth]{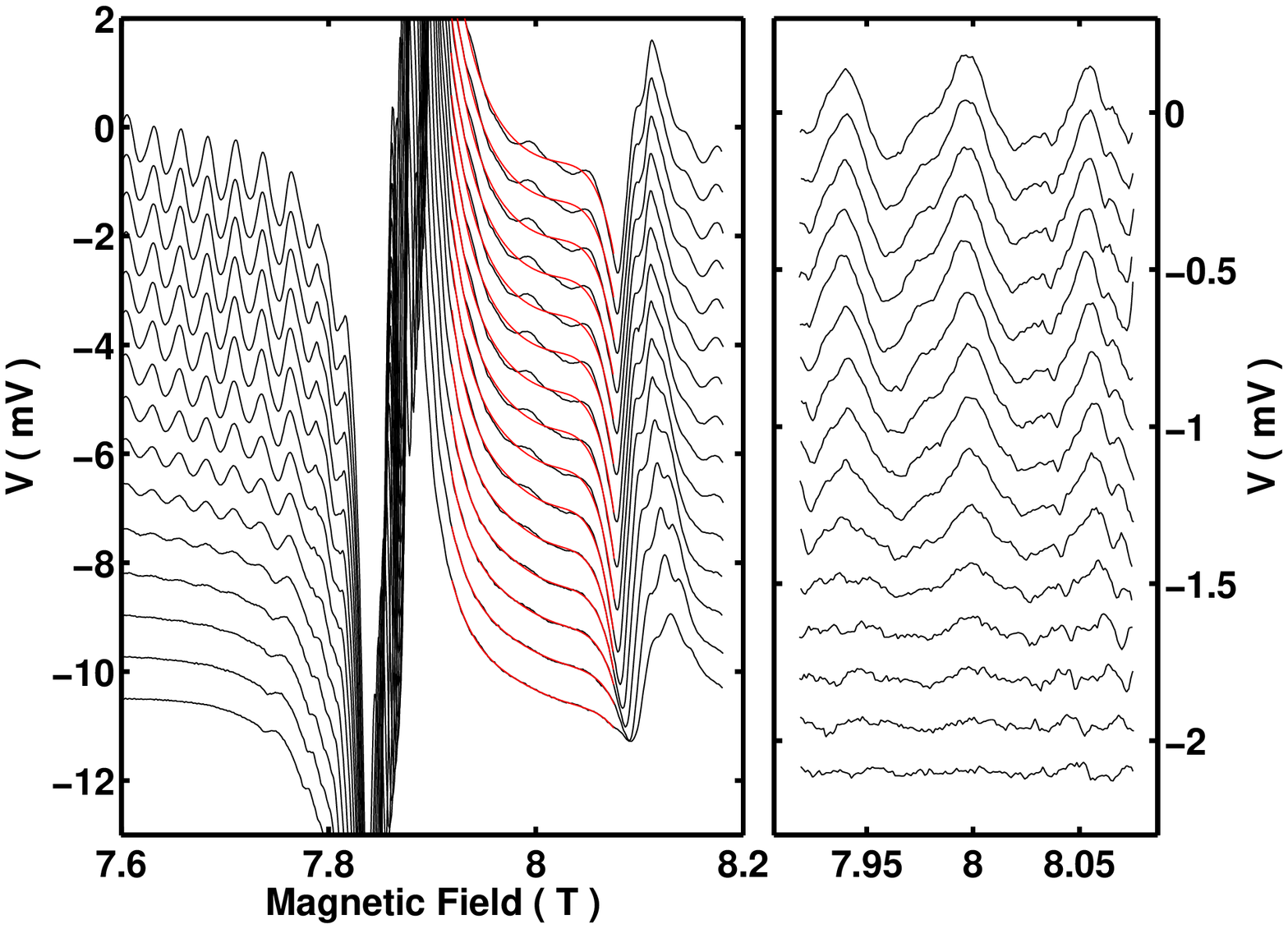}
		\includegraphics[width=\columnwidth]{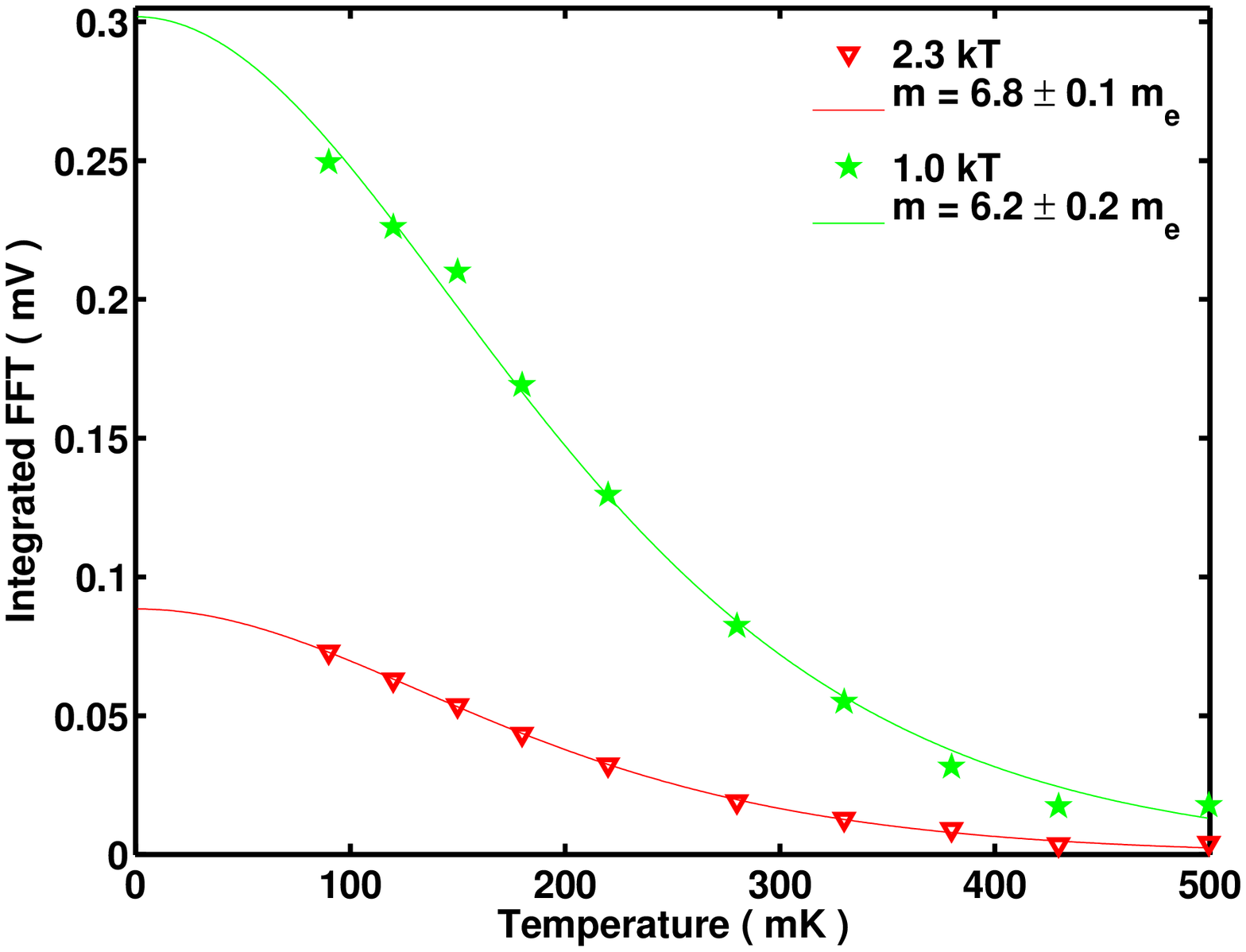}
		\end{center}
	\end{minipage}
	\caption[Oscillations in the nematic phase in sample C698I]{$Top$ $left$ Raw second harmonic data for sample C698I, with the $c$-axis aligned with the magnetic field, at 15 different temperatures from 25 to 500~mK. $Top$ $right$ The background substraction scheme is shown in red on top of the raw data, on the left, and the signal after background substraction is presented on the right. $Bottom$ $left$ Fourier transforms of the data in the top right plot, at the various measured temperatures. $Bottom$ $right$ Square root of the integral of the peaks in the power spectra as a function of temperature, along with two parameter LK fits, in solid lines. Fit results are given in the legend.}
	\label{fig: NematicFFT_c-axis_a}
\end{figure}

\begin{figure}[!t]
	\begin{minipage}[t]{.5\columnwidth}
		\begin{center}
		\includegraphics[width=\columnwidth]{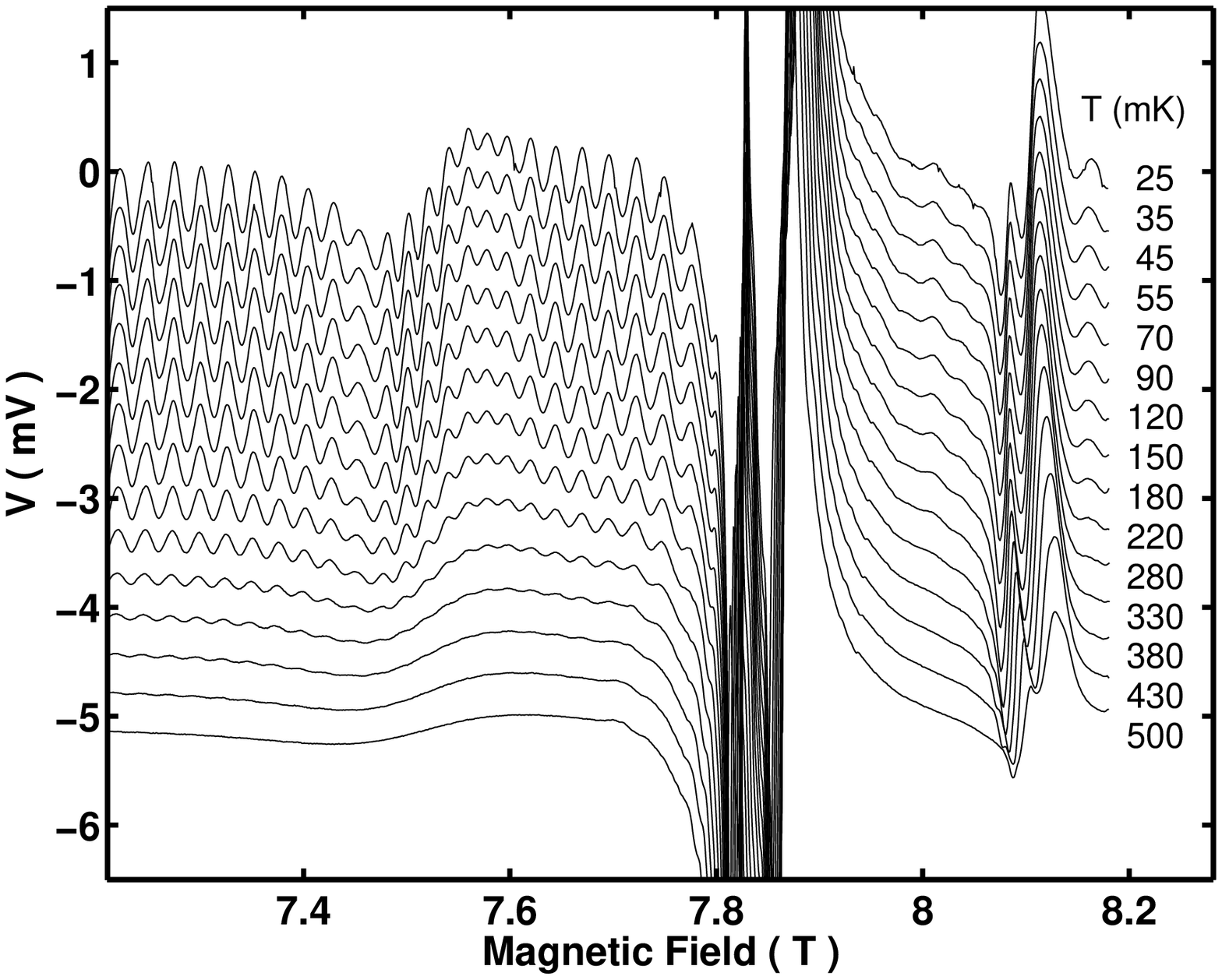}
		\includegraphics[width=\columnwidth]{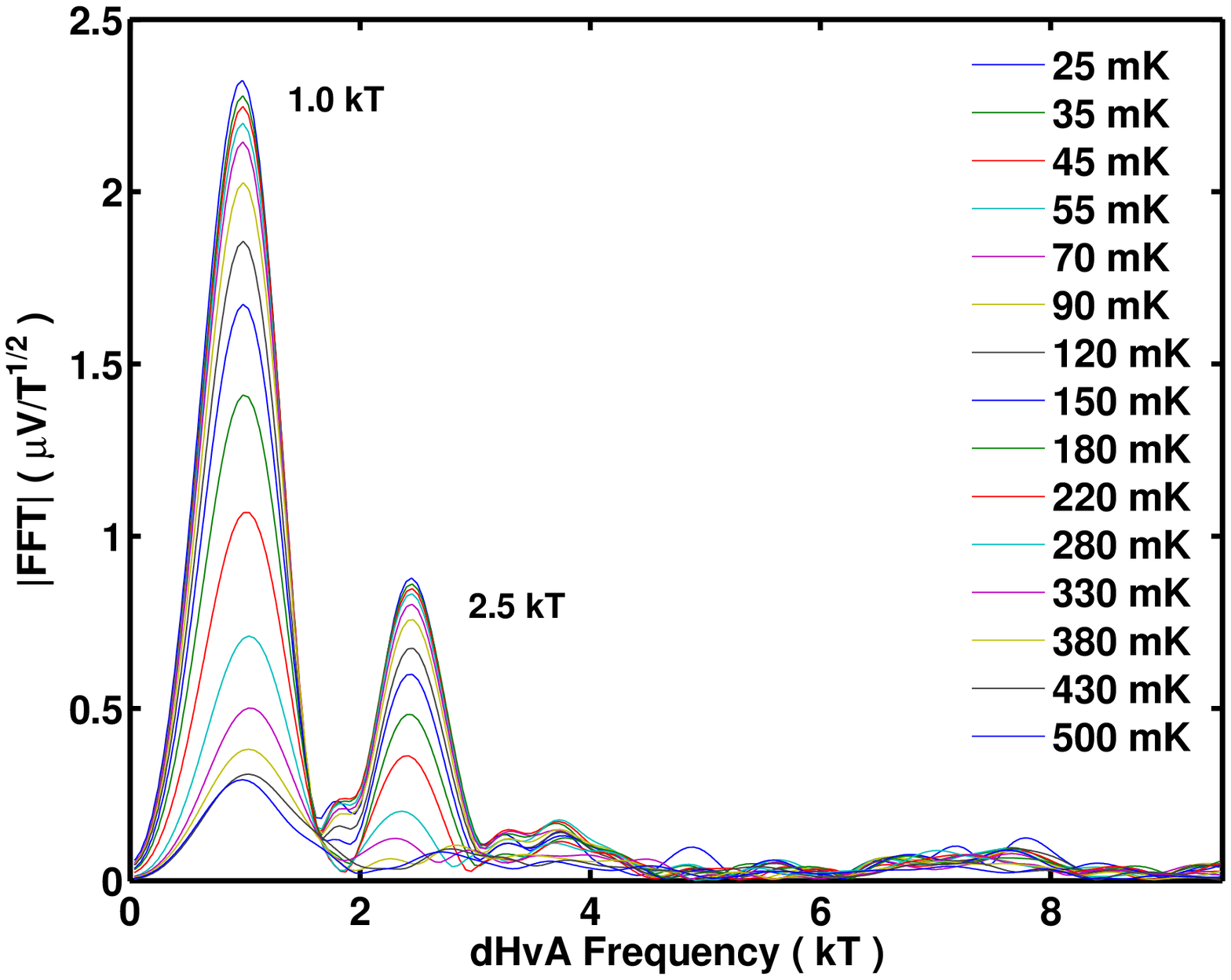}
		\end{center}
	\end{minipage}
	\hfill
	\begin{minipage}[t]{.5\columnwidth}
		\begin{center}
		\includegraphics[width=\columnwidth]{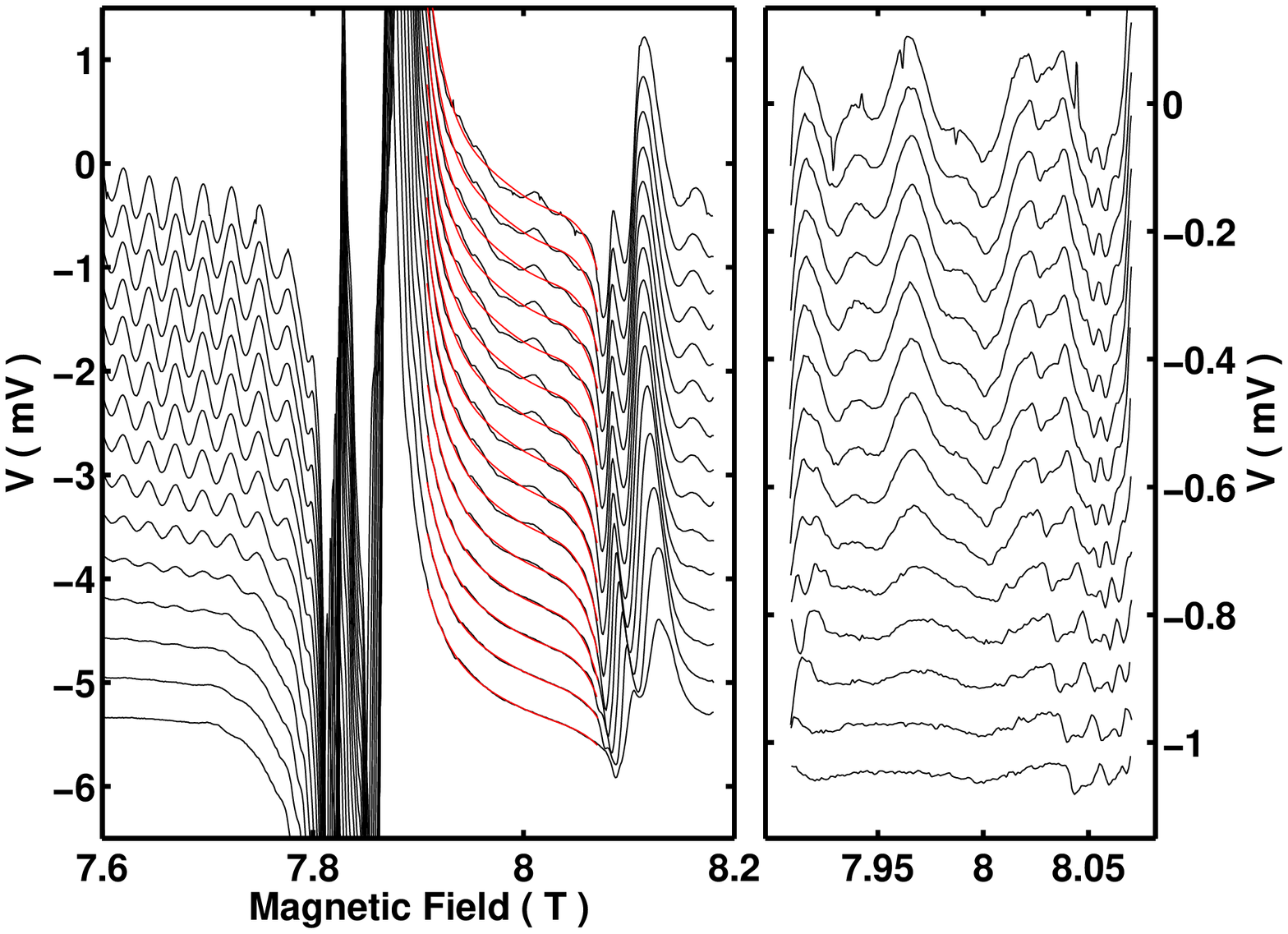}
		\includegraphics[width=\columnwidth]{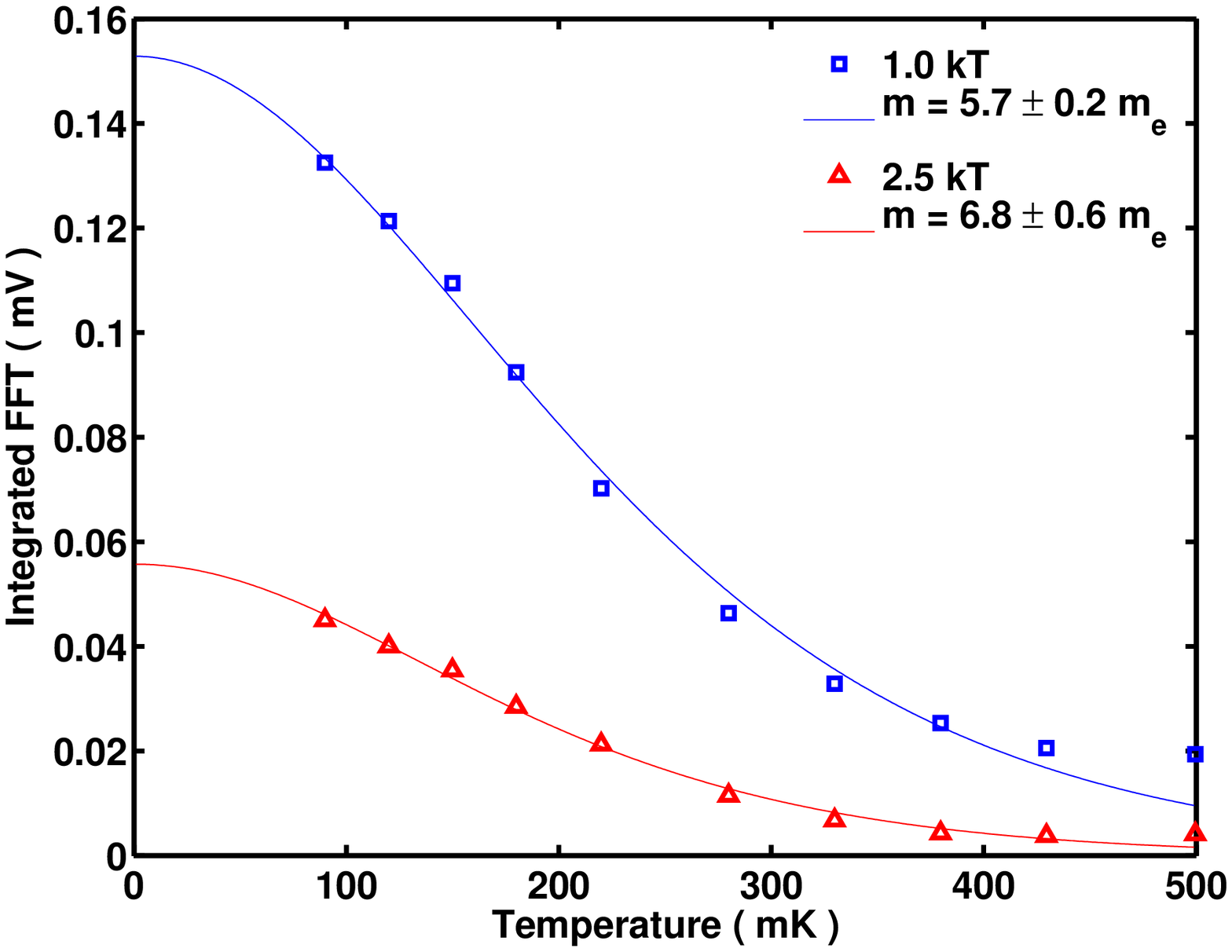}
		\end{center}
	\end{minipage}
	\caption[Oscillations in the nematic phase in sample C698A]{Same analysis as in figure \ref{fig: NematicFFT_c-axis_b} for the data taken with sample C698A.}
	\label{fig: NematicFFT_c-axis_b}
\end{figure}

Figure \ref{fig: NematicFFT_c-axis_a} presents the temperature dependence of the dHvA oscillations in the nematic phase for sample C698I, with its $c$-axis well aligned with the magnetic field.\footnote{The error is of about $\pm 1^{\circ}$ in the [110] direction, the plane of rotation, and $\pm 3^{\circ}$ in the [100] direction.} These were obtained at 15 temperatures between 25 and 500~mK. The data are of high quality and the noise was hardly visible at all in comparison with the oscillations outside the phase.

The top left plot of figure \ref{fig: NematicFFT_c-axis_a} presents the raw data for all temperatures. One can see the metamagnetic transition signal, first discussed in section \ref{sect:dataharmonics}. We mentioned that the first feature at 7.5~T on the left is a metamagnetic crossover, and the second and third are the first order transitions delimiting the nematic phase. We observed between these a few cycles of oscillations, which repeated with identical phase at different temperatures, and their amplitude decreased when the temperature increased. The top right plot presents how the background substraction was done, in red, where a fifth order polynomial was adjusted to the data between appropriate field values such that it properly reproduced the magnetic background, and was subsequently substracted from the data. Any lower polynomial did not yield the appropriate shape, while higher orders reproduced partly the oscillatory component, which in turn resulted in removing part of the signal. The result after removing the background in shown on the right, using the same temperature ordering as in the figure on the left.

From this data set, we carried out Fourier transforms, shown in the lower left plot. Two frequencies were found, at 1.0 and 2.3 kT, at all temperatures. We integrated the peaks and plotted the result as a function of temperature, shown in the lower right graph. The distributions followed the LK function, and two parameter LK fits were performed, using data at temperatures above 90~mK \footnote{This was done for the same considerations as in the previous section regarding low temperature data.}, where we obtained masses of 6.2$\pm$0.2 and 6.8$\pm$0.1 electron masses for the 1.0 and 2.3~kT peaks respectively. Figure \ref{fig: NematicFFT_c-axis_b} shows the same measurements and analysis performed on sample C698A. Very similar results were obtained, with frequencies of 1.0 and 2.5 kT, and masses of 5.7$\pm$0.2 and 6.8$\pm$0.6 electron masses, respectively.

We may speculate about an origin for the measured frequencies and masses. We know already from the field dependence of the dHvA spectra, figure \ref{fig:FvsBCam}, section \ref{sect:CamSpectra}, that the dHvA peak at 1.8~kT increases in frequency to about 2.5~kT near the metamagnetic transition, but the group of peaks near 0.9~kT does not evolve significantly. It is therefore appropriate to associate the peak at 2.3 or 2.5~kT to the one at 1.8~kT in the low field side, and the one at 1.0~kT to the one at 0.9~kT. Although the masses do not agree extremely well, since the 0.9~kT group has a mass of about 8~$m_e$ and the 1.8~kT peak a mass of about 7~$m_e$, we argue that systematic errors due to the difficult background substraction might have led to underestimating the mass within the phase.  
 
\subsection{Angular dependence of the oscillations}

\begin{figure}[!t]
	\begin{minipage}[t]{.5\columnwidth}
		\begin{center}
		\includegraphics[width=\columnwidth]{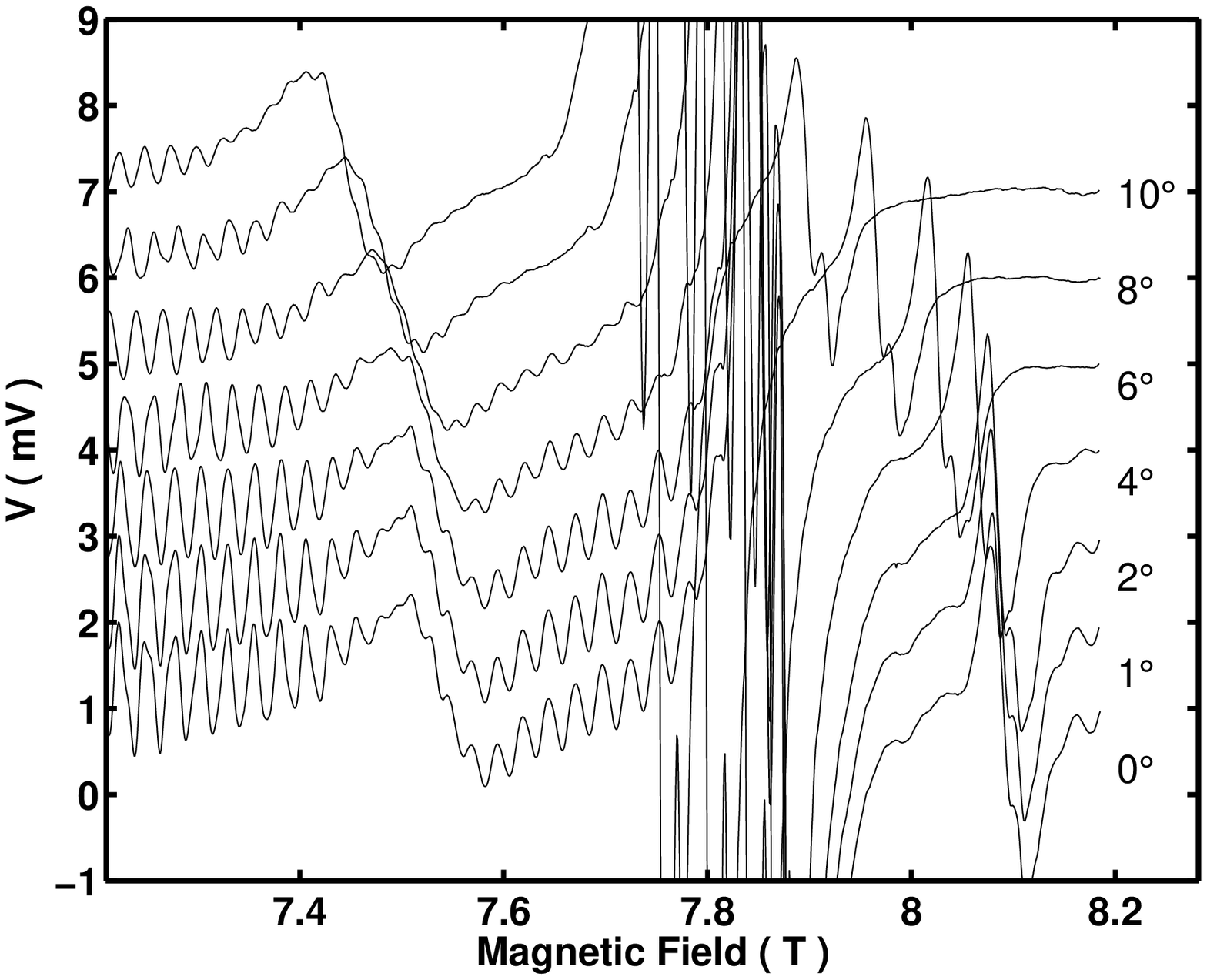}
		\includegraphics[width=\columnwidth]{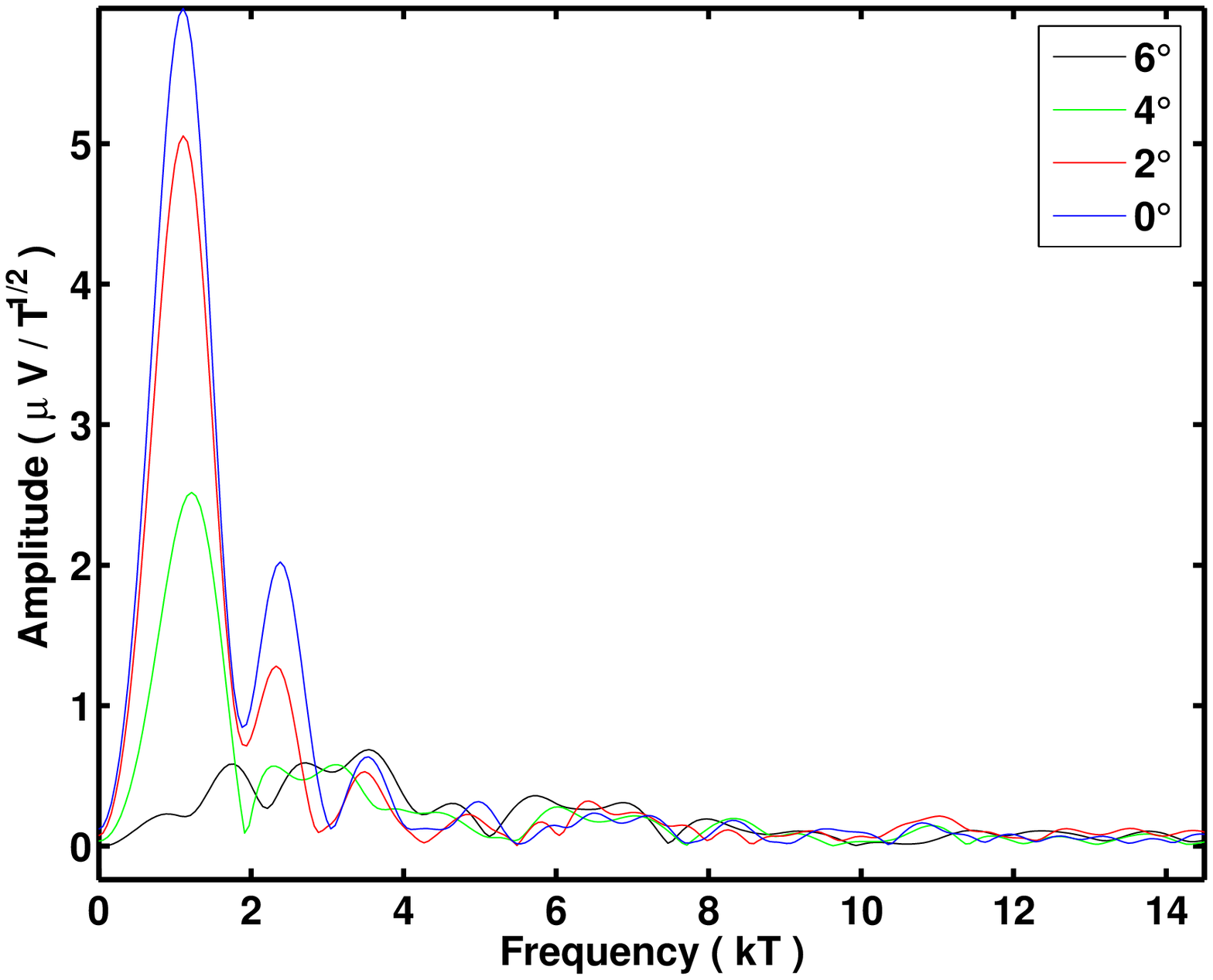}
		\end{center}
	\end{minipage}
	\hfill
	\begin{minipage}[t]{.5\columnwidth}
		\begin{center}
		\includegraphics[width=\columnwidth]{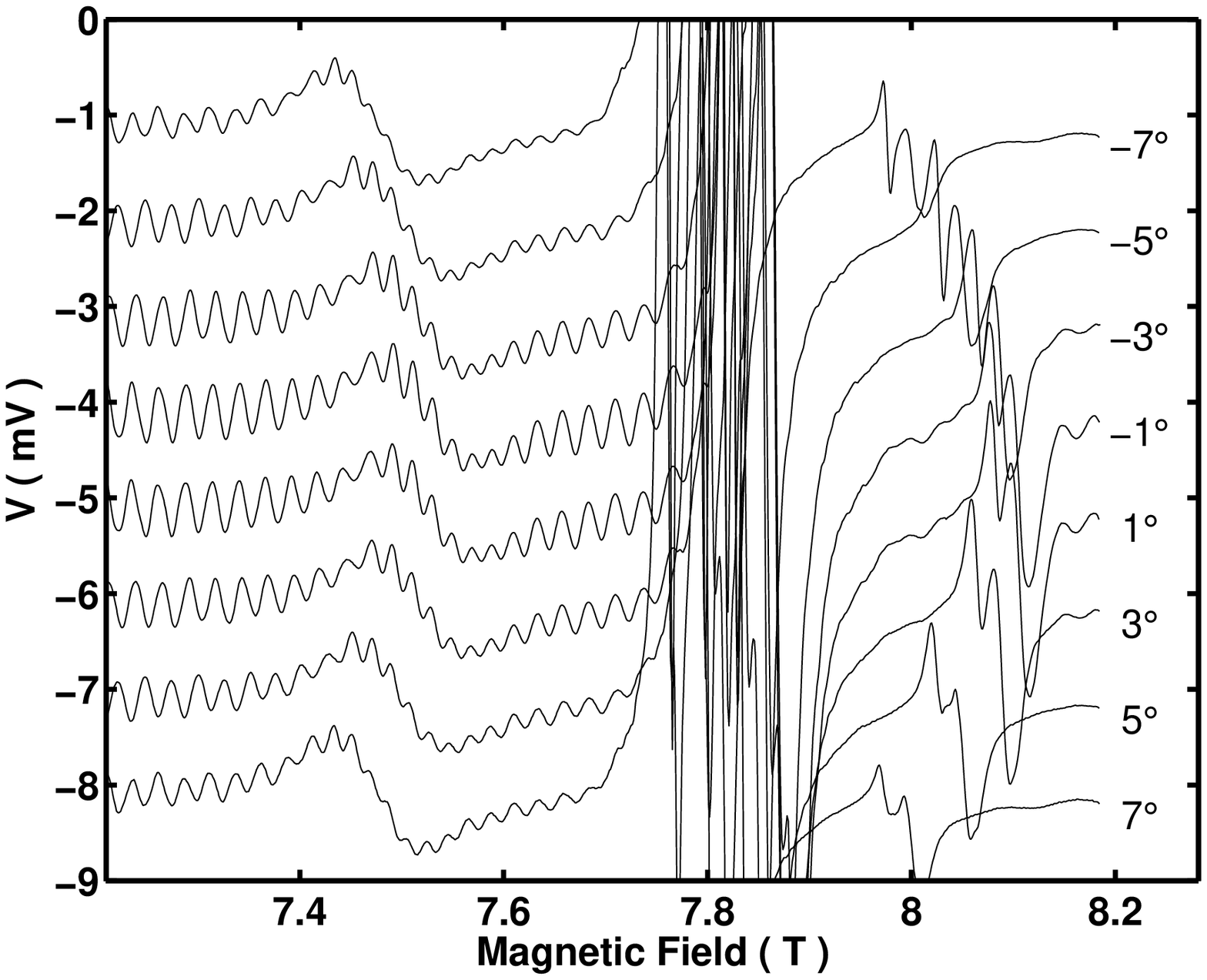}
		\includegraphics[width=\columnwidth]{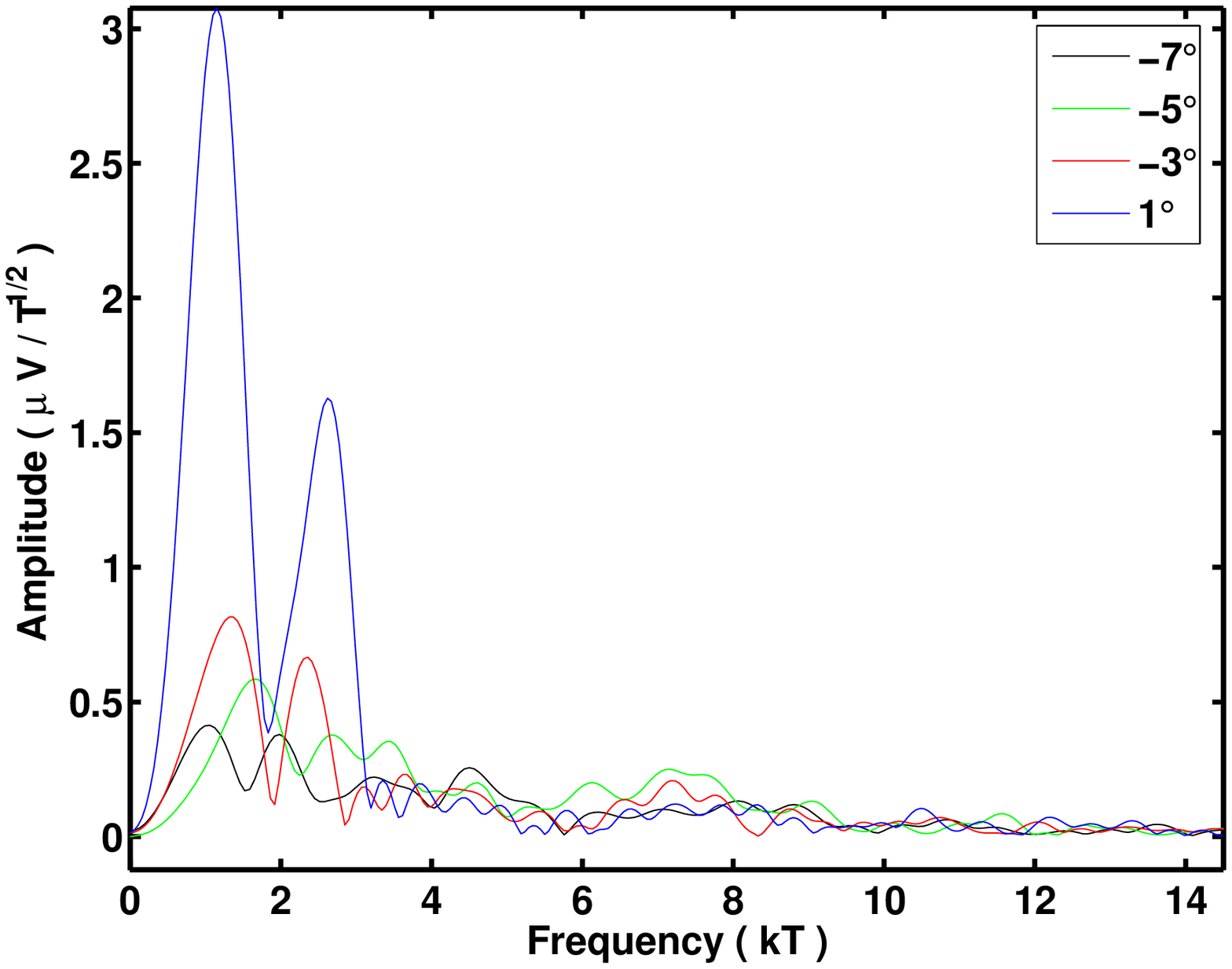}
		\end{center}
	\end{minipage}
	\caption[Angle dependence of the oscillations in the nematic phase]{$Top$ $left$ Raw second harmonic data taken on sample C698I, as a function of the angle between its $c$-axis and the magnetic field, at a temperature of 25~mK. The field is rotated in the crystallographic direction of [110]. $Top$ $right$ Same type of data taken on C698A, and the rotation is performed towards [100]. $Bottom$ Fourier transforms of the data in the nematic phase as a function of angle for sample C698I ($left$) and C698A ($Bottom$).}
	\label{fig: RotNematic_Raw_a}
\end{figure}

dHvA oscillations were measured at a base temperature of 25~mK, at 16 different angles between 0$^{\circ}$ and 25$^{\circ}$ for sample C698I, and between -14$^{\circ}$ and 16$^{\circ}$ for sample C698A, with an angle step of about 1.6$^{\circ}$. The difference in angle spans originated from a misalignment of the samples with respect to the field (see section \ref{sect:FirstRotation} for details). Figure \ref{fig: RotNematic_Raw_a} presents the angular dependence of dHvA obtained on both samples. The top left plot presents data from C698I, where the magnetic field was rotated from $c$-axis towards [110], while the top right plot relates to C698A, where the field was rotated towards [100]. We observed that the oscillations were only present near 0$^{\circ}$, and were significantly suppressed as we rotated away the magnetic field, where effectively, dHvA vanished at an angle of about 5-6$^{\circ}$. In order to emphasise this observation, we performed Fourier transforms using the same background substraction method as described previously, shown in the bottom plots. Both peaks at 1.0 and 2.3~kT decreased in intensity as we rotated away from $c$-axis.

Although the temperature dependence and the frequency values for the dHvA peaks that we found in the nematic phase lead us to associate these to the low field side dHvA spectra of \TTS, this strong angular dependence could be related to the high field dHvA, where we observed that frequencies disappear and appear as a function of angle. If the peak at 2.3 or 2.5~kT really corresponds to the 1.8 - 1.6~kT frequency, then it is consistent to find that it vanishes as a function of angle inside the nematic phase. It is not the same for the peak at 1.0 kT, since we know that in the high field side, it is not present around the $c$-axis, but appears at around 5$^{\circ}$ (see figure \ref{fig: FFT10to18T_A}). It is then, at this point, difficult understand what we observed inside the nematic phase.

\chapter{Discussion \label{chap:discussion}}
\markright{Chapter~\ref{chap:discussion}: Discussion}

We present in this chapter an analysis and an interpretation of the data presented in the preceding chapter. We provide a complete model for the Fermi surface of \TTS, as well as a an interpretation of its properties near the QCEP at the metamagnetic transition. With our data alone, we could not produce a unique model for the FS, but this has become possible by the use of a combination of our data, ARPES measurements that were performed by A. Tamai $et$ $al.$ and magnetocaloric oscillations experiments carried out by A. Rost \cite{tamai, rost}. The model introduced by A. Tamai $et$ $al.$ suggests that metamagnetism may be produced by a peak in the DOS situated near a new small FS pocket which was subsequently discovered by A. Rost in magnetocaloric oscillations. This suggestion is consistent with our data showing that no quasiparticle mass enhancement is present for any of the previously known FS pockets. We unfortunately provide no definite proof but only partial evidence indicating that the new FS pocket is responsible for metamagnetism and the previously observed quantum critical properties.

We moreover make use of various models in order to interpret some unusual properties of our dHvA data, the anomalous beat patterns in the low field side and the highly split spectra in the high field side. For the first problem, we introduce a simple generic model of metamagnetism that features a peak in the DOS, which we use to simulate the field and angle dependence of the amplitude of the 1.8~kT frequency. We obtain only partial agreement with the data, which indicates that the physics taking place are more complex. For the second problem, we use the theory for magnetic breakdown in order to show that a large number of almost degenerate orbits can arise from only a few symmetrical breakdown points in the BZ. We provide there the most probable hot spots in the BZ where magnetic breakdown could occur, and calculate the size of the new expected cyclotron orbits. The two models presented are far from exact. They were only developed to point to probable sources for these anomalous properties; detailed calculations are beyond the scope of this thesis. 

Finally, we discuss the absence of an enhancement of the quasiparticle masses near the QCEP and the appearance of dHvA inside the nematic phase. For the first of these phenomena, we provide a mechanism which was probably responsible for producing spurious mass enhancements through systematic errors in the non-linear LK fits of the previous work of Borzi $et$ $al.$ \cite{borzi}. We give examples of an artificial mass enhancement, but we provide detailed evidence in appendix~\ref{App:F} using numerical simulations. We furthermore discuss the implication of the absence of a mass enhancement by comparing with electronic specific heat data and measurements of the $A$ coefficient of the resistivity. Regarding the nematic phase oscillations, we provide an interpretation for their presence, and argue that their suppression with angle may be related to putative nematic domains. 
 
 \section{Zero field Fermi surface \label{sect:ZeroFieldFS}}
 
In a dHvA experiment on a two-dimensional material, one cannot determine the in-plane shape of the FS (see section \ref{sect:BergemanAnalysis}). In such a case, one knows the sizes of the various FS pockets, but not their position in $\bf{k}$-space or their form and orientation (the terms with $\mu \neq 0$ in the expansion of eq. \ref{eq:kfexpansion}), nor whether they consist of electrons or holes. In order to orient ourselves in this respect, ARPES experiments were highly desirable.

Such experiments were performed by Tamai and co-workers at the Stanford Synchrotron Radiation Laboratory, using single crystals of \TTS\ provided by the author, originating from the ensemble studied in section \ref{sect:search} \cite{tamai}. High quality data were obtained, and a complete interpretation of the low field side FS was constructed, using both ARPES and dHvA data from this work. These results are part of this thesis work only inasmuch as they used low field dHvA frequencies and quasiparticle masses from this project, which had first been measured by Borzi $et$ $al.$ \cite{borzi}, and subsequently remeasured with more precision by the author. They are presented here for the reason that they are essential to the interpretation of the main results of this project. We present in this section the photoemission data and its interpretation, along with a comparison to dHvA data.

Moreover, quantum oscillations in the magnetocaloric effect were discovered in \TTS\ by A. Rost\cite{rost}, featuring an additional low frequency orbit that was not detected by the author of this work using dHvA, due to lack of sensitivity in this frequency region\footnote{Second harmonic dHvA possesses a sensitivity that vanishes at zero frequency as $F^2$, while the magnetocaloric oscillations feature good sensitivity at low, but not at high frequencies.}. This information has been important for the interpretation of our results and is presented as well. Therefore, we first describe the measurements performed with the magnetocaloric effect, followed by those using ARPES. We then introduce the model constructed by A. Tamai $et$ $al.$ and an interpretation for the zero field FS of \TTS.

\subsection{Oscillations in the magnetocaloric effect \label{sect:magnetocaloric}}
 
 \begin{figure}[t]
	\begin{minipage}[t]{.5\columnwidth}
		\begin{center}
		\includegraphics[width=\columnwidth]{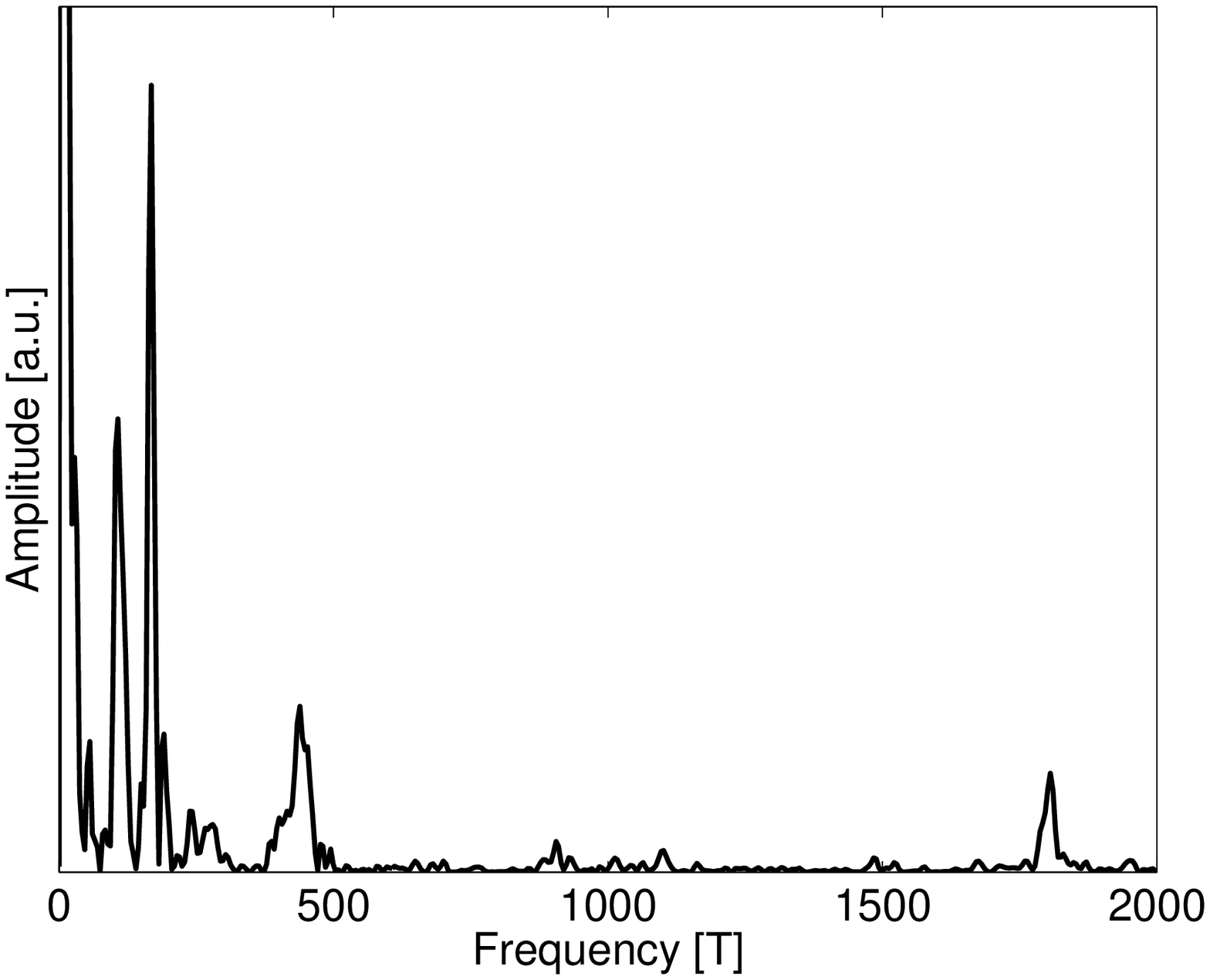}
		\end{center}
	\end{minipage}
	\hfill
	\begin{minipage}[t]{.5\columnwidth}
		\begin{center}
		\includegraphics[width=\columnwidth]{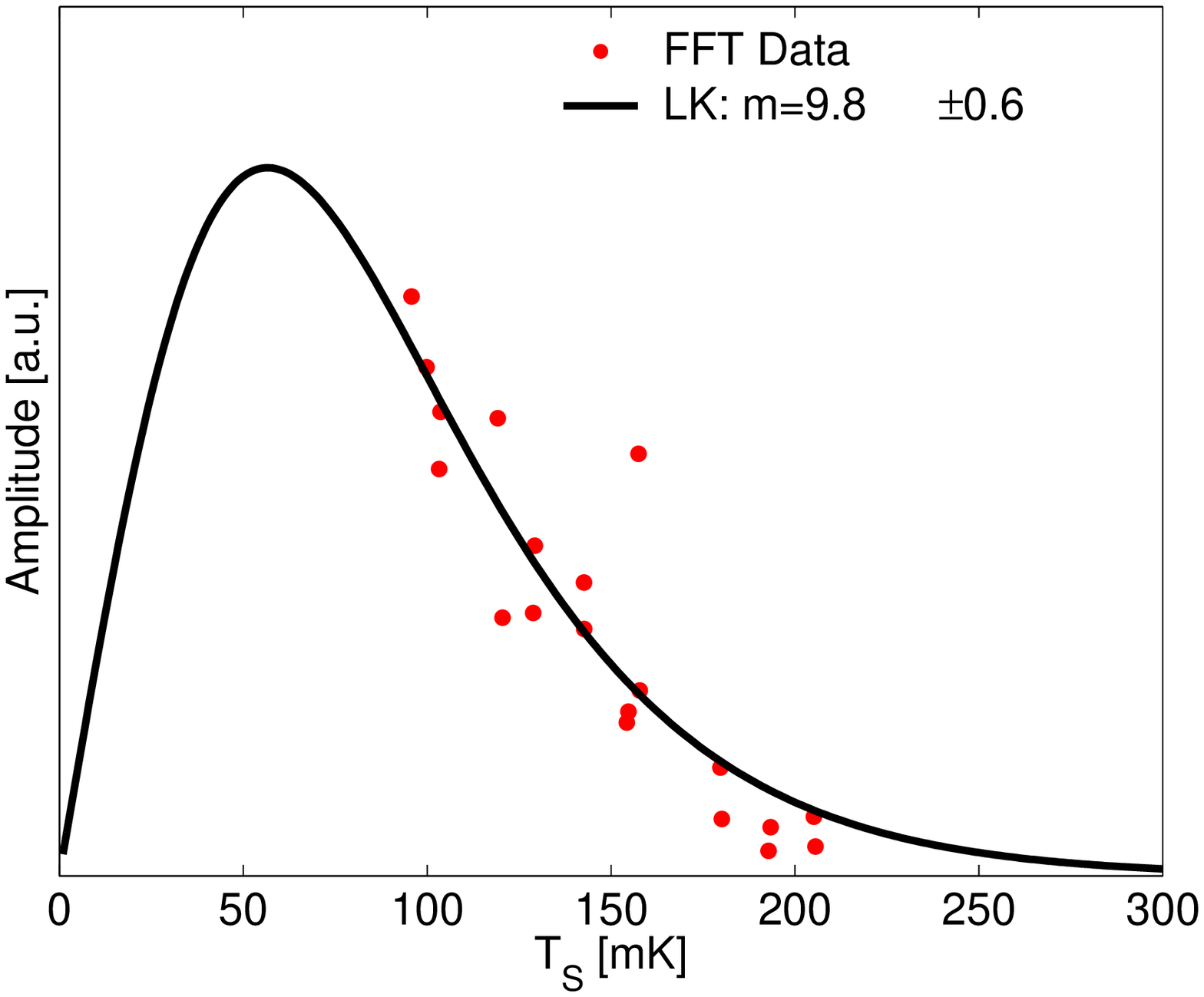}
		\end{center}
	\end{minipage}
	\caption[Quantum oscillations in the magnetocaloric effect]{Quantum oscillations in the magnetocaloric effect. $Left$ Fourier transform of the oscillations between 4 and 7 T. $Right$ Temperature dependence of the amplitude of the peak at 110 T, circles, along with a fit of $\partial LK(T) /\partial T$, solid line. Taken from A. Rost \cite{rost}.}
	\label{fig:newmass}
\end{figure}

Quantum oscillations in the magnetocaloric effect in \TTS\ were found by A. Rost\cite{rost}, using samples that were grown by R. S. Perry and characterised by the author of this project, as described in section \ref{sect:search}. Two high purity samples were used, labelled C698J and C698K. Oscillations were found both above and below the metamagnetic transition field value. In the low field side, the transition, all of the frequencies measured in this project were reproduced except that at 4.2~kT, and the extracted quasiparticle masses were in agreement as well. An additional very low frequency peak was discovered, which in second harmonic dHvA possessed an amplitude that was too low for detection, but became crucial for the interpretation of physics of \TTS.

Figure \ref{fig:newmass}, left panel, shows the Fourier transform of the magnetocaloric oscillatory signal between 4 and 7 T, where one can see peaks at 0.11, 0.15, 0.43 and 1.8~kT, and a faint group near 0.9~kT. Excluding the peak at 0.11~kT, the other ones correspond to the same FS sheets as those observed in this work (see table \ref{tab:freqmassesCambridge}, p. \pageref{tab:freqmassesCambridge}) except for the peak at 4.2~kT, which was not observed in the magnetocaloric effect. As presented in the work of Rost\cite{rost}, these oscillations are damped with increasing frequency due to phase smearing related to the heat relaxation time constant of the system. Figure \ref{fig:newmass} presents the quasiparticle mass analysis of the new peak at 0.11~kT. As the magnetocaloric effect involves the temperature derivative of the magnetisation, the temperature dependence of the oscillations follows the temperature derivative of the LK function rather than the LK function itself. By adjusting $\partial LK(T) /\partial T$, Rost obtained a quasiparticle mass of 9.8$\pm$0.6$m_e$. This discovery will be of relevance in the next section in order to obtain a complete set of quantum oscillations peaks in \TTS.

\subsection{Angle Resolved Photoemission Spectroscopy}
 

\begin{figure}[p]
\begin{center}
	\includegraphics[width=1\columnwidth]{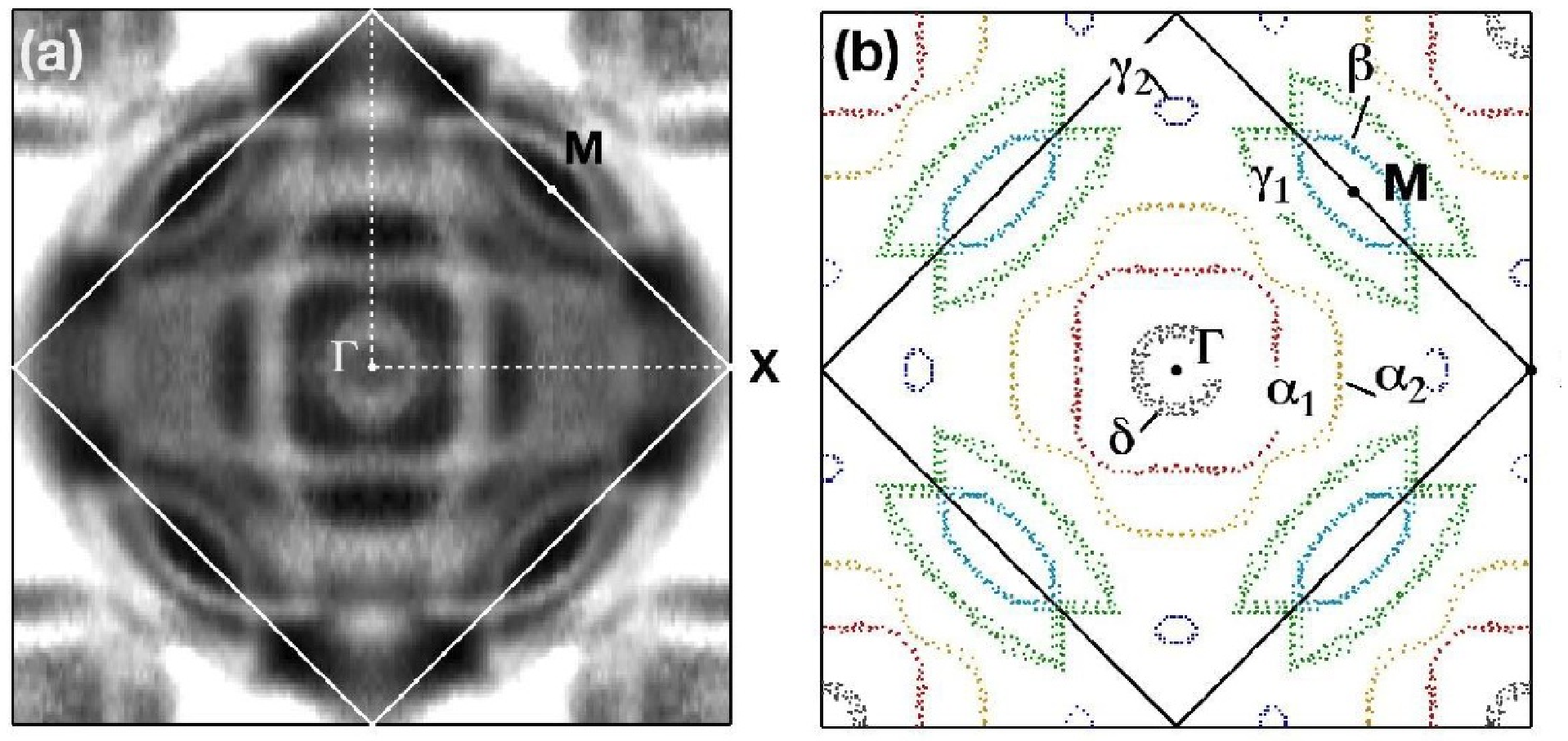}
	\includegraphics[width=1\columnwidth]{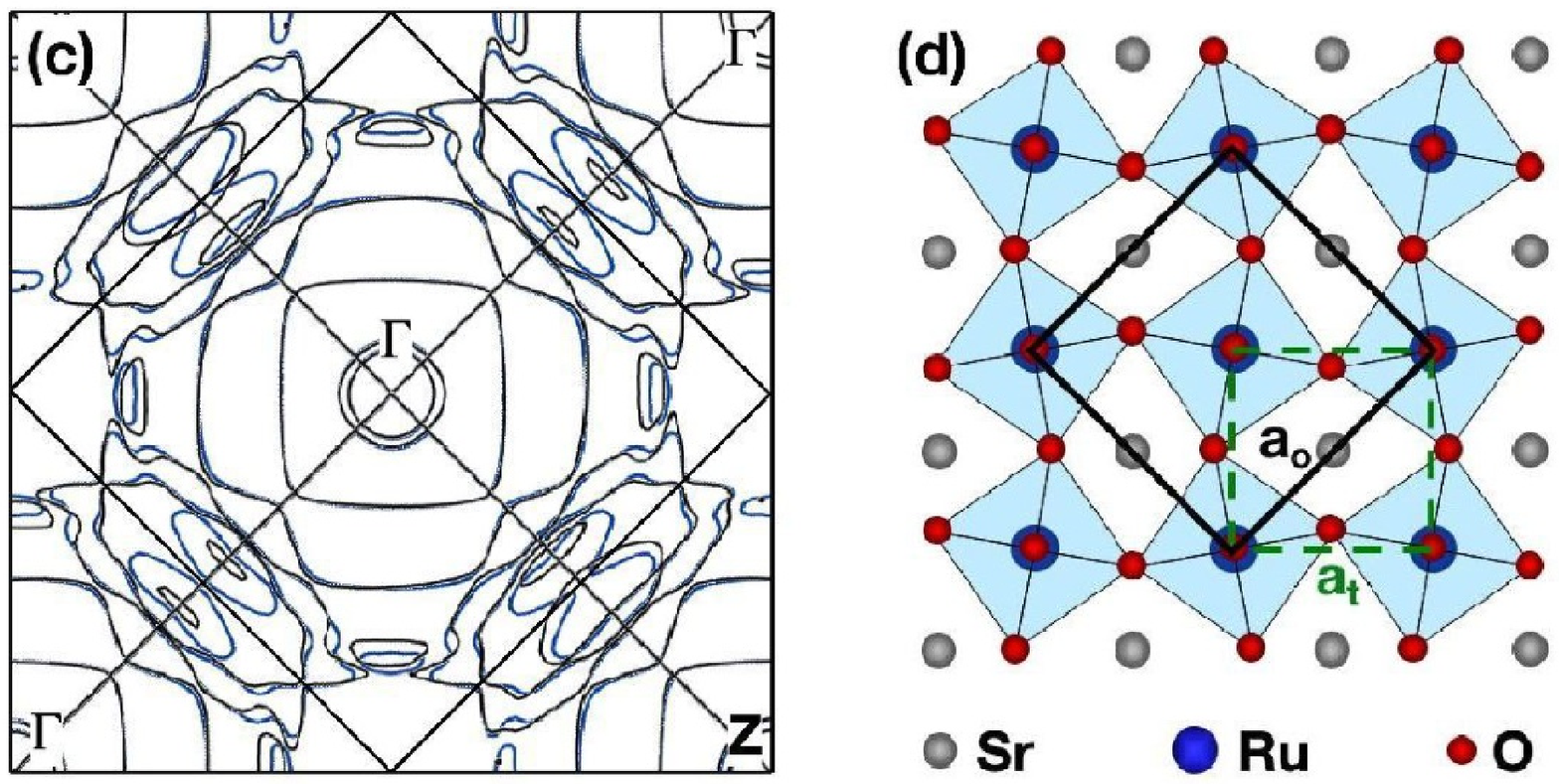}
		\begin{tabular*}{1\textwidth}{@{\extracolsep{\fill}}r c c c c c c}
			\hline
			\hline
			dHvA&$\alpha_1$&$\alpha_2$&$\beta$&$\gamma_1$&$\gamma_2$&$\delta$\\
			\hline
			$F$ (kT) & 1.78 & 4.13 & 0.15 & 0.91 & $-$& 0.43\\
			Area (\% BZ) & 13.0 & 30.1 & 1.09 & 6.64 &$-$& 3.14\\
			$m^*/m_e$ & $6.9\pm0.1$ & $10.1\pm0.1$  & $5.6\pm0.3$ & $7.7 \pm 0.3$ & $-$ & $8.4\pm0.7$\\
			\hline
			ARPES&$\alpha_1$&$\alpha_2$&$\beta$&$\gamma_1$&$\gamma_2$&$\delta$\\
			\hline
			Area (\% BZ) & $14.1\pm2$ & $31.5\pm3$ & $2.6\pm1$ & $8.0\pm2$  & $<1$& $2.1\pm1$\\
			$m^*/m_e$ & $8.6\pm3$ & $18\pm8$ & $4.3 \pm 2$ & $9.6\pm3$ & $10\pm4$ & $8.6\pm3$\\
			\hline
			\hline
		\end{tabular*}
	\caption[ARPES data and comparison with dHvA]{$(a) $ARPES FS data measured by Tamai and co-workers \cite{tamai}. The image was symmetrised from the first quadrant. $(b)$ FS contours extracted from $(a)$. $(c)$ FS calculated by Singh. $(d)$ Schematic structure of a single ruthenium oxide plane. $Table$ dHvA FS areas and quasiparticle masses obtained by the author (see table \ref{tab:freqmassesCambridge}), compared with the same quantities obtained with ARPES. The error on the dHvA frequency and areas are of 10~T and 0.1\%.}
	\label{fig: fig1_tamai3}
\end{center}
\end{figure}

ARPES data were obtained over several series of experiments using samples that were characterised by the author, using the full analysis method described in section \ref{sect:search}. The crystals used were the ones labelled C6103A, C6103ZA, C698C, C698DA and C698DB, and their properties can be found in tables \ref{tab:327Samples1} and \ref{tab:327Samples2}. These possessed very low disorder and had impurity phase volume fractions of less than 1\%, except for C698DA and C698DB, which contained 4\% of \TOF. For more details on the technical aspects of the ARPES experiments, the reader is referred to the review on the subject by Damascelli \cite{damascelli} and the article by A. Tamai $et$ $al.$ \cite{tamai}.

Figure \ref{fig: fig1_tamai3} presents the measured ARPES FS. The plot labelled $(a)$ presents the FS photoemission data in the undistorted zone, with the $\sqrt{2} \times \sqrt{2}$ reconstructed zone indicated as a white diamond. From this image, FS contours have been extracted, shown in $(b)$. Six FS pockets have been identified and labelled $\alpha_1$, $\alpha_2$, $\gamma_1$, $\gamma_2$, $\beta$ and $\delta$. The first two were so labelled for their common origin with the $\alpha$ pocket of \TOF, which possesses $d_{xz}$, $d_{yz}$ orbital character (see section \ref{sect:estructure} and figure \ref{fig: 327FS2}). The next two were labelled $\gamma_1$, $\gamma_2$ for their relation with the $\gamma$ pocket of \TOF, or the $d_{xy}$ orbital, although they are also hybridised partly with a band parent to the $\beta$ pocket of \TOF, and it is not presently known with certainty whether $\gamma_2$ actually crosses $\epsilon_F$ or exists below the Fermi energy. The $\beta$ pocket is, on the other hand, related to that with the same name in \TOF. Finally, the $\delta$ pocket originates from a new band pushed down from the $e_g$ manifold above the Fermi level, of $d_{x^2-y^2}$ character, as inferred from LDA calculations of Mazin and Singh, \cite{singh}. The third part of figure \ref{fig: fig1_tamai3}, $(c)$, presents LDA calculations performed by Singh, for an octahedral rotation of 6.8$^{\circ}$. The blue lines represent the FS at the basal plane, $k_z = 0$, and the black lines are for $k_z c = \pi/2$, showing the extent of the $k_z$ dispersion, or corrugation of the FS sheets. 

The bottom table in figure \ref{fig: fig1_tamai3} shows a comparison between dHvA and ARPES data. The first row presents values of the dHvA frequencies $F$ associated with the Fermi surface contours identified by ARPES, with the same notation as in $(b)$, followed by the fraction of the BZ area each extend to, and their quasiparticle masses. The cross sectional areas in the $c$-axis direction from dHvA data and those from ARPES all agree very well, except for the $\gamma_2$ pocket, which possesses a small area and was not detected by dHvA.  There is also a good agreement between the dHvA quasiparticle masses and those extracted from ARPES Fermi velocities. 

This data set provides us with a robust in-plane picture of the FS of \TTS. ARPES moreover tells us, from dispersion data, whether the pockets consist of electrons or holes\footnote{A. Tamai $et$ $al.$, private communication.}. It was discovered that the circular pocket near $\Gamma$ ($\delta$) is electron-like, the square and cross shaped bands ($\alpha_1$, $\alpha_2$) centred on $\Gamma$ are hole-like, the two surfaces at the $\bf{M}$ point ($\gamma_1$, $\beta$) are electron-like, and the very small pocket near the corner $\bf{X}$ of the zone ($\gamma_2$) is hole-like.

\subsection{Luttinger sum rule and specific heat \label{sect:Luttingersum}}

As was described in sections \ref{sect:OscZeroT} (eq. \ref{eq:LuttingerSum}) and \ref{sect:LK} (eq. \ref{eq:SpecificHeatSum}), the total number of electrons detected with dHvA can be counted, as can be the total number of degrees of freedom, which may be compared with, respectively, the expected number of electrons per formula unit and the electronic specific heat at zero field. For this purpose one requires a model of the FS, with knowledge of whether the various sheets are composed of electrons or holes. 

One such model was proposed by A. Tamai $et$ $al.$, where these two sums were performed when using the ARPES areas and quasiparticle masses. It is the following: $\alpha_1$ and $\alpha_2$ consist of holes and count as negative. $\gamma_1$ and $\beta$ are composed of electrons, but from energy dispersion data, Tamai and co-workers have demonstrated that $\gamma_1$ is double, a phenomenon attributable to bilayer splitting of the $d_{xy}$ band, and so counts twice. The $\beta$ band is not double, however, since as one can see in figure \ref{fig: 327FS2} f., the bilayer splitting was already taken into account and leads to another band situated elsewhere\footnote{The same argument applies to the $\alpha_1$, $\alpha_2$ bands.}. Moreover, $\beta$ and $\gamma_1$ are situated at the $\bf{M}$ point, and one is required to count them both twice, such that $\beta$ is counted twice and $\gamma_1$ four times. Finally, the putative pocket $\gamma_2$ is very small and does not make a very large contribution to the total area, but possesses a mass that is not negligible, and appears four times inside the BZ. It is not presently known but rather expected from DFT calculations that this pocket should also be bilayer split, increasing its count to eight times. Finally, the $\delta$ pocket was predicted to be bilayer split too, and hence will be counted twice. 

Since ARPES does not provide a definite answer as to whether $\gamma_2$ is part of the FS or not, we tried to find out through the Luttinger sum and the total specific heat. One should account for 16 electrons in \TTS, since four electrons originate from four Ru atoms in the distorted lattice. We mentioned in section \ref{sect:estructure} how the FS could be constructed from simple arguments, and presented a schematic plot of the result, figure \ref{fig: 327FS2}. In this model, the back-folding of the four outermost FS sheets seen in figure \ref{fig: 327FS2} $(d)$ should give rise to eight electron filled BZ, one per spin flavour per sheet. With this FS model, ARPES and dHvA data, the Luttinger sum excluding $\gamma_2$ is
\bea
A_{tot} &=& 2(1-A_{\alpha_1}+1-A_{\alpha_2}+4A_{\gamma_1} + 2A_{\beta} + 2A_{\delta}) + 8\nn\\
 &=& 12 - .08\pm.09 \, \mathrm{electrons} \quad (\mathrm{ARPES}),\nn\\
 &=& 12 - .162\pm .006 \, \mathrm{electrons} \quad (\mathrm{dHvA}),\nn\\
 \eea
where the factor of two refers to spin flavours and the additional eight corresponds to filled zones. In both results, there are four electrons missing, which could be hidden in additional filled Brillouin zones. In this view, the ARPES result is consistent with 16 electrons, but the dHvA result misses almost a fifth of an electron. Including $\gamma_2$, assuming that it covers an area of 1\% with an error of 100\% of its value for ARPES, and using the frequency of 0.11 measured by A. Rost for dHvA, the sum becomes
\bea
A_{tot} &=& 2(1-A_{\alpha_1}+1-A_{\alpha_2}+4A_{\gamma_1} + 2A_{\beta}+2 - 8A_{\gamma_2} + 2A_{\delta}) + 8\nn\\
 &=& 16 - .24\pm.12 \, \mathrm{ electrons} \quad (\mathrm{ARPES}),\nn\\
 &=& 16 - .291\pm.008 \, \mathrm{ electrons} \quad (\mathrm{dHvA}).\nn\\
 \eea
 In this case, we obtain the appropriate number short of a quarter of an electron for both experiments.
 
With the same two models, the total specific heat is counted by adding the quasiparticle masses, where electrons and holes contribute positively, and one should reproduce the value of the zero field specific heat of 110 mJ/mol Ru K$^2$ measured by Ikeda $et$ $al.$ \cite{Ikeda2000}. Excluding $\gamma_2$, we obtain
\bea
\sum m^* &=& m_{\alpha_1} + m_{\alpha_2} + 4m_{\gamma_1} + 2m_{\beta} + 2m_{\delta} \nn\\
&=& 91\pm15 \, m_e \Rightarrow \gamma = 67 \pm 21\, \mathrm{mJ/mol Ru K}^2 \quad (\mathrm{ARPES}),\nn\\
&=& 75.8 \pm1.2 \, m_e, \Rightarrow \gamma = 56 \pm 1 \, \mathrm{mJ/mol Ru K}^2 \quad (\mathrm{dHvA}),\nn\\
\eea
and find that around half of the specific heat is missing.
If we now include $\gamma_2$ into the sum, taking for the missing value in dHvA that measured by A. Rost with the magnetocaloric effect, we obtain
\bea
\sum m^* &=& m_{\alpha_1} + m_{\alpha_2} + 4m_{\gamma_1} + 2m_{\beta} + 2m_{\delta} + 8m_{\gamma_2}\nn\\
&=& 174 \pm29 \, m_e \Rightarrow \gamma = 129 \pm 21\, \mathrm{mJ/mol Ru K}^2 \quad (\mathrm{ARPES}),\nn\\
&=& 159 \pm 11 \, m_e, \Rightarrow \gamma = 118 \pm 8 \, \mathrm{mJ/mol Ru K}^2 \quad (\mathrm{dHvA}),\nn\\
\eea
and recover the correct specific heat within experimental error. 

We conclude from the calculation of the specific heat that the band giving rise to $\gamma_2$ most probably crosses the Fermi level and that it should be included in the FS, whereas if we do not, half of the specific heat is not accounted for. However, the Luttinger sum does not seem to agree with this statement, though, as we do not obtain a integer number of electrons in either cases whether we include or not the $\gamma_2$ pocket, but the difference is greater when we do include it.

\begin{figure}[!t]
\begin{center}
	\includegraphics[width=1\columnwidth]{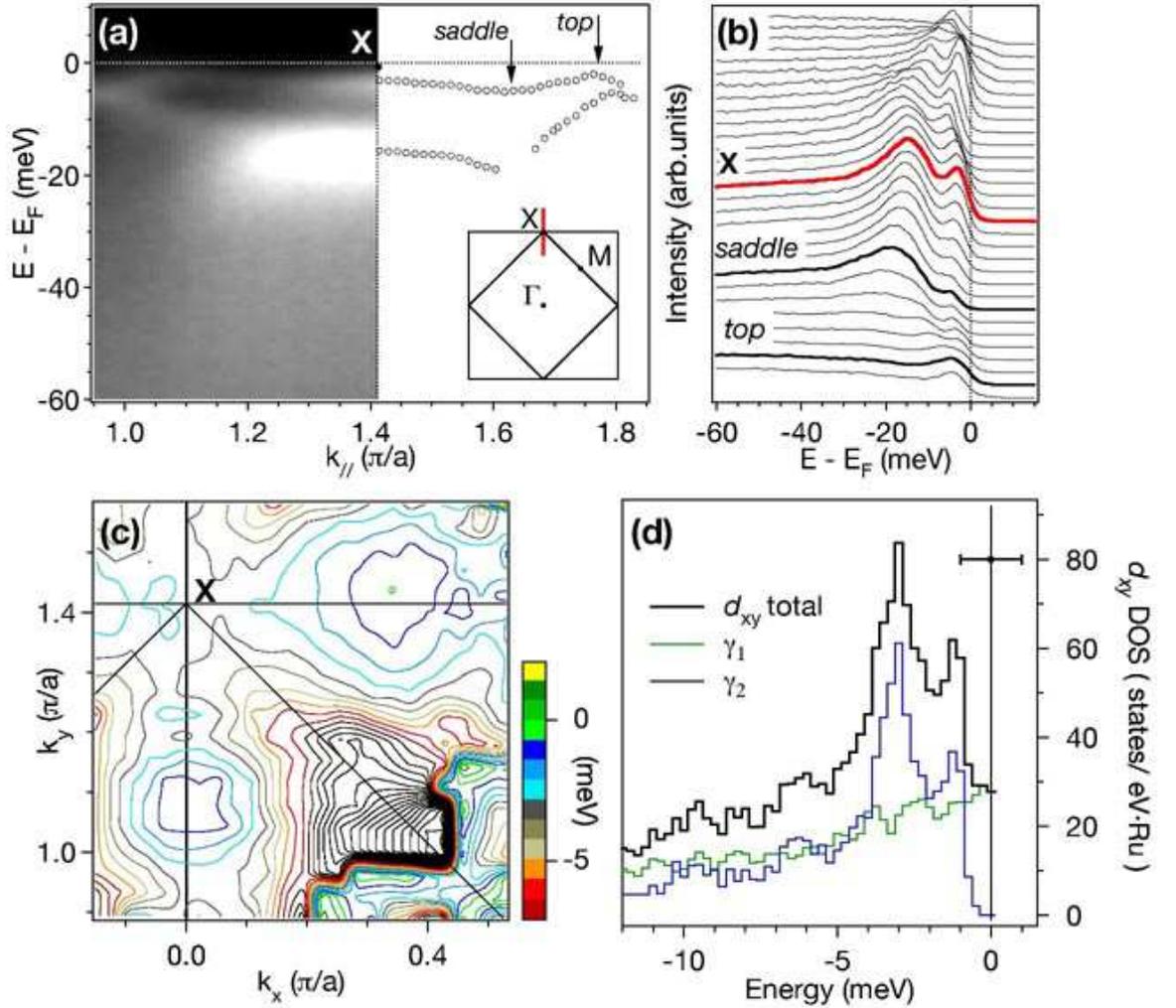}
	\caption[van Hove singularity in \TTS]{$(a)$ Dispersion near the corner $X$ of the BZ and the $\gamma_2$ pocket, in the $k_y$ (or $k_x$) direction. $(b)$ Raw energy distribution curves extracted from $(a)$. $(c)$ Topographic map of the dispersion in the plane near $X$, where a saddle point exists. $(d)$ Histogram of the dispersion in $(c)$ as a function of energy.}
	\label{fig: fig3_tamai3}
\end{center}
\end{figure}

Finally, let us mention that A. Tamai $et$ $al.$ also discovered the existence of a saddle point very close to the Fermi level, in the region surrounding the $\gamma_2$ band. A dispersion of this type corresponds to a van Hove singularity (vHS), which produces a peak in the DOS if it is located at the Fermi level, and, if it is located within a few meV of the Fermi level, could be responsible for metamagnetism in \TTS\ when it is reached by Zeeman splitting. Tamai and co-workers found evidence for such a peak in the DOS below the Fermi level by calculating a histogram of the number of states as a function of energy from a topographic map of the dispersion in $k_x$,$k_y$ space near the corner $X$ of the BZ, around the $\gamma_2$ band. Figure \ref{fig: fig3_tamai3} $(a)$, shows the dispersion near the $\gamma_2$ pocket along the $k_y$ direction. One can observe that this band is very flat and of hole type. The $(b)$ panel shows the extracted energy distribution curves, which we do not discuss here. Panel $(c)$ presents the level curves in the vicinity of the corner of the BZ $X$, where the saddle point is located, between the band $\gamma_2$, in dark blue, and $X$. From these level curves was extracted a contribution to the DOS that features two peaks, shown in $(d)$. Tamai and co-workers suggested that the energy scale of 1-3 meV makes these peaks good candidates for explaining metamagnetism in \TTS. This vHS would be reached with magnetic field as the $\gamma_2$ band expands towards the corner of the zone. They also suggested that this could explain the changes in dHvA frequencies observed by Borzi $et$ $al.$ at the metamagnetic transition, which could correspond to a Fermi surface reconstruction involving transfers of quasiparticles between bands.

 \section{Low field Fermi surface \label{sect:LowFieldSideModel}}
 
The low field FS of \TTS\ possesses a structure that is more complex than that predicted in a two dimensional system like \TOF, as we saw in section \ref{sect:FirstRotation}. In \TOF, the anisotropy of the crystal lattice leads to FS sheets that are connected in the $k_z$ direction, the shape of which can be expanded in cylindrical coordinates (section \ref{sect:BergemanAnalysis}). In the work of Bergemann \cite{bergemann}, the beat patterns of the quantum oscillations in field and field angle were calculated very accurately in outstanding agreement with dHvA measurements. In \TTS, we saw that the beat pattern of the 1.8~kT peak is not accounted for by any combination of terms in the expansion of that model. An additional contribution plays a role here, most probably related to metamagnetism. 

We propose in this section a model where a peak in the total DOS, combined with charge transfers between bands, produces a non-linear variation of the $k$ space area of the Zeeman split bands, resulting in interference in the dHvA oscillations. Moreover, we present evidence that the gyromagnetic factor of the quasiparticles $g$ is spatially anisotropic and produces an additional angle dependence in the dHvA. We do not, however, reproduce exactly the complex beat pattern of the 1.8~kT peak, but this model explains relatively well some of the features of the data. A more elaborate and accurate theory proved too difficult to construct for the scope of this project.
 
 \subsection{Anisotropic $g$ factor}
 
In materials with strong crystalline field, the coupling in energy of the electron's magnetic moment in a magnetic field can be anisotropic. The energy term of Pauli paramagnetism is expressed with a  tensorial form for the $g$ factor, $\bf{S} \cdot \bf{g} \cdot \bf{B}$, where ${\bf g} = {\bf i} g_{xx} {\bf i} + {\bf j} g_{yy} {\bf j} + {\bf k} g_{zz} {\bf k}$, and the directions $\bf i,j,k$ are defined with respect to the crystal field symmetry. The $g$ factor has three main axes with different values, and the shift in energy due to the Zeeman splitting changes with the direction of $B$. In spherical coordinates, it is expressed as 
\beq
g(\theta,\phi) = \sqrt{g_{xx}^2 \sin^2\theta \cos^2\phi + g_{yy}^2 \sin^2\theta \sin^2\phi + g_{zz}^2 \cos^2\theta },\nn
\eeq
 where $\theta$ is the angle between $B$ and the $z$ direction and $\phi$ that in the $xy$ plane. The $g$ factor respects the crystal field symmetry, and we take $g_{zz}$ as that in the $c$-axis direction, and, $g_{xx}, g_{yy}$ in the $ab$-plane. For the sake of simplicity, we assume that $g_{xx}$ and $g_{yy}$ are of very similar value, which we will find is approximately but not absolutely true in \TTS. With this approximation, the expression for $g$ reduces to $g(\theta) = \sqrt{g_{ab}^2 \sin^2\theta + g_{c}^2 \cos^2\theta }$.

 \begin{figure}[t]
	\begin{minipage}[t]{7cm}
		\begin{center}
		\includegraphics[width=7cm]{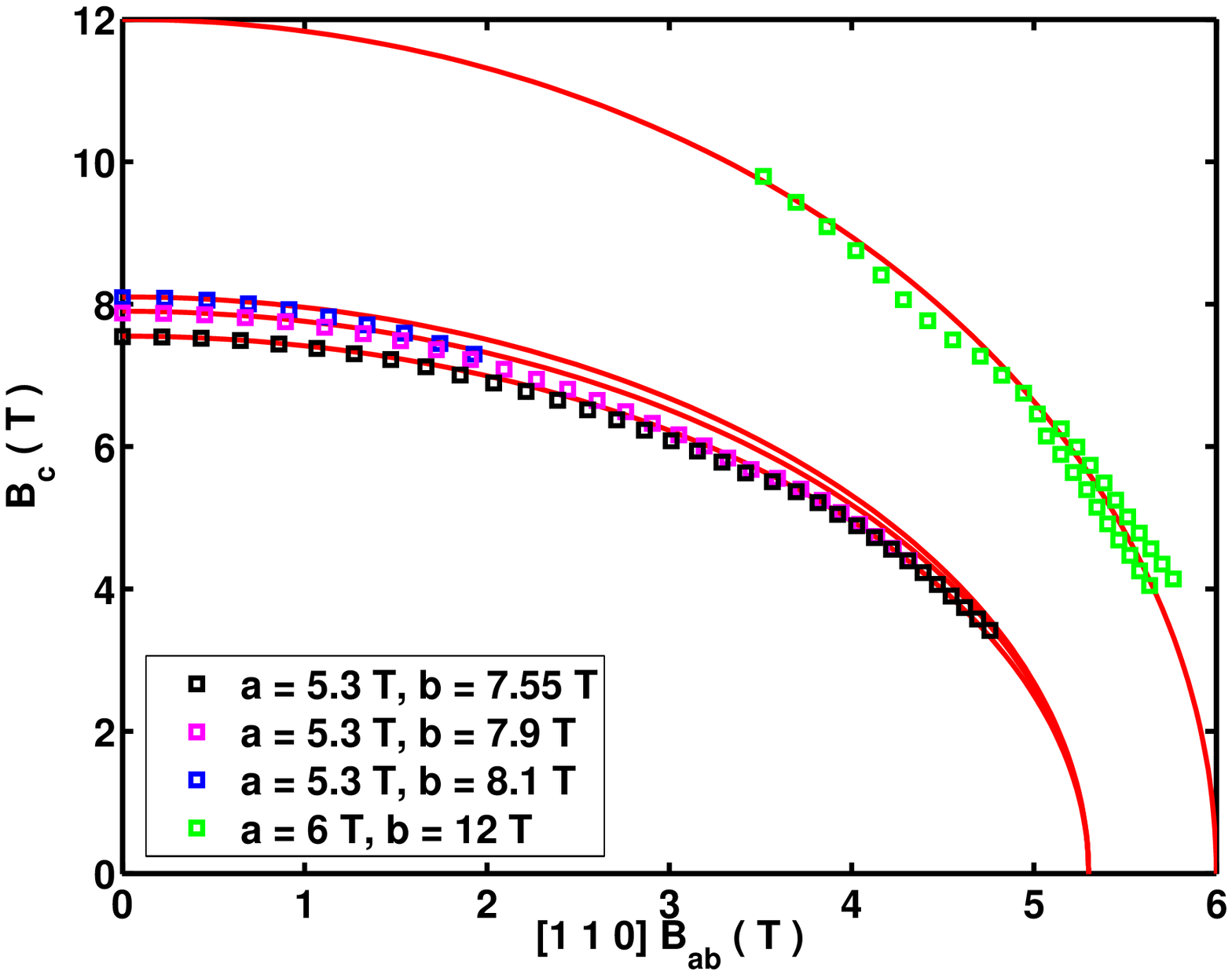}
		\end{center}
	\end{minipage}
	\hfill
	\begin{minipage}[t]{7cm}
		\begin{center}
		\includegraphics[width=7cm]{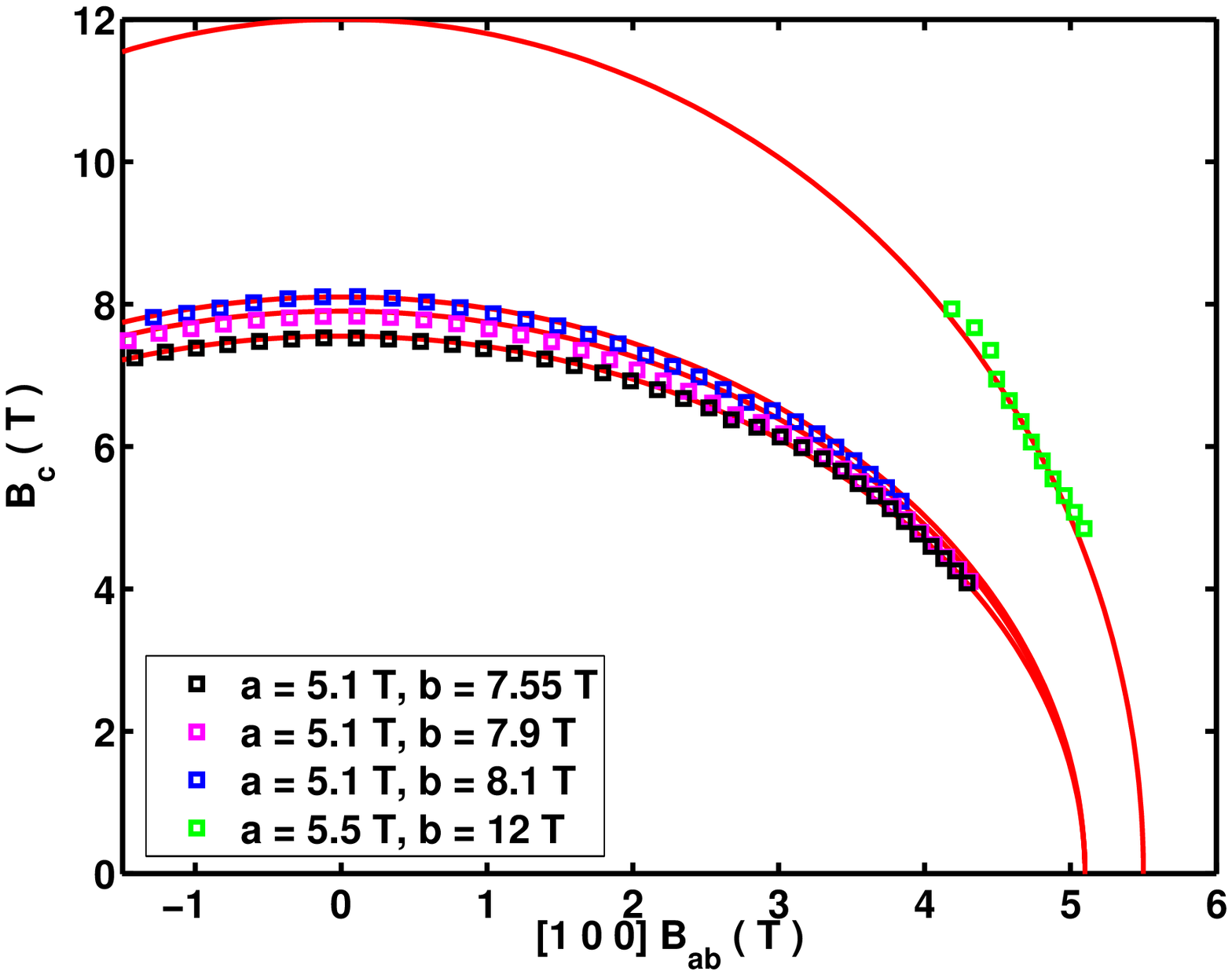}
		\end{center}
	\end{minipage}
	\caption[Orientation dependent $g$ factor]{Out-of-plane magnetic field component as a function of the in-plane magnetic field component of the metamagnetic transitions and crossovers, taken from the positions of peaks in the second harmonic AC susceptibility. The data is well modelled with elliptical curves, of the type ${B_{ab}^2\over a^2} + {B_c^2\over b^2} = 1$, shown in red. The plots present data for rotations performed from [001] towards [110], left, and towards [100], right. The legends show the pairs of parameters $a$ and $b$ for each curve, which reveal a small anisotropy in the plane.}
	\label{fig: Ellipses-a}
\end{figure}

The consequence of the angular dependence of $g$ is that properties driven by the magnetic field should exhibit an additional variation with angle. It is indeed the case with the metamagnetic transition, as we saw from figure \ref{fig:MMTpositions}. In that case, if we assume that the magnetisation follows an empirical function $f$ of the type
\beq
M(B) \propto f\big({1\over2}\mu_B g(\theta) B - \epsilon_c\big),\nn
\eeq
where $\epsilon_c$ is the critical energy, then it follows that at the metamagnetic transition, the product $g(\theta)B$ is constant and we have a critical field $B^*$ that follows 
\beq
B^*(\theta) = {\epsilon_c \over \mu_B\sqrt{g_{ab}^2 \sin^2\theta + g_{c}^2 \cos^2\theta }}.\nn
\eeq
This can be expressed with the in-plane and out of plane critical fields $B_c$ and $B_{ab}$ in the form of an ellipse,
\beq
g_{ab}^2B_{ab}^2 + g_{c}^2B_{c}^2 = {\epsilon_c^2 \over \mu_B^2}.\nn
\eeq
From the work of Grigera\footnote{S. A. Grigera, private communication. Complete mapping of the metamagnetic transition in the field-angle plane, as seen from various physical properties of \TTS.}, that of Perry \cite{perry1, perry2, perryJPSJ} and this thesis, we know that the critical field in the $c$-axis direction is between 7.9 and 8.1 T, and that in the $ab$-plane is between 5.1 and 5.3~T \footnote{The in-plane anisotropy of the critical field was never investigated in detail; we know from this work and from that of A. Rost (in preparation) that differences exist in the $ab$-plane between 5.1 and 5.4, which indicates that $g_{xx} \neq g_{yy}$.}, and that it evolves continuously between the two sets of values. Figure \ref{fig: Ellipses-a} shows the position of the $c$-axis component of the critical fields as a function of the $ab$-plane component, along with elliptical curves taken from this description. They agree remarkably well. Note that we do not presently know the exact value of $g$, except for the fact that it is smaller in the $c$-axis direction compared to the in-plane value by a factor of about 5.2/7.9 $\simeq$ 0.66, which corresponds to the ratio of the critical field values for these directions. 

\subsection{Evolution of frequencies with magnetic field \label{sect:electrontransfer}}

\begin{figure}[p]
\begin{center}
	\includegraphics[width=1\columnwidth]{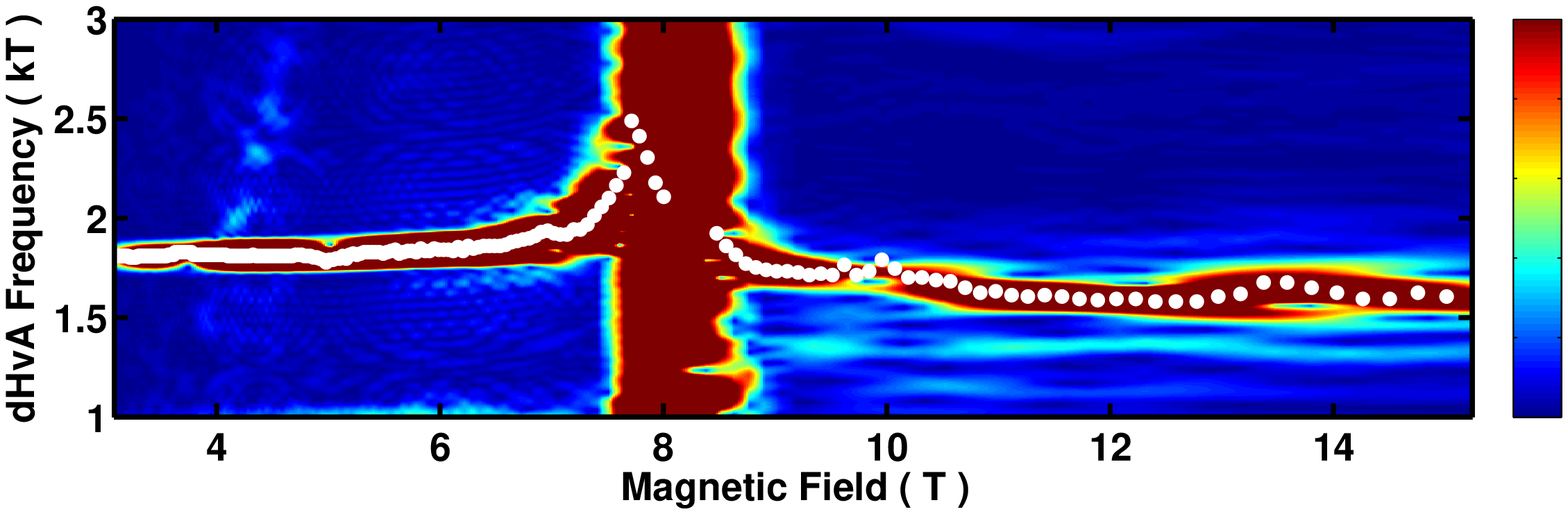}
	\caption[Field dependence of the 1.8~kT frequency]{Field dependence of the $\alpha_1$ pocket. The colours correspond to the intensity of the Fourier transforms (see figure \ref{fig:FvsBCam}) and the white circles to the maximum of the peak near 1.8-1.6~kT for each field value, using 360 windows of width 0.005 T$^{-1}$.}
	\vspace{24pt}
	\label{fig: Freq2kTc-axis}
	\includegraphics[width=1\columnwidth]{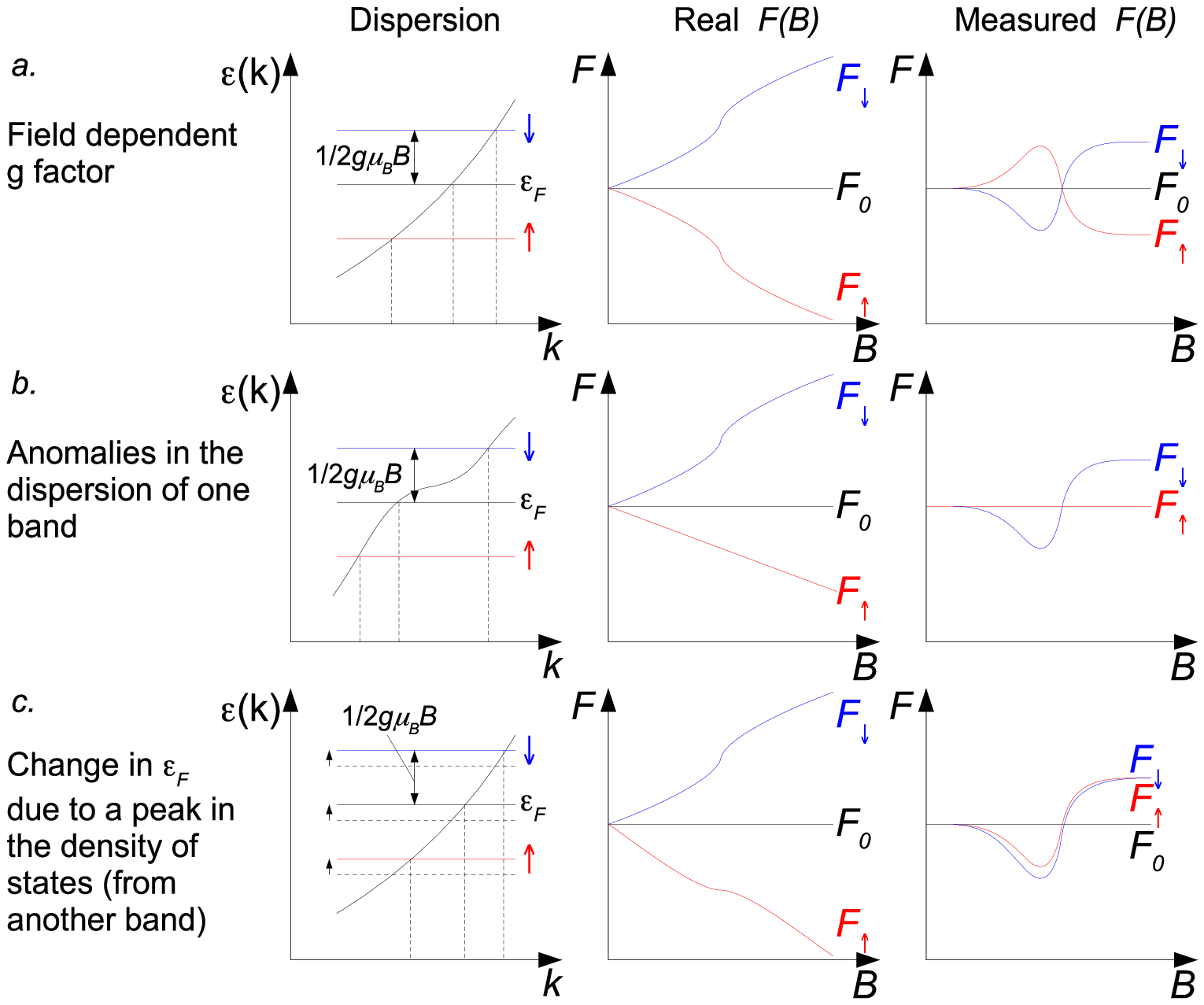}
	\caption[Scenarios for changes in dHvA frequencies]{Three scenarios that can lead to changes in measured dHvA frequencies. Blue denotes energies and dHvA frequencies related to spin down quasiparticles, red to spin up and black to zero field values.}
	\label{fig: BackProjection}
\end{center}
\end{figure}

We presented in section \ref{sect:CamOscSpectra}, figure \ref{fig:FvsBCam}, evidence for an evolution of several of the dHvA frequencies near the metamagnetic transition, where in particular, the 1.8-1.6 and 4.2-4.0~kT frequencies change by 0.2~kT (shown again in figure \ref{fig: Freq2kTc-axis}, for the 1.8-1.6~kT frequency). These changes lead to anomalous interference patterns in the amplitude of the 1.8~kT frequency in the low field side of the transition, as we saw in section \ref{sect:CamEnvelopes}, figure \ref{fig: LF1p8kT-A}. Several physical phenomena may produce such changes in dHvA frequency, and we analyse here three of these. We conclude that, as was suggested by A. Tamai $et$ $al.$, charge transfers indeed occur in \TTS\ at the metamagnetic transition, probably due to a peak in the DOS, which means that the Fermi level\footnote{The Fermi level here means the energy at mid-point between the Fermi levels for the up and down spin species.} changes in that region of the phase diagram.

The first case that we study here is that of a field dependence of the $g$ factor, sketched figure \ref{fig: BackProjection} ($a$). In metallic systems, changes in the energy coupling between the paramagnetic moment of the electrons and the magnetic field may arise, which produces non-linear changes in the magnetisation, sketched in the top row of figure \ref{fig: BackProjection}. In such a situation, Zeeman splitting in a normal dispersion accelerates, producing non-linear changes in the size of FS pockets, and consequently in their associated real dHvA frequency. However, one does not measure the real size of the FS pockets with dHvA but rather their $back$-$projected$ frequency, as described in section \ref{sect:spinsplitting}. Equation \ref{eq:backprojection} means that one measures the real frequency minus its linear part, which corresponds to the phase of the oscillations,
\beq
F_{obs} = F - B{dF\over dB}.\nn
\eeq
One therefore observes peaks in the frequency, described in figure \ref{fig: SpinSplitting3}. In the case of a field dependent $g$ factor, these peaks appear with opposite sign for the spin species, or in other words, one measures both a sharp peak and a sharp dip. This is however not what was observed in the data, as only a positive increase in frequency was measured. 

The second case which leads to evolutions in the frequency is that where anomalies are present in the dispersion, depicted in figure \ref{fig: BackProjection} ($b$). In that situation, the anomalies do not likely appear symmetrically above and below the Fermi level, and it is probable that only one spin split pocket undergoes a non-linear change. In that case the dHvA peak becomes split and the measured frequency features a peak, positive or negative, for one of the spin species, and none for the other, which remains equal to the zero field frequency. Moreover, the frequency featuring a peak also possesses a field dependent quasiparticle mass, since the gradient of the dispersion changes with energy value, and the phenomenon of spin dependent mass is observed. In our data, no evidence was found for strong spin splitting with one of the peaks featuring a field dependent mass\footnote{No strictly horizontal lines of intensity are present in figure \ref{fig:FvsBCam} at 1.8 or 4.2~kT.}. Since the change in frequency between the low and high field side is of about 0.2~kT, much more than the peak width, such a phenomenon would have been detected if it had been present. However, if anomalies are placed symmetrically around $\epsilon_F$, one should observe a behaviour similar to that of the field dependent $g$ factor, a possibility we have already eliminated.

The third case, which we think is of relevance for \TTS, is where a change in Fermi level produces non-linearities in the frequencies, shown in figure \ref{fig: BackProjection} ($c$). According the ARPES data of A. Tamai $et$ $al.$ presented in the previous section, a vHS was found in the vicinity of the $\gamma_2$ pocket, which produces peak in the DOS below the Fermi level. If the $\gamma_2$ pocket suddenly increases its share of the DOS, the required charge will be taken from the whole system, resulting in, to first order, an increase of the Fermi level. When measuring the frequency of a band other than $\gamma_2$, one should observe that one of the spin split frequencies features a non-linear increase, while the other one exhibits a slowdown, even though the dispersion does not possess any sharp change in gradient. Consequently, both measured frequencies possess peaks, but not necessarily of the same magnitude, as this depends strongly on the band structure. For electron-like bands, these peaks will be negative (as in figure \ref{fig: BackProjection} ($c$)), while for hole-like bands, they will be positive, as in the data of figure \ref{fig: Freq2kTc-axis}.

We did not measure the field dependence of the $\gamma_2$ frequency, but we did for all the other bands. We believe that these possess normal dispersions since we measured that their quasiparticle mass does not change. For some of these, $\alpha_1$ and $\alpha_2$, non-linear changes in the real frequency arise for both spin species, of the same sign. This third model is consistent with our observations in figure \ref{fig:FvsBCam}, where only a positive peak in the frequency was measured for both the $\alpha_1$ and $\alpha_2$ frequencies, at 1.8-1.6 and 4.2-4.0~kT respectively. We conclude that electron transfers occur between these bands and $\gamma_2$. 

However, in this simple picture a change in Fermi level should affect all the bands, and in particular, electron bands should feature dHvA frequencies that increase across the metamagnetic transition. The data does not seem to suggest that, but instead the $\gamma_1$ and $\delta$ pockets vanish at $c$-axis\footnote{However, $\gamma_1$ and $\delta$ seem to appear again at higher angles, see figure \ref{fig: FFT10to18T_A} and the discussion in section \ref{sect:Fdisappear}.}. It is possible that we have misinterpreted high field side data, and that these bands simply possess very different frequencies there compared to the low field side. We argue that the metamagnetic transition might affect some bands more than others, such that $\alpha_1$, $\alpha_2$ and $\beta$ remain essentially intact while the other two evolve more dramatically.

\subsection{Interference patterns from non-linear magnetisation and spin splitting}

Anomalous beat patterns may occur in dHvA when frequencies are not constants of the magnetic field. In such a case, the interference occurs between spin split bands, and the resulting pattern is more complex than the usual ``spin zero'' structure (eq. \ref{eq:spinzero}, section \ref{sect:spinsplitting}). In our case of study, we believe that a change in Fermi level leads to non-linear changes in the $k$-space size of the spin split bands, and that this evolution produces the intricate interference that we observed in the 1.8-1.6~kT band (the $\alpha_1$ band) calculated in section \ref{sect:CamEnvelopes}, shown in figure \ref{fig: LF1p8kT-A}. We present in this section a simple model that reproduces some of the observed features, but that fails to reproduce some others. This indicates that the real physics is more complicated than that which is presented here. However, we believe that comparison of the model and the data, along with the analysis shown in the last section, provides partial evidence for theories of metamagnetism involving a peak in the DOS.

Interference in dHvA can be written in the following manner, using subscript symbols of plus and minus to represent spin up and spin down:
\bea
M(\theta,B) &\propto& \cos\bigg({F_0 + \Delta F_{+} \over B \cos(\theta)}\bigg) + \cos\bigg({F_0 + \Delta F_{-} \over B \cos\theta}\bigg)\nn\\
&=&
\cos\bigg({\Delta F_+ - \Delta F_- \over 2 B \cos\theta}\bigg)\cos\bigg({2 F_0 + \Delta F_+ + \Delta F_- \over 2 B \cos \theta}\bigg),
\label{eq:AnomalousInterf}
\eea
where $F_0$ represents the zero field frequency, $\Delta F_{\pm}$ the respective changes in frequency for the spin species, and $\theta$ is the angle between the field and the $c$-axis. The sum of the signal for both spins results in an interference factor which involves the difference in frequency changes $\Delta F_+ - \Delta F_-$ \footnote{Note that $\Delta F_+$ is negative.}. We are interested in modelling the field dependence of $\Delta F_{\pm}(B)$ which correspond to changes in electron density of the up and down spin sheets of specific FS pockets as the system approaches the metamagnetic transition. If these were simply linear functions of $B$ as it is the case with simple spin splitting, one would obtain ordinary spin zeros, but we know from the data that this is not the case.

In our model, the $\gamma_2$ band crosses a putative vHS at the metamagnetic field, producing a peak in the DOS, which affects all the other bands through a change in Fermi level in order to conserve the number of quasiparticles. Essentially, if the peak is located below the Fermi level (as proposed by A. Tamai $et$ $al.$), all bands are required near the metamagnetic field to donate a certain amount of holes, raising the Fermi level. Consequently, hole bands like the $\alpha_1$ and $\alpha_2$ undergo a reduction of size without changes in their quasiparticle masses, since their band structure does not feature sharp changes in their gradient. This produces the changes in frequency discussed in the last section, where both spin up and down electron densities for these bands are affected. 

\begin{figure}[t]
	\begin{minipage}[t]{7cm}
		\begin{center}
		\includegraphics[width=7cm]{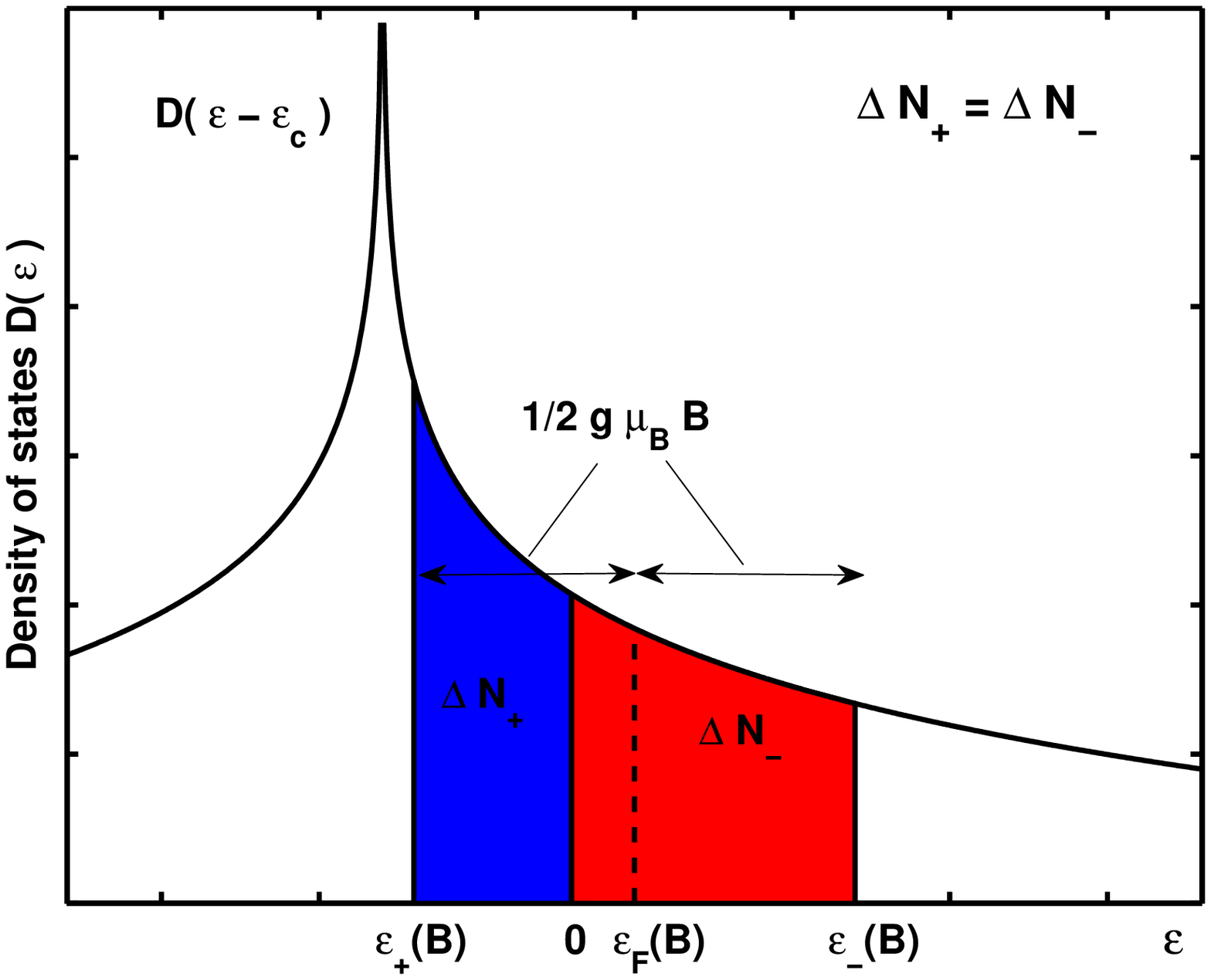}
		\end{center}
	\end{minipage}
	\hfill
	\begin{minipage}[t]{7cm}
		\begin{center}
		\includegraphics[width=7cm]{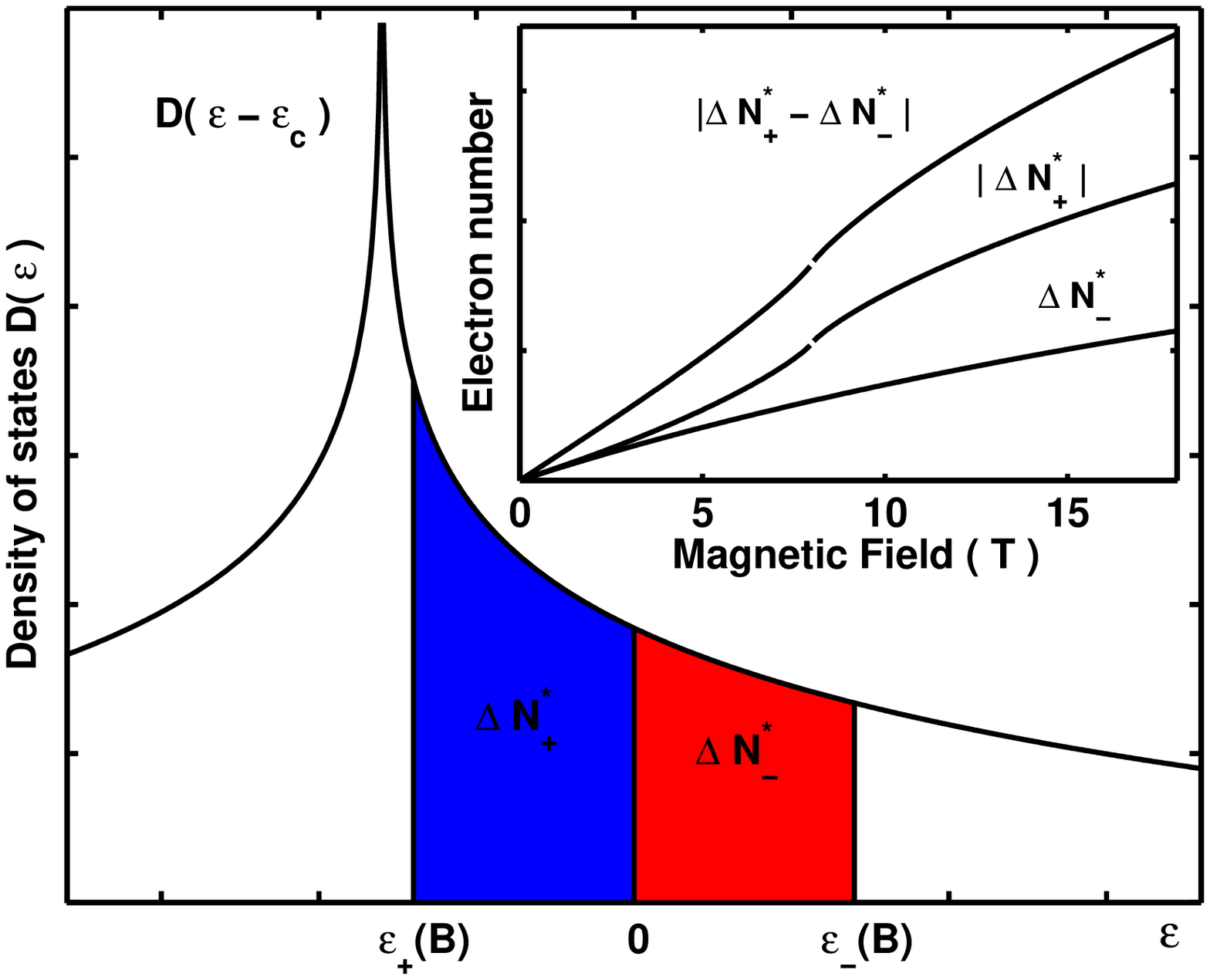}
		\end{center}
	\end{minipage}
	\caption[DOS near a vHS, with or without particle conservation]{$Left$ Illustration of the DOS near a van Hove singularity. Number differences for up and down spins ($\Delta N_+$ and $\Delta N_-$) calculated when conserving the total number of quasiparticles are obtained from integrals of the DOS, shown in blue and red coloured areas, and the number differences are equal, but the limits of integration are not easily found. The Fermi level is required by particle conservation to move upwards as the magnetic field increases, shown with a dashed line. $Right$ Similar illustration for the case where particle conservation is not required. In such a case, $\Delta N_+^* \neq \Delta N_-^*$, but the limits of integration are known, and $\Delta N_\pm^*$ are simple to calculate. $Inset$ Number differences calculated as functions of magnetic field from the integrals illustrated in the main graph. The total magnetisation is proportional to $\Delta N_+^*$-$\Delta N_-^*$.}
	\label{fig:vHsDensity2}
\end{figure}

Figure \ref{fig:vHsDensity2}, left panel, illustrates the situation presented here. In two-dimensional materials, a peak in the DOS due to a vHS follows a logarithmic form \cite{Binz}\footnote{The form of this function is not critical, but it must meet two criteria: it diverges at $\epsilon_c$ but its integral does not. A function of the form $A_1|\epsilon - \epsilon_c|^{-A_2}$, $0 < A_2 < 1$ could also be used, and produces similar results.},
\beq
D(\epsilon) = A_1 \log\bigg| {A_2 \over \epsilon - \epsilon_c}\bigg|.\nn
\eeq
Taking the zero at the zero field Fermi energy, in such a model, when one applies a magnetic field the change in the number of quasiparticles with up ($\Delta N_+$) and down spins ($\Delta N_-$), shown as blue and red areas in the left graph of figure \ref{fig:vHsDensity2}, must conserve the total number of particles, $\Delta N_- = -\Delta N_+$. The Fermi level, situated midway between $\epsilon_+$ and $\epsilon_-$, cannot remain at its zero field value but must move upwards as the magnetic field increases, in a way which is complex to calculate. The number differences are therefore difficult to evaluate since the limits of integration are not easily found. 

However, if one does not conserve the total number of particles, one has number differences, denoted with a star, which are not equal in absolute value, $\Delta N_-^* \neq -\Delta N_+^*$, but the integrals for the number differences are known, a situation shown in the right hand side panel of figure \ref{fig:vHsDensity2}. One finds $\Delta N_\pm^*$ by evaluating the following integrals of the DOS,
\bea
&&\Delta N_{\pm}^* = \int^{\pm {1\over2}\mu_B g(\theta) B}_{0} D(\epsilon) d\epsilon \nn\\
&&= \pm A_1\bigg[{1\over2}\mu_B g(\theta)\log A_2 - \bigg({1\over2}\mu_B g(\theta) B \pm \epsilon_c\bigg)\bigg(\log|{1\over2}\mu_B g(\theta) B \pm \epsilon_c| - 1\bigg) \nn\\
&&\quad\quad \mp (\epsilon_c \log \epsilon_c - \epsilon_c)\bigg].
\label{eq:DeltaNstar}
\eea
$\Delta N_+^*$ features a superlinear rise near $.5\mu_B g(\theta) B = \epsilon_c$, where $g(\theta)$ is the orientation dependent $g$ factor discussed earlier, while $\Delta N_-^*$ does not. The magnetisation corresponds to the difference between the two, $M \propto \Delta N_+^* - \Delta N_-^*$,
\beq
M(B) = A_1 \sum_\pm \bigg[{1\over2}\mu_B g(\theta)\log A_2 - \bigg({1\over2}\mu_B g(\theta) B \pm \epsilon_c\bigg)\bigg(\log|{1\over2}\mu_B g(\theta) B \pm \epsilon_c| - 1\bigg)\bigg],\nn
\eeq
and also features a superlinear rise near $\epsilon_c$, since only $\Delta N_+^*$ undergoes a sharp increase (see the inset of figure \ref{fig:vHsDensity2}). 

In order to construct a simple model, we have assumed that the $\gamma_2$ pocket produces the sharp increase in the DOS below the Fermi level, and when this is reached by magnetic field, the missing number of electrons are provided by all the other bands through a change in Fermi level. However, we have kept the total number differences of eq. \ref{eq:DeltaNstar}. For these other bands (for instance, $\alpha_1$ and $\alpha_2$), their frequency differences $\Delta F_+ - \Delta F_-$ are related to unknown fractions of the total number differences, $\Delta N_{\alpha \pm}$. We supposed that the change in Fermi level produces evolutions of these numbers such that 
\beq
\Delta N_{\alpha +} - \Delta N_{\alpha -} \propto \Delta N_+^* - \Delta N_-^*.\nn
\eeq
Such a statement means that the argument of the beat pattern factor is simply
\beq
\Delta F_+ - \Delta F_- \propto M(B).\nn
\eeq
It does not mean, however, that the share of the DOS produced by these bands ($\alpha_1$,$\alpha_2$)  possesses a peak, as we know from field resolved mass studies that this is not the case. It is through the change in Fermi level and a share of the DOS that is not constant, but which evolves slowly, that these fractional number differences $\Delta N_{\alpha \pm}$ undergo non-linear changes \footnote{If the structure of these bands was perfectly parabolic, and their share of the DOS constant, there would be no difference between the up and down spin dHvA frequencies, and therefore no beat patterns, even though both would feature a non-linear change near 8 T.}.

These assumptions are reasonable in regard to the data, which seems to indicate that the frequencies evolve in a way which is roughly consistent with this model (see figures \ref{fig: Freq2kTc-axis} and \ref{fig: BackProjection} ($c$) of the last section, compared to the curves in the inset of the left panel of figure \ref{fig:vHsDensity2}). However, this calculation does not take into account properly particle number conservation, and is not perfectly adequate. It also neglects particle-particle interactions, which could lead to a discontinuous jump across the peak in the DOS, corresponding to a first order jump in the magnetisation. An appropriate calculation is outside the scope of this thesis, and the reader is referred to the work of Binz and Sigrist for a model that respects particle conservation \cite{Binz} .

 \begin{figure}[p]
 	\begin{center}
 		\includegraphics[width=7cm]{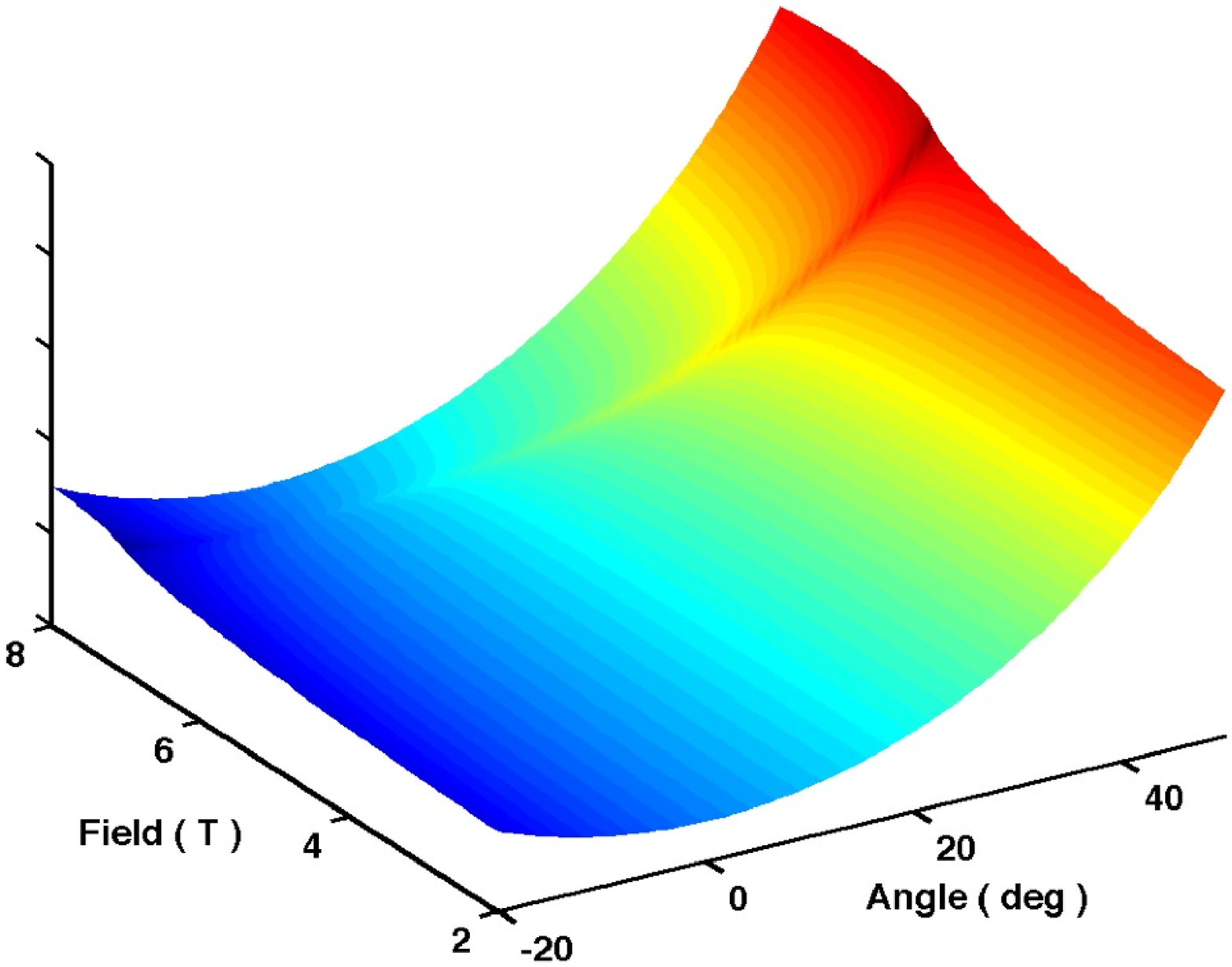}
	\end{center}
	\begin{minipage}[t]{.5\columnwidth}
		\begin{center}
		\includegraphics[width=1\columnwidth]{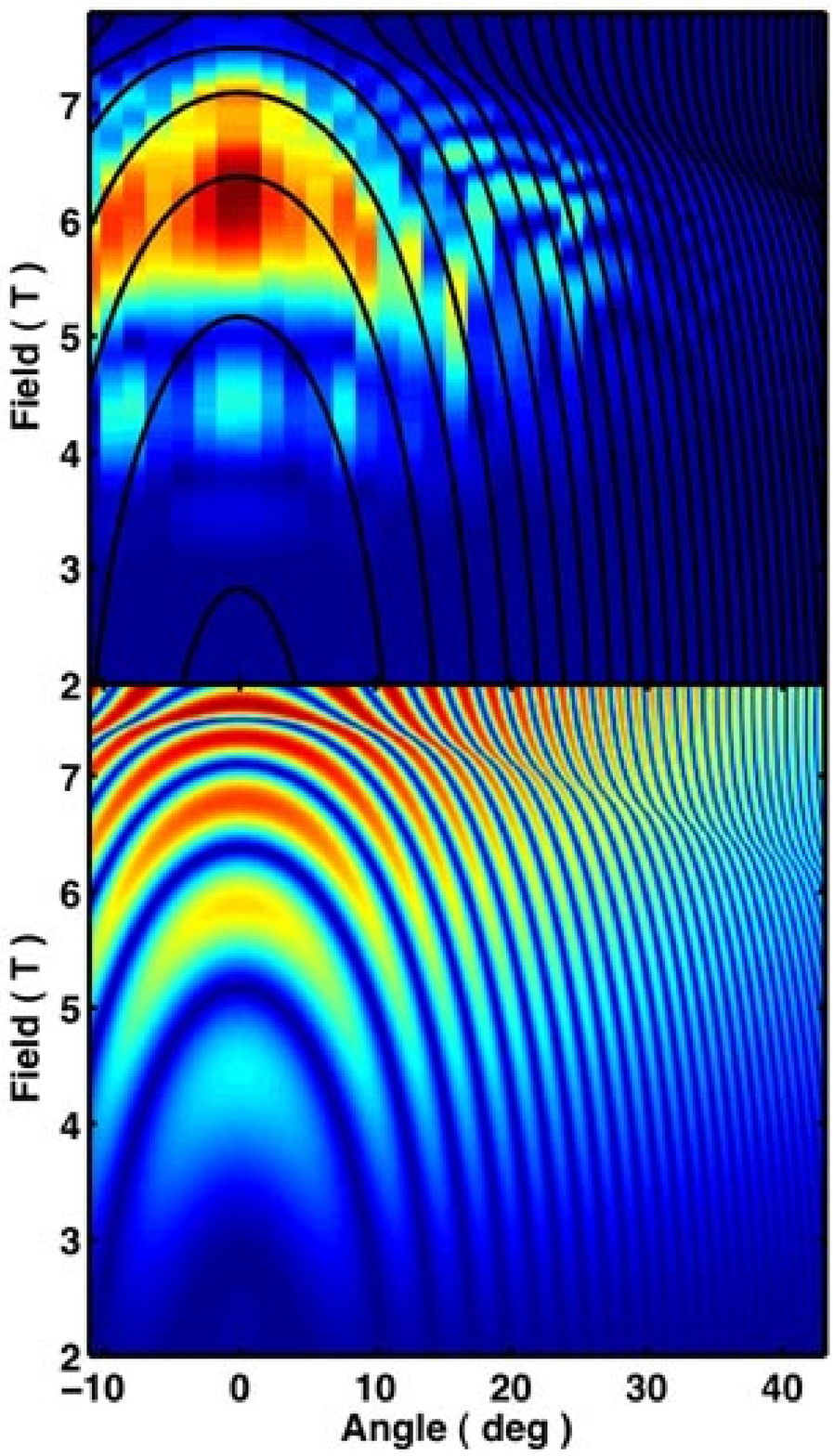}
		\end{center}
	\end{minipage}
	\hfill
	\begin{minipage}[t]{.5\columnwidth}
		\begin{center}
		\includegraphics[width=1\columnwidth]{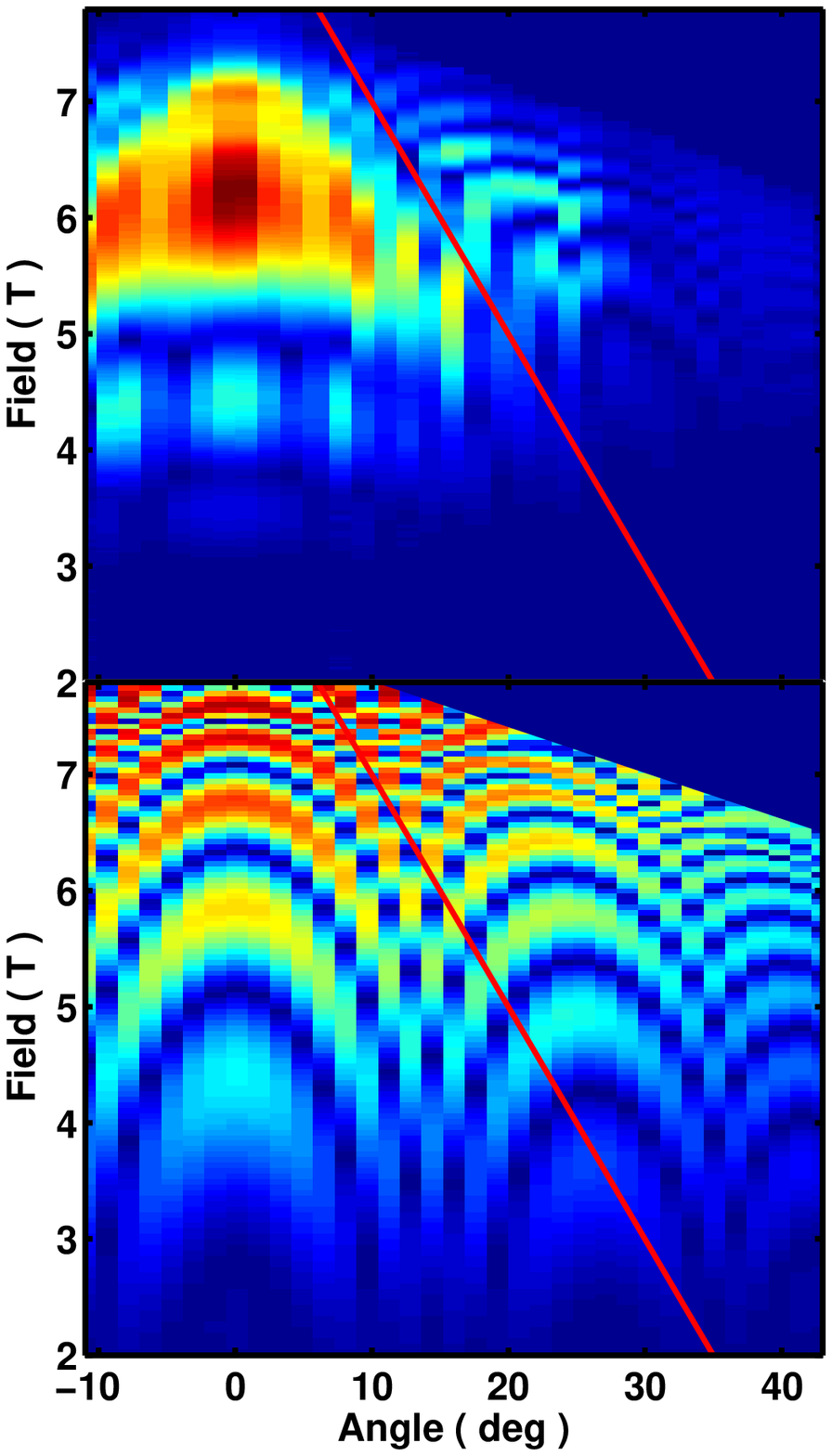}
		\end{center}
	\end{minipage}
	\caption[Simulations of anomalous beat patterns]{$Top$ Surface plot of $(\Delta F_+ - \Delta F_-)/B\cos \theta$ as a function of field $B$ and angle $\theta$. $Middle$ $left$ Amplitude of the 1.8~kT frequency for the rotation towards [110], in colour, overlaid with beat pattern nodes, solid black lines. $Bottom$ $left$ Simulation of the beat pattern produced by the model of this section. $Middle$ $right$ Amplitude of the 1.8~kT frequency for the rotation towards [110] for comparison purposes. $Bottom$ $right$ Same simulation of the beat pattern, deliberately under-sampled with similar angle steps as the data.}
	\label{fig: LF1p8_b_AnisG2}
\end{figure}

Figure \ref{fig: LF1p8_b_AnisG2}, top graph, shows a surface of the argument
\beq
(\Delta F_+ - \Delta F_-)/B \cos\theta,\nn
\eeq
calculated with this model. The nodes (or zeros) of the interference can be calculated by drawing a contour plot of this expression at values of $\pi/2 + n \pi$, $n$ an integer. As level curves do not cross but close onto themselves, the same properties are associated with these beat pattern nodes. Moreover, these oscillations undergo a drastic change at ${1\over 2} g(\theta) \mu_B B = \epsilon_c$, but also, since $F(B)$ is highly non-linear, we expect that the beat patterns accelerate as the system comes near the metamagnetic transition on the low field side, as was observed in the data. Note that the modulation produced by the corrugation of the Fermi cylinder is independent of that resulting from non-linear magnetisation since, as one can see from eq. \ref{eq:simplewarping}, it can be expressed as a separate factor to the oscillatory term.

Figure \ref{fig: LF1p8_b_AnisG2}, bottom left graph, shows the result of the calculation of the beat pattern due to non-linear susceptibility combined with the anisotropy of $g$, using values for $A_1$ and $A_2$ of 0.4~kT and 80 T, and the DOS in the form $A_1 \log| A_2/(.5g(\theta) B-B_c)|$. It produces a pattern where the original spin zeros curve towards the horizontal at high fields, and come closer together at high angles. The region at high fields where the zeros change curvature corresponds to the metamagnetic transition, along the line $.5g\mu_B B = \epsilon_c$. The middle left graph shows the zeros of the calculation plotted on top of the data for the rotation towards [110], at 1.8kT, One can see that this model produces the appropriate behaviour at high angles\footnote{Unfortunately, the data is highly under-sampled. One would need measurements performed with finer angular steps, which was clearly out of reach with the time that was allowed for this experiment.}. However, near $c$-axis, the agreement not good, and on axis, there is no resemblance at all between the simulation and the data. In order to convince the reader of the qualitative agreement of the model and the data at high angles, the bottom right graph presents the same calculation as in the bottom left graph, using a sampling identical to that of the data, shown middle right graph. The data and simulation on the right of the red lines agree qualitatively, but do not agree on the left. The effect of sampling is to repeat the modulation periodically with angle, as in a moiré pattern, which is what appears in the data\footnote{Note that the damping of the intensity in the data as a function of angle is not correctly reproduced in the simulations, but is related to the experimental system and is not the focus here.}.

We know from figure \ref{fig: LF1p8kT_Sim2_A2}, section \ref{sect:CamEnvelopes}, that the low angle part of the data (left of the red lines in figure \ref{fig: LF1p8_b_AnisG2}, top right) is well reproduced by a simple warping parameter $k_{01}$ of 4.4$\times 10^6$m$^{-1}$, except very near the metamagnetic transition, but not at high angles, where the Yamaji angle region of high intensity is not observed. Combining the two contributions to the beat patterns, warping and spin splitting, does not improve the agreement, since in that case both beat patterns are superimposed over the whole field and angle range. The modulation produced by simple warping, given by eq. \ref{eq:simplewarping} (p. \pageref{eq:simplewarping}), originates from a factor independent of Zeeman splitting, and does not influence the position of the zeros produced by the effect discussed here\footnote{Even when adding Zeeman splitting into the value for $k_{00}$ used in the calculation of the simple warping, an unrealistically high amplitude for the splitting has to be used in order to significantly change the beat pattern.}. Consequently, this simple model does not fully explain the anomalous beat pattern observed in the data.

The angular dependence of the $g$ factor in this model produces a simple displacement of the peak in the DOS as a function of field. Since the anomalous modulation is of less intensity near $c$-axis, we propose that in this scenario, the peak in the DOS changes its width with angle, such that the magnetisation begins its superlinear rise at a higher field interval from the critical field at high angles than at $c$-axis, where it becomes very small\footnote{Less than .5~T at $c$-axis compared to 3~T at 45$^{\circ}$, see figure \ref{fig: LF1p8kT-A}, p. \pageref{fig: LF1p8kT-A}.}. Such a model is difficult to construct, and is beyond the scope of this thesis. We nevertheless conclude that the combination of an anisotropic $g$ factor, a peak in the DOS and non-linear spin splitting is the most probable origin of the anomalous beat pattern of the 1.8~kT peak in field and angle, and that the same can probably be said of the other frequencies in the low field side.

 \section{High field Fermi surface}
 
The dHvA effect at fields higher than about that of the metamagnetic transition possesses an enhanced complexity compared to low field regions. Two phenomena take place: all the dHvA peaks split into a large number of frequencies closely spaced, but frequencies also appear and disappear as a function of field angle. We propose in this section a model for the explanation of the first of these two phenomena. The field dependence of the dHvA frequencies calculated in section \ref{sect:CamSpectra}, figure \ref{fig:FvsBCam}, suggests that the splitting of the peaks increase with magnetic field. This type of behaviour is generally indicative of the occurrence of magnetic breakdown, where electrons tunnel between FS sheets. However, we have not found any reasonable explanation for the anomalous angular dependence of the high field side frequencies, but we argue that it is magnetically driven structural evolutions that lie at the origin of the phenomenon.
 
\subsection{Magnetic breakdown}

Magnetic breakdown occurs in a material when the the energy spacing between Landau levels, $\Delta \epsilon_L$, becomes comparable to the gap energy between two bands in a certain region of $k$ space, $\epsilon_g$ \cite{shoenberg}:
\beq
\Delta \epsilon_L \simeq {\epsilon_g \over \epsilon_F^2},\nn
\eeq
where $\epsilon_F$ is the Fermi energy. In such situation, electrons may jump between FS sheets, and the hopping probability $P_H$ is exponential,
\beq
P_H \propto e^{-H_B / H},\nn
\eeq
where $H_B$ is the magnetic breakdown field, given by
\beq
H_B = {m^* \over e \hbar } {\epsilon_g^2 \over \epsilon_F}.
\label{eq:breakdownfield}
\eeq
Due to simple conservation of the number of electrons, the intensity of the signal from the original band decreases as magnetic breakdown increases, and the intensity from each orbit changes with the value of the magnetic field. As was derived initially by Pippard \cite{pippard1, pippard2}, if one takes $p$ as the quantum mechanical wave function amplitude for an electron that hops between bands, where
\beq
P_H = p^2 \propto e^{-H_B \over H},\nn
\eeq
and $q$, the amplitude of an electron that does not hop, such that $q^2 = 1 - p^2$, one can calculate the total probability of specific paths through $k$ space by counting the number of inter-band hops $m$ and non-hops $n$, and the amplitude reduction factor of dHvA amplitude for this orbit is 
\beq
R_M \propto d(ip)^mq^n,\nn
\eeq
where $d$ accounts for the degeneracy of the orbit, and $m$ and $n$ are even numbers. Since different paths possess different probabilities, one expects complexity to arise with magnetic field, until it is high enough that only the breakdown orbits are populated.

\begin{figure}[t]
	\begin{minipage}[hbt]{0.5\columnwidth}
		\begin{flushright}
		\includegraphics[width=1\columnwidth]{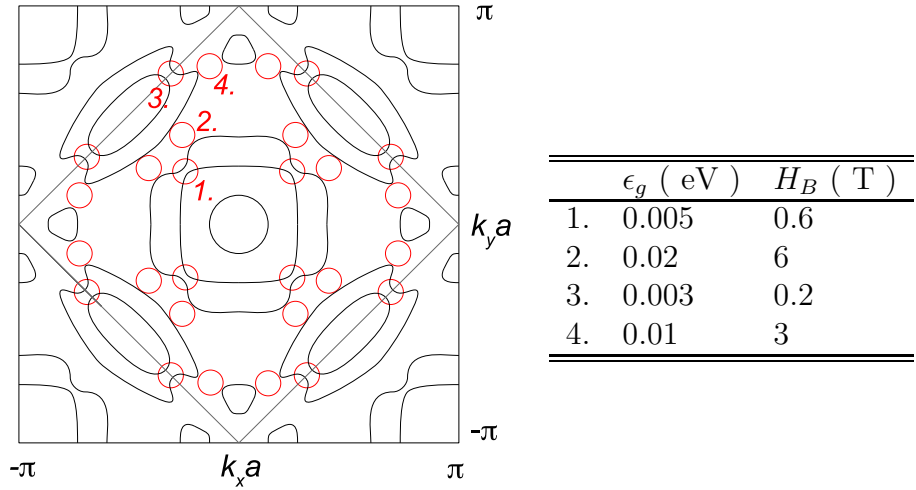}
		\end{flushright}
	\end{minipage}
	\begin{minipage}[hbt]{0.3\columnwidth}
		\begin{flushleft}
		\begin{tabular}{r l l}
			\hline
			\hline
			&$\epsilon_g$ ( eV )& $H_B$ ( T )\\
			\hline
			1.& 0.005 & 0.6\\
			2.& 0.02 & 6\\
			3.& 0.003 & 0.2\\
			4.& 0.01 & 3\\
			\hline
			\hline
		\end{tabular}
		\end{flushleft}
	\end{minipage}
	\caption[Regions of possible magnetic breakdown and breakdown field values]{$Left$ Regions of the BZ where magnetic breakdown may be possible. Four different points, numbered 1-4 in the plot, are repeated by symmetry all over the BZ. $Right$ Table of calculated gap energies between bands for the four magnetic breakdown points, and corresponding magnetic breakdown fields. The error on these numbers is of about $\pm$20\%.}
	\label{fig:MagBreakdown2-2}
\end{figure}

\begin{figure}[p]
	\begin{minipage}[hbt]{0.55\linewidth}
		\begin{flushright}
		\begin{tabular}{r l}
			\hline
			\hline
			&Frequency ( kT )\\
			\hline
			a.&4.2, 4.26, 4.32, 4.38, 4.44, ..., 0.064\\
			b.&12.1, 11.65, 11.43\\
			c.&1.80, 2.40, 3.00, 3.60, ..., 0.60\\
			d.&4.2, 5.1, 6.0, 6.9, 7.8, ..., 0.9\\
			e.& 2.5\\
			f. & 0.43, .48, .9, 1.32\\
			\hline
			\hline
		\end{tabular}
		\end{flushright}
	\vspace{.001cm}
	\end{minipage}
	\begin{minipage}[hbt]{0.45\linewidth}
		\begin{flushleft}
		\includegraphics[width=4.5cm]{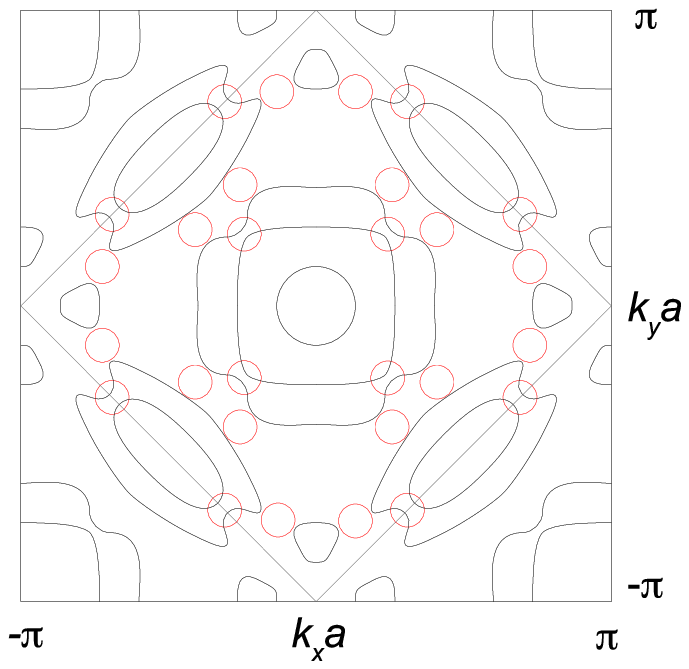}
		\end{flushleft}
	\end{minipage}

	\begin{center}
	\includegraphics[width=0.8\columnwidth]{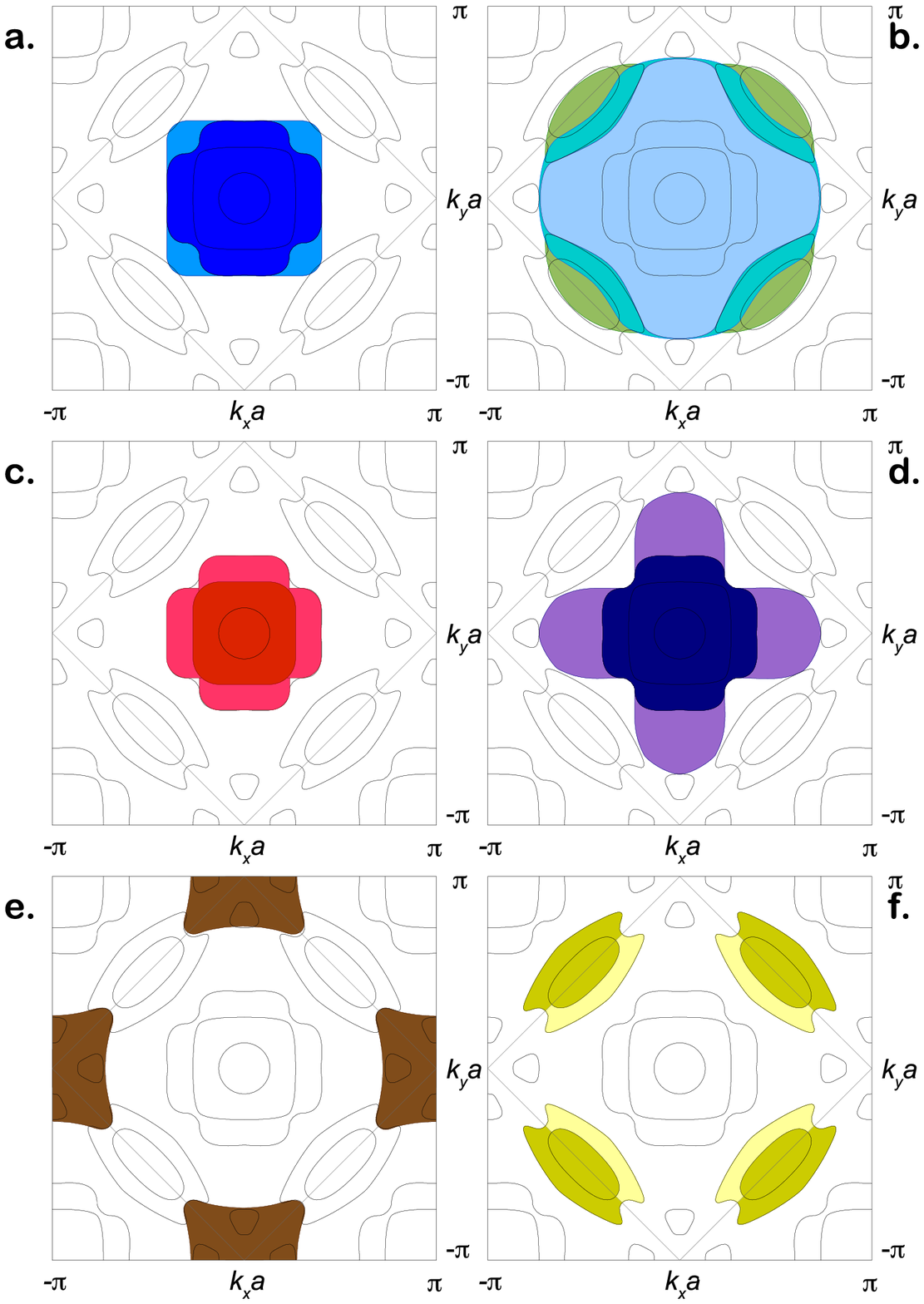}
	
	\caption[Possible new orbits due to magnetic breakdown]{Schematic possibilities of magnetic breakdown in \TTS. $Top$ $right$ Most probable magnetic breakdown points, where quasiparticle could tunnel in between FS sheets. $Top$ $left$ Table of roughly calculated additional frequencies starting from low field side orbits (ignoring electron transfers). $a.$ $to$ $f.$ Sketches of possible new quasiparticle orbits.}
	\label{fig: MagBreakdown}
	\end{center}
\end{figure}

In \TTS, we effectively have a situation where the complexity of the dHvA spectrum increases with magnetic field (see the field dependence of the spectrum, figure \ref{fig:FvsBCam}, p. \pageref{fig:FvsBCam}). Figure \ref{fig:MagBreakdown2-2}, left panel, presents four regions in the low field side BZ of \TTS\ where magnetic breakdown may potentially occur, which repeat all over the BZ due to symmetry. Using average Fermi velocities, calculated using average values of $k_F$ and $m^*$, we evaluated the energy gaps at these points by taking approximate $k$-space distances ($\Delta k$) and using the equation
\beq
\epsilon_g \simeq {\Delta k \over 2 \hbar} \big(\nabla \epsilon_1(k) + \nabla \epsilon_2(k)\big) = {\hbar^2 \over 2} \Delta k \bigg({k_1\over m_1} + {k_2\over m_2}\bigg).
\label{eq:breakdowngaps}
\eeq
We converted these results into magnetic breakdown fields using eq. \ref{eq:breakdownfield}, and tabulated these along with the gap values in the table of figure \ref{fig:MagBreakdown2-2}. All the breakdown fields were found to be within experimental range \footnote{The absolute values should not be considered accurate, as they were produced using average values for $k$ and $m^*$, while the absolute value of these parameters vary in $k$-space. More accurate values could be obtained using ARPES Fermi velocities, which was not possible during this project.}. The error value of 20\% was obtained from the maximum and minimum possible values the Fermi velocities used in eq. \ref{eq:breakdowngaps} are expected to take.

From the four breakdown regions of figure \ref{fig:MagBreakdown2-2}, a number of breakdown orbits were considered, shown in the main panel of figure \ref{fig: MagBreakdown}, using the low field side FS. Taking these into account, we were able to calculate approximate frequencies for the resulting peak splitting for various FS sheets, using geometrical considerations. Unfortunately, these do not take into account the electron transfers occurring at the metamagnetic transition, since we do not know with certainty the details of the high field side FS. The table in figure \ref{fig: MagBreakdown} shows the results.

The first situation we considered is that of panel $a.$, figure \ref{fig: MagBreakdown}, where electrons hop from bands $\alpha_2$ to $\gamma_1$. It gives rise to eight possible hopping points, and consequently four regions where the orbit can be expanded. If $a$ is the original $k$ space area and $b$ the additional area due to magnetic breakdown, we expect to find, to first order, that the $\alpha_2$ pocket gives rise to five orbits at $a$, $a+b$, $a+2b$, $a+3b$ and $a+4b$, with amplitudes corresponding to $q^8$, $p^2q^6$, $p^4q^4$, $p^6q^2$ and $p^8$. However, it is possible that the small orbit of area $b$ is observed by itself, or that it is added in degenerate fashion, such that areas of $a+5b$, for instance, are possible. The number of ways equivalent areas are generated can be high in some cases, for example with $a+2b$, where it can appear at probabilities of $p^4q^4$, $p^4q^6$ and $p^4q^8$, with degeneracy of 6, 12 and 10 respectively. This degeneracy may enhance the amplitude of some of the possibilities. 

The complete list of possibilities is of course infinite, but we were able to identify all orbits of area up to $a+8b$, and 278 different paths were counted, in 46 different values of probability, for 9 possible areas, excluding the orbit of area $b$ and its harmonics. In this specific case, $b$ is of size of about 0.06~kT, and frequencies between 4.2 and 4.7~kT are expected. If we assume that the electron transfer of the metamagnetic transition reduced the size of $\alpha_2$ to around 3.9~kT, then we expect frequencies between 3.9 and 4.4~kT. In dHvA experiments, frequencies between 3.9 and 4.7 were observed, which gives credibility to this model (see figure \ref{fig:FvsBCam}, p. \pageref{fig:FvsBCam}).

The second situation we analysed is that depicted in panel $b.$ of figure \ref{fig: MagBreakdown}, where an orbit similar to the $\beta$ pocket of \TOF\ is recovered, which lies around 12~kT, along with small variants. These were not observed. Next, in $c.$, we have a situation similar to that in $a.$, which gives rise to multiple frequencies between 1.6 and 4.0~kT spaced by around 0.6~kT, which could explain the peak seen near 3.2~kT, which in combination with the situation in $a.$, could furthermore explain why this peak near 3.2~kT is multiple. We do not see, however, other peaks between 4.0 and 1.6~kT, except at higher angles (see figure \ref{fig: FFT5to6p5T_A}). Another similar situation is that in $d.$, where frequencies between 4.2 and 8.9~kT were predicted, spaced by around 1~kT. Combined with the possibilities in $a.$, could potentially explain the group of peaks near 8~kT. The situation in $e.$ gives rise to one orbit of around 2.5~kT, which was not clearly observed, although it cannot be ruled out. And finally, hopping between $\gamma_1$ and $\beta$ gives rise to frequencies between 0.43 and 1.32~kT, which are clearly observed.

We conclude this section by stating that the situations depicted in $a.$, $c.$, $d.$ and $f.$ and combinations between these could potentially explain the full complexity of the dHvA spectrum at $c$-axis. In particular, the situation in $a.$ is a very likely explanation to the complex splitting of the 4.0~kT pocket. Moreover, it seems quite accidental that magnetic breakdown appears to be only observed above the metamagnetic transition. We argue that it is actually not the case, and that it starts to occur in the low field side of the transition. One can observe in figure \ref{fig:FvsBCam} that the 4.2~kT peak splits into several above 5 T, a phenomenon that is also visible in the spectra taken from the data interval between 5 to 6.5 T, shown in figure \ref{fig: FFT_C698I}, top graph.

 \subsection{Strongly angle-dependent amplitudes \label{sect:Fdisappear}}
 

The amplitude of the dHvA signal in field and angle in the high field side possesses an intriguing feature that is difficult to understand, that some frequencies disappear in specific angular spans. In particular, we have that the 1.8~kT peak vanishes completely above an angle between 5 and 10$^{\circ}$, and the 0.9~kT peak vanishes below a similar angle. Moreover, the 0.43~kT peak appears at an angle of around 15-20$^{\circ}$, and other peaks appear briefly in the region between 2.5 and 4.0~kT at low angles. We also have that those for which we can observe the angular dependence do not follow the $F_0/\cos \theta$ form. These types of behaviour are not related to anything in the usual dHvA theory for 2D systems, and we do not present in this thesis any model to explain it. It remains to this point a rather mysterious phenomenon, but we suggest here a few ideas about its possible origin.

The simplest explanation appears to be that angle changes the condition for magnetic breakdown. However, its origin does not lie with a combination of Zeeman spin splitting and magnetic breakdown. The reason is that the effect of magnetic breakdown is not directly related with Zeeman splitting\footnote{Many thanks to C. A. Hooley, who pointed that out.}. Magnetic breakdown stems from Landau diamagnetism, or the orbital motion of electrons in a magnetic field, without reference to their spin. In an imaginary world where charged fermions without spin existed, the dHvA effect could still be observed, whereas in a spin split band structure, there is no process by which the two spin species can interact  except through spin orbit coupling. Consequently, the effect of the anisotropy of $g$ discussed in section \ref{sect:LowFieldSideModel} does not change with angle the condition for magnetic breakdown. 

Thus there must be another process which changes the magnetic breakdown field values, and we suggest two here. The first and simplest is that the gap between bands may possess a $k_z$ dependence, which can be smaller in certain regions than others, which at certain angles, would become by coincidence extremely broken down extremal orbits. In such a situation, one may find that new orbits appear briefly as a function of angle, and that original ones disappear as well, both of which have been observed. The second process we suggest, which appears the most probable, is that the strong magneto-structural coupling in \TTS\ produces small changes in the crystalline structure with field angle, which distorts the FS itself \cite{shaked, friedt}. Consequently, the gaps between bands may change as the FS evolves and thus change the condition for magnetic breakdown. 

Finally, another process has been suggested by S. R. Julian\footnote{Private communication.}, where stress induced changes in samples occur due to the experimental mounting procedure, an effect which we have not been able to rule out due to time constraints. In this scenario, the layer of silver epoxy used for the heat sinking contacts could induce, with magnetostriction effects at the metamagnetic field, inhomogeneous stress across the samples. Due to the strong magnetostriction properties of \TTS, these could deform the FS inhomogeneously and produce various dHvA frequencies from different parts of the sample. We do not believe that this is possible, due to the very thin nature of the epoxy layers. Moreover, as we mentioned earlier, magnetic breakdown occurs not only above but also below the metamagnetic transition, indicating that it is probably not related with the metamagnetic transition and magnetostriction occurring there. We could not however rule out this possibility based on the current data.

\section{Absence of quasiparticle mass \newline enhancements as seen from dHvA}

The absence of an enhancement of the quasiparticle masses in \TTS\ is an extremely unexpected phenomenon. Not only since it was reported otherwise in the past \cite{borzi}, but also through the fact that other metamagnets, CeRu$_2$Si$_2$, CePd$_2$Si$_2$, CeRhIn$_5$ \cite{Aoki,Sheikin,settai} do feature mass enhancements and \TTS\ now figures as an exception. Secondly, as we saw in section \ref{sect:LK}, in 2D systems one can calculate the electronic specific heat coefficient by summing the quasiparticle masses, and this should be true at all fields. We described in section \ref{sect:QC} that the electronic specific heat of \TTS\ is greatly enhanced near the QCEP, and so should the sum of the quasiparticle masses be, a fact that was confirmed again recently down to 100~mK by A. Rost \cite{rost}. Thirdly, in a Fermi liquid the $A$ coefficient to the quadratic temperature dependence of the resistivity is, as we saw in section \ref{sect:FermiLiquid}, eq. \ref{eq:resistivity}, proportional to the square of the quasiparticle mass. This parameter was also reported strongly enhanced (see section \ref{sect:QC}).

We are facing an important discrepancy between measurements if we assume that we have determined a complete set of FS sheets and know all of the FS quasiparticle masses. However, we cannot make such a statement with certainty, but can only instead present very stringent constraints to the origin of the ``missing" mass. Moreover, in order to strengthen our position, we are required to investigate how it came to be that the opposite conclusion to ours was inferred in the past. We thus present here first an explanation of how a divergence of the quasiparticle mass may be erroneously deduced from field dependent low signal to noise ratio and the use of non-linear LK fits with three parameters. We discuss here the process by which it can happen, and a give a detailed numerical analysis in appendix~\ref{App:F}. We moreover show there that our quasiparticle mass extraction method is able to well detect an enhancement of the mass as long as the signal to noise ratio is larger that 0.5, a condition which is fulfilled for all the bands that we analysed in section \ref{sect:MassNonDiv}. Finally, we discuss the implication of our result and argue that the $\gamma_2$ pocket may possess the missing mass.

\subsection{Source of systematic errors in previous work \label{sect:MassSystematic}}

\begin{figure}[p]
         \begin{center}
	\includegraphics[width=7cm]{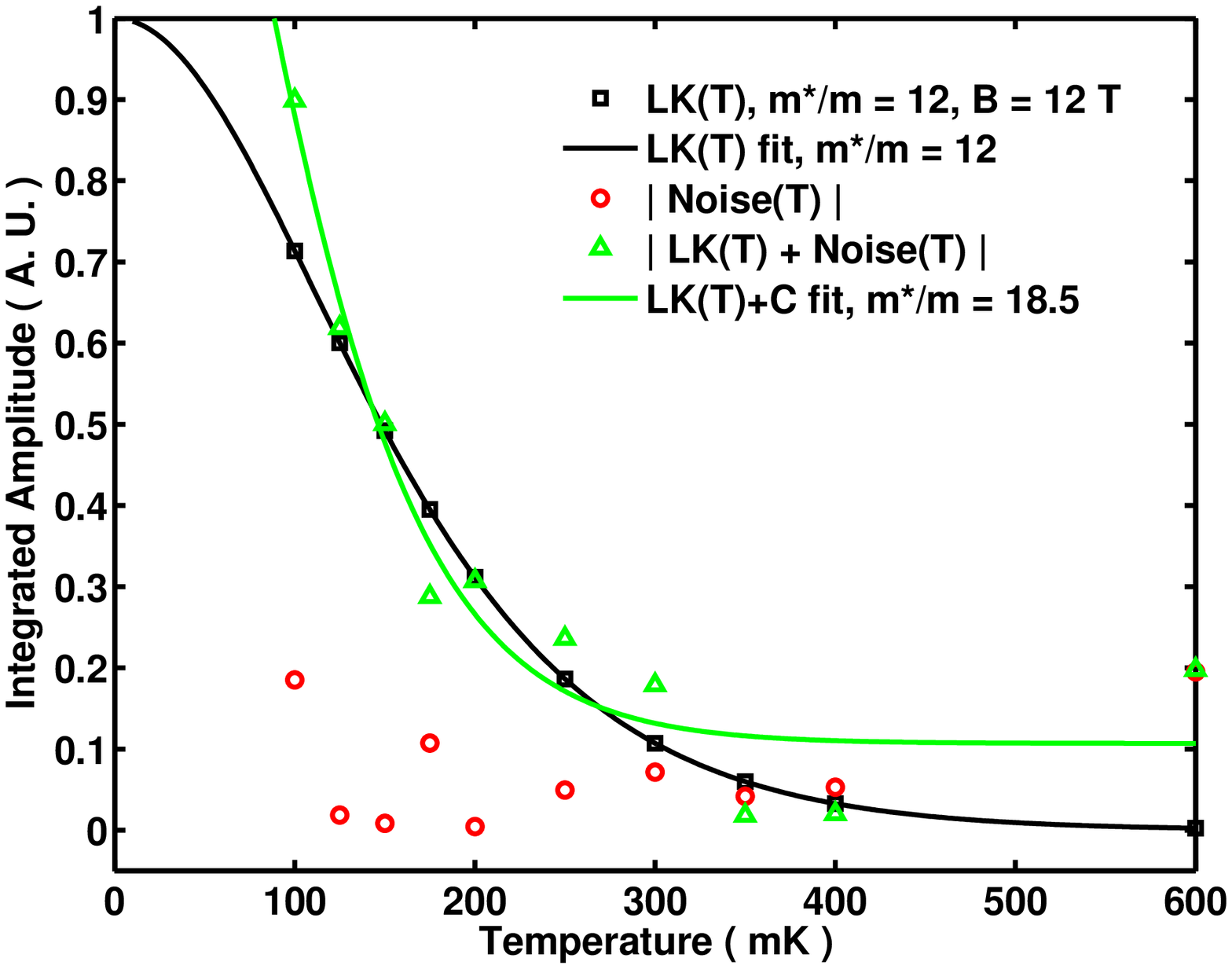}
	\end{center}
	\begin{minipage}[t]{7cm}
		\begin{center}
		\includegraphics[width=1\columnwidth]{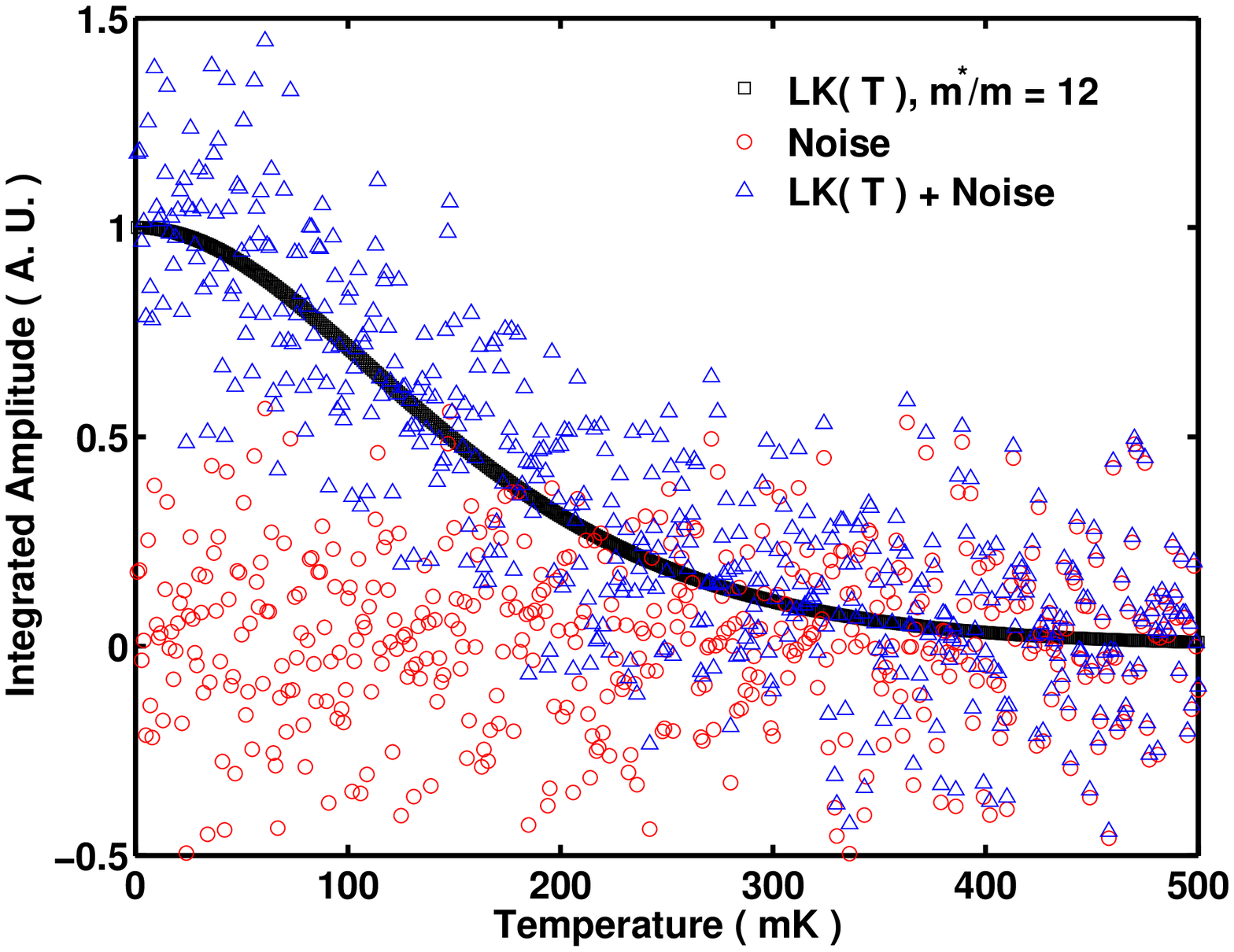}
		\end{center}
	\end{minipage}
	\hfill
	\begin{minipage}[t]{7cm}
		\begin{center}
		\includegraphics[width=1\columnwidth]{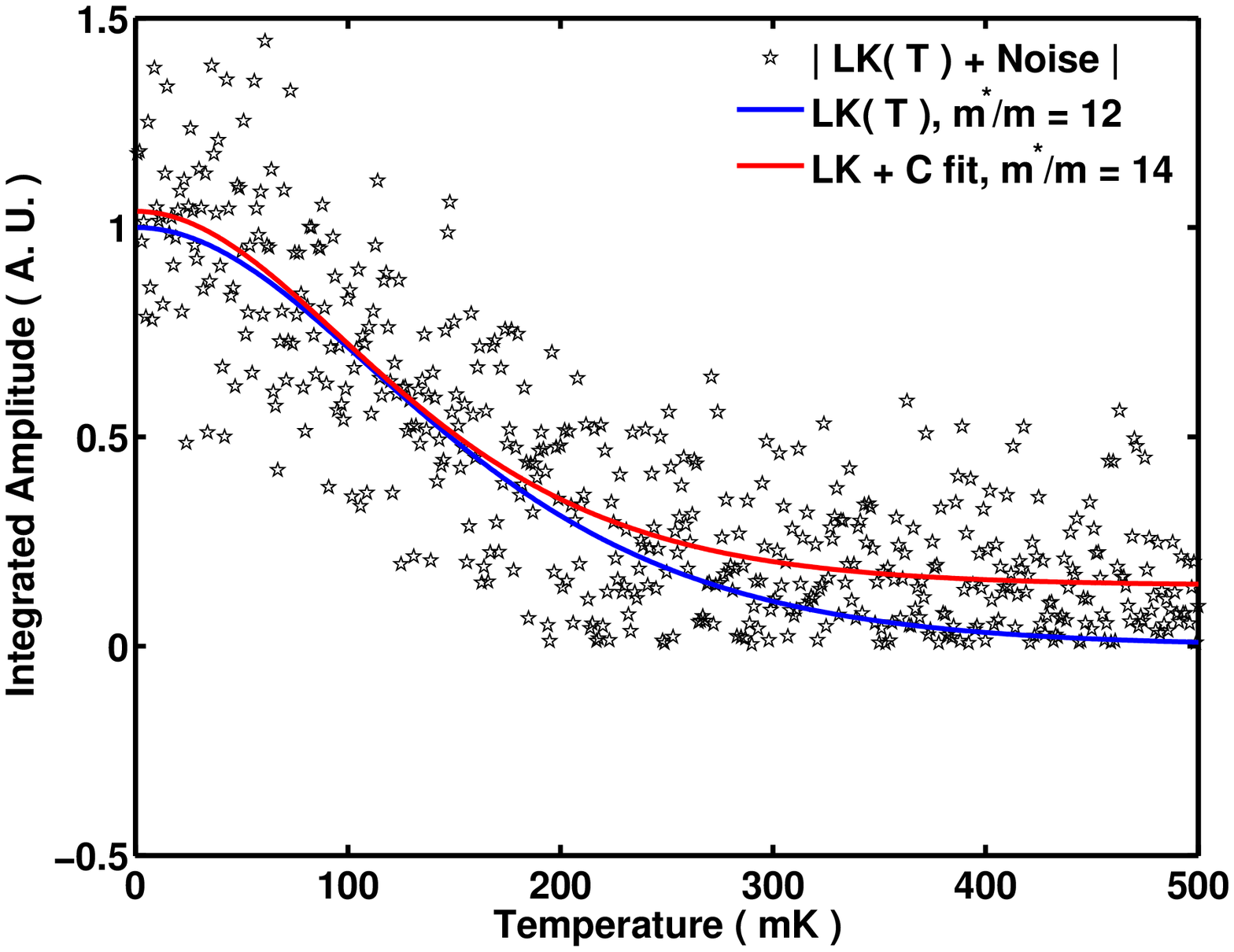}
		\end{center}
	\end{minipage}
	\caption[Systematic errors due to three parameter LK fits]{Two illustrations of a situation where a three parameter fit is used with noisy data and results in a systematic error for the mass. $Top$ Realistic situation with only a few data points. The data without noise is shown in black squares, with associated LK function of mass $m^*=12 m_e$ at a field of 12 T, shown with a solid line. In red circles appears the absolute value of noise created with a random number generator of gaussian distribution, and in green triangles, the absolute value of the sum of the LK function and the noise. In solid green appears the best three parameter LK fit, with a spurious enhanced mass of $m^* = 18.5 \pm 2 m_e$. Note that the same temperatures as those in the experiment of Borzi were used. $Bottom$ $left$ $and$ $right$ Same calculation using 500 uniformly distributed data points and the same noise level. On the left is shown the original LK function in black, the noise in red and the sum of both in blue. On the right is shown the LK function, in blue, the absolute value of the sum of the noise and the LK function, in black, and the three parameter LK fit, in red.}
	\label{fig: LKNoiseFit2}
\end{figure}

Previous dHvA work on \TTS\ by Borzi $et$ $al.$ featured the use of three parameter non-linear LK fits to the temperature dependence of the dHVA amplitude \footnote{For a definition, see eq. \ref{eq:3param}, page \pageref{eq:3param}. This information was obtained through a private communication with R. A. Borzi.}. We suggest here how such a procedure can generate an increase of the quasiparticle mass when the signal to noise ratio is low and varies systematically with magnetic field. The signal one has in such a case is of the form of a modulated oscillating function to which random noise $f$ is added. In a hypothetical case where only the LK function modulated the oscillations, one has that
\beq
M(X) = LK(X,T)\sin(F_0 X) + f(X,T).\nn
\eeq
In the normal analysis procedure of dHvA, in order to calculate the oscillatory amplitude, one performs a Fourier transform over an inverse field width $\Delta X$, centred at $X_0$, 
\beq
\tilde{M}(F) = \tilde{LK}(F,X_0,T) + \tilde{f}(F,X_0,T).\nn
\eeq
Without noise, one integrates the power spectrum, or modulus square, of $\tilde{LK}$ over $F$ in the region where $\tilde{LK}(F)$ takes the shape of a peak, and take a square root, leading to a value of $\tilde{LK}$ proportional to $\langle LK(X)\rangle$, an average of $LK(X)$ over the range $\Delta X$, centred at $X_0$ \footnote{Using Parseval's theorem.}. With noise added, one has
\beq
\sqrt{\int |\tilde{M} (F,X_0)|^2 dF} = |\langle LK(X,T) + f(F,X,T)\rangle|,\nn
\eeq
which is the average of the sum of the noise and the amplitude function, in absolute value. This $does$ $not$ correspond to the average of $LK(X,T)$ plus the average of $f(X,T)$, and even less to the sum of the absolute value of the averages:
\bea
&&|\langle LK(X,T) + f(F,X,T)\rangle| \leq |\langle LK(X_0,T)\rangle + \langle f(F_0,X,T)\rangle|\nn\\
&&\leq |\langle LK(X_0,T)\rangle| + |\langle f(F_0,X,T)\rangle|.
\label{eq:LKnoise}
\eea
We stress that the result is smaller than just the LK function added to the modulus of random noise. As one performs calculations using the absolute value of the average of the sum of the signal and the noise, one finds that at high temperatures where the LK function is very small, the result is positive and non-zero, of which one tries to take account of by adding an offset to the LK fit in the three parameter scheme. However, the signal of amplitude $\tilde{LK}$ is not simply added to a constant, as we showed in eq. \ref{eq:LKnoise}, but to noise which may be positive or negative and the average and absolute value are taken $after$ the sum. As noise increases in magnitude, the LK function becomes ``buried" into noise, only the low temperature part rising above. Using a third parameter in the fit appears to work, but leads to sharper LK functions than reality, which means the quasiparticle mass, inversely proportional to the width of the LK function, is estimated larger. 

Moreover, $\langle LK(X,T)\rangle$ and $\langle f(X,T)\rangle$ are averages of the originals over a width $\Delta X$ (see section \ref{sect:FFTwindowing} for details), such that the noise function $\bar{f}(X_0,T) = \langle f(X,T)\rangle$ is not completely random, but has a correlation in $X$ of length $\Delta X$. It should not be surprising, then, to find that when the noise is large compared to the signal, the deviations that one measures on the mass are smooth and continuous curves as a function of inverse field.

Figure \ref{fig: LKNoiseFit2} presents two numerical examples of this systematic effect, using a quasiparticle mass of 12 $m_e$ at a field of 12 T. The noise was created with a random number generator of normal distribution, and had an rms amplitude of about 20\% of the signal at the lowest temperature. In the first example (top graph), the temperature values were the same as in the experiment by Borzi. We found that when we adjusted the three parameter LK function to the absolute value of the data added to the noise, in this specific instance we obtained a mass of $18.5 \pm 2$ $m_e$. When we performed this calculation several times by regenerating the noise, we systematically obtained masses between 1 and 1.5 times larger than the original value of 12 $m_e$. This is changes with differing noise amplitudes, and the systematic error on average becomes larger the smaller the signal to noise ratio. In the second case (bottom two graphs), 500 data points were used with the same scenario. An enhanced mass of 14 $m_e$ was obtained. Hence, combined with the fact  that the noise has a correlation length, the field dependent mass calculated with three parameter fits may feature an apparent continuous enhancement if the signal to noise ratio is low and has itself a field dependence, a condition which was fulfilled in the previous work of Borzi $et$ $al.$\footnote{This information was obtained by an analysis of the original data of Borzi $et$ $al.$. Note that we reproduced the reported enhancement when using three parameter fits with the original data.}.

Appendix~\ref{App:F} presents detailed calculations that we performed in order to verify the self-consistency of our calculation method for the quasiparticle masses, and to demonstrate how the masses can feature enhancements following the argument of this section. The method we used was to simulate dHvA oscillations with noise added to them, and extracted the quasiparticle mass as a function of magnetic field in the same way as was done in section \ref{sect:MassNonDiv}. We found that our calculation method is appropriate, and that we effectively produced a continuous mass enhancement when using three parameter fits applied to data where a low signal to noise ratio evolved with magnetic field.

\subsection{Interpretation of the absence of an enhancement}

We present here strong constraints to parameters of a putative band, from the viewpoint of the dHvA experiment, that possesses the missing, enhanced quasiparticle mass that we require in order to reconcile the field dependent specific heat and the sum of the dHvA quasiparticle masses. It must possess parameters that make it difficult to measure with second harmonic dHvA, and there are four possibilities. Firstly, this band could possess such a large in-plane $k$ space area that the Dingle reduction factor damps it enough to escape detection. Secondly, it can have a very large quasiparticle mass even away from the metamagnetic transition, where it is expected to diverge, and in that case it is the LK reduction factor that decreases its amplitude. Thirdly, it can be so small in area that, due to our use of  second harmonic detection, it is the Bessel function modulation factor that damps its low frequency amplitude. Finally, the field dependent missing mass observed in specific heat and resistivity could be caused by exotic non-fermionic degrees of freedom, not by the band structure, and so would be invisible to dHvA\footnote{Such theories are beyond the scope of this thesis and are not discussed.}.

In the first case, one can see in the ARPES data of figure \ref{fig: fig1_tamai3} that it is difficult to place an orbit in the BZ that is much larger than 4.2~kT, which we detected with good signal to noise ratio. Part of this large band would need to be situated in the narrow region between $\alpha_2$ and $\gamma_1$, which according to ARPES dispersion data along the $\Gamma$M line \footnote{A. Tamai $et$ $al.$, private communication.}, no evidence was found for such a band.

In the second case, a very high, field dependent mass for a single band would be consistent with the scenario where $\gamma_2$ is not included in the FS. In that case, near zero field this band would be required to possess a quasiparticle mass which accounts for half of the total specific heat, of around 74 $m_e$. However, such high masses have been observed in the past with the Cambridge system \footnote{See for instance the work of Taillefer $et$ $al.$\cite{taillefer}}, and if this band was small, it would give rise to an extremely flat dispersion near the Fermi level, which would most probably have been observed by ARPES. From ARPES data, there is no evidence of any other band than $\delta$ located at $\Gamma$ within the area covered by $\alpha_1$. Finally, if such a band was to be located elsewhere than in the centre of the zone, by symmetry argument, it would be required to be present in in two or four equivalent locations, depending whether it is situated at the $M$ point or elsewhere. If it was at the M point, the required mass per part of the orbit would be of 37 $m_e$, but no evidence for a corresponding very flat dispersion was found there. If it was located elsewhere, then symmetry requires it to be present four times and the required mass per part of the orbit would be around 19 $m_e$, low enough to be observed.

Consequently, if one dismisses theories involving non-fermionic degrees of freedom, all evidence points by elimination towards $\gamma_2$ being part of the FS and possessing the missing mass enhancement and the mass responsible for half of the zero field electronic specific heat. As stated earlier, band structure calculations by Mazin and Singh \cite{singh} predict that this band should be bilayer split, and consequently, counted eight times. Moreover, the finding by A. Rost, presented in section \ref{sect:magnetocaloric}, of a quantum oscillations peak around 110~T with a quasiparticle mass of about 9 $m_e$ accounts for nearly the right number of degrees of freedom, 72 $m_e$ when counted eight times. A. Tamai $et$ $al.$ also reported that a vHS is present in its vicinity, which could be reached by Zeeman splitting in a magnetic field, and lead to a peak in the DOS. In conclusion, we argue that it is the small $\gamma_2$ band observed in ARPES that holds the key to the reconciliation of our measurements to the field dependent specific heat.

This model does not, however, explain the divergence of the $A$ coefficient of the power law of the resistivity, measured by Grigera $et$ $al.$ \cite{science1}. Even though this parameter is proportional to the square of the quasiparticle mass, in a multi-band material, one obtains the resistivity by adding the conductivity of the various bands and inverting the result. In other words, the total resistivity is added as in a parallel circuit,
\beq
{1\over \rho_{tot}} = \sum_k {1 \over \rho_{0k} + A_k T^2} ... \nn,
\eeq
where $k$ refers to specific bands. A divergence of $A$ for one band does not show as a divergence of $A$ in the total resistivity. It is therefore required that the $A$ coefficients for more than one band diverge simultaneously. We do not hold any reasonable solution to this conflict between measurements. 

\section{dHvA in the nematic phase}
\subsection{Interpretation}

The observation of quantum oscillations inside the nematic phase was an extremely interesting discovery which provided us with important new information. The amount of established facts about this phase of the electronic system in \TTS\ is still small at this point, and the observation of dHvA may help moving towards an explanation of the phenomenon.

The most important fact that this measurement established is that the phase is most probably composed of Landau quasiparticles. Essentially, quantum oscillations with an LK temperature dependence stem from basic Fermi liquid theory, where fermionic charged particles follow the the Fermi-Dirac distribution function, and the LK function originates from the Fourier transform of the derivative of this distribution (see section \ref{sect:LK}, where the calculation is given). 

These oscillations also told us more about the nature of the electronic structure of this phase. We observed two quasiparticle orbits, at 1.0 and 2.5~kT, with masses of around 6-7 $m_e$. We have argued in section \ref{sect:nematicCambridge} that these frequencies, due to back-projection issues, corresponded to 0.9 and 1.8~kT, two of the low field side orbits, of which the masses were not entirely inconsistent. This suggested that although we did not observe any other frequencies in the nematic phase, its Fermi surface appeared to possess some similarity with that of the low field side. We therefore argue that the nematic behaviour in \TTS\ is produced by a specific subset of the low field side FS. Such a statement is in accordance with the suggestion by A. Tamai $et$ $al.$ from ARPES (section \ref{sect:Luttingersum}) measurements that the metamagnetic behaviour could be produced by the $\gamma_2$ pocket or the saddle point in the dispersion situated nearby \cite{tamai}. 

Moreover, one can extract a length scale from these frequencies, since the area within a $k$ space orbit is proportional to that of the real space orbit through the factor $(eH/\hbar)^2$. From eq. \ref{eq:period}, the frequency of 1.8~kT corresponds to a $k$ space orbit of average radius of 3.57$\times$10$^9$ m$^{-1}$, which through eq. \ref{eq:scaling}, in a field of 8 T, equals to an average real space diameter of approximately 580 nm. This value may constitute an upper boundary for the size of any real space orbit, and a lower boundary for the distance between nematic phase domain walls. Moreover, the interaction of the domain walls with the cyclotron orbits may explain the extreme reduction in amplitude of the oscillations between the low field side and the nematic phase.

Finally, the strong angular dependence of the amplitude of the quantum oscillations in the nematic phase is an aspect that we do not understand at this point. The similarity with the behaviour at higher fields than the metamagnetic region reminds us that we face the same problem in these two situations: we do not know of a process that damps the amplitude as a function of angle as dramatically as what is observed. But we may still suggest a process that could have this result, related to the anisotropic transport properties that are triggered in \TTS\ by the application of a small in-plane magnetic field. If, as was suggested by Grigera $et$ $al.$ \cite{science2}, the nematic phase domain walls are aligned in some way by the in-plane magnetic field, resulting in anisotropic transport, the distance between walls could also be affected. If that distance is reduced, it may be more difficult for the system to fit quasiparticle orbits anywhere in real space, and lead to suppression of dHvA.

\addcontentsline{toc}{chapter}{\sffamily Conclusion}
 
 \chapter*{Conclusion}
 \markright{Conclusion}
 
 We review in this conclusion the facts that have been established in this thesis regarding the FS near the QCEP in \TTS. We had set out to study three aspects of this problem: the low and high field FS, that in the nematic phase and the evolution of the quasiparticle properties in the region surrounding the QCEP. We have, in collaboration with ARPES experiments by A. Tamai $et$ $al.$ \cite{tamai}, determined with some degree of confidence the topology of the FS at low fields. We discovered that dHvA oscillations exist inside the nematic phase, and finally, we established that none of the orbits for which we were able to obtain high resolution dHvA signals undergo quasiparticle mass enhancements near the QCEP. We present here a brief unified picture of the physics of the FS of \TTS\ based on what is known at the present time. 
 
\vspace{18pt}

 The dHvA experiments at low fields revealed the proof that the FS of \TTS\ is quasi-two-dimensional. We showed that all the FS pockets except one exhibited a behaviour consistent with a cylindrical topology connected to the top of the BZ. Following this, A. Tamai $et$ $al.$ suggested a model of the in-plane FS that is consistent both with ARPES and our dHvA data, respecting measured cross-sectional areas and quasiparticle masses. One pocket was missing in our dHvA data, but was found by A. Rost in magneto-caloric quantum oscillations \cite{rost}. This model was extremely successful in reproducing the zero field electronic specific heat measured by Ikeda $et$ $al.$ \cite{Ikeda2000}, but the Luttinger counting revealed a quarter of an electron missing over the 16 required by the crystal structure.
 
 We moreover determined that a non-linear change in Fermi level occurs at the metamagnetic transition, such that some FS pockets change size abruptly, as inferred from dHvA frequencies. Through this frequency evolution, we found that the metamagnetic transition influenced the dHvA oscillations at low fields, starting from 3 T away from the critical region at high angles and from 0.5 T at $c$-axis. The gradual difference arising in the number of electrons between the spin species of the $\alpha_1$ pocket produced an additional interference pattern to that inherent to its corrugation along the $k_z$ direction. We were not able to reproduce fully this pattern, but we performed a simulation that agreed reasonably well with high angle data, suggesting that the effect gradually changed magnitude with angle, reflecting the fact that the new pattern modulated the data from a higher field interval down to lower fields at high angles compared to the $c$-axis direction. We interpreted that as a change of width of the peak in the density of states responsible for metamagnetism, which could be due to small evolutions of the crystal structure with field angle. This is effectively expected from the known magneto-structural coupling. We assume that the other frequencies present in the low field side possess the same property as the $\alpha_1$ pocket, but their field and angular amplitude modulation was too complex to analyse due to the close proximity of different peaks, related to bilayer splitting or possible magnetic breakdown even at such low fields.
 
 We studied the FS at high fields as well, above the metamagnetic transition, and found surprising complexity in the spectra, at all angles. The frequencies were all split into a high number, much more than what is predicted by spin splitting due to super-linear changes in the magnetisation. As this complexity increased with magnetic field, we associated it with magnetic breakdown, which starts at fields as low as 5 T in one case. We discovered another unexpected phenomenon, which is not fully explained yet, that the frequencies possessed highly unusual angular dependence of their amplitude, where some disappeared and others appeared at specific angles. We suggested that it could be related with small evolutions in the crystal structure with the value of the in-plane field, resulting in changes in the gap value between bands, modifying the condition for magnetic breakdown. In such a model, new orbits appear while others lose intensity, controlled by the different hopping probabilities included in each $k$-space path. The frequency values also increased with angle, but were not completely consistent with the normal quasi-two-dimensional angular dependence.
 
\vspace{18pt}

Turning our attention to the nematic phase, we made the interesting discovery that dHvA orbits exist within the first order phase transition boundaries. We found a few cycles of two frequencies which were consistent with the $\alpha_1$ and $\gamma_1$ pockets. They possessed normal LK temperature dependence, and their quasiparticle masses did not agree perfectly with those of $\alpha_1$, $\gamma_1$ but possessed nearly the right value. We moreover discovered that these oscillations exist only within 5$^{\circ}$ of the $c$-axis, and are rapidly suppressed at higher angles. 

The existence of quantum oscillations inside the nematic phase is a surprising result since the resistivity is highly enhanced and one expects high scattering of charge carriers. However, if the scattering is produced by domain walls, it is possible that only orbits enclosing real space areas of average diameter larger than the average distance between walls are suppressed by that process. We calculated that such a distance is of the order of 600 nm, comparable to the value of the mean free path of 300 nm.  Moreover, we reflected on the fact that with an in-plane field, as the domains become aligned, the average distance between domain walls may decrease significantly and suppress all cyclotron orbits altogether, producing the observed strong angular dependence of their amplitude.

\vspace{18pt}

The most important result of this thesis is related to quantum criticality and metamagnetism. We calculated the field dependence of the effective quasiparticle mass of five of the six orbits observed by ARPES, and discovered no enhancement near the QCEP, in complete contradiction with previous dHvA experiments \cite{borzi}. We infer that quantum critical properties are not present in any of the bands observed in dHvA, putting strong constraints on many theoretical models. Moreover, since in two-dimensional materials, the sum of the quasiparticle masses must correspond to the electronic specific heat, this conclusion is in complete contradiction with past measurements \cite{perry1,zhou} except if the enhancement is produced by the missing pocket, labelled $\gamma_2$ by A. Tamai $et$ $al.$. It is also in contradiction with existing transport measurements for the same reason, since the $A$ coefficient of the quadratic temperature dependence of the resistivity is proportional to the square of the quasiparticle mass, and features a strong enhancement \cite{science1}.

We proposed that the specific heat enhancement is either produced by the $\gamma_2$ pocket or by non-fermionic degrees of freedom. In the first case, we suggested that metamagnetism is produced by the van Hove singularity identified by A. Tamai $et$ $al.$ near the corner of the orthorhombic BZ, reached with magnetic field by the $\gamma_2$ pocket, which already possesses half the total electronic specific heat. It would also explain the electron transfers that we detected in the $\alpha_1$ and $\alpha_2$ pockets, as moving through a large peak in the density of states with field should force a change in Fermi level, as well as a non-linear increase in magnetisation. We hold, however, only indirect proofs of this inference, and cannot completely rule out other explanations related to non-fermionic modes. Direct quasiparticle mass measurements of the $\gamma_2$ pocket are required in order to decide which explanation is the right one, and they have not been possible during this project, the signal being too small to detect with dHvA.
 
\vspace{18pt}

 We end this conclusion by mentioning two types of experiments that could be done in the future to complete this work. The first class corresponds to those which are required to close all the uncertainties in the work to date, and the second to new experiments which could help gather more information about the $\gamma_2$ pocket. 
 
 Three measurements should be done using the current dHvA system in order to clear a number of issues in the interpretation of the data that was obtained in this project. The first regards questions about the sample mounting procedure, and to whether, as was suggested to us by S. R. Julian, it produces inhomogeneous strain in the samples, leading to the high peak splitting of the oscillations in the high field side. The second is to improve the knowledge of the thermometry or thermal equilibrium of our mounting procedure used in the Cambridge system, for instance by using another known Fermi liquid system using the same procedure. This should enable us to understand better the low temperature deviations of our data compared to the LK function. The third is to verify into even more detail whether a modulation field higher than that which was used by Borzi $et$ $al.$ did not suppress an enhancement of the quasiparticle mass, and would involve measuring the first harmonic dHvA at very low modulation fields. 
 
 A number of new experiments could be done in order to determine with certainty whether $\gamma_2$ is responsible for the specific heat enhancement and metamagnetism in \TTS. Unfortunately, we have already used most of the methods that provide information related to specific branches of the FS, as only ARPES and quantum oscillations can resolve individual FS pockets. There exists still one type of quantum oscillations experiment that has not been performed yet, using torque magnetometry. However, the signal from such experiments possesses an amplitude that is proportional to the derivative of the dHvA frequency $F$ as a function of angle $\theta$, which equals zero at $c$-axis, the direction of interest. Consequently, one cannot study the $\gamma_2$ pocket at the QCEP, but valuable information could still be obtained near the metamagnetic transition, since it exists at all angles. 
 
 However, more experiments could be done using dHvA. All our measurements were performed with second harmonic, the amplitude of which is proportional to the square of the dHvA frequency. The first harmonic could be used, since is it linear in dHvA frequency and better signal could be obtained for $F < 1$ kT. Time constraints have not allowed us to perform these experiments in this project. One could also perform dHvA using samples under pressure, which could tune the metamagnetic transition to a different field value. However, in both of these cases, the hope to resolve the field dependence of the quasiparticle mass is slim, as good signal to noise will be difficult to achieve. In that respect, experiments using the magnetocaloric effect could produce better signal than dHvA for small FS pockets if performed at lower temperatures than the recent measurements by A. Rost \cite{rost}, using for instance the Cambridge dilution refrigerator.
 
 Other types of experiments could also provide more information about $\gamma_2$. In particular, ARPES experiments performed with doped samples of \TTS\ may allow the observation of the behaviour of $\gamma_2$ as one crosses the peak in the density of states. But according to our model, the band responsible for metamagnetism is of hole type, which means that hole doping is necessary. This was confirmed by J. Farrell, who doped electrons in \TTS\ and observed an increase in the metamagnetic field value\footnote{Private communication, to be published.}. Hole doping has not been performed in \TTS\ to our knowledge, and is probably difficult to achieve\footnote{It involves replacing the Sr atoms by one of the alkaline elements, which are extremely reactive.}, but would provide an excellent tool to study metamagnetism and the QCEP, as it would bring the metamagnetic transition towards zero field.

\appendix

\chapter{Fourier transform of the derivative of the Fermi function \label{App:A}} 
\markright{Appendix~\ref{App:A}}

This Fourier transform is more easily done backwards, by applying the back Fourier transform on the LK function, as a function of $\lambda$, with reciprocal variable $\psi = \phi / \lambda$. The LK function is
\beq
\tilde{g}(\lambda) = {\pi \lambda \over \sinh( \pi \lambda)}\nn
\eeq
The back Fourier transform is
\beq
g(\psi) = {1 \over \sqrt{2\pi}} \int_{-\infty}^\infty {\pi \lambda e^{i \lambda \psi}\over \sinh( \pi \lambda)} d\lambda \nn
\eeq
This integral converges, since $\sinh(\lambda)$ diverges faster than $\lambda$ for $\lambda \rightarrow \pm \infty$. One applies Cauchy's residue theorem to solve it, by setting $\lambda = R e^{i \theta}$ and performing an integral over a semicircle contour in the complex plane, the linear part on the real axis. The arc is chosen to lie either on the positive or negative half complex plane, depending on the sign of $\psi$. The integral involves two terms, one for the section on the real axis, one for the arc in the complex plane. By taking the limit of an infinite radius $R$, the sum of these integrals becomes equal to $2 \pi i$ the sum of the residues at the poles,
\beq
\int_{-\infty}^{\infty} {\pi \lambda e^{i \lambda \psi}\over \sinh(\pi \lambda)} d\lambda + \lim_{R \rightarrow \infty} \int_0^{2 \pi} {\pi Re^{i \theta} e^{i Re^{i \theta} \psi}\over \sinh(\pi Re^{i \theta})} Rd\theta  = 2 \pi i \sum (residues),\nn
\eeq
where the second term is zero everywhere. 

One is required to locate the poles and determine their order. The integrand diverges on the imaginary axis, where the $\sinh(i\pi R)$ function becomes $-\sin(\pi R)$, except at $R = 0$, where the integrand is zero. The poles are evenly spaced, appearing at
\beq
R = n,\nn
\eeq
with $n$ an integer. Their order corresponds to that of the first term in a Laurent series,
\beq
{1 \over \sin(\pi R)} \simeq {1\over \pi R} + ...,\nn
\eeq
and is equal to one. The residues are calculated in the following manner
\beq
\lim_{R \rightarrow n} (i\pi R-in\pi) {-i\pi R e^{- R \psi}\over \sin(\pi R)} = - (-1)^nin\pi e^{- \psi n}\nn
\eeq  
As indicated earlier, one chooses the side to take the integral along the arc that defines the sign of $n$, with that of $\psi$, such that the argument of the exponential in the numerator of the integrand is always negative when on the imaginary axis. The result of the integral is
\beq
I = \int_{-\infty}^{\infty} {\pi \lambda e^{i \lambda \psi}\over \sinh(\pi \lambda)} d\lambda = - 2 \pi i\sum_{n=1}^\infty (-1)^nin\pi e^{- |\psi n|} = 2 \pi \sum_{n=1}^\infty (-1)^nn\pi e^{- |\psi n|}.\nn
\eeq
Since this is an analytic function of $\psi$ except at zero, one uses its derivative in order to have a more tractable form:
\beq
I =  2 \pi {d\over d\psi}\sum_{n=1}^\infty (-1)^n e^{- |\psi n|}.\nn
\eeq
This is a geometric series, where setting 
\beq
r = -e^{-|\psi|},\nn
\eeq
results in
\beq
\sum_{n=0}^\infty {1\over r^n} = {1 \over 1-r}.\nn
\eeq
The integral becomes
\beq
I = 2 \pi {d\over d\psi} {1\over 1 + e^{- |\psi|}} = 2 {- \pi e^{- |\psi|} \over (1 + e^{-|\psi|})^2} = 2 \pi {1 \over 1 + \cosh|\psi|},\nn
\eeq
and the result of the Fourier transform is
\beq
g(\psi) = {1 \over \sqrt{2\pi}} {1 \over 1 + \cosh|\psi|},\nn
\eeq
the derivative of the Fermi function.

 \chapter{Arbitrary cross-sectional area of a cylindrical expansion \label{App:B}} 
\markright{Appendix~\ref{App:B}}

\begin{figure}[t]
         \begin{center}
	\includegraphics[width=0.4\columnwidth]{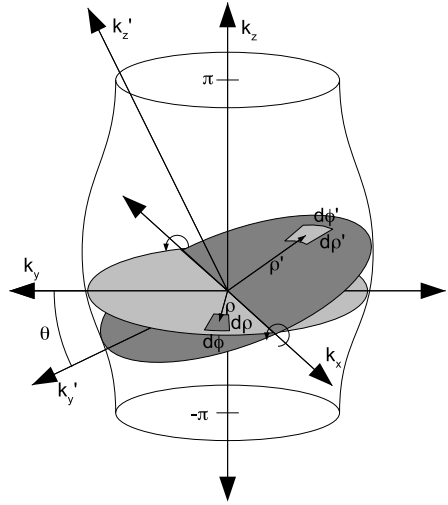}
	\caption[Intersection of a plane and a cylindrical expansion of the FS]{Schematic Fermi surface, with a plane intersecting at an arbitrary angle $\theta$ and height in $k_z$. Two systems of cylindrical coordinates are shown, the normal one with $k_z$, $\rho$ and $\phi$ as variables, and the tilted one with primed variables $k_z'$, $\rho'$ and $\phi'$. Two infinitesimal areas are drawn, one in the normal cylindrical coordinates, the other in the tilted system. The infinitesimal units of area are $\rho d\rho d\phi$ and $\rho' d\rho' d\phi'$. In order to calculate the area of the tilted ellipse using $\rho'$ and $\phi$, one has to rescale the area unit by $\cos \theta$.}
	\label{fig: FSArea}
	\end{center}
\end{figure}

We carry out in this appendix the calculation from the point it was left off in section \ref{sect:BergemanAnalysis}, where we required to calculate the cross-sectional area $A(\kappa_z,\theta, \phi_0 )$ in order to perform the integral
\beq
M \propto \int_{-\pi}^{\pi} d\kappa_z \sin\bigg({\hbar \over e} A(\kappa_z)X \bigg),
\label{eq:IntOscAppendix}
\eeq
with a cylindrical expansion of $A$ with parameters $\mu$ and $\nu$. The in-plane angle $\phi_0$ was added in order to orient the in-plane shape with the direction of rotation of the magnetic field. The expansion of the Fermi wave vector is 
\beq
k_F(\kappa_z, \phi, \phi_0) = k_{00} + \sum_{\mu = 1, \nu = 1} k_{\mu \nu} \cos \nu \kappa_z \cos (\mu\phi + \phi_0),
\label{eq:kfexpansionAppendix}
\eeq
where $\kappa_z$ corresponds to $k_z \times 2\pi / h_{bz}$, and $h_{bz}$ is the height of the Brillouin zone $4 \pi /c$, with $c$ the $c$-axis lattice parameter.

We define two cylindrical coordinate systems, one with $\hat{k_z}$ direction parallel to the $c$-axis, the other rotated by $\theta$ in the direction of $\phi_0$, as shown in figure \ref{fig: FSArea}, using prime notations for the tilted system. Using a cartesian system with the direction $\hat{k_x}$ in the axis of rotation of the tilted system, we have a set of equations for the transformation between the two coordinate systems:
\bea
k_y' &=& k_y\cos \theta + k_z\sin \theta, \nn\\
k_z' &=& -k_y\sin \theta + k_z\cos \theta, \nn\\
k_y &=& k_y'\cos \theta - k_z'\sin \theta, \nn\\
k_z &=& k_y'\sin \theta + k_z'\cos \theta, \nn\\
k_x &=& k_x'.\nn
\eea
In relation to the cylindrical coordinates
\bea
&&\rho^2 = k_x^2 + k_y^2, \quad k_x = \rho \cos \phi, \quad k_y = \rho \sin \phi, \nn\\
&&\rho'^2 = k_x'^2 + k_y'^2, \quad k_x' = \rho' \cos \phi', \quad k_y' = \rho' \sin \phi',\nn
\eea
one can put these equations together to get a new set for $k_z$ and $k_z'$:
\bea
k_z &=& \rho' \sin \phi' \sin \theta + k_z' \cos \theta,\nn\\
k_z' &=& - \rho \sin \phi \sin \theta + k_z \cos \theta.\nn
\eea
As shown in figure \ref{fig: FSArea}, we are required to integrate the ellipse that corresponds to the intersection of the plane at constant $k_z'$ and the arbitrary shape of $k_F(k_z)$. We then evaluate the infinitesimal unit of area on the tilted ellipse that lies at constant $k_z'$, $\rho d\rho d\phi$, using variables that are easier to work with. For that, we differentiate the last set of equations on this plane, where $dk_z' = 0$:
\bea
dk_z &=& d \rho' \sin \phi' \sin \theta + \rho' \cos \phi' \sin \theta d\phi'\nn\\
0 &=& - d\rho \sin \phi \sin \theta + dk_z \cos \theta - \rho \cos \phi \sin \theta d\phi,\nn
\eea
which, by eliminating $dk_z$ leads to
\beq
d\rho \sin \phi \tan \theta + d\phi \rho \cos \phi \tan \theta = d\rho' \sin \phi' \sin \theta + \rho' \cos \phi' \sin \theta d\phi'.\nn
\eeq
The radius $\rho'$ only varies when $\rho$ varies as well, and when only the angles change, using the fact that $k_x' = k_x = \rho \cos \phi = \rho' \cos \phi'$, we obtain
\beq
d\phi = \cos \theta d\phi',\nn
\eeq
which results in an area unit of
\beq
\rho'd\rho'd\phi' = {\rho' d\rho' d\phi \over \cos \theta},\nn
\eeq
and enables us to perform the calculation using the variable $\rho'$. We now calculate the area of the ellipse:
\bea
A &=& \int da' = \int \rho' d\rho' d\phi' = {1\over \cos \theta} \int_0^{2\pi} d\phi \int_0^{k_F(k_z')} \rho' d\rho'\nn\\ 
&=& {1\over 2\cos \theta} \int_0^{2\pi} d\phi k_F(k_z')^2\nn
\eea
We replace $k_z'$ for its value in terms of variables of the normal system, substituting $\rho$ by its average value $k_{00}$, and multiplying by $2\pi/h_{BZ}$,
\beq
\kappa_z' = \kappa_z \cos \theta - \kappa_{00} \sin \phi \sin \theta,\nn
\eeq
and normalising $\kappa_z'$ to a value such that $\kappa_z$ varies between $-\pi$ to $\pi$, as in equation \ref{eq:IntOscAppendix}, i.e. by $\cos \theta$ \footnote{The calculation is at the same point as in Bergemann, eq 9, p. 655, \cite{bergemann}.}: 
\beq
A = {1\over 2\cos \theta} \int_0^{2\pi} d\phi k_F^2(\kappa_z - \rho \sin \phi \tan \theta)\nn
\eeq
From here, we may check our calculation with a circular cylindrical Fermi surface $k_F = k_{00}$, which appropriately yields $A = \pi k_{00}^2 / \cos \theta$. 

We carry on with the expansion of the square of $k_F$ to first order (dropping terms in $k_{\mu \nu} k_{\sigma \tau}$, and the orientation angle $\phi_0$ for now):
\bea
k_F^2(k_z,\phi) &=& (k_{00} + ... + k_{\mu \nu} \cos \nu \kappa_z \cos \mu \phi)^2\nn\\ 
&=& k_{00}^2 + ... + 2k_{00} k_{\mu \nu} \cos \nu \kappa_z \cos \mu \phi + ...\nn\\
&\simeq& k_{00}^2 + 2k_{00}\sum_{\mu \nu} k_{\mu \nu} \cos \nu \kappa_z \cos \mu \phi.\nn
\eea
We substitute this in the area integral:
\beq
A = {1\over 2\cos \theta} \sum_{\mu \nu} \int_0^{2\pi} d\phi \bigg( k_{00}^2 + 2 k_{00} k_{\mu \nu} \cos \mu \phi \cos (\nu \kappa_z - \nu \kappa_{00} \sin \phi \tan \theta) \bigg).\nn
\eeq
The cosine of a sine in an integral leads to Bessel functions, where
\beq
J_\mu(\lambda) = {1\over 2 \pi} \int_{-\pi}^{\pi} \cos(\mu x - \lambda \sin(x)) dx.\nn
\eeq
We transform our integral in such a form. Expanding and regrouping, using trigonometric identities, we have
\bea
&& \cos \mu \phi \cos(\nu \kappa_z - \kappa_{00} \sin \phi \tan \theta) \nn\\
&=& {1\over2} \bigg(\cos(\mu\phi + \nu \kappa_z - \nu \kappa_{00} \tan \theta \sin \phi) + \cos(\mu\phi - \nu \kappa_z + \nu \kappa_{00} \tan\theta\sin\phi)\bigg)\nn\\
&=& {1\over2}\bigg(\cos(\mu\phi - \nu \kappa_{00} \tan \theta \sin \phi) \cos \nu \kappa_z - \sin(\mu\phi - \nu \kappa_{00} \tan \theta \sin \phi)\sin \nu \kappa_z \nn\\
&+& \cos(\mu \phi + \nu \kappa_{00} \tan \theta \sin \phi)\cos \nu \kappa_z + \sin(\mu \phi + \nu \kappa_{00} \tan \theta \sin \phi)\sin \nu \kappa_z \bigg).\nn\\
\eea
The $\phi$ integral of the second and last terms will be zero, and we are left with two Bessel functions:
\beq
\int_0^{2\pi} d\phi \cos(\mu \phi \mp \nu \kappa_{00} \tan \theta \sin \phi)\cos \nu \kappa_z = J_\mu( \pm \nu \kappa_{00} \tan \theta)\cos\nu \kappa_z\nn
\eeq
The Bessel functions are either symmetric or antisymmetric, depending whether $\mu$ is respectively even or odd. We rewrite the sum of the integrals
\bea
J_\mu(\nu \kappa_{00} \tan \theta) + J_\mu(-\nu \kappa_{00} \tan \theta) &=& 2 J_\mu(\nu \kappa_{00} \tan \theta), \quad \mu \quad \textrm{even},\nn\\
&=& 0 \quad \mu \quad \textrm{odd},\nn
\eea
Finally, the area is
\beq
A = {\pi k_{00}^2 \over \cos \theta} + 2\pi \sum_{\mu \nu}{k_{00} k_{\mu \nu} \over \cos \theta} J_\mu (\nu \kappa_{00} \tan \theta) \cos \nu \kappa_z, \quad \mu \quad even\nn
\eeq
This is the result when imposing an orientation angle $\phi_0$ of zero. When using $\phi_0 \neq 0$, the calculation is more complicated but essentially the same and yields the result of Bergemann $et$ $al.$, eq. 13, p. 657 \cite{bergemann},
\bea
&&{\pi k_{00}^2 \over \cos \theta} + 2\pi \sum_{\mu \nu}{k_{00} k_{\mu \nu} \over \cos \theta} J_\mu (\nu \kappa_{00} \tan \theta) \cos \mu \phi_0 (-1)^{\mu/2}\cos\nu \kappa_z, \quad even \quad \mu, \nn\\
&&{\pi k_{00}^2 \over \cos \theta} + 2\pi \sum_{\mu \nu}{k_{00} k_{\mu \nu} \over \cos \theta} J_\mu (\nu \kappa_{00} \tan \theta) \cos \mu \phi_0 (-1)^{(\mu-1)/2}\sin\nu \kappa_z, \quad odd \quad \mu,\nn\\
\eea

With this equation, adding all the different $A_{\mu\nu}$ to the area will give the total area of the intersection of a plane tilted by $\theta$ and the Fermi surface with a shape defined by the $k_{\mu \nu}$. We cannot go any further unless we choose one $k_{\mu \nu}$ and set all the others to zero, except $k_{00}$. 

We are finally required to calculate dHvA oscillations from a single warping parameter. We replace the last equation with $\phi_0 = 0$ into the equation \ref{eq:IntOscAppendix}, and obtain
\bea
M &\propto& \int_{-\pi}^{\pi} dk_z \sin\bigg[{\hbar X \over e} \bigg({\pi _{00}^2 \over \cos \theta} +  2\pi \sum_{\mu \nu}{k_{00}k_{\mu\nu} \over \cos \theta} J_\mu(\nu \kappa_{00} \tan \theta) \cos \nu \kappa_z \bigg)\bigg]\nn\\
&=& \int_{-\pi}^{\pi} dk_z \sin ( a +  b_{\mu \nu} \cos \nu \kappa_z)\nn\\
&=& \int_{-\pi}^{\pi} dk_z \sin a \cos ( b_{\mu \nu} \cos \nu \kappa_z) + \cos a \sin( b_{\mu \nu} \cos \nu \kappa_z),\nn
\eea
where the second term is zero by symmetry. We have once more an integral of the form of a Bessel function, of order zero. Putting back the symbols into $a$ and $b$, we obtain (with $\mu$ even):
\bea
M &\propto& \sin \bigg({\hbar X\over e}{ \pi k_{00}^2 \over \cos \theta}\bigg) {1\over \nu} J_0 \bigg({2\pi \hbar X\over e}{k_{00} k_{\mu \nu} \over \cos \theta} J_\mu(\nu \kappa_{00} \tan \theta)\bigg)\nn\\
&=& \sin\bigg({2 \pi F_0 X \over \cos \theta} \bigg) {1\over \nu} J_0\bigg( {2 \pi \Delta F_{\mu \nu} X \over \cos \theta}J_\mu(\nu \kappa_{00} \tan \theta)\bigg),\nn
\eea
where $F_0$ is the average dHvA frequency, and $\Delta F_{\mu \nu}$ is the difference which produces a beat pattern.

 \chapter{Fourier series of the functions $\sin\big(\lambda \sin(\omega t)\big)$ and $\cos\big(\lambda \sin(\omega t)\big)$ \label{App:C}} 
 \markright{Appendix~\ref{App:C}}

We begin with the function $\cos\big(\lambda \sin(\omega t)\big)$. The Fourier series requires to calculate
\beq
\cos\big(\lambda \sin(\omega t)\big) = a_0 + \sum_{k=1}^{\infty} a_k \cos(k\omega t) + b_k \sin(k\omega t),\nn
\eeq 
with
\bea
a_k &=& {1\over 2 \pi} \int_{-\pi}^{\pi} \cos\big(\lambda \sin(\omega t)\big) \cos(k \omega t) d(\omega t), \nn\\
b_k &=& {1\over 2 \pi} \int_{-\pi}^{\pi} \cos\big(\lambda \sin(\omega t)\big) \sin(k \omega t) d(\omega t), \nn\\
a_0 &=& {1\over 2 \pi} \int_{-\pi}^{\pi} \cos\big(\lambda \sin(\omega t)\big) d(\omega t).\nn
\eea
Using trigonometric properties, the first two equations can be rewritten
\bea
a_k &=& {1\over 2 \pi} \int_{-\pi}^{\pi} \bigg(\cos\big(k \omega t + \lambda \sin(\omega t)\big) + \cos\big(k \omega t - \lambda \sin(\omega t)\big) \bigg)d(\omega t), \nn\\
b_k &=& 0,\nn
\eea
the second being zero by symmetry. The equation for $a_k$ corresponds to Bessel's first integral
\beq
J_k(\lambda) = {1\over 2 \pi} \int_{-\pi}^{\pi} \cos(kx - \lambda \sin(x)) dx,\nn
\eeq
and thus 
\beq
a_k = J_k(\lambda) + J_k(-\lambda).\nn
\eeq
The constant offset $a_0$ also possesses the form of Bessel's first integral, with $k=0$:
\beq
a_0 = J_0(\lambda).\nn
\eeq
A similar calculation can be done with the function $\sin\big(\lambda \sin(\omega t)\big)$. In this case, $a_k$ and $a_0$ equal zero, by symmetry, and the remaining term is
\bea
b_k &=& {1\over 2 \pi} \int_{-\pi}^{\pi} \bigg(\cos\big(k \omega t + \lambda \sin(\omega t)\big) - \cos\big(k \omega t - \lambda \sin(\omega t)\big) \bigg)d(\omega t), \nn\\
&=& J_k(\lambda) - J_k(\lambda).\nn
\eea
The Bessel functions are either symmetric or antisymmetric, depending whether $k$ is respectively even or odd. One can rewrite these results as
\bea
J_k(\lambda) + J_k(-\lambda) &=& 2 J_k(\lambda), \quad k \quad \textrm{even},\nn\\
&=& 0 \quad k \quad \textrm{odd},\nn
\eea
and
\bea
J_k(\lambda) - J_k(-\lambda) &=& 2 J_k(\lambda), \quad k \quad \textrm{odd},\nn\\
&=& 0 \quad k \quad \textrm{even},\nn
\eea

The Fourier series are then:
\bea
\cos\big(\lambda \sin(\omega t)\big) &=& J_0(\lambda) + 2 \sum_{k=1}^{\infty} J_{2k} \cos(2k\omega t)\nn\\
\sin\big(\lambda \sin(\omega t)\big) &=& 2 \sum_{k=1}^{\infty} J_{2k-1} \cos\big((2k-1)\omega t \big)\nn
\eea

Finally, one may  approximate $J_k(\lambda)$ for $\lambda<1$. The Bessel function is defined as
\bea
J_k(\lambda) &=& \sum_{n=0}^\infty {(-1)^n\over n! (n+k+1)!}\bigg({\lambda \over 2}\bigg)^{2n+k}\nn\\
&=& \bigg({\lambda \over 2}\bigg)^{k} \sum_{n=0}^\infty {(-1)^n\over n! (n+k)!}\bigg({\lambda \over 2}\bigg)^{2n}\nn
\eea
When $\lambda$ is 1 or smaller, the terms of the sum become very rapidly small with increasing $n$, and keeping only the first one, $n=0$, which is equal to $1/k!$, one obtains
\beq
J_k(\lambda) \simeq {\lambda^k \over 2^k k!}.\nn
\eeq

 \chapter{Cooling power of the adiabatic demagnetisation refrigerator \label{App:D}} 
 \markright{Appendix~\ref{App:D}}

The principle of adiabatic demagnetisation refrigeration uses the entropy of an isolated paramagnetic material under a magnetic field. If the spins have a total angular momentum of $J = L + S$, their interaction with the field is $\epsilon = -g(J,L,S)\mu_B\bs{J} \cdot \bs{H}$. From this, one can calculate the free energy, and derive the magnetisation and the entropy \footnote{Ashcroft and Mermin \cite{ashcroft}, chapter 3}:
\bea
M = -\bigg({\partial F \over \partial H}\bigg)_T &=& N g\mu_B J B_J(gJx ), \quad x = {\mu_BH \over k_BT}\nn\\
S = -\bigg({\partial F \over \partial T}\bigg)_H &=& Nk_B\bigg[\ln{\sinh g(J+{1 \over 2})x \over \sinh{gx \over 2}} - gJxB_J(gJx)\bigg] - k_B\ln N!,\nn\\
\eea
with $B_J$ the Brillouin function,
\beq
B_J(y) = (1+{1\over 2J}) \coth(1+{1\over 2J})y - {1 \over 2 J} \coth {y \over 2 J}.\nn
\eeq
Both the entropy and the magnetisation are functions of $H$ and $T$ through $x = {\mu_BH \over k_BT}$ only. Consequently, for an adiabatic process where $H$ varies, $T$ evolves such that $x$ remains constant. In such a scenario, reducing $H$ to zero should produce zero temperature. It is not the case, since below a certain field, $H_0$, the intrinsic internal field of the paramagnetic material prevents reaching zero field, from very small but non-vanishing spin interactions. When the system reaches the minimum field of $H_0$, the cooling power vanishes and one obtains a base temperature $T_{BT}$.

\begin{figure}[t]
\begin{center}
	\includegraphics[width=.7\columnwidth]{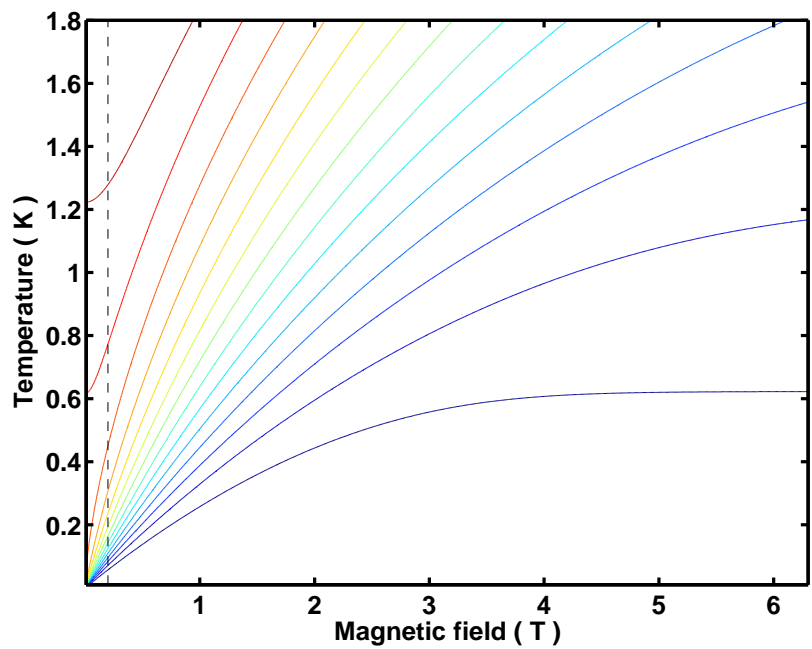}
	\caption[Contour lines of the entropy in the ADR]{Contour lines of the total entropy of the LTS of the ADR while in the adiabatic state. The red lines correspond to high entropy, which decreases towards the blue curves. The dashed line represents the intrinsic ordering magnetic field of the paramagnetic material, at which the cooling process stops.}
	\label{fig:HvsTcontour}
	\end{center}
\end{figure}

This analysis works when only the spin system is held in adiabatic conditions in the cryostat and no significant load of heat capacity is thermally connected. In the ADR system, the heat capacity of the LTS is not negligible. As temperature is reduced, heat is transferred from the LTS to the paramagnetic material, and the entropy of the spin system $S$ is not constant. As one reduces the magnetic field by $\Delta H$, the temperature reduces by $\Delta T$, but their ratio also changes by $\Delta x$, as does the entropy of the spin system by $\Delta S$, and 
\beq
- T\Delta S = C_M^{LTS}(T) \Delta T,
\label{eq:TdS_ADF}
\eeq
where $C_M^{LTS}$ is the heat capacity load of the LTS. The LTS is made of non-magnetic metal:
\beq
C_M^{LTS}(T) = N (\gamma T + \beta T^3)\nn,\nn
\eeq
where $N$ is a number of moles. One takes a path during demagnetisation where the rate of temperature decrease is constant. Integrating eq. \ref{eq:TdS_ADF} over this path:
\beq
- S(T',H)  = \int_{T_{1K}}^{T} {C_M^{LTS}(T') \over T'} dT' = N (\gamma T + \beta {T^3\over 3}) + C.\nn
\eeq
Figure \ref{fig:HvsTcontour} shows curves of $T$ versus $H$ when appropriate constants are used, calculated by equating the entropy values. It consists of a contour plot of the total conserved entropy of the paramagnetic system and LTS. One observes that the rate of decrease of the LTS temperature accelerates during cooldown with a constant rate of demagnetisation, which corresponds to the behaviour of the ADR, and is due to the decrease in heat capacity of the LTS with $T$.

 \chapter{Raw data from the rotation study \label{App:E}} 
 \markright{Appendix~\ref{App:E}}

 This appendix presents raw dHvA data for the reader to appreciate and get a feeling for. The complexity of the data can be seen just by looking at the patterns formed by the raw data, before even performing any Fourier transforms. The data which is presented here is only the rotation study of dHvA measured in Cambridge, which was performed at a base temperature $30\pm5$ mK, during around 20 days.
 
 The angle was measured using a rotating knob potentiometer, which was turned with the mechanism by a computer controlled electrical motor. The resistance is then simply proportional to the angle, and was measured in k$\Omega$. The conversion from resistance to angle values was later performed by simply taking photos of the coil system at various potentiometer values, method that turned out to be the simplest.\footnote{I thank Michael Sutherland and Swee K. Goh for this careful measurement, which was done in my absence.} We obtained conversion values of $8.23^{\circ}/k\Omega \times R - 8.23^{\circ}$ for coil A, which held sample C698I, and $8.45^{\circ}/k\Omega \times R - 17.8^{\circ}$ for the other one, determined with a linear regression of angles measured on photos as a function of measured resistances.\footnote{The $conversion$ numbers come from a series of photos, but the zeros are not from the linear regression, but rather from the symmetry of the data, i.e. the $F = F_0/\cos\theta$ law.} The rotation was performed in steps of 0.2 k$\Omega$, which corresponds to approximately 1.6$^{\circ}$. Moreover, it was found during the experiments that the first few data sets had rotation steps of less than that value, probably due to torsion building within the rotation mechanism before the rotation actually started at the sample stage.
 
 The graphs presented in this section show data at resistance values of 0.6 to 7.6k$\Omega$, although the measurements started at zero resistance, and this is due to the fact that the first four were almost identical. The angles extend between -3.3$^{\circ}$ to 54.3$^{\circ}$ for C698I, and from -12.7$^{\circ}$ to 46.4$^{\circ}$ for sample C698A. The difference in those angle intervals, for the same position in rotator, is due to the an error in the installation of the coil onto the bobbin system (see figure \ref{fig:Camprobe}, section \ref{sect:Camprobe}), and the error in cutting the sample with respect to their c-axis.
 
 \begin{figure}[p]
         \begin{center}
	\includegraphics[width=1\columnwidth]{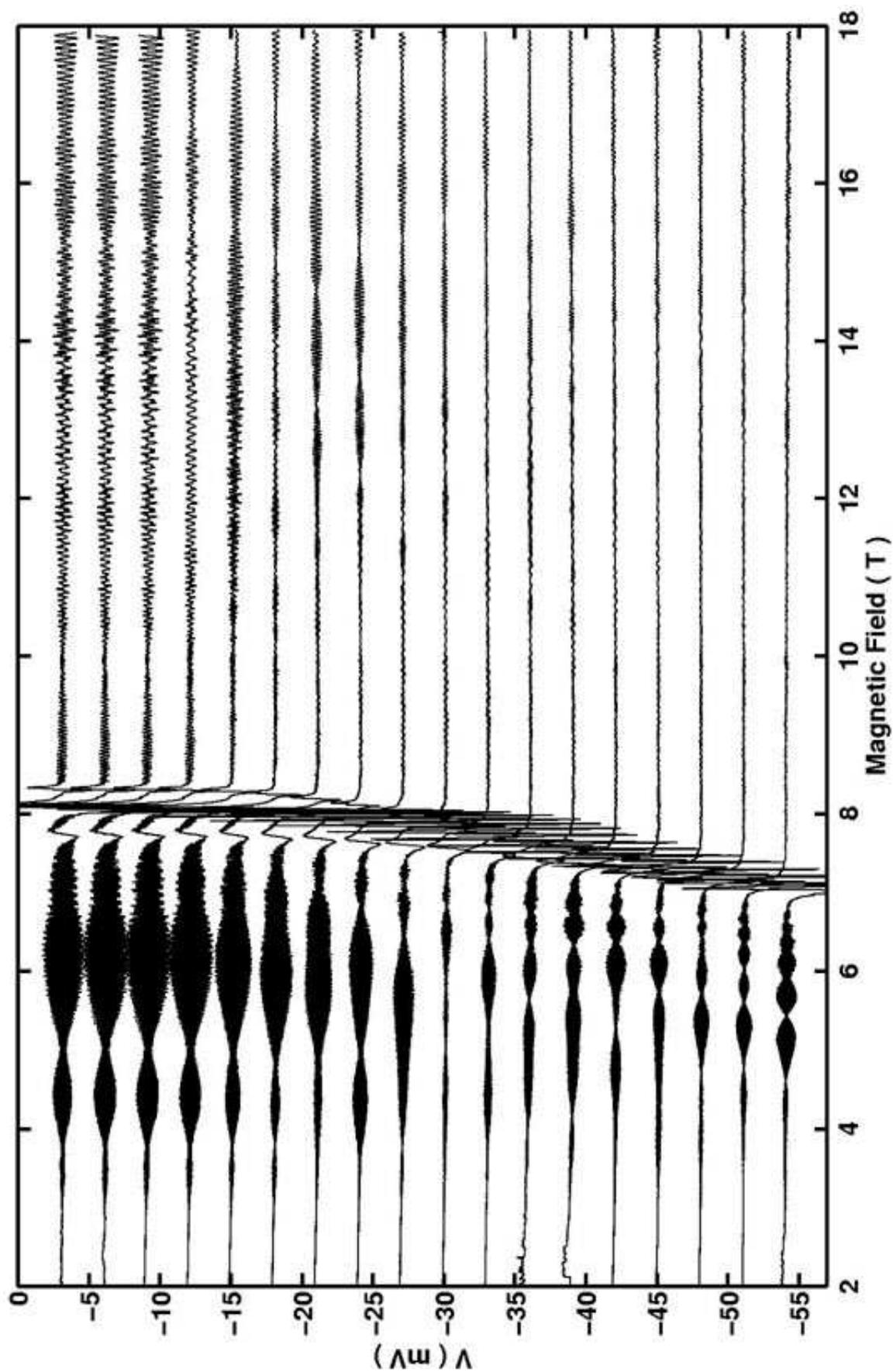}
	\caption[Raw dHvA data from the rotation study of sample C698I]{Raw data from the rotation study of sample C698I. This first half of the measurements were performed at potentiometer values between 0.6 and 4 k$\Omega$, corresponding to angles of -3.3$^{\circ}$ to 24.7$^{\circ}$.}
	\label{fig: LIAa3to20}
	\end{center}
\end{figure}

 \begin{figure}[p]
         \begin{center}
	\includegraphics[width=1\columnwidth]{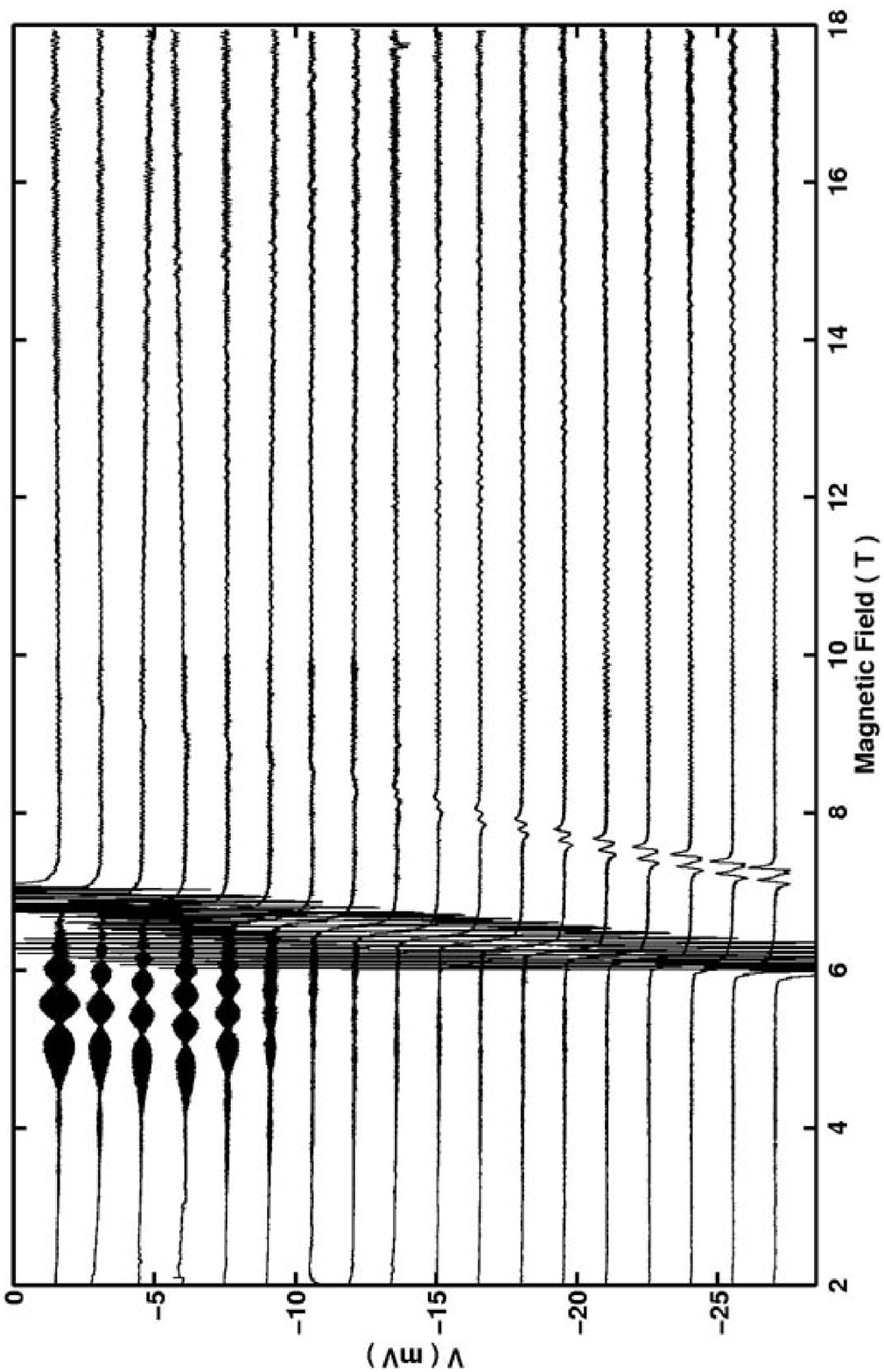}
	\caption[Raw dHvA data from the rotation study of sample C698I, continued]{Raw data from the rotation study of sample C698I. The second half of the measurements were performed at potentiometer values between 4.2 and 7.6 k$\Omega$, corresponding to angles of 26.3$^{\circ}$ to 54.3$^{\circ}$.}
	\label{fig: LIAa21to38}
	\end{center}
\end{figure}

 \begin{figure}[p]
         \begin{center}
	\includegraphics[width=1\columnwidth]{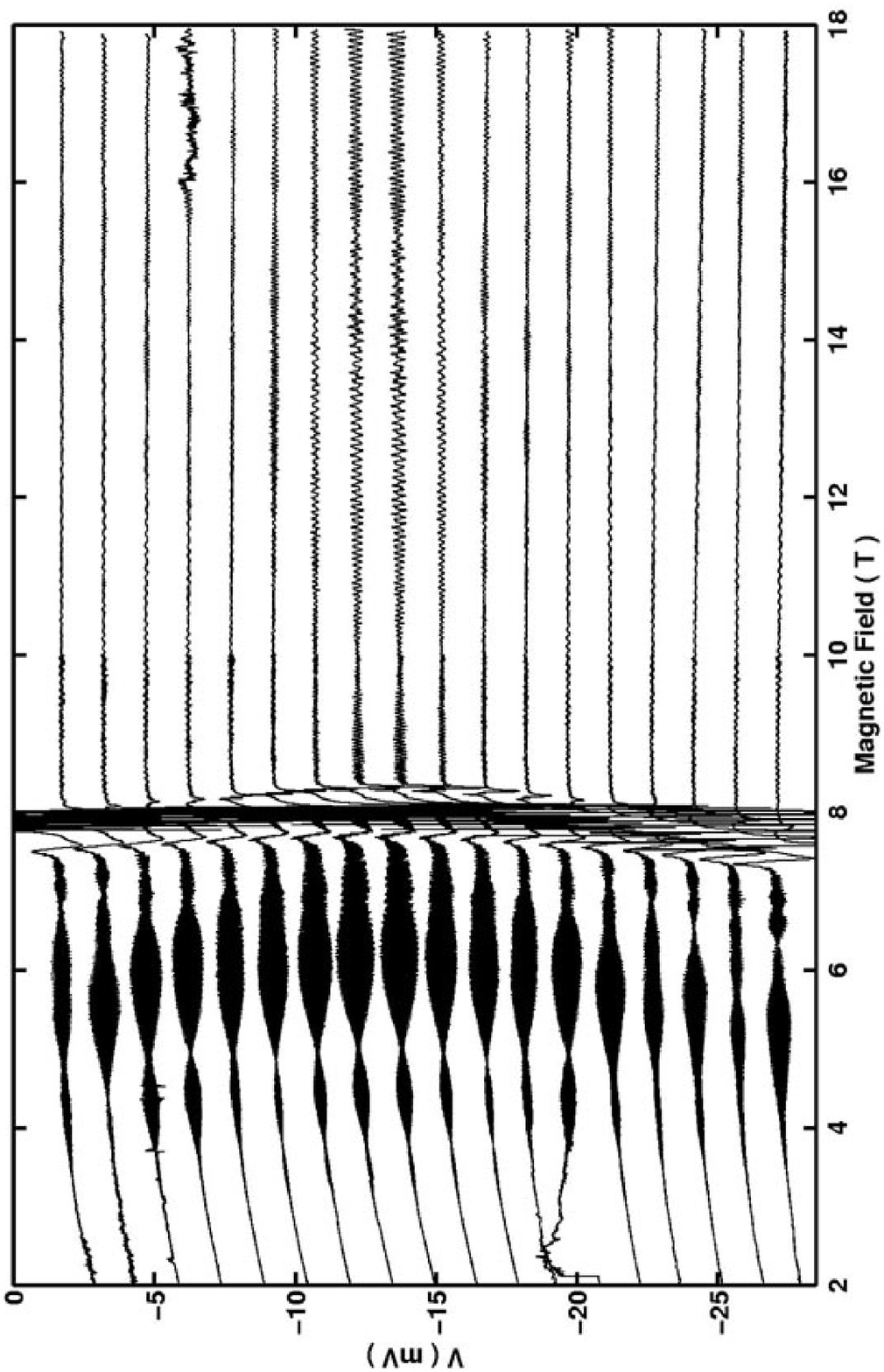}
	\caption[Raw dHvA data from the rotation study of sample C698A]{Raw data from the rotation study of sample C698A. This first half of the measurements were performed at potentiometer values between 0.6 and 4 k$\Omega$, corresponding to angles of -12.7$^{\circ}$ to 16.0$^{\circ}$.}
	\label{fig: LIAb3to20}
	\end{center}
\end{figure}

 \begin{figure}[p]
         \begin{center}
	\includegraphics[width=1\columnwidth]{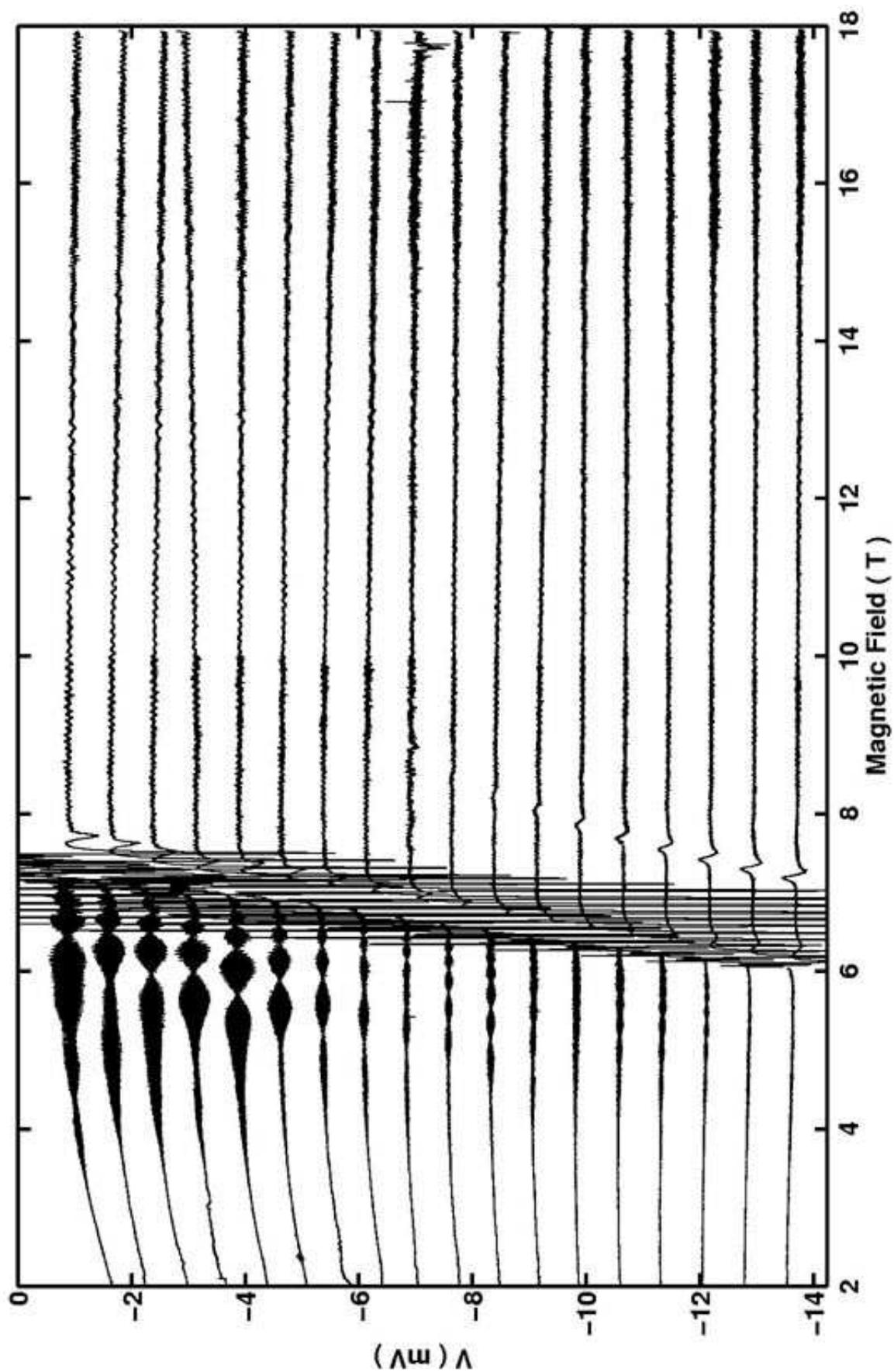}
	\caption[Raw dHvA data from the rotation study of sample C698A, continued]{Raw data from the rotation study of sample C698A. The second half of the measurements were performed at potentiometer values between 4.2 and 7.6 k$\Omega$, corresponding to angles of 17.7$^{\circ}$ to 46.4$^{\circ}$.}
	\label{fig: LIAb21to38}
	\end{center}
\end{figure}

 \chapter{Numerical self-consistency verifications for calculations of the quasiparticle mass \label{App:F}} 
 \markright{Appendix~\ref{App:F}}
 
We present in this appendix simulations of dHvA data sets along with their analysis for the quasiparticle mass. We have performed two type of calculations, presented in two separate sections. The first type was done with a simulated mass enhancement, using the same calculation procedure as was used in section \ref{sect:MassNonDiv} with two parameter fits of the LK relation, in order to verify the validity of the method. The second was done with a constant mass using a three parameter fit, in order to demonstrate how systematic errors can arise when the signal to noise ratio is poor and a free offset is used in the LK fit. Gaussian noise of various amplitudes was deliberately added to the signal, such that the simulations imitate real life experiments. We observed that the method we used to analyse the dHvA data from the Cambridge experiments produced the appropriate results, and that spurious mass enhancements can be produced when using the same method as that which was used in the work of Borzi $et$ $al.$ \cite{borzi}, confirming our discussion of section \ref{sect:MassSystematic}.
 
 \subsection{Calculation of quasiparticle mass enhancements with two parameters fits}
 \begin{figure}[p]
         \begin{center}
	\includegraphics[width=1\columnwidth]{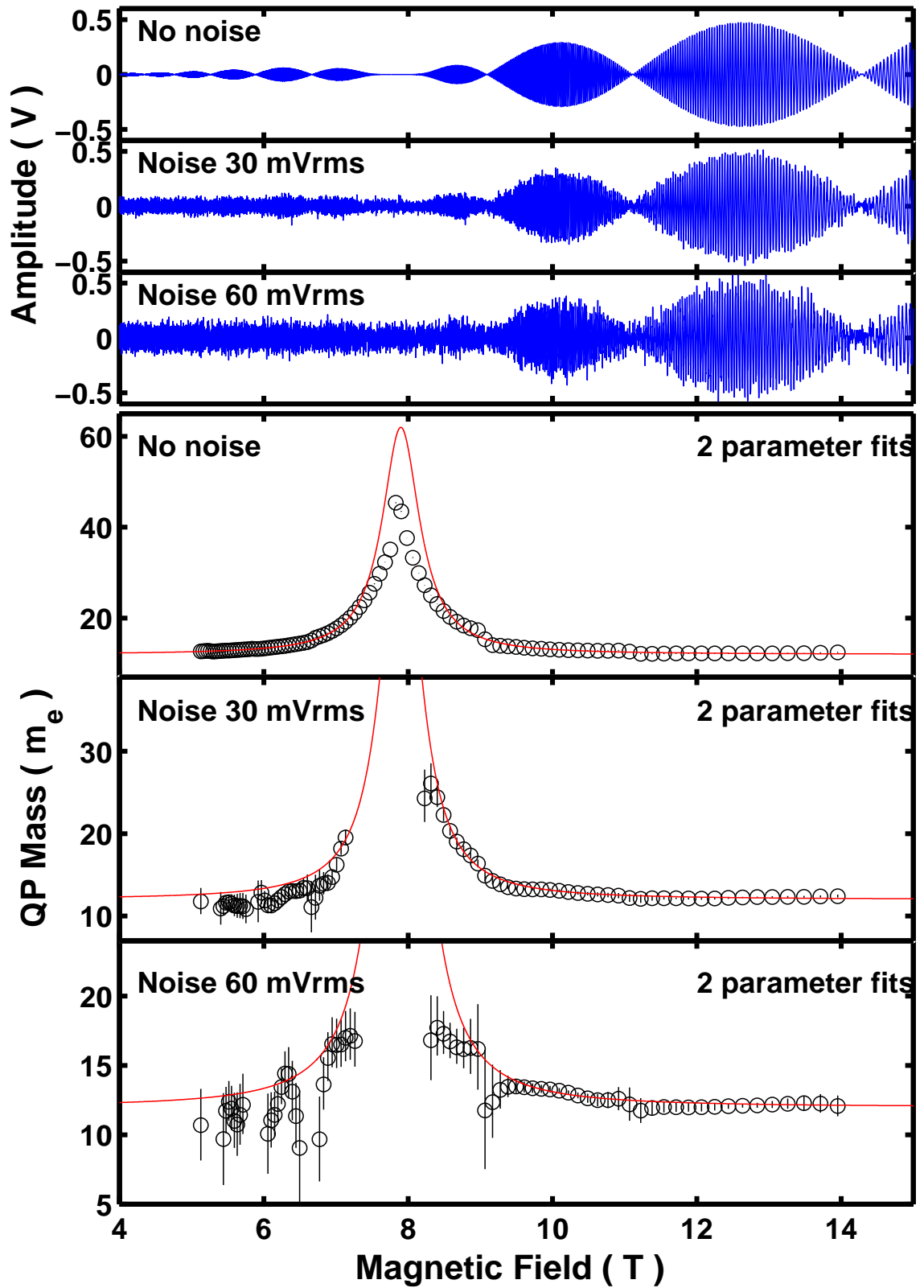}
	\caption[Numerical self-consistency verification for a mass enhancement]{Simulations of dHvA data incorporating a strong mass enhancement around 7.9 T, and quasiparticle mass values extracted from the simulations using the normal method. Gaussian white noise was added to the second and third simulations, of standard deviation 10 and 20 mVrms. The data shown were calculated with a temperature of 90 mK.}
	\label{fig: MassDivSim}
	\end{center}
\end{figure}

The top part of figure \ref{fig: MassDivSim} presents simulated dHvA oscillations, produced by evaluating the following function,
\beq
M(X) \propto LK\big(X,T,m(X)\big) D\big(\ell,F,X\big) \bigg[ \cos ( 2\pi F_1 X) + \cos ( 2\pi F_2 X)\bigg] + f(X),\nn
\eeq
where $LK(X)$ is the LK function, D(X) the Dingle factor (see section \ref{sect:exprosc}) and $f(X)$ represents white noise with a Gaussian distribution. The data was calculated for a field range of 5 to 15 T, using two slightly different frequencies, $F_1 = 4.20 kT$ and $F_2 = 4.25 kT$, in order to produce ordinary interference patterns, a mean free path of $\ell = 1\mu$m and a field dependent quasiparticle mass $m^* = m^*(X)$. The function was evaluated at ten temperatures between 90 and 500 mK \footnote{These were the same as those in the real experiments carried out in Cambridge: $T$ = 90, 120, 150, 180, 220, 280, 330, 380, 430 and 500 mK.}. The mass was defined as constant and equal to 12 $m_e$ away from 7.9 T and increases up to 60 $m_e$ following a Lorentzian function, centred at 7.9 T, of half-width at half-height of 0.3 T. No particular reason lies behind this choice of this function, the only requirement being its reproduction when calculating the mass using the simulations. One may observe that at 7.9 T, the oscillations amplitude is very close to zero.

The second and third data sets presented in figure \ref{fig: MassDivSim} feature noise, which was obtained using a random number generator, and subsequently weighed such that its probability distribution was Gaussian. The noise amplitudes quoted in the graphs, 10 and 20 mVrms, refer to the standard deviation $\sigma$, even though the visible amplitude appears larger. The noise in real experiments may not be necessarily of that type. However, from lack of better knowledge over its distribution and its correlation and due to time constraints, we used the simplest realistic type of noise.

The method used to analyse the simulated dHvA data was exactly the same as that employed for the extraction of the quasiparticle masses presented in section \ref{sect:MassNonDiv}, figures \ref{fig: p9p45kT_c-axis_a} and \ref{fig: p9p45kT_c-axis_b}, which we describe again here. The data was processed using 100 sections of length of 0.01 T$^{-1}$, which were allowed to overlap, their centres evenly spaced in inverse field. Fourier transforms were performed on these for all temperatures, and the peak in the power spectrum corresponding to the oscillations was integrated. The result was fitted using the normal two parameter LK function, eq. \ref{eq:LKfit} (see section \ref{sect:mass}). Each fit was inspected by eye, and the results for which the data distribution was too noisy to obtain a reasonable fit were discarded.

The lower part of figure \ref{fig: MassDivSim} shows the results of the mass analysis for the three situations described, with two different levels of noise and without noise at all, and the error bars correspond to the fit error given by the least squares method. In red is plotted the field dependent mass $m^*(B)$ which was used in the creation of the oscillations. Without noise, the calculation procedure reproduces the original mass properly except where its value is very high (note that error bars are much smaller than the size of the symbols). This is partly due to the width of the data sets used in the Fourier transform, which leads to a convolution of the mass function with the Fourier transform window of width of about half the window size, 0.01 T$^{-1}$. But also, it is due to systematic errors related to the lowest temperature point, which was deliberately chosen high, 90 mK, in order to reflect the real data. Very high masses lead to temperature distributions in which only the lowest data point possesses a non-zero amplitude, leading to errors much larger than those given by the fitting algorithm. Nevertheless, these results indicate that without noise, any mass enhancement should be properly detected by our data analysis procedure.

With 30 mVrms of noise added to the oscillations, we performed the same analysis and obtained a similar mass enhancement, but cut off near 26 electron masses. The gap in the data reflects the relative size of the oscillatory amplitude and the noise in the region where the mass is very high, and corresponds to inverse field windows where the LK fit was not possible at all. In the low field side of the mass enhancement, it is recovered, but the results are noisier due to the exponential decrease of the Dingle factor. Effectively, large underestimates arise at nodes of the beat pattern, where the signal to noise vanishes. We conclude however that with such a signal to noise ratio, a mass enhancement can still be detected.

Finally, using a noise amplitude of 60 mVrms, we performed the same analysis again, and different results were obtained. In the enhanced mass region, LK fits were much poorer and systematic errors were observed, especially in the low field side. The mass was mostly underestimated at every node of the beat pattern, but was accurate otherwise. However, the mass enhancement was cut off at around 18 electron masses, reflecting how the noise amplitude may hinder its detection.

Quantitatively, we found that it is when the signal to noise ratio of the 90 mK temperature point decreases below 0.5 that the mass analysis becomes difficult. We then compared this with the signal to noise ratios for all the frequencies of which masses were extracted and presented in figures \ref{fig: p9p45kT_c-axis_a} and \ref{fig: p9p45kT_c-axis_b}, for the $c$-axis direction. For the low field side frequencies at 4.2, 1.8, 0.9, 0.43 and 0.15 kT, the lowest signal to noise ratios recorded very near the transition (away from any beat pattern nodes) were respectively of 8, 20, 1.8 1.4 and 0.8. In the high field side, we presented data only for the 4.0 and 1.6 kT peaks, and they possessed signal to noise ratios of respectively 4 and 12. Consequently, none of the frequencies analysed were in the situation depicted here, where the noise hinders the mass analysis. We conclude that if a mass enhancement had been present for any of the pockets we analysed, we should have detected it.

 \subsection{Systematic errors with three parameters fits}
 \begin{figure}[p]
         \begin{center}
	\includegraphics[width=1\columnwidth]{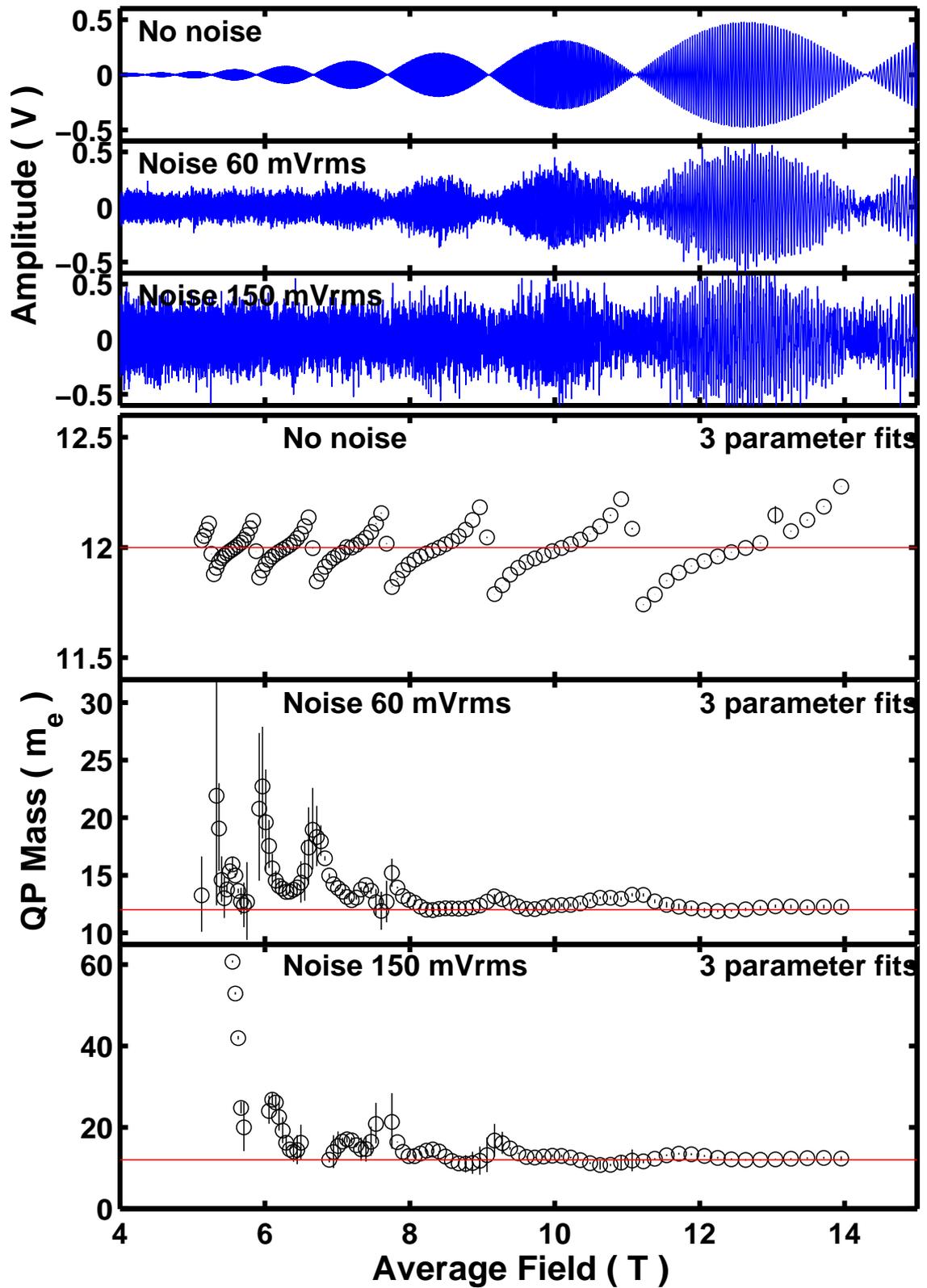}
	\caption[Simulation of spurious mass enhancement]{Simulations of dHvA data using a constant quasiparticle mass, without noise in the first case, and with different noise levels for the other two. Quasiparticle mass values extracted from the simulations using three parameter fits produce strong spurious mass enhancements where the signal to noise ratio is low.  The data shown were calculated with a temperature of 90 mK.}
	\label{fig: MassNonDivSim}
	\end{center}
\end{figure}

We discussed in section \ref{sect:MassSystematic} how the use of three parameter fits in order to accommodate data of poor signal to noise ratio could lead to spurious continuous mass enhancements as the signal to noise ratio evolve with magnetic field. Such a phenomenon is relatively easy to simulate, and we present such a calculation in this section, giving further support to our suggestion that it is the phenomenon which led to the mass enhancement measured by Borzi $et$ $al.$ \cite{borzi}.

Simulations similar to those of the previous section were performed, without an enhancement of the quasiparticle mass, which was defined as constant and equal to 12 $m_e$. All other parameters were the same, except for higher noise levels, which were of 60 and 150 mVrms. The quasiparticle mass analysis was performed in exactly the same way as previously, with the exception that the non-linear LK fits were performed using three parameters, as defined in section \ref{eq:3param} and discussed in section \ref{sect:MassSystematic}. 

The top three graphs in figure \ref{fig: MassNonDivSim} present the simulated data, the first featuring data without noise, while the other two possess noise levels of 60 and 150 mVrms. The bottom three graphs present the mass extracted with the three parameter LK fits. In the first case, without noise, we observed that the quasiparticle mass oscillated around the original value of 12, deviating at each beat pattern node. As noise was added to the oscillations, we found that these deviations evolved into strong mass enhancements as the signal amplitude became comparable with the noise level. With a noise level of 150 mVrms, stronger mass enhancements were calculated, of up to 60 at the lowest signal to noise ratio. Note that again here, all the error bars that we present correspond to the fit errors given by the algorithm. These seem inconsistent in places, being small where strong enhancements were observed. We argue that this points strongly to how misleading three parameter fits can be: very good fits can be obtained where noise is the dominant component of the data. 

The mass enhancements appear at beat pattern nodes, where the signal to noise ratio vanishes, as we discussed in section \ref{sect:MassSystematic}. However, here, one can observe that the increase is more and more pronounced the lower the signal to noise ratio is away from beat pattern nodes, and that the system becomes unstable to small changes in signal amplitude. From the mass extracted from data with 60 mVrms of noise, we determined that it is when the signal to noise ratio away from beat pattern nodes is of around 3 or less that the system becomes susceptible to large mass changes. This essentially means that for a system where the signal to noise ratio is not good where the signal amplitude is at its maximum, strong mass enhancements can be expected if one uses three parameter fits in regions where beat pattern nodes arise. It is possible that in such a case, two parameter fits are not possible either, due to the bad quality of the data. Such data should then simply not be used for a field resolved mass analysis.

We believe that it is the assignation of an additive constant to a random value added to the LK function that lies behind this phenomenon. Although in many fits, this seems very appropriate, and small error bars are obtained, the information extracted can be very inaccurate. We conclude by suggesting that only LK fits with two parameters should be used, and only with data that possesses a signal to noise ratio higher than 0.5.

\markright{Bibliography}
\bibliographystyle{unsrt}
\bibliography{PhDBib.bib}
\end{document}